\documentclass[final,1p,times]{elsarticle}
\usepackage{amssymb}
\usepackage{epsfig}
\usepackage{graphicx,amsmath}
\usepackage[english]{babel}
\usepackage{bm}
\journal{Physics Reports}

\graphicspath{{Figures/}}

\begin{document}

\begin{frontmatter}



\title{Scaling Behavior of Heavy Fermion Metals}

\author[label1,label2]{V.R.
Shaginyan\corref{cor1}}\ead{vrshag@thd.pnpi.spb.ru}\cortext[cor1]{Corresponding
author at: Petersburg Nuclear Physics Institute, RAS, Gatchina,
188300, Russia; Phones: (office) 7-813-714-6096, (fax) 7-813-713-1963}
\address[label1]{Petersburg Nuclear Physics Institute, RAS, Gatchina,
188300, Russia}
\address[label2]{CTSPS, Clark Atlanta University,
Atlanta, Georgia 30314, USA}
\author[label3,label5]{M.Ya. Amusia}\address[label3]{Racah Institute
of Physics, Hebrew University, Jerusalem 91904, Israel}
\address[label5]{Ioffe Physical –Technical Institute, RAS, St. Petersburg 194021, Russia}
\author[label2]{A.Z. Msezane}
\author[label4]{K.G. Popov}
\address[label4]{Komi Science Center, Ural Division, RAS, 3a, Chernova
str. Syktyvkar, 167982, Russia}

\begin{abstract}

Strongly correlated Fermi systems are fundamental systems in physics
that are best studied experimentally, which until very recently have
lacked theoretical explanations. This review discusses the
construction of a theory and the analysis of phenomena occurring in
strongly correlated Fermi systems such as heavy-fermion (HF) metals
and two-dimensional (2D) Fermi systems. It is shown that the basic
properties and the scaling behavior of HF metals
can be described within the framework of a fermion
condensation quantum phase transition (FCQPT) and extended
quasiparticle paradigm that allow us to explain the non-Fermi liquid
behavior observed in strongly correlated Fermi systems. In contrast
to the Landau paradigm stating that the quasiparticle effective mass
is a constant, the effective mass of new quasiparticles strongly
depends on temperature, magnetic field, pressure, and other
parameters. Having analyzed collected facts on strongly correlated
Fermi systems with quite different microscopic nature, we find these
to exhibit the same non-Fermi liquid behavior at FCQPT. We show both
analytically and using arguments based entirely on the experimental
grounds that the data collected on very different strongly
correlated Fermi systems have a universal scaling behavior, and
materials with strongly correlated fermions can unexpectedly be
uniform in their diversity. Our analysis of strongly correlated
systems such as HF metals and 2D Fermi systems is in the context of
salient experimental results. Our calculations of the non-Fermi
liquid behavior, the scales and thermodynamic, relaxation and
transport properties are in good agreement with experimental facts.
\end{abstract}


\end{frontmatter}

\tableofcontents


\section{Introduction} \label{INTR}

Strongly correlated Fermi systems, such as heavy fermion (HF)
metals, high-$T_c$ superconductors, and two-dimensional (2D) Fermi
liquids, are among the most intriguing and best experimentally
studied fundamental systems in physics. However until very recently
lacked theoretical explanations. The properties of these materials
differ dramatically from those of ordinary Fermi systems
\cite{ste,varma,vojta,voj,belkop,obz,cust,sen,senth1,senth2,senth3,col11}.
For instance, in the case of metals with heavy fermions, the strong
correlation of electrons leads to a renormalization of the effective
mass of quasiparticles, which may exceed the ordinary, "bare", mass
by several orders of magnitude or even become infinitely large. The
effective mass strongly depends on the temperature, pressure, or
applied magnetic field. Such metals exhibit NFL behavior and unusual
power laws of the temperature dependence of the thermodynamic
properties at low temperatures . Ideas based on quantum and thermal
fluctuations taking place at a quantum critical point (QCP) have
been put forward and the fascinating behavior of these systems known
as the non-Fermi liquid (NFL) behavior was attributed to the
fluctuations \cite{ste,vojta,col2,geg1,geg,huy,steg}. Suggested to
describe one property, the ideas failed to do the same with the
others and there was a real crisis and a new quantum phase
transition responsible for the observed behavior was required
\cite{cust,sen,senth1,senth2,senth3,col2,col1}.

The Landau theory of the Fermi liquid has a long history and
remarkable results in describing a multitude of properties of the
electron liquid in ordinary metals and Fermi liquids of the $^3$He
type \cite{landau,lanl1,PinNoz}. The theory is based on the
assumption that elementary excitations determine the physics at low
temperatures. These excitations behave as quasiparticles, have a
certain effective mass, and, judging by their basic properties,
belong to the class of quasiparticles of a weakly interacting Fermi
gas. Hence, the effective mass $M^*$ is independent of the
temperature, pressure, and magnetic field strength and is a
parameter of the theory.

The Landau Fermi liquid (LFL) theory fails to explain the results of
experimental observations related to the dependence of $M^*$ on the
temperature $T$, magnetic field $B$, pressure, etc.; this has led to
the conclusion that quasiparticles do not survive in strongly
correlated Fermi systems and that the heavy electron does not retain
its identity as a quasiparticle excitation
\cite{cust,sen,senth1,senth2,senth3,col11,col2,col1}.

\subsection{Quantum phase transitions
and the non-Fermi liquid behavior of correlated Fermi systems}

The unusual properties and NFL behavior observed in high-$T_c$
superconductors, HF metals and 2D Fermi systems are assumed to be
determined by various magnetic quantum phase transitions
\cite{ste,varma,vojta,voj,belkop,obz,cust,sen,senth1,senth3,col11,col2,geg1}.
Since a quantum phase transition occurs at the temperature $T=0$,
the control parameters are the composition, electron (hole) number
density $x$, pressure, magnetic field strength $B$, etc. A quantum
phase transition occurs at a quantum critical point, which separates
the ordered phase that emerges as a result of quantum phase
transition from the disordered phase. It is usually assumed that
magnetic (e.g., ferromagnetic and antiferromagnetic) quantum phase
transitions are responsible for the NFL behavior. The critical point
of such a phase transition can be shifted to absolute zero by
varying the above parameters.

Universal behavior can be expected only if the system under
consideration is very close to a quantum critical point, e.g., when
the correlation length is much longer than the microscopic length
scale, and critical quantum and thermal fluctuations determine the
anomalous contribution to the thermodynamic functions of the metal.
Quantum phase transitions of this type are so widespread
\cite{varma,vojta,voj,senth1,senth2,senth3,col11,col2} that we call
them ordinary quantum phase transitions \cite{shag3}. In this case,
the physics of the phenomenon is determined by thermal and quantum
fluctuations of the critical state, while quasiparticle excitations
are destroyed by these fluctuations. Conventional arguments that
quasiparticles in strongly correlated Fermi liquids "get heavy and
die" at a quantum critical point commonly employ the well-known
formula based on the assumptions that the $z$-factor (the
quasiparticle weight in the single-particle state) vanishes at the
points of second-order phase transitions \cite{col1}. However, it
has been shown that this scenario is problematic \cite{khodb,x18}.

The fluctuations in the order parameter developing an infinite
correlation and the absence of quasiparticle excitations is
considered the main reason for the NFL behavior of heavy-fermion
metals, 2D fermion systems and high-$T_c$ superconductors
\cite{vojta,voj,senth1,col11,col2,col3}. This approach faces certain
difficulties, however. Critical behavior in experiments with metals
containing heavy fermions is observed at high temperatures
comparable to the effective Fermi temperature $T_k$. For instance,
the thermal expansion coefficient $\alpha(T)$, which is a linear
function of temperature for normal LFL, $\alpha(T)\propto T$,
demonstrates the $\sqrt{T}$ temperature dependence in measurements
involving CeNi$_2$Ge$_2$ as the temperature varies by two orders of
magnitude (as it decreases from 6 K to at least 50 mK) \cite{geg1}.
Such behavior can hardly be explained within the framework of the
critical point fluctuation theory. Obviously, such a situation is
possible only as $T\to0$, when the critical fluctuations make the
leading contribution to the entropy and when the correlation length
is much longer than the microscopic length scale. At a certain
temperature $T_k$, this macroscopically large correlation length
must be destroyed by ordinary thermal fluctuations and the
corresponding universal behavior must disappear.

Another difficulty is in explaining the restoration of the LFL
behavior under the application of magnetic field $B$, as observed in
HF metals and in high-$T_c$ superconductors \cite{ste,geg,cyr}. For
the LFL state as $T\to0$, the electric resistivity
$\rho(T)=\rho_0+AT^2$, the heat capacity $C(T)=\gamma_0T$, and the
magnetic susceptibility $\chi=const$. It turns out that the
coefficient $A(B)$, the Sommerfeld coefficient $\gamma_0(B)\propto
M^*$, and the magnetic susceptibility $\chi(B)$ depend on the
magnetic field strength B such that $A(B)\propto\gamma_0^2(B)$ and
$A(B)\propto\chi^2(B)$, which implies that the Kadowaki-Woods
relation $K=A(B)/\gamma_0^2(B)$ \cite{kadw} is $B$-independent and
is preserved \cite{geg}. Such universal behavior, quite natural when
quasiparticles with the effective mass $M^*$ playing the main role,
can hardly be explained within the framework of the approach that
presupposes the absence of quasiparticles, which is characteristic
of ordinary quantum phase transitions in the vicinity of QCP.
Indeed, there is no reason to expect that $\gamma_0$, $\chi$ and $A$
are affected by the fluctuations in a correlated fashion.

For instance, the Kadowaki-Woods relation does not agree with the
spin density wave scenario \cite{geg} and with the results of
research in quantum criticality based on the renormalization-group
approach \cite{mill}. Moreover, measurements of charge and heat
transfer have shown that the Wiedemann-Franz law holds in some
high-$T_c$ superconductors \cite{cyr,cyr1} and HF metals
\cite{pag,pag2,ronn1,ronn}. All this suggests that quasiparticles do
exist in such metals, and this conclusion is also corroborated by
photoemission spectroscopy results \cite{koral,fujim}.

The inability to explain the behavior of heavy-fermion metals while
staying within the framework of theories based on ordinary quantum
phase transitions implies that another important concept introduced
by Landau, the order parameter, also ceases to operate (e.g., see
Refs \cite{senth1, senth3, col1, col2}). Thus, we are left without
the most fundamental principles of many-body quantum physics
\cite{landau,lanl1,PinNoz}, and many interesting phenomena
associated with the NFL behavior of strongly correlated Fermi
systems remain unexplained.

\begin{figure} [! ht]
\begin{center}
\vspace*{-0.2cm}
\includegraphics [width=0.60\textwidth]{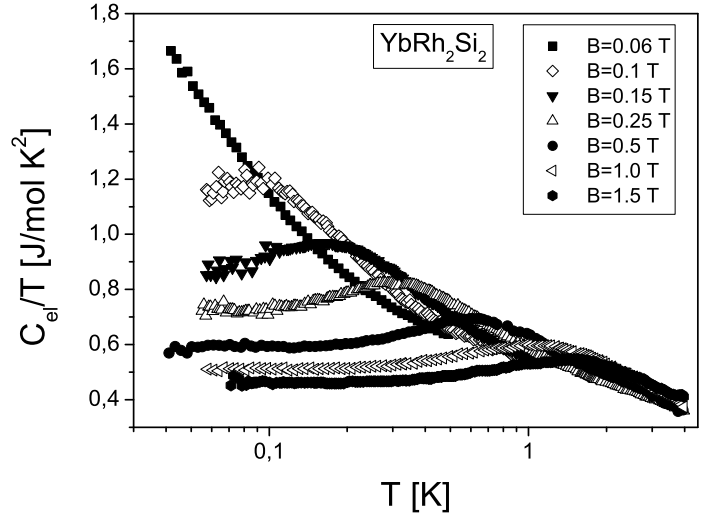}
\end{center}
\vspace*{-0.3cm} \caption{Electronic specific heat of $\rm
YbRh_2Si_2$, $C/T$, versus temperature $T$ as a function of magnetic
field $B$ \cite{oesb} shown in the legend.}\label{YBRHSI}
\end{figure}
NFL behavior manifests itself in the power-law behavior of the
physical quantities of strongly correlated Fermi systems located
close to their QCPs, with exponents different from those of a Fermi
liquid \cite{oesb,oesbs}. It is common belief that the main output
of theory is the explanation of these exponents which are at least
depended on the magnetic character of QCP and dimensionality of the
system. On the other hand, the NFL behavior cannot be captured by
these exponents as seen from Fig. \ref{YBRHSI}.
Indeed, the specific heat $C/T$ exhibits a behavior that is to be
described as a function of both temperature $T$ and magnetic $B$
field rather than by a single exponent. One can see that at low
temperatures $C/T$ demonstrates  the LFL behavior which is changed
by the transition regime at which $C/T$ reaches its maximum and
finally $C/T$ decays into NFL behavior as a function of $T$ at fixed
$B$. It is clearly seen from Fig. \ref{YBRHSI} that, in particularly
in the transition regime, these exponents may have little physical
significance.

\begin{figure} [! ht]
\begin{center}
\vspace*{-0.2cm}
\includegraphics [width=0.60\textwidth]{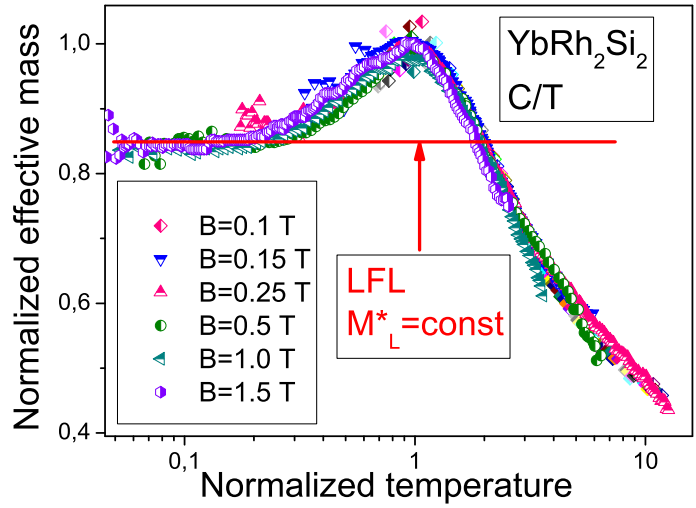}
\end{center}
\vspace*{-0.3cm} \caption{The normalized effective mass $M^*_N$
versus normalized temperature $T_N$.  $M^*_N$ is extracted from the
measurements of the specific heat $C/T$ on $\rm YbRh_2Si_2$ in
magnetic fields $B$ \cite{oesb} listed in the legend. Constant
effective mass $M^*_L$ inherent in normal Landau Fermi liquids is
depicted by the solid line.}\label{YBRHSIN}
\end{figure}
In order to show that the behavior of $C/T$ displayed in Fig.
\ref{YBRHSI} is of generic character, we remember that in the
vicinity of QCP it is helpful to use "internal" scales to measure
the effective mass $M^*\propto C/T$ and temperature $T$
\cite{dft373,dftjtp}. As seen from Fig. \ref{YBRHSI}, a maximum
structure in $C/T\propto M^*_M$ at temperature $T_M$ appears under
the application of magnetic field $B$ and $T_M$ shifts to higher $T$
as $B$ is increased. The value of the Sommerfeld coefficient
$C/T=\gamma_0$ is saturated towards lower temperatures decreasing at
elevated magnetic field. To obtain the normalized effective mass
$M^*_N$, we use $M^*_M$ and $T_M$ as "internal" scales: The maximum
structure in $C/T$ was used to normalize $C/T$, and $T$ was
normalized by $T_M$. In Fig. \ref{YBRHSIN} the obtained
$M^*_N=M^*/M^*_M$ as a function of normalized temperature
$T_N=T/T_M$ is shown by geometrical figures. Note that we have
excluded the experimental data taken in magnetic field $B=0.06$ T.
In that case, as will be shown in Subsections \ref{HCEL3} and
\ref{HCEL9}, $T_M\to0$ and the corresponding $T_M$ and $M^*_M$ are
unavailable. It is seen that the LFL state and NFL one are separated
by the transition regime at which $M^*_N$ reaches its maximum value.
Figure \ref{YBRHSIN} reveals the scaling behavior of the normalized
experimental curves - the curves at different magnetic fields $B$
merge into a single one in terms of the normalized variable
$y=T/T_M$. As seen from Fig. \ref{YBRHSIN}, the normalized effective
mass $M^*_N(y)$ extracted from the measurements is not a constant,
as would be for a LFL, and shows the scaling behavior over three
decades in normalized temperature $y$. It is seen from Figs.
\ref{YBRHSI} and \ref{YBRHSIN} that the NFL behavior and the
associated scaling extend at least to temperatures up to few
Kelvins. Scenario where fluctuations in the order parameter of an
infinite (or sufficiently large)  correlation length and an infinite
correlation time (or sufficiently large) develop the NFL behavior
can hardly match up such high temperatures.

Thus, we conclude that a challenging problem for theories considering
the critical behavior of the HF metals is to explain the scaling
behavior of $M^*_N(y)$. While the theories calculating only the
exponents that characterize $M^*_N(y)$ at $y\gg 1$ deal with a part
of the observed facts related to the problem and overlook, for
example, consideration of the transition regime.
Another part of the problem is the remarkably large temperature ranges
over which the NFL behavior is observed.

As we will see below, the large temperature ranges are precursors of
new quasiparticles, and  it is the scaling behavior of the
normalized effective mass that allows us to explain the
thermodynamic, transport and relaxation properties of HF metals at
the transition and NFL regimes.

Taking into account the simple behavior shown in Fig. \ref{YBRHSIN},
we ask the question: what theoretical concepts can replace the
Fermi-liquid paradigm with the notion of the effective mass in cases
where Fermi-liquid theory breaks down? To date such a concept is not
available \cite{vojta}. Therefore, in our review we focus on a
concept of fermion condensation quantum phase transition (FCQPT)
preserving quasiparticles and intimately related to the unlimited
growth of $M^*$. We shall show that it is capable revealing the
scaling behavior of the effective mass and delivering an adequate
theoretical explanation of a vast majority of experimental results
in different HF metals. In contrast to the Landau paradigm based on
the assumption that $M^*$ is a constant as shown by the solid line
in Fig. \ref{YBRHSIN}, in FCQPT approach the
effective mass $M^*$ of new quasiparticles strongly depends on
$T$, $x$, $B$ etc. Therefore, in accord with numerous experimental
facts the extended quasiparticles paradigm is to be introduced. The
main point here is that the well-defined quasiparticles determine as
before the thermodynamic, relaxation and transport properties of strongly
correlated Fermi-systems in large temperature ranges
(see Section \ref{HCEL} and Subsection \ref{HCEL4}),
while $M^*$ becomes a function of $T$,
$x$, $B$ etc. The FCQPT approach had been already successfully
applied to describe the thermodynamic properties of such different
strongly correlated systems as $^3$He on the one hand and
complicated heavy-fermion (HF) compounds on the other
\cite{obz,khodb,prl3he}.

\subsection{Limits and goals of the review}

The purpose of this review is to show that diverse strongly
correlated Fermi systems such three dimensional (3D) and 2D
compounds as HF metals and 2D strongly correlated Fermi liquids
exhibit a scaling behavior, which can be described within a single
approach based on FCQPT theory \cite{obz,ks,ksk}. We discuss the
construction of the theory and show that it delivers theoretical
explanations of the vast majority of experimental results in
strongly correlated systems such as HF metals and 2D systems. Our
analysis is in the context of salient experimental results. Our
calculations of the non-Fermi liquid behavior, the scales and
thermodynamic, relaxation and transport properties are in good
agreement with experimental facts. We shall also focus on the
scaling behavior of the thermodynamic, transport and relaxation
properties that can be revealed from experimental facts and
theoretical analysis. As a result, we do not discuss the specific
features of strongly correlated systems in full; instead, we focus
on the universal behavior of such systems. For instance, we ignore
the physics of Fermi systems such as neutron stars, atomic clusters
and nuclei, quark plasma, and ultra-cold gases in traps, in which we
believe fermion condensate (FC) induced by FCQPT can exist
\cite{khod3,amsh3,khod4,volovik1,volovik2,vol_1}. Ultra-cold gases
in traps are interesting because their easy tuning allows selecting
the values of the parameters required for observations of quantum
critical point and FC. We do not discuss also microscopic mechanisms
of quantum criticality related to FCQPT. Such mechanisms can be
developed within FC theory. For example, the mechanism of quantum
criticality as observed in f-electron materials can take place in
systems when the centers of merged single-particle levels "get
stuck" at the Fermi surface. One observes that this could provide a
simple mechanism for pinning narrow bands in solids to the Fermi
surface \cite{vol_1}. On the other hand, we consider high-$T_c$
superconductors within a coarse-grained model based on the FCQPT
theory in order to illuminate their generic relationships with HF
metals.

Experimental studies of the properties of quantum phase transitions
and their critical points are very important for understanding the
physical nature of high-$T_c$ superconductivity and HF metals. The
experimental data that refer to different strongly correlated Fermi
systems complement each other. In the case of high-$T_c$
superconductors, only few experiments dealing with their QCPs have
been conducted, because the respective QCPs are in the
superconductivity range at low temperatures and the physical
properties of the respective quantum phase transition are altered by
the superconductivity. As a result, high magnetic fields are needed
to destroy the superconducting state. But such experiments can be
conducted for HF metals. Experimental research has provided data on
the behavior of HF metals, shedding light on the nature of critical
points and phase transitions (e.g., see Refs
\cite{geg,cyr,cyr1,pag2,ronn1,koral,fujim}). Hence, a key issue is
the simultaneous study of high-$T_c$ superconductors and the NFL
behavior of HF metals.

Since we are concentrated on properties that are
non-sensitive to the detailed structure of the system we
avoid difficulties associated with the anisotropy generated by
the crystal lattice of solids, its special features, defects, etc.,
We study the universal behavior of high-$T_c$ superconductors, HF
metals, and 2D Fermi systems at low temperatures using the model of
a homogeneous HF liquid \cite{dft373,dftjtp}. The model is quite
meaningful because we consider
the scaling behavior exhibited by these materials at low
temperatures, a behavior related to the scaling of
quantities such as the effective mass, the heat capacity, the
thermal expansion, etc. The scaling properties of the normalized
effective mass that characterizes them, are determined by momentum
transfers that are small compared to momenta of the order of the
reciprocal lattice length. The high momentum contributions can
therefore be ignored by substituting the lattice for the jelly
model. While the values of the scales like the maximum $M^*_M$ of the
effective mass and $T_M$ at which $M^*_M$ takes place are determined
by a wide range of momenta and thus these scales are controlled by
the specific properties of the system.

We analyze the universal properties of strongly correlated Fermi
systems using the FCQPT theory \cite{obz,ks,ksk,shag4}, because the
behavior of heavy-fermion metals already suggests that their unusual
properties can be associated with the quantum phase transition
related to the unlimited increase in the effective mass at the
critical point. Moreover, we shall see that the scaling behavior
displayed in Fig. \ref{YBRHSIN} can be quite naturally captured
within the framework of the quasiparticle extended paradigm
supported by FCQPT which gives explanations of the NFL behavior
observed in strongly correlated Fermi systems.

\section{ Landau theory of Fermi liquids}\label{FLFC1}

One of the most complex problems of modern condensed matter physics
is the problem of the structure and properties of Fermi systems with
large inter particle coupling constants. Theory of Fermi liquids,
later called "normal", was first proposed by Landau as a means for
solving such problems by introducing the concept of quasiparticles
and amplitudes that characterize the effective quasiparticle
interaction \cite{landau, lanl1}. The Landau theory can be regarded
as an effective low-energy theory with the high-energy degrees of
freedom eliminated by introducing amplitudes that determine the
quasiparticle interaction instead of the strong inter particle
interaction. The stability of the ground state of the Landau Fermi
liquid is determined by the Pomeranchuk stability conditions:
stability is violated when at least one Landau amplitude becomes
negative and reaches its critical value \cite{lanl1,pom}. We note
that the new phase in which stability is restored can also be
described, in principle, by the LFL theory.

We begin by recalling the main ideas of the LFL theory \cite{landau,
lanl1,PinNoz}. The theory is based on the quasiparticle paradigm,
which states that quasiparticles are elementary weakly excited
states of Fermi liquids and are therefore specific excitations that
determine the low-temperature thermodynamic and transport properties
of Fermi liquids. In the case of the electron liquid, the
quasiparticles are characterized by the electron quantum numbers and
the effective mass $M ^*$. The ground state energy of the system is
a functional of the quasiparticle occupation numbers (or the
quasiparticle distribution function) $n({\bf p},T)$, and the same is
true of the free energy $F(n({\bf p},T))$, the entropy $S(n({\bf
p},T))$, and other thermodynamic functions. We can find the
distribution function from the minimum condition for the free energy
$F=E-TS$ (here and in what follows $k_B=\hbar=1$)
\begin{equation}\label{FL1} \frac{\delta(F-\mu N)}{\delta n({\bf p}, T)}=\varepsilon({\bf
p}, T) -\mu (T)-T\ln\frac{1-n({\bf p},T)}{n({\bf p},T)}=0.
\end{equation}
Here $\mu$ is the chemical potential fixing the number density
\begin{equation}\label{NUMX}
x=\int n({\bf p},T)\frac{d{\bf p}}{(2\pi)^3},
\end{equation} and
\begin{equation} \varepsilon({\bf p},T) \ = \ \frac {\delta
E(n({\bf p},T))} {\delta n({\bf p},T)}\,\label{FL2} \end{equation}
is the quasiparticle energy. This energy is a functional of $n({\bf
p},T)$, in the same way as the energy $E$ is: $\varepsilon({\bf
p},T,n)$. The entropy $S(n({\bf p},T))$ related to quasiparticles is
given by the well-known expression \cite{landau,lanl1}
\begin{eqnarray}
S(n({\bf p},T))&=& -2\int[n({\bf p},T) \ln (n({\bf p},T))+(1-n({\bf p},T))\nonumber \\
&\times&\ln (1-n({\bf p},T))]\frac{d{\bf p}}{(2\pi) ^3},\label{FL3}
\end{eqnarray}
which follows from combinatorial reasoning. Equation \eqref{FL1} is
usually written in the standard form of the Fermi-Dirac
distribution,
\begin{equation} n({\bf p},T)=
\left\{1+\exp\left[\frac{(\varepsilon({\bf p},T)-\mu)}
{T}\right]\right\}^{-1}.\label{FL4} \end{equation} At $T\to 0 $,
(\ref{FL1}) and (\ref{FL4}) have the standard solution
$n(p,T\to0)\to\theta(p_F-p)$ if the derivative $\partial
\varepsilon(p\simeq p_F)/\partial p$ is finite and positive. Here
$p_F$ is the Fermi momentum and $\theta (p_F-p)$ is the step
function. The single particle energy can be approximated as
$\varepsilon(p\simeq p_F)-\mu\simeq p_F(p-p_F)/M^*_L$, and $M ^*_L$
inversely proportional to the derivative is the effective mass of
the Landau quasiparticle,
\begin{equation}
\frac1{M^*_L}=\frac1p\, \frac{d\varepsilon(p,T=0)}{dp}|_{p=p_F}\
\label{FL5}.\end{equation} In turn, the effective mass $M^*_L$
is related to the bare electron mass $m$ by the well-known Landau
equation \cite{landau,lanl1,PinNoz} \begin{eqnarray}\label{LANDM}
\frac{1}{M^*_L} &=& \frac{1}{m}+\sum_{\sigma_1}\int \frac{{\bf
p}_F{\bf p_1}}{p_F^3} F_{\sigma,\sigma_1}({\bf p_F},{\bf p}_1)
\nonumber \\ &\times & \frac{\partial n_{\sigma_1}({\bf
p}_1,T)}{\partial {p}_1} \frac{d{\bf p}_1}{(2\pi)^3}.
\end{eqnarray}
where $F_{\sigma,\sigma_1}({\bf p_F},{\bf p}_1)$ is the Landau
amplitude, which depends on the momenta ${\bf p_F}$ and ${\bf p}$
and the spins $\sigma$. For simplicity, we ignore the spin
dependence of the effective mass, because $M^*_L$ is almost
completely spin-independent in the case of a homogeneous liquid and
weak magnetic fields. The Landau amplitude $F$ is given by
\begin{equation}\label{AMPL} F_{\sigma,\sigma_1}({\bf p},{\bf p}_1,n)
=\frac{\delta^2E(n)}{\delta n_{\sigma}({\bf p})\delta
n_{\sigma_1}({\bf p}_1)}.\end{equation} The stability of the ground
state of LFL is determined by the Pomeranchuk stability conditions:
stability is violated when at least one Landau amplitude becomes
negative and reaches its critical value \cite{lanl1,PinNoz,pom}
\begin{equation}\label{POM}
F^{a,s}_L=-(2L+1).
\end{equation}
Here $F^{a}_L$ and $F^{s}_L$ are the dimensionless spin-symmetric
and spin-antisymmetric Landau amplitudes, $L$ is the angular
momentum related to the corresponding Legendre polynomials $P_L$,
\begin{equation}\label{LEG}
F({\bf p\sigma},{\bf
p}_1\sigma_1)=\frac{1}{N}\sum^{\infty}_{L=0}P_L(\Theta)
\left[F^{a}_L\sigma,\sigma_1
+F^{s}_L\right].
\end{equation}
Here $\Theta$ is the angle between momenta ${\bf p}$ and ${\bf p}_1$
and the density of states $N=M^*_Lp_F/(2\pi^2)$. It follows from Eq.
\eqref{LANDM} that
\begin{equation}\label{EFFM1}
\frac{M^*_L}{m}=1+\frac{F^s_1}{3}.
\end{equation}
In accordance with the Pomeranchuk stability conditions it is seen
from Eq. \eqref{EFFM1} that $F^s_1>-3$, otherwise the effective mass
becomes negative leading to unstable state when it is energetically
favorable to excite quasiparticles near the Fermi surface. In what
follows, we shall omit the spin indices $\sigma$ for simplicity.

To deal with the transport properties of Fermi systems, one needs a
transport equation describing slowly varying disturbances of the
quasiparticle distribution function $n_{\bf p}({\bf r},t)$ which
depends on position ${\bf r}$ and time $t$. As long as the
transferred energy $\omega$ and momentum $q$ of the quanta of
external field are much smaller than the energy and momentum of the
quasiparticles, $qp_F/(T M^*_L)\ll 1$ and $\omega/T\ll 1$, the
quasiparticle distribution function $n({\bf q},\omega)$ satisfies
the transport equation \cite{landau,lanl1,PinNoz}
\begin{equation}\label{TREQ}
\frac{\partial n_{\bf p}}{\partial t}+\nabla_{\bf p}\varepsilon_{\bf
p}\nabla_{\bf r}n_{\bf p}-\nabla_{\bf r}\varepsilon_{\bf
p}\nabla_{\bf p}n_{\bf p}=I[n_{\bf p}].
\end{equation}
The left-hand side of Eq. \eqref{TREQ} describes the dissipationless
dynamic of quasiparticles in phase space. The quasiparticle energy
$\varepsilon_{\bf p}({\bf r},t)$ now depends on its position and
time, and the collision integral $I[n_{\bf p}]$ measures the rate of
change of the distribution function due to collisions. The transport
equation \eqref{TREQ} allows one to derive all the transport
properties and collective excitations of a Fermi system.

It is common belief that the equations of this subsection are
phenomenological and inapplicable to describe Fermi systems
characterized by the effective mass $M^*$ strongly dependent on
temperature, external magnetic fields $B$, pressure $P$ etc. On the
other hand, facts collected on HF metals demonstrate the specific
behavior when the effective mass strongly depends on temperature
$T$, doping (or the number density) $x$ and applied magnetic fields
$B$, while the effective mass $M^*$ itself can reach very high
values or even diverge, see e.g. \cite{vojta,voj}. As we have seen
in Section \ref{INTR} such a behavior is so unusual that the
traditional Landau quasiparticles paradigm fails to describe it.
Therefore, in accord with numerous experimental facts the extended
quasiparticles paradigm is to be introduced with the well-defined
quasiparticles determining as before the thermodynamic and transport
properties of strongly correlated Fermi-systems, $M^*$ becomes a
function of $T$, $x$, $B$, while the dependence of the effective
mass on $T$, $x$, $B$ gives rise to the NFL behavior
\cite{obz,khodb,dft373,dkss,dkss1,ckhz}.

As we shall see in the following Section \ref{POM_M}, Eq.
\eqref{LANDM} can be derived microscopically and it becomes
compatible with the extended paradigm.

\section{Equation for the effective mass and the scaling
behavior}\label{POM_M}

To derive the equation determining the effective mass, we consider the
model of a homogeneous HF liquid and employ the density functional
theory for superconductors (SCDFT) \cite{gross} which allows us to
consider $E$ as a functional of the occupations numbers $n({\bf p})$
\cite{dft373,dft,dft269,sh}. As a result, the ground state energy of
the normal state $E$ becomes the functional of the occupation
numbers and the function of the number density $x$, $E=E(n({\bf
p}),x)$, while Eq. \eqref{FL2} gives the single-particle spectrum.
Upon differentiating both sides of Eq. \eqref{FL2} with respect
to ${\bf p}$ and after some algebra and integration by parts, we
obtain \cite{khodb,dft373,dft,dft269}
\begin{equation}\label{EM}
\frac{\partial\varepsilon({\bf p})}{\partial {\bf p}}=\frac{{\bf
p}}{m}+\int F({\bf p},{\bf p}_1,n)\frac{\partial n({\bf
p}_1)}{\partial{\bf p}_1}\frac{d{\bf p}_1}{(2\pi)^3}.
\end{equation}
To calculate the derivative $\partial\varepsilon({\bf p})/\partial
{\bf p}$, we employ the functional representation
\begin{eqnarray}\label{ENP}
E(n)&=&\int\frac{p^2}{2m}n({\bf p})\frac{d{\bf
p}}{(2\pi)^3}\nonumber
\\&+&\frac{1}{2}\int F({\bf p},{\bf p}_1,n)_{|_{n=0}}\,n({\bf p})n({\bf
p}_1)\frac{d{\bf p}d{\bf p}_1}{(2\pi)^6} +...
\end{eqnarray}
It is seen directly from Eq. \eqref{EM} that the effective mass is
given by the well-known Landau equation \begin{equation}\label{FLL}
\frac{1}{M^*} = \frac{1}{m}+\int \frac{{\bf p}_F{\bf p_1}}{p_F^3}
F({\bf p_F},{\bf p}_1,n)\frac{\partial n(p_1)}{\partial p_1}
\frac{d{\bf p}_1}{(2\pi)^3}.
\end{equation}
For simplicity, we ignore the spin dependencies. To calculate $M^*$
as a function of $T$, we construct the free energy $F=E-TS$, where
the entropy $S$ is given by Eq. \eqref{FL3}. Minimizing $F$ with
respect to $n({\bf p})$, we arrive at the Fermi-Dirac distribution,
Eq. \eqref{FL4}.  Due to the above derivation, we conclude that Eqs.
\eqref{EM} and \eqref{FLL} are exact ones and allow us to calculate
the behavior of both $\partial\varepsilon({\bf p})/\partial {\bf p}$
and $M^*$ which now is a function of temperature $T$, external
magnetic field $B$, number density $x$ and pressure $P$ rather than a
constant. As we will see it is this feature of $M^*$ that forms both
the scaling and the NFL behavior observed in measurements on HF
metals.

In LFL theory it is assumed that $M^*_L$ is positive, finite  and
constant. As a result, the temperature-dependent corrections to $M
^*_L$, the quasiparticle energy $\varepsilon({\bf p})$ and other
quantities begin with the term proportional to $T^2$ in 3D systems
and with the term proportional to $T$ in 2D one \cite{chub}. The
effective mass is given by Eq. \eqref{LANDM}, and the specific heat
$C$ is \cite{landau}
\begin{equation}\label{HEAT}
C=\frac{2\pi^2NT}{3}=\gamma_0T=T\frac{\partial S}{\partial T},
\end{equation}
and the spin susceptibility
\begin{equation}\label{SPINS}
\chi=\frac{3\gamma_0\mu_B^2}{\pi^2(1+F^a_0)},
\end{equation}
where $\mu_B$ is the Bohr magneton and $\gamma_0\propto M^*_L$. In
the case of LFL, upon using the transport Eq. \eqref{TREQ} one finds
for the electrical resistivity at low $T$ \cite{PinNoz}
\begin{equation}\label{RESIS}
\rho(T)=\rho_0+AT^{\alpha_R},
\end{equation}
where $\rho_0$ is the residual resistivity, the exponent
$\alpha_R=2$ and $A$ is the coefficient determining the charge
transport. The coefficient is proportional to the
quasiparticle-quasiparticle scattering cross-section. Equation
\eqref{RESIS} symbolizes and defines the LFL behavior observed in
normal metals.

Equation (\ref{FLL}) at $T=0$, combined with the fact that $n({\bf
p},T=0)$ becomes $\theta (p_F-p) $, yields the well-known result
\cite{vollh,pfw,vollh1}
$$ \frac{M^*}{m}=\frac{1}{1-F^1/3}.$$
where $F^1=N_0f^1$, $N_0=mp_F/(2\pi^2)$ is the density of states of
a free Fermi gas and $f^1(p_F,p_F)$ is the $p$-wave component of the
Landau interaction amplitude. Because $x=p_F^3/3\pi^2$ in the Landau
Fermi-liquid theory, the Landau interaction amplitude can be written
as $F^1(p_F,p_F)=F^1(x)$. Provided that at a certain critical point
$x_{FC}$, the denominator $(1-F^1(x)/3)$ tends to zero, i.e.,
$(1-F^1(x)/3)\propto(x-x_{FC})+a(x-x_{FC})^2 + ...\to 0$, we find
that \cite{khod1,shag1}
\begin{equation}
\frac{M^*(x)}{m}\simeq a_1+\frac{a_2}{x-x_{FC}}\propto\frac{1}{r}.
\label{FL7}\end{equation} where $a_1$ and $a_2$ are constants and
$r=(x-x_{FC})/x_{FC}$ is the ``distance'' from QCP $x_{FC}$ at which
$M^*(x\to x_{FC})\to\infty$. We note that the divergence of the
effective mass given by Eq. \eqref{FL7} does preserve the
Pomeranchuk stability conditions for $F^1$ positive, see Eq.
\eqref{POM}. Equations \eqref{EFFM1} and \eqref{FL7} seem to be
different but it is not the case since $F^1\propto m$, while
$F^s_1\propto M^*$ and Eq. \eqref{EFFM1} represents an implicit
formula for the effective mass.

\begin{figure} [! ht]
\begin{center}
\includegraphics [width=0.60\textwidth] {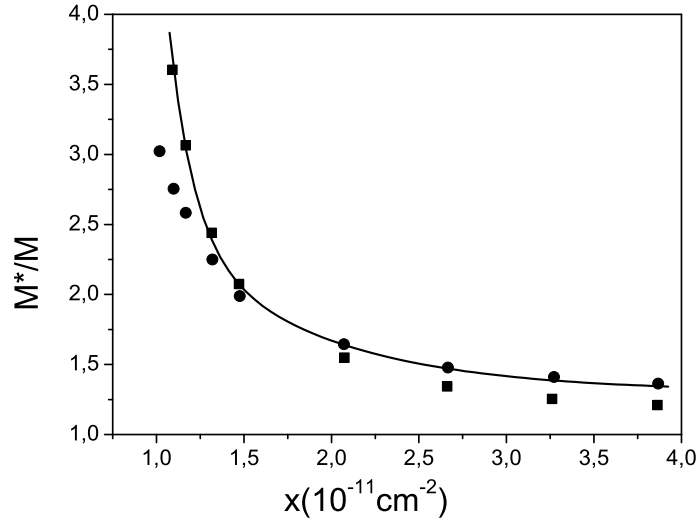}
\end{center}
\caption {The ratio $M ^*/M$ in a silicon MOSFET as a function of
the electron number density $x$. The black squares mark the
experimental data on the Shubnikov-de Haas oscillations. The data
obtained by applying a parallel magnetic field are marked by black
circles \cite{skdk,skdk1,krav}. The solid line represents the function
(\ref{DL4}).} \label{Fig5}
\end{figure}

The behavior of $M^*(x)$ described by formula (\ref{FL7}) is in good
agreement with the results of experiments \cite{cas1,skdk,skdk1} and
calculations \cite{krot,sarm1,sarm2}. In the case of electron
systems, Eq. \eqref{FL7} holds for $x>x_{FC}$, while for 2D $^3$He
we have $x<x_{FC}$ so that always $r>0$ \cite{ksk,ksz} (see also
Section \ref{FCDL}). Such behavior of the effective mass is observed
in HF metals, which have a fairly flat and narrow conductivity band
corresponding to a large effective mass, with a strong correlation
and the effective Fermi temperature $T_k\sim p_F^2/M ^*(x)$ of the
order of several dozen degrees kelvin or even lower (e.g., see Ref.
\cite{ste}).

\begin{figure} [! ht]
\begin{center}
\includegraphics [width=0.60\textwidth] {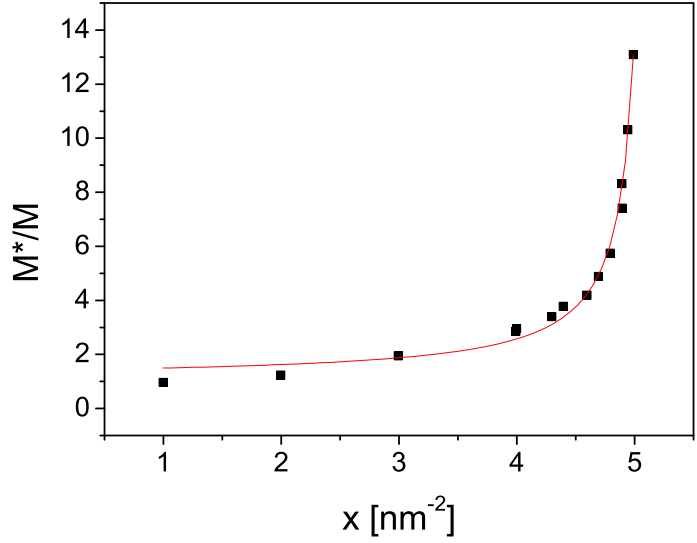}
\end{center}
\caption {The ratio $M ^*/M$ in 2D $^3$He as a function of the
density $x$ of the liquid, obtained from heat capacity and
magnetization measurements. The experimental data are marked by
black squares \cite{cas1,cas}, and the solid line represents the
function given by Eq. \eqref{FL7}, where $a_1$=1.09, $a_2=1.68\,\rm
nm^{-2}$, and $x_{FC}=5.11$ $\rm nm^{-2}$.} \label {Fig6}
\end{figure}

The effective mass as a function of the electron density $x$ in a
silicon MOSFET (Metal Oxide Semiconductor Field Effect Transistor),
approximated by Eq. (\ref{FL7}), is shown in Fig. \ref{Fig5}. The
parameters $a_1$, $a_2$ and $x_{FC}$ are taken as fitting. We see
that Eq. \eqref{FL7} provides a good description of the experimental
results.

The divergence of the effective mass  $M^*(x)$ discovered in
measurements involving 2D $^3$He \cite{cas1,cas} is illustrated in
Fig. \ref{Fig6}. Figures \ref{Fig5} and \ref{Fig6} show that the
description provided by Eq. \eqref{FL7} does not depend on
elementary Fermi particles constituting the system and is in good
agreement with the experimental data.

It is instructive to briefly explore the scaling behavior of $M^*$
in order to illustrate the ability of the quasiparticle extended
paradigm to capture the scaling behavior, while more detailed
consideration is reserved for Section \ref{HCEL}. Let us write
the quasiparticle distribution function as $n_1({\bf p})=n({\bf
p},T)-n({\bf p})$, with $n({\bf p})$ being the step function, and Eq.
\eqref{FLL} then becomes
\begin{equation}
\frac{1}{M^*(T)}=\frac{1}{M^*}+\int\frac{{\bf p}_F{\bf
p_1}}{p_F^3}F({\bf p_F},{\bf p}_1)\frac{\partial n_1(p_1,T)}
{\partial p_1}\frac{d{\bf p}_1}{(2\pi) ^3}. \label{LF1}
\end{equation}
At QCP $x\to x_{FC}$, the effective mass $M^*(x)$ diverges and Eq.
\eqref{LF1} becomes homogeneous determining $M^*$ as a function of
temperature while the system exhibits the NFL behavior. If the
system is located before QCP, $M^*$ is finite, at low temperatures
the integral on the right hand side of Eq. \eqref{LF1} represents a
small correction to $1/M^*$ and  the system demonstrates the LFL
behavior seen in Figs. \ref{YBRHSI} and \ref{YBRHSIN}. The LFL
behavior assumes that the effective mass is independent of
temperature, $M^*(T)\simeq const$, as shown by the horizontal line
in Fig. \ref{YBRHSIN}. Obviously, the LFL behavior takes place only
if the second term on the right hand side of Eq. \eqref{LF1} is
small in comparison with the first one. Then, as temperature rises
the system enters the transition regime: $M^*$ grows, reaching its
maximum $M^*_M$ at $T=T_M$, with subsequent diminishing. As seen
from Fig. \ref{YBRHSIN}, near temperatures $T\geq T_M$ the last
"traces" of LFL regime disappear, the second term starts to
dominate, and again Eq. \eqref{LF1} becomes homogeneous, and the NFL
behavior is restored, manifesting itself in decreasing $M^*$ as a
function of $T$.

\section{Fermion condensation quantum phase transition}\label{FLFC}

As shown in Section \ref{POM_M}, the Pomeranchuk stability
conditions do not encompass all possible types of instabilities and
that at least one related to the divergence of the effective mass
given by Eq. \eqref{FL7} was overlooked \cite{ks}. This type of
instability corresponds to a situation where the effective mass, the
most important characteristic of quasiparticles, can become
infinitely large. As a result, the quasiparticle kinetic energy is
infinitely small near the Fermi surface and the quasiparticle
distribution function $n({\bf p})$ minimizing $E(n({\bf p}))$ is
determined by the potential energy. This leads to the formation of a
new class of strongly correlated Fermi liquids with FC
\cite{ks,ksk,vol_1,vol}, separated from the normal Fermi liquid by
FCQPT \cite{shag3,ms,shb}.

It follows from (\ref{FL7}) that at  $T=0$ and as $r\to0$ the
effective mass diverges, $M^*(r)\to\infty$. Beyond the critical
point $x_{FC}$, the distance $r$ becomes negative and,
correspondingly, so does the effective mass. To avoid an unstable
and physically meaningless state with a negative effective mass, the
system must undergo a quantum phase transition at the critical point
$x=x_{FC}$, which, as we will see shortly, is FCQPT \cite{ms,shb,shag3}.
Because the kinetic energy of quasiparticles that are near the Fermi
surface is proportional to the inverse effective mass, the potential
energy of the quasiparticles near the Fermi surface determines the
ground-state energy as $x\to x_{FC}$. Hence, a phase transition
reduces the energy of the system and transforms the quasiparticle
distribution function. Beyond QCP $x=x_{FC}$, the quasiparticle
distribution is determined by the ordinary equation for a minimum of
the energy functional \cite{ks}:
\begin{equation} \frac{\delta E(n({\bf p}))}{\delta
n({\bf p},T=0)}=\varepsilon({\bf p})=\mu; \, p_i\leq p\leq p_f.
\label{FL8}\end{equation}

\begin{figure} [! ht]
\begin{center}
\includegraphics [width=0.60\textwidth]{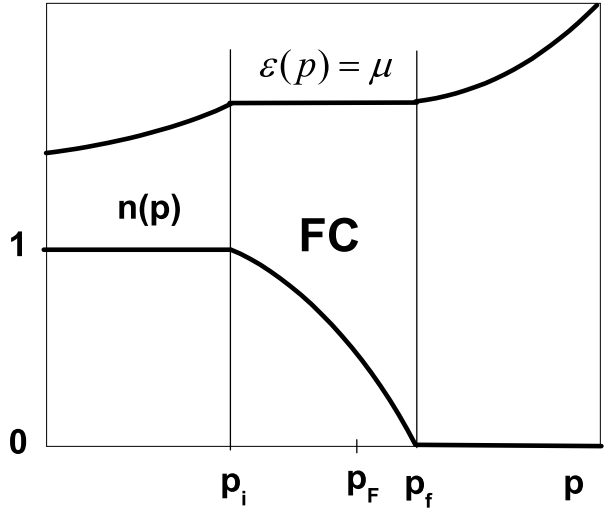}
\end{center}
\caption {The single-particle spectrum $\varepsilon(p)$ and the
quasiparticle distribution function $n_0(p)$. Because $n_0(p)$ is a
solution of Eq. (\ref{FL8}), we have $n_0(p<p_i)=1$,
$0<n_0(p_i<p<p_f)<1$, and $n_0(p>p_f)=0$, while $\varepsilon
(p_i<p<p_f)=\mu$. The Fermi momentum $p_F$  satisfies the condition
$p_i<p_F<p_f$.} \label{Fig1}
\end{figure}

Equation (\ref{FL8}) yields the quasiparticle distribution function
$n_0({\bf p}) $ that minimizes the ground-state energy $E$. This
function found from Eq. (\ref{FL8}) differs from the step function
in the interval from $p_i$ to $p_f$, where $0<n_0({\bf p})<1$, and
coincides with the step function outside this interval. In fact, Eq.
\eqref{FL8} coincides with Eq. \eqref{FL2} provided that the Fermi
surface at $p=p_F$ transforms into the Fermi volume at $p_i\leq
p\leq p_f$ suggesting that the single-particle spectrum is
absolutely ``flat'' within this interval. A possible solution
$n_0({\bf p})$ of Eq. (\ref{FL8}) and the corresponding
single-particle spectrum $\varepsilon({\bf p})$ are depicted in Fig.
\ref{Fig1}. Quasiparticles with momenta within the interval
$(p_f-p_i)$ have the same single-particle energies equal to the
chemical potential $\mu$ and form FC, while the distribution
$n_0({\bf p})$ describes the new state of the Fermi liquid with FC
\cite{ks,ksk,vol}.
In contrast to the Landau, marginal, or Luttinger Fermi liquids
\cite{varma,var,varm1}, which exhibit the same topological structure
of the Green's function, in systems with FC, where the Fermi surface
spreads into a strip, the Green's function belongs to a different
topological class. The topological class of the Fermi liquid is
characterized by the invariant \cite{volovik1,volovik2,vol}
\begin{equation} N=tr\oint_C\frac{dl}{2\pi i}G(i\omega,{\bf p})
\partial_lG^{-1}(i\omega,{\bf p})\label{FLVOL},\end{equation}
where ``tr'' denotes the trace over the spin indices of the Green's
function and the integral is taken along an arbitrary contour $C$
encircling the singularity of the Green's function. The invariant
$N$ in (\ref{FLVOL}) takes integer values even when the singularity
is not of the pole type, cannot vary continuously, and is conserved
in a transition from the Landau Fermi liquid to marginal liquids and
under small perturbations of the Green's function. As shown by
Volovik \cite{volovik1,volovik2,vol}, the situation is quite
different for systems with FC, where the invariant $N$ becomes a
half-integer and the system with FC transforms into an entirely new
class of Fermi liquids with its own topological structure.

\subsection{The order parameter of FCQPT}\label{FCPD}

We start with visualizing the main properties of FCQPT. To this end,
again consider SCDFT. SCDFT states that the thermodynamic potential
$\Phi$ is a universal functional of the number density $n({\bf r})$
and the anomalous density (or the order parameter) $\kappa({\bf
r},{\bf r}_1)$, providing a variational principle to determine the
densities. At the superconducting transition temperature $T_c$ a
superconducting state undergoes the second order phase transition.
Our goal now is to construct a quantum phase transition which
evolves from the superconducting one.

Let us assume that the coupling constant $\lambda_0$ of the BCS-like
pairing interaction \cite{bcs} vanishes, with $\lambda_0\to0$ making
vanish the superconducting gap at any finite temperature. In that
case, $T_c\to0$ and the superconducting state takes place at $T=0$
while at finite temperatures there is a normal state. This means
that at $T=0$ the anomalous density
\begin{equation}\label{ANOM}\kappa({\bf
r},{\bf r_1})=\langle\Psi\uparrow({\bf r})\Psi\downarrow({\bf
r_1})\rangle\end{equation} is finite, while the superconducting gap
\begin{equation}\label{DEL}\Delta({\bf
r})=\lambda_0\int\kappa({\bf r},{\bf r_1})d{\bf r_1}\end{equation}
is infinitely small \cite{obz,shag1}. In Eq. \eqref{ANOM}, the field
operator $\Psi_{\sigma}({\bf r})$ annihilates an electron of spin
$\sigma, \sigma=\uparrow,\downarrow$ at the position ${\bf r}$. For
the sake of simplicity, we consider the model of  homogeneous HF
liquid \cite{obz}. Then at $T=0$, the thermodynamic potential $\Phi$
reduces to the ground state energy $E$ which turns out to be a
functional of the occupation number $n({\bf p})$ since in that case
the order parameter $\kappa({\bf p})=v({\bf p})u({\bf
p})=\sqrt{n({\bf p})(1-n({\bf p}))}$. Indeed,
\begin{equation} n({\bf p})=v^2({\bf
p}); \, \, \, \kappa({\bf p})=v({\bf p})u({\bf p}),\label{SC2}
\end{equation}
where $u({\bf p})$ and $v({\bf p})$ are normalized parameters such
that $v^2({\bf p})+u^2({\bf p})=1$ and $\kappa({\bf p})=\sqrt{n({\bf
p})(1-n({\bf p}))}$, see e.g. \cite{lanl1}.

Upon minimizing $E$ with respect to $n({\bf p})$, we obtain Eq.
\eqref{FL8}. As soon as Eq. \eqref{FL8} has nontrivial solution
$n_0({\bf p})$ then instead of the Fermi step, we have $0<n_0({\bf
p})<1$ in certain range of momenta $p_i\leq p\leq p_f$ with
$\kappa({\bf p})=\sqrt{n_0({\bf p})(1-n_0({\bf p}))}$ being finite in
this range, while the single particle spectrum $\varepsilon({\bf
p})$ is flat. Thus, the step-like Fermi filling inevitably undergoes
restructuring and forms FC when Eq. \eqref{FL8} possesses for the
first time the nontrivial solution at $x=x_c$ which is QCP of FCQPT.
In that case, the range vanishes, $p_i\to p_f\to p_F$, and the
effective mass $M^*$ diverges at QCP \cite{obz,khodb,ks}
\begin{equation}\label{EFM}
\frac{1}{M^*(x\to x_c)}=\frac{1}{p_F}\frac{\partial\varepsilon({\bf
p})}{\partial{\bf p}}|_{p\to p_F;\,x\to x_c}\to 0.\end{equation} At
any small but finite temperature the anomalous density $\kappa$ (or
the order parameter) decays and this state undergoes the first order
phase transition and converts into a normal state characterized by
the thermodynamic potential $\Phi_0$. Indeed, at $T\to0$, the
entropy $S=-\partial \Phi_0/\partial T$ of the normal state is given
by Eq. \eqref{FL3}. It is seen from Eq. \eqref{FL3} that the normal
state is characterized by the temperature-independent entropy $S_0$
\cite{obz,khodb,yakov}. Since the entropy of the superconducting
ground state is zero, we conclude that the entropy is discontinuous
at the phase transition point, with its discontinuity $\delta
S=S_0$. Thus, the system undergoes the first order phase transition.
The heat $q$ of transition from the asymmetrical to the symmetrical
phase is $q=T_cS_0=0$ since $T_c=0$. Because of the stability
condition at the point of the first order phase transition, we have
$\Phi_0(n({\bf p}))=\Phi(\kappa({\bf p}))$. Obviously the condition
is satisfied since $q=0$.

\subsection{Quantum protectorate related to FCQPT}\label{FCQP}

With FCQPT  (as well as with other phase transitions), we have to
deal with strong particle interaction, and there is no way in which
a theoretical investigation based on first principles can provide an
absolutely reliable solution. Hence, the only way to verify that FC
exists is to study this state by exactly solvable models and to
examine the experimental facts that could be interpreted as direct
confirmation of the existence of FC. Exactly solvable models
unambiguously suggest that Fermi systems with FC exist (e.g., see
Refs \cite{khv,dzyal,lid,irk}). Taking the results of topological
investigations into account, we can state that the new class of
Fermi liquids with FC is nonempty, actually exists, and represents
an extended family of new states of Fermi systems
\cite{volovik1,volovik2,vol}.

We note that the solutions  $n_0({\bf p})$ of Eq. (\ref{FL8}) are
new solutions of the well-known equations of the Landau Fermi-
liquid theory. Indeed, at $T=0$, the standard solution given by a
step function, $n({\bf p},T\to0)\to\theta(p_F-p)$, is not the only
possible one. Anomalous solutions $\varepsilon({\bf p})=\mu$ of Eq.
(\ref{FL1}) can exist if the logarithmic expression on its
right-hand side is finite. This is possible if $0<n_0({\bf p})<1$
within a certain interval $(p_i\leq p \leq p_f)$. Then, this
logarithmic expression remains finite within this interval as
$T\to0$, the product $T\ln[(1-n_0({\bf p}))/n_0({\bf
p})]_{|T\to0}\to0$, and we again arrive at Eq. (\ref{FL8}).

Thus, as $T\to0$, the quasiparticle distribution function $n_0({\bf
p})$, which is a solution of Eq. (\ref{FL8}), does not tend to the
step function $\theta(p_F-p)$ and, correspondingly, in accordance
with Eq. (\ref{FL3}), the entropy $S(T)$ of this state tends to a
finite value $S_0$ as $T\to0$:
\begin{equation}
S(T\to0)\to S_0.\label{snfl} \end{equation}

As the density $x\to x_{FC}$ (or as the interaction force
increases), the system reaches QCP at which FC is formed. This means
that $p_i\to p_f\to p_F$ and that the deviation $\delta n({\bf p})$
from the step function is small. Expanding the function $E(n({\bf
p}))$ in Taylor series in $\delta n({\bf p})$ and keeping only the
leading terms, we can use Eq. (\ref{FL8}) to obtain the following
relation that is valid within the interval $p_i\leq p\leq p_f$:
\begin{equation} \mu=\varepsilon({\bf p})=\varepsilon_0({\bf
p})+\int F({\bf p},{\bf p}_1)\delta n({\bf p_1})\frac{d{\bf
p}_1}{(2\pi)^2}.\label{FL9}\end{equation} Both quantities, the
Landau amplitude $F({\bf p},{\bf p}_1)$ and the single-particle
energy $\varepsilon_0({\bf p})$, are calculated at $n({\bf
p})=\theta(p_F-p)$. Equation (\ref{FL9}) has nontrivial solutions
for densities $x\leq x_{FC}$ if the corresponding Landau amplitude,
which is density-dependent, is positive and sufficiently large for
the potential energy to be higher than the kinetic energy. For
instance, such a state is realized in a low-density electron liquid.
The transformation of the Fermi step function $n({\bf
p})=\theta(p_F-p)$ into a smooth function determined by Eq.
(\ref{FL9}) then becomes possible \cite{ks,ksk,ksz}.

It follows from Eq. (\ref{FL9}) that the quasiparticles of FC form a
collective state, because their state is determined by the
macroscopic number of quasiparticles with momenta $p_i<p<p_f$. The
shape of the single-particle spectrum related to FC is independent
of the Landau interaction, which is in general determined by the
properties of the system as a whole, including the collective
states, irregularities of structure, the presence of impurities, and
composition. The length of the interval from $p_i$ to $p_f$ where FC
exists is the only characteristic determined by the Landau
interaction; of course, the interaction must be strong enough for
FCQPT to occur. Therefore, we conclude that spectra related to FC
have a universal shape. In Sections \ref{FLSH} and \ref{SCFC} we
show that these spectra are dependent on the temperature and the
superconducting gap and that this dependence is also universal. The
existence of such spectra can be considered a characteristic feature
of a "quantum protectorate", in which the properties of the
material, including the thermodynamic properties, are determined by
a certain fundamental principle \cite{rlp,pa}. In our case, the
state of matter with FC  is also a quantum protectorate, since the
new type of quasiparticles of this state determines the special
universal thermodynamic and transport properties of Fermi liquids
with FC .

\subsection{The influence of FCQPT at finite temperatures} \label{FLSH}

According to Eq. (\ref{FL1}), the single-particle energy
$\varepsilon({\bf p},T)$ is linear in $T$ for $T\ll T_f$ within the
interval $(p_f-p_i)$ \cite{kcs}. Expanding $\ln((1-n({\bf
p}))/n({\bf p}))$ in a series in $n({\bf p})$ at $p\simeq p_F$, we
can write the expression
\begin{equation} \frac{\varepsilon({\bf p},T)-\mu(T)}{T}= \ln\frac{1-n({\bf
p})}{n({\bf p})}\simeq \frac{1-2n({\bf p})}{n({\bf
p})}\bigg|_{p\simeq p_F}. \label{FL10}
\end{equation}
where $T_f$ is the temperature above which the effect of FC is
insignificant \cite{dkss}:
\begin{equation} \frac {T_f}{\varepsilon_F} \sim
\frac{p_f^2-p_i^2}{2M\varepsilon_F}\sim\frac
{\Omega_{FC}}{\Omega_F}. \label{FL11}\end{equation} with
$\Omega_{FC}$ being the volume occupied by FC, $\varepsilon_F$ being
the Fermi energy, and $\Omega_F$ being the volume of the Fermi
sphere. We note that for $T\ll T_f$, the occupation numbers $n({\bf
p})$ obtained from Eq. (\ref{FL8}) are almost perfectly independent
of $T$ \cite{dkss,dkss1,kcs}. At finite temperatures, according to
Eq. (\ref{FL10}), the dispersionless plateau $\varepsilon ({\bf
p})=\mu$ shown in Fig. \ref{Fig1} is slightly rotated
counterclockwise in relation to $\mu$. As a result, the plateau is
slightly tilted and rounded off at its end points. According to Eqs.
(\ref{FL5}) and (\ref{FL10}), the effective mass $M^*_{FC}$ that
refers to the FC quasiparticles is given by
\begin{equation} M^*_{FC}\simeq p_F\frac{p_f-p_i}{4T}.
\label{FL12}\end{equation} In deriving (\ref{FL12}), we approximated
the derivative as $dn(p)/dp\simeq -1/(p_f-p_i)$. Equation
(\ref{FL12}) clearly shows that for $0<T\ll T_f$, the electron
liquid with FC  behaves as if it were placed at a quantum critical
point, since the electron effective mass diverges as $T\to0$.
Actually, as we shall see in Subsection \ref{PHDFC} the system is at
a quantum critical line, because critical behavior is observed
behind QCP with $x=x_{FC}$ of FCQPT as $T\to0$. In Sections
\ref{HFL} and \ref{SCEL}, we show that the behavior of such a system
differs dramatically from that of a system at a quantum critical
point.

Upon using Eqs. \eqref{FL11} and \eqref{FL12}, we estimate the
effective mass $M^*_{FC}$ as
\begin{equation}
\frac{M^*_{FC}}{M}\sim\frac{N(0)}{N_0(0)}\sim\frac{T_f}T,
\label{FL13} \end{equation} where $N_0(0)$ is the density of states
of a noninteracting electron gas and $N(0)$ is the density of
states on the Fermi surface. Equations (\ref{FL12}) and
(\ref{FL13}) yield the temperature dependence of $M^*_{FC}$.

Multiplying both sides of Eq. \eqref{FL12} by $(p_f-p_i)$, we obtain
an expression for the characteristic energy,
\begin{equation} E_0\ \simeq\ 4T,\label{FL14}\end{equation}
which determines the momentum interval $(p_f-p_i)$ with the
low-energy quasiparticles characterized by the energy
$|\varepsilon({\bf p})-\mu|\leq E_0/2$ and the effective mass
$M^*_{FC}$. The quasiparticles that do not belong to this momentum
interval have an energy $|\varepsilon({\bf p})-\mu|>E_0/2$ and
an effective mass $M^*_{L}$ that is weakly temperature-dependent
\cite{ms, shb,ars}. Equation (\ref{FL14}) shows that $E_0$ is
independent of the condensate volume. We conclude from Eqs.
(\ref{FL12}) and (\ref{FL14}) that for $T\ll T_f$, the
single-electron spectrum of FC quasiparticles has a universal shape
and has the features of a quantum protectorate.

Thus, a system with FC  is characterized by two effective masses,
$M^*_{FC}$ and $M ^*_L$. This fact manifests itself in a break or an
abrupt change in the quasiparticle dispersion law, which for
quasiparticles with energies $\varepsilon({\bf p})\leq \mu$ can be
approximated by two straight lines intersecting at $E_0/2\simeq 2T$.
Figure \ref{Fig1} shows that at $T=0$, the straight lines intersect
at $p=p_i$. This break also occurs when the system is in its
superconducting state at temperatures $T_c\leq T\ll T_f$, where
$T_c$ is the critical temperature of the superconducting phase
transition, which agrees with the experimental data in \cite{blk}
and, as we will see in Section \ref{SC}, this behavior agrees with the
experimental data at $T\leq T_c$. At $T>T_c$, the quasiparticles are
well-defined, because their width $\gamma$ is small compared to
their energy and is proportional to the temperature, $\gamma\sim T$
\cite{koral,dkss}. The quasiparticle excitation curve (see Section
\ref{DISP}) can be approximately described by a simple Lorentzian
\cite{ars}, which also agrees with the experimental data
\cite{blk,krc,vall,vall1}.

We estimate the density $x_{FC}$ at which FCQPT occurs. We show in
Section \ref{FCDL} that an unlimited increase in the effective mass
precedes the appearance of a density wave or a charge density wave
formed in electron systems at $r_s=r_{cdw}$, where $r_s=r_0/a_B$,
$r_0$ is the average distance between electrons, and $a_B$ is the
Bohr radius. Hence, FCQPT certainly occurs at $T=0$ when $r_s$
reaches its critical value $r_{FC}$ corresponding to $x_{FC}$, with
$r_{FC}<r_{cdw}$ \cite{ksz}. We note that the increase in the
effective mass as the electron number density decreases was observed
in experiments, see Figs. \ref{Fig5} and \ref{Fig6}.

Thus, the formation of FC  can be considered a general property of
different strongly correlated systems rather than an exotic
phenomenon corresponding to the anomalous solution of Eq.
\eqref{FL8}. Beyond FCQPT, the condensate volume is proportional to
$(r_s-r_ {FC})$, with $T_f/\varepsilon_F\sim (r_s-r_{FC})/r_{FC}$,
at least when $ (r_s-r_{FC})/r_{FC} \ll 1 $. This implies that
\cite{obz}
\begin{equation}\frac {r_s-r_{FC}}{r_{FC}}\sim \frac{p_f-p_{i}}{p_{F}}
\sim \frac {x_{FC}-x}{x_{FC}}. \label{FL15}
\end{equation}
Because a state of a system with FC  is highly degenerate, FCQPT
serves as a stimulator of phase transitions that could lift the
degeneracy of the spectrum. For instance, FC  can stimulate the
formation of spin density waves, antiferromagnetic state and
ferromagnetic state etc., thus strongly stimulating the competition
between phase transitions eliminating the degeneracy. The presence
of FC strongly facilitates a transition to the superconducting
state, because both phases have the same order parameter.

\subsection{Phase diagram of Fermi system with FCQPT}\label{PHDFC}

At $T=0$, a quantum phase transition is driven by a nonthermal
control parameter, e.g. the number density $x$. As we have seen, at
QCP, $x=x_{FC}$, the effective mass diverges. It follows from Eq.
\eqref{FL7} that beyond QCP, the effective mass becomes negative. To
avoid an unstable and physically meaningless state with a negative
effective mass, the system undergoes FCQPT leading to the formation of
FC.

\begin{figure} [! ht]
\begin{center}
\vspace*{-0.5cm}
\includegraphics [width=0.60\textwidth]{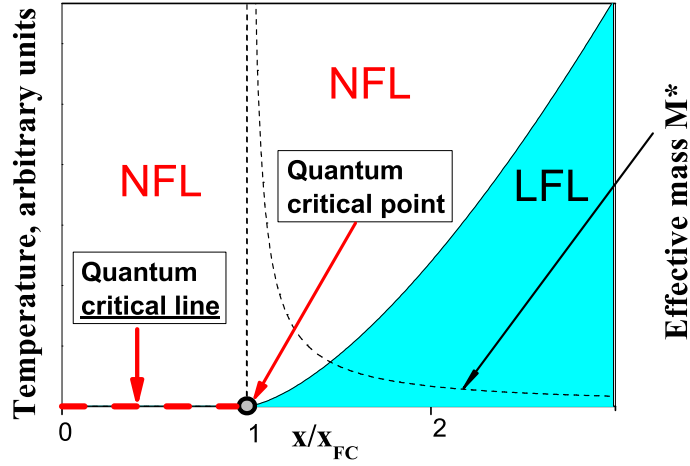}
\end{center}
\vspace*{-0.8cm} \caption{Schematic phase diagram of system with FC.
The number density $x$ is taken as the control parameter and
depicted as $x/x_{FC}$. The dashed line shows $M^*(x/x_{FC})$ as the
system approaches QCP, $x/x_{FC}=1$, of FCQPT which is denoted by
the arrow. At $x/x_{FC}>1$ and sufficiently low temperatures, the
system is in the LFL state as shown by the shadow area. At $T=0$ and
beyond the critical point, $x/x_{FC}<1$, the system is at the
quantum critical line depicted by the dashed line and shown by the
vertical arrow. The critical line is characterized by the FC state
with finite superconducting order parameter $\kappa$. At any finite
low temperature $T>T_c=0$, $\kappa$ is destroyed, the system
undergoes the first order phase transition, possesses finite entropy
$S_0$ and exhibits the NFL behavior at any finite temperatures
$T<T_f$.}\label{fig1}
\end{figure}

A schematic phase diagram of the system which is driven to the FC
state by variation of $x$ is reported in Fig. \ref{fig1}.  Upon
approaching the critical density $x_{FC}$ the system remains in the
LFL region at sufficiently low temperatures as it is shown by the
shadow area. The temperature range of the shadow area shrinks as the
system approaches QCP, and $M^*(x/x_{FC})$ diverges as shown by the
dashed line and Eq. \eqref{FL7}. At QCP $x_{FC}$ shown by the arrow in
Fig. \ref{fig1}, the system demonstrates the NFL behavior down to
the lowest temperatures. Beyond the critical point at finite
temperatures the behavior remains the NFL and is determined
by the temperature-independent entropy $S_0$ \cite{obz,yakov}. In
that case at $T\to 0$, the system is approaching a quantum critical
line (shown by the vertical arrow and the dashed line in Fig.
\ref{fig1}) rather than a quantum critical point. Upon reaching the
quantum critical line from the above at $T\to0$ the system undergoes
the first order quantum phase transition, which is FCQPT taking
place at $T_c=0$. While at diminishing temperatures, the systems
located before QCP do not undergo a phase transition and their
behavior transits from NFL to LFL.

It is seen from Fig. \ref{fig1} that at finite temperatures there is
no boundary (or phase transition) between the states of systems
located before or behind QCP shown by the arrow. Therefore, at
elevated temperatures the properties of systems with $x/x_{FC}<1$ or
with $x/x_{FC}>1$ become indistinguishable. On the other hand, at
$T>0$ the NFL state above the critical line and in the vicinity of
QCP is strongly degenerate, therefore the degeneracy stimulates the
emergence of different phase transitions lifting it and the NFL
state can be captured by the other states such as superconducting
(for example, by the superconducting state (SC) in $\rm CeCoIn_5$
\cite{shag1,yakov}) or by antiferromagnetic (AF) state (e.g. AF one
in $\rm YbRh_2Si_2$ \cite{dft373}) etc. The diversity of phase
transitions occurring at low temperatures is one of the most
spectacular features of the physics of many HF metals. Within the
scenario of ordinary quantum phase transitions, it is hard to
understand why these transitions are so different from one another
and their critical temperatures are so extremely small. However,
such diversity is endemic to systems with a FC \cite{khodb}.

Upon using nonthermal tuning parameters like the number density,
pressure or magnetic field, the NFL behavior is destroyed and the LFL
one is restored as we shall see in Sections \ref{HCEL} and
\ref{SCEL}. For example, the application of magnetic field
$B>B_{c0}$ drives a system to QCP and destroys the AF state restoring
the LFL behavior. Here, $B_{c0}$ is a critical magnetic field, such
that at $B>B_{c0}$ the system is driven towards its LFL state. In
some cases as in the HF metal $\rm CeRu_2Si_2$, $B_{c0}=0$, see e.g.
\cite{takah}, while in $\rm YbRh_2Si_2$, $B_{c0}\simeq 0.06$ T
\cite{geg}.

\section{The superconducting state
with FC } \label{SC}

In this section we discuss the superconducting state of a 2D liquid
of heavy electrons, since high-$T_c$ superconductors are
represented mainly by 2D structures. On the other hand, our study
can easily be generalized to the 3D case. To show that there is no
fundamental difference between the 2D and 3D cases, we derive
Green's functions for the 3D case in Section \ref{FC}.

\subsection{The superconducting state at $T=0$}
\label{SCFC}

As we have seen in Subsection \ref{FCPD}, the ground-state energy
$E_{gs}(\kappa({\bf p}),n({\bf p}))$ of a 2D electron liquid is a
functional of the superconducting state order parameter $\kappa({\bf
p})$ and of the quasiparticle occupation numbers $n({\bf p})$. This
energy is determined by the well-known Bardeen-Cooper-Schrieffer
(BCS) equations and in the weak-coupling superconductivity theory is
given by \cite{gross,bcs,til}
\begin{eqnarray}
 \nonumber
E_{gs}(\kappa({\bf p}),n({\bf p}))
&=&E(n({\bf p}))+\lambda_0\int V({\bf p}_1,{\bf p}_2) \\
& \times & \kappa ({\bf p}_1)\kappa^*({\bf p}_2) \frac{d{\bf
p}_1d{\bf p}_2}{(2\pi)^4}. \label{SC1}
\end{eqnarray}
It is assumed that the constant $\lambda_0$, which determines the
magnitude of the pairing interaction $\lambda_0V({\bf p}_1,{\bf
p}_2)$, is small. We define the superconducting gap as
\begin{equation} \Delta({\bf p})=-\lambda_0\int V({\bf p},{\bf
p}_1)\kappa({\bf p}_1) \frac{d{\bf
p}_1}{4\pi^2}.\label{SC3}\end{equation} Minimizing $E_{gs}$ in
$v({\bf p})$ and using (\ref{SC3}), we arrive at equations that
relate the single-particle energy $\varepsilon({\bf p})$ to
$\Delta({\bf p})$ and $E({\bf p})$
\begin{equation} \varepsilon ({\bf p})-\mu = \Delta ({\bf
p})\frac{1-2v^2({\bf p})}{2\kappa({\bf p})},\,\, \frac{\Delta({\bf
p})}{E({\bf p})}=2\kappa({\bf p}).\label{SC4}\end{equation} Here the
single-particle energy $\varepsilon ({\bf p})$ is determined by Eq.
\eqref{FL2}, and
\begin{equation}E({\bf p})=\sqrt{\xi^2({\bf
p})+\Delta^2({\bf p})},\label{SC3.1}\end{equation} with $\xi({\bf
p})=\varepsilon({\bf p})-\mu$. Substituting the expression for
$\kappa ({\bf p})$ from (\ref{SC4}) in Eq. \eqref{SC3}, we obtain
the well-known equation of the BCS theory for $\Delta ({\bf p})$
\begin{equation}
\Delta({\bf p})=-\frac{\lambda_0}{2}\int V({\bf p},{\bf p}_1)
\frac{\Delta({\bf p}_1)}{E({\bf p}_1)} \frac{d{\bf p}_1}{4\pi^2}.
\label{SC5}\end{equation} As $\lambda_0\to0$, the maximum value of
$\Delta_1$ of the superconducting gap $\Delta({\bf p})$ tends to
zero and each equation in (\ref{SC4}) reduces to Eq. (\ref{FL8})
\begin{equation} \frac{\delta E(n({\bf p}))}{\delta n({\bf p})}=
\varepsilon({\bf p})-\mu=0, \label{SC6} \end{equation} if $0<n({\bf
p})<1,\,\, \mbox{or}\,\, \kappa({\bf p})\neq 0$, in the interval
$p_i\leq p\leq p_f$. Equation (\ref{SC6}) shows that the function
$n_0({\bf p})$ is determined from the solution to the standard
problem of finding the minimum of the functional $E(n({\bf p}))$
\cite{ks,dkss,dkss1}. Equation (\ref{SC6}) specifies the
quasiparticle distribution function $n_0({\bf p})$ that ensures the
minimum of the ground-state energy $E(\kappa({\bf p}),n({\bf p}))$.
We can now study the relation between the state specified by Eq.
\eqref{SC6} or Eq. \eqref{FL8} and the superconducting state.

At $T=0$, Eq. (\ref{SC6}) determines the specific state of a Fermi
liquid with FC, the state for which the absolute value of the order
parameter $|\kappa({\bf p})|$ is finite in the momentum interval
$p_i\leq p\leq p_f$ as $\Delta_1\to0$. Such a state can be
considered superconducting with an infinitely small value of
$\Delta_1$. Hence, the entropy of this state at $T=0$ is zero.
Solutions $n_0({\bf p})$ of Eq. (\ref{SC6}) constitute a new class
of solutions of both the BCS equations and the Landau Fermi-liquid
equations. In contrast to the ordinary solutions of the BCS
equations \cite{bcs}, the new solutions are characterized by an
infinitely small superconducting gap $\Delta_1\to0$, with the order
parameter $\kappa({\bf p})$ remaining finite. On the other hand, in
contrast to the standard solution of the Landau Fermi-liquid
theory, the new solutions $n_0({\bf p})$ determine the state of a
heavy-electron liquid with a finite entropy $S_0$ as $T\to0$ (see
Eq. (\ref{snfl})). We arrive at an important conclusion that the
solutions of Eq. (\ref{SC6}) can be interpreted as the general
solutions of the BCS equations and the Landau Fermi-liquid theory
equations, while Eq. (\ref{SC6}) itself can be derived either from
the BCS theory or from the Landau Fermi-liquid theory. Thus, as
shown in Subsection \ref{FCPD} both states of the system coexist as
$T\to0$. As the system passes into a state with the order parameter
$\kappa({\bf p})$, the entropy suddenly vanishes, with the system
undergoing the first-order transition near which the critical
quantum and thermal fluctuations are suppressed and the
quasiparticles are well- defined excitations (see also Section
\ref{SCEL}). It follows from Eq. (\ref{FLVOL}) that FCQPT is
related to a change in the topological structure of the Green's
function and belongs to Lifshitz's topological phase transitions,
which occur at absolute zero \cite{vol}. This fact establishes a
relation between FCQPT and quantum phase transitions under which
the Fermi sphere splits into a sequence of Fermi layers
\cite{asp,pogshag} (see Sections \ref{HFL} and \ref{TFT}). We note
that in the state with the order parameter $\kappa({\bf p})$, the
system entropy $S=0$ and the Nernst theorem holds in systems with
FC.

If $\lambda_0\neq0$, the gap $\Delta_1$ becomes finite, leading to
a finite value of the effective mass $M^*_{FC}$, which may be
obtained from Eq. (\ref{SC4}) by taking the derivative with respect
to the momentum $p$ of both sides and using Eq. (\ref{FL5})
\cite{ms,shb,ars}:
\begin{equation} M^*_{FC}\simeq
p_F\frac{p_f-p_i}{2\Delta_1}.\label{SC7}\end{equation} It follows
from Eq. \eqref{SC7} that in the superconducting state the effective
mass is always finite. As regards the energy scale, it is determined
by the parameter $E_0$:
\begin{equation} E_0=\varepsilon({\bf p}_f)- \varepsilon({\bf
p}_i)\simeq p_F \frac{(p_f-p_i)}{M^*_{FC}}\simeq
2\Delta_1.\label{SC8}\end{equation}

\subsection{Green's function of the superconducting state
with FC at $T=0$}\label{FC}

We write two equations for the 3D case, the Gor'kov equations
\cite{gorkov}, which determine the Green's functions $F^+({\bf
p},\omega)$ and $G({\bf p},\omega)$ of a superconductor (e.g., see
Ref.  \cite{lanl1}):
\begin{eqnarray}
\nonumber
F^+&=&\frac{-\lambda_0\Xi^*}{(\omega -E({\bf p})+i\,0)(\omega +E({\bf p})-i\,0)};\\
G&=&\frac{u^2({\bf p})}{\omega -E({\bf p})+i\,0}+\frac{v^2({\bf
p})}{\omega +E({\bf p})-i\,0},\label{zui2}
\end{eqnarray}
The gap  $\Delta$ and the function $\Xi$ are given by
\begin{equation}\label{zui3}
\Delta=\lambda_0|\Xi|,\quad i\Xi= \int\int_{-\infty }^{\infty
}F^+({\bf p},\omega)\frac{d\omega d{\bf p} }{(2\pi)^4}.
\end{equation}
We recall that the function $F^+({\bf p},\omega)$ has the meaning of
the wave function of Cooper pairs and $\Xi$ is the wave function of
the motion of these pairs as a whole and is just a constant in a
homogeneous system \cite{lanl1}. It follows from Eqs. \eqref{SC4}
and \eqref{zui3} that
\begin{equation}\label{zui7}
i\Xi=\int_{-\infty }^{\infty }F_0^+({\bf p},\omega )\frac{d\omega
d{\bf p}}{(2\pi)^4}=i\int\kappa({\bf p})\frac{d{\bf p}}{(2\pi)^3}.
\end{equation}
Taking Eqs. \eqref{zui3} and \eqref{SC4} into account, we can write
Eqs. \eqref{zui2} as
\begin{eqnarray}
\nonumber F^+&=&-\frac{\kappa({\bf p})}{\omega -E({\bf
p})+i\,0}+\frac{\kappa({\bf p})}{\omega +E({\bf p})-i\,0};\\
G&=&\frac{u^2({\bf p})}{\omega -E({\bf p})+i\,0}+\frac{v^2({\bf
p})}{\omega +E({\bf p})-i\,0}.\label{zui8}
\end{eqnarray}
As $\lambda_0\to0$, the gap $\Delta\to0$, but $\Xi$ and
$\kappa({\bf p})$ remain finite if the spectrum becomes flat,
$E({\bf p})=0$, and Eqs. (\ref{zui8}) become
\begin{eqnarray}
\nonumber F^+({\bf p},\omega )&=&-\kappa({\bf p})\left[
\frac{1}{\omega +i\,0}
-\frac{1}{\omega -i\,0}\right];\\
G({\bf p},\omega )&=&\frac{u^2({\bf p})}{\omega
+i\,0}+\frac{v^2({\bf p})}{\omega -i\,0}.\label{zui9}
\end{eqnarray}
in the interval $p_i\leq p\leq p_f$. The parameters $v({\bf p})$
and $u({\bf p})$ are determined by the condition that the spectrum
be flat: $\varepsilon({\bf p})=\mu$. If we take the Landau equation
\eqref{FL2} into account, this condition again reduces to Eqs.
(\ref{FL8}) and (\ref{SC6}) for determining the minimum of the
functional $E(n({\bf p}))$.

We construct the functions $F^+({\bf p},\omega)$ and  $G({\bf
p},\omega)$ in the case where the constant $\lambda_0$ is finite but
small, such that $v({\bf p})$ and $\kappa({\bf p})$ can be found on
the basis of the FC solutions of Eq. (\ref{FL8}). Then $\Xi$,
$E({\bf p})$ and $\Delta$ are given by Eqs. (\ref{zui7}),
(\ref{zui3}), and (\ref{SC4}) respectively. Substituting the functions
constructed in this manner into (\ref{zui8}), we obtain $F^+({\bf
p},\omega)$ and $G({\bf p},\omega)$ \cite{shagstep}. We note that
Eqs. (\ref{zui3}) imply that the gap $\Delta$ is a linear function
of $\lambda_0$ under the adopted conditions. As we shall see in
Subsection \ref{SCFT}, this gives rise to high-$T_c$ at common
values of the superconducting coupling constant

\subsection{The superconducting state at finite
temperatures}\label{SCFT}

We assume that the region occupied by FC is small:
$(p_f-p_i)/p_F\ll1$ and $\Delta_1\ll T_f$. Then, the order parameter
$\kappa({\bf p})$ is determined primarily by FC, i.e., the
distribution function $n_0({\bf p})$ \cite{ms,shb}. To be able to
solve Eq. \eqref{SC5} analytically, we adopt the BCS approximation
for the interaction \cite{bcs}: $\lambda_0V ({\bf p},{\bf
p}_1)=-\lambda_0$ if $|\varepsilon({\bf p})-\mu |\leq\omega_D$ and
the interaction is zero outside this region, with $\omega_D$ being a
certain characteristic energy. As a result, the superconducting gap
depends only on the temperature, $\Delta({\bf p})=\Delta_1(T)$, and
Eq. \eqref{SC5} becomes
\begin{eqnarray}
 \nonumber
  1 &=& N_{FC}\lambda_0\int\limits_0^{E_0/2}
\frac{d\xi}{\sqrt{\xi^2 +\Delta^2_1 (0)}} \\
  &+& N_{L}\lambda_0\int\limits_{E_0/2}^{\omega_D} \frac
{d\xi}{\sqrt{\xi^2+\Delta^2_1(0)}}.\label{SC9}
\end{eqnarray}
where we introduced the notation $\xi=\varepsilon({\bf p})-\mu$ and
the density of states $N_{FC}$ in the interval $(p_f-p_i)$ or in the
$E_0$-energy interval. It follows from Eq. (\ref{SC7}) that $N
_{FC}=(p_f-p_F)p_F/(2\pi\Delta_1)$. Within the energy interval
$(\omega_D-E_0/2)$, the density of states $N_{L}$ has the standard
form $N_{L}=M ^*_{L}/2\pi$. As $E_0\to0$, Eq. (\ref{SC9}) becomes
the BCS equation. On the other hand, assuming that
$E_0\leq2\omega_D$ and discarding the second integral on the
right-hand side of Eq. (\ref{SC9}), we obtain
\begin{eqnarray}
\nonumber
  \Delta_1 (0)&=& \frac
{\lambda_0p_F(p_f-p_F)}{2\pi}\ln\left(1+\sqrt2\right) \\
&=& 2\beta\varepsilon_F\frac{p_f-p_F}{p_F}\ln\left(1+\sqrt2\right),
\label{SC10}\end{eqnarray} where $\varepsilon_F=p_F^2/2M ^*_L$ is
the Fermi energy and $\beta=\lambda_0 M^*_L/2\pi$ is the
dimensionless coupling constant. Using the standard value of $\beta$
for ordinary superconductors, e.g., $\beta\simeq0.3$, and assuming
that $(p_f-p_F)/p_F\simeq 0.2$, we obtain a large value
$\Delta_1(0)\sim0.1\varepsilon_F$ from Eq. (\ref{SC10}); for
ordinary superconductors, this gap has a much smaller value:
$\Delta_1(0)\sim10^{-3}\varepsilon_F$. With the integral discarded
earlier taken into account, we find that
\begin{eqnarray}
\nonumber
 \Delta_1 (0)&\simeq & 2\beta\varepsilon_F \frac
{p_f-p_F} {p_F} \ln\left(1 +\sqrt2\right)\\
&+&\Delta_1(0)\beta\ln\left(\frac {2\omega_D}{\Delta_1(0)}\right).
\label{SC11}\end{eqnarray} On the right-hand side of Eq.
(\ref{SC11}), the value of $\Delta_1$ is given by (\ref{SC10}). As
$E_0\to0$ and $p_f\to p_F$, the first term on the right-hand side of
Eq. (\ref{SC9}) is zero, and we obtain the ordinary BCS result with
$\Delta_1\propto\exp{(-1/\lambda_0)}$. The correction related to the
second integral in \eqref{SC9} is small because the second term on
the right-hand side of Eq. \eqref{SC11} contains the additional
factor $\beta$. In what follows, we show that $2T_c\simeq
\Delta_1(0)$. The isotopic effect is small in this case, because
$T_c$ depends on $\omega_D$ logarithmical, but the effect is
restored as $E_0\to0$.

At $T\simeq T_c$, Eqs. (\ref{SC7}) and (\ref{SC8}) are replaced by
Eqs. (\ref{FL12}) and (\ref{FL14}), which also hold for $T_c\leq
T\ll T_f$:
\begin{equation}
M^*_{FC}\simeq p_F\frac{p_f-p_i}{4T_c},\,\,\, E_0\simeq
4T_c,\,\,\mbox {\rm if}\,\,T\simeq T_c; \label{SC12}\end{equation}
\begin{equation} M^*_{FC}\simeq
p_F\frac{p_f-p_i}{4T},\,\,\, E_0\simeq 4T,\,\, \mbox {\rm
at}\,\,T<T_c.\label{SC13}
\end{equation}
Equation \eqref{SC9} is replaced by its standard generalization
valid for finite temperatures:
\begin{eqnarray} \label{SC14}
&1&=N_{FC}\lambda_0\int\limits_0^{E_0/2} \frac{d\xi}
{\sqrt{\xi^2+\Delta^2_1}}\tanh\frac{\sqrt{\xi^2+\Delta^2_1}}{2T}\
\nonumber\\
&+&N_{L}\lambda_0\int\limits_{E_0/2}^{\omega_D}
\frac{d\xi}{\sqrt{\xi^2+\Delta^2_1}}
\tanh\frac{\sqrt{\xi^2+\Delta^2_1}}{2T}.
\end{eqnarray}
Because $\Delta_1(T\to T_c)\to0$, Eq. (\ref{SC14}) implies a
relation that closely resembles the BCS result \cite{belkop},
\begin{equation}
2T_c\simeq\Delta_1(0)\label{SC15},\end{equation} where
$\Delta_1(T=0)$ is found from Eq. (\ref{SC11}). Comparing
(\ref{SC7}) and (\ref{SC8}) with (\ref{SC12}) and (\ref{SC13}), we
see that both $M ^*_{FC}$ and $E_0$ are temperature-independent for
$T\leq T_c$.

\subsection{Bogoliubov quasiparticles}

Equation \eqref{SC5} shows that the superconducting gap depends on
the single-particle spectrum $\varepsilon({\bf p})$. On the other
hand, it follows from Eq. \eqref{SC4} that $\varepsilon({\bf p})$
depends on $\varepsilon({\bf p})$ if Eq. \eqref{SC6} has a solution that
determines the existence of FC  as $\lambda_0\to 0$. We assume that
$\lambda_0$ is so small that the pairing interaction
$\lambda_0V({\bf p},{\bf p}_1)$ leads only to a small perturbation
of the order parameter $\kappa({\bf p})$. Equation \eqref{SC7}
implies that the effective mass and the density of states
$N(0)\propto M^*_{FC}\propto 1/\Delta_1$ are finite.
Thus, in contrast to the spectrum in the standard superconductivity
theory, the single-particle spectrum $\varepsilon({\bf p})$ depends
strongly on the superconducting gap, and Eqs. \eqref{FL2} and
\eqref{SC5} must be solved by a self-consistent method.

We assume that Eqs. \eqref{FL2} and \eqref{SC5} have been solved
and the effective mass $M ^*_{FC}$ has been found. This means that
we can find the quasiparticle dispersion law $\varepsilon({\bf p})$
by choosing the effective mass $M^*$ equal to the obtained value of
$M^*_{FC}$ and then solve Eq. (\ref{SC5}) without taking (\ref{FL2})
into account, as is done in the standard BCS superconductivity
theory \cite{bcs}. Hence, the superconducting state with FC  is
characterized by Bogoliubov quasiparticles \cite{bogol} with
dispersion \eqref{SC3.1} and the normalization condition $v^2({\bf
p})+u^2 ({\bf p})=1$ for the coefficients $v({\bf p})$ and $u({\bf
p})$. Moreover, quasiparticle excitations of the superconducting
state in the presence of FC  coincide with the Bogoliubov
quasiparticles characteristic of the BCS theory, and
superconductivity with FC  resembling the BCS superconductivity,
which points to the applicability of the BCS formalism to the
description of the high-$T_c$ superconducting state \cite{asjetpl}.
At the same time, the maximum value of the superconducting gap set
by Eq. (\ref{SC11}) and other exotic properties are determined by
the presence of FC. These results are in good agreement with the
experimental facts obtained for the high-$T_c$ superconductor
Bi$_2$Sr$_2$Ca$_2$Cu$_3$O$_{10+\delta}$ \cite{mat}.

In constructing the superconducting state with FC, we returned to
the foundations of the LFL theory, from which the high-energy
degrees of freedom had been eliminated by the introduction of
quasiparticles. The main difference between the LFL, which forms the
basis for constructing the superconducting state, and the Fermi
liquid with FC is that in the latter case we must increase the
number of low-energy degrees of freedom by introducing the new type of
quasiparticle with the effective mass $M^*_{FC}$ and the
characteristic energy $E_0$ given by Eq. \eqref{SC8}. Hence, the
dispersion law $\varepsilon({\bf p})$ is characterized by two types
of quasiparticles with the effective masses $M^*_L$ and $M^*_{FC}$
and the scale $E_0$. The extended paradigm and new quasiparticles
determine the properties of the superconductor, including the
lineshape of quasiparticle excitations \cite{ms,shb,ams}, while the
dispersion of the Bogoliubov quasiparticles has the standard form.

We note that for $T<T_c$, the effective mass $M^*_{FC}$ and the
scale $E_0$ are temperature-independent \cite{ams}. For $T>T_c$, the
effective mass $M ^*_{FC}$ and the scale $E_0$ are given by Eqs.
(\ref{FL12}) and (\ref{FL14}). Obviously, we cannot directly relate
these new quasiparticles (excitations) of the Fermi liquid with FC
to excitations (quasiparticles) of an ideal Fermi gas, as is done in
the standard LFL theory, because the system is beyond FCQPT. The
properties and dynamics of quasiparticles are given by the extended
paradigm and closely related to the properties of the
superconducting state and are of a collective nature, formed by
FCQPT and determined by the macroscopic number of FC quasiparticles
with momenta in the interval $(p_f-p_i)$. Such a system cannot be
perturbed by scattering on impurities and lattice defects and,
therefore, has the features of a quantum protectorate and
demonstrates universal behavior \cite{ms,shb,rlp,pa}.

Several remarks concerning the quantum protectorate and the
universal behavior of superconductors with FC  are in order.
Similarly to the Landau Fermi liquid theory, the theory of
high-$T_c$ superconductivity based on FCQPT deals with
quasiparticles that are elementary low-energy excitations. The
theory provides a qualitative general description of the
superconducting and the normal states of superconductors and HF
metals. Of course, with phenomenological parameters (e.g., the
pairing coupling constant) chosen, we can obtain a quantitative
description of superconductivity, in the same way as this can be
done in the Landau theory when describing a normal Fermi liquid,
e.g., $^3$He. Hence, any theory capable of describing FC  and
compatible with the BCS theory gives the same qualitative picture of
the superconducting and normal states as the picture based on FCQPT.
Obviously, both approaches may be coordinated on the level of
numerical results by choosing the appropriate parameters. For
instance, because the formation of FC is possible in the Hubbard
model \cite{irk}, it allows reproducing the results of the theory
based on FCQPT. It is appropriate to note here that the
corresponding description restricted to the case of $T=0$ has been
obtained within the framework of the Hubbard model \cite{rand,pwa}.

\subsection{The pseudogap}
\label{pseudogap}

We now discuss some features of the superconducting state with FC
\cite{sh,ars,ms1} considering two possible types of the
superconducting gap $\Delta({\bf p})$ determined by Eq. (\ref{SC5})
and the interaction $\lambda_0V({\bf p},{\bf p}_1)$. If the
interaction is caused by attraction, occurring, for instance, as a
result of an exchange of phonons or magnetic excitations, the
solution of Eq. (\ref{SC5}) with an $s$-wave or $s+d$-mixed waves
has the lowest energy. If the pairing interaction $\lambda_0V({\bf
p}_1,{\bf p}_2)$ is a combination of an attractive interaction and a
strongly repulsive interaction, $d$-wave superconductivity may occur
(e.g., see Refs \cite{kug,abr}). However, both the $s$- and $d$-wave
symmetries lead to approximately the same result for the size of the
gap $\Delta_1$ in Eq. \eqref{SC11}. Hence, $d$-wave
superconductivity is not a universal and necessary property of
high-$T_c$ superconductors. This conclusion agrees with the
experimental evidence described in Refs.
\cite{skin,bis,skin1,skin2,chen}.

\begin{figure} [! ht]
\begin{center}
\includegraphics [width=0.60\textwidth] {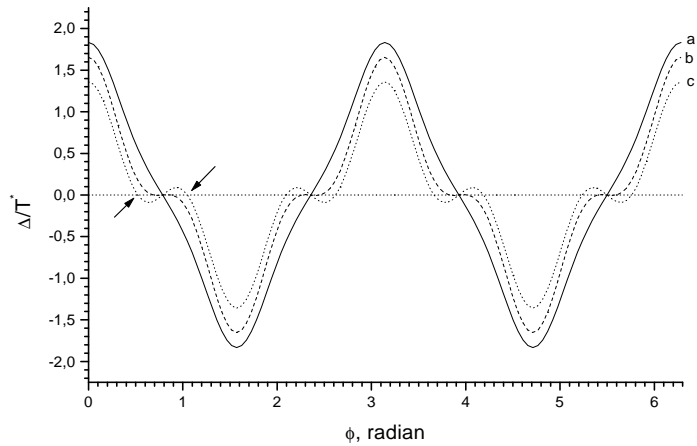}
\end{center}
\caption {The gap $\Delta(p_F,\phi)$ as a function of $\phi$
calculated for three values of the temperature expressed in units of
$T_{node} \simeq T_c$. The solid curve (a) represents the function
$\Delta(p_F,\phi)$ calculated for the temperature $0.9\,T_{node}$.
The dashed curves (b) represents the same function at $T=T_{node}$,
and the dotted curve (c) depicts the function calculated at
$1.2\,T_{node}$. The arrows indicate the region $\theta_c$ limited
by the two new zeros that emerge at $T>T_{node}$.} \label{Fig2}
\end{figure}

We can define the critical temperature $T^*$ as the temperature at
which $\Delta_1(T^*)\equiv0$. For $T\geq T^*$, Eq. \eqref{SC14} has
only the trivial solution $\Delta_1\equiv 0$. On the other hand, the
critical temperature $T_c$ can be defined as the temperature at
which superconductivity disappears and the gap occupies only a part
of the Fermi surface. Thus, there are two different temperatures
$T_c$ and $T^*$, which may not coincide in the case of the $d$-wave
symmetry of the gap. As shown in Refs \cite{sh,ars}, in the presence
of FC , Eq. (\ref{SC14}) has nontrivial solutions at $T_c\leq T\leq
T^*$, when the pairing interaction $\lambda_0V({\bf p}_1,{\bf p}_2)$
consists of attraction and strong repulsion, which leads to $d$-wave
superconductivity. In this case, the gap $\Delta({\bf p})$ as a
function of the angle $\phi$, or $\Delta({\bf p})=\Delta(p_F,\phi)$,
has new nodes at $T>T_{node}$, as shown in Fig. \ref{Fig2}
\cite{ars}.

Figure \ref{Fig2} shows the ratio $\Delta (p_F,\phi)/T^*$
calculated for three temperatures: $0.9\,T_{node}$, $T_{node}$ and
$1.2 \,T_{node}$. In contrast to curve (a), curves (b) and (c) have
flat sections. Clearly, the flattening occurs because of the two
new zeros that emerge at $T=T_{node}$. As the temperature
increases, the region $\theta_c$ between the zeros (indicated by
arrows in Fig. \ref{Fig2}) increases in size. It is also clear that
the gap $\Delta$ is very small within the interval $\theta_c$. It
was found in \cite{nm,th} that the magnetism and the
superconductivity affect each other, which leads to suppression of
the magnetism at temperatures below $T_c$. In view of this, we can
expect suppression of superconductivity due to magnetism.

\begin{figure} [! ht]
\begin{center}
\includegraphics [width=0.60\textwidth] {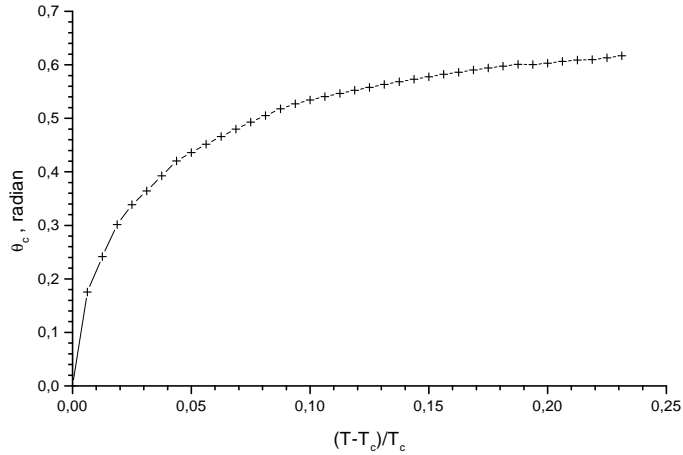}
\end{center}
\caption {. The result of a numerical calculation of the angle
$\theta_c$ separating two zeros as a function of $(T-T_c)/T_c$.}
\label {Fig3}
\end{figure}

Thus, we may conclude that the gap in the vicinity of $T_c$ can be
destroyed by strong antiferromagnetic correlations (or spin density
waves), impurities, and sizable inhomogeneities existing in
high-$T_c$ superconductors \cite{kresin}. Because the
superconducting gap is destroyed in a macroscopic region of the
phase space, $\theta_c$, superconductivity is also destroyed, and
therefore $T_c\simeq T_{node}$. The exact value of $T_c$ is
determined by the competition between the antiferromagnetic state
(or spin density waves) and the superconductivity in the interval
$\theta_c$. The behavior and the shape of the pseudogap closely
resemble the similar characteristics of the superconducting gap, as
Fig. \ref{Fig2} shows. The main difference is that the pseudogap
disappears in the segment $\theta_c$ of the Fermi surface, while
the gap disappears at isolated nodes of the $d$-wave. This result is
in accord with observations \cite{pbar}. Our estimates
show that for small values of the angle $\psi$, the function
$\theta_c(\psi)$ rapidly increases,
$\theta_c(\psi)\simeq\sqrt{\psi}$. These estimates agree with the
results of numerical calculations of the function
$\theta_c([T-T_c]/T_c)$, (Fig. \ref{Fig3}). Hence, we may conclude
that $T_c$ is close to $T_{node}$.
Thus, the pseudogap state appears at $T\geq T_c\simeq T_{node}$ and
disappears at temperatures $T\geq T^*$ at which Eq. \eqref{SC14} has
only the trivial solution $\Delta_1\equiv0$. Obviously, $\Delta_1$
determines $T ^*$ and not $T_c$, with the result that Eq.
(\ref{SC15}) should be rewritten as
\begin{equation}
2T^* \simeq\Delta_1(0).\label{SC16}
\end{equation}
The temperature $T^*$ has the physical meaning of the temperature of
the BCS transition between the state with the order parameter
$\kappa\neq0$ and the normal state.

At temperatures below $T<T_c $, the quasiparticle excitations of the
superconducting state are characterized by the presence of sharp
peaks. When the temperature becomes high ($T>T_c$) and
$\Delta(\theta)\equiv0 $ in the interval $\theta_c$, normal
quasiparticle excitations with a width $\gamma$ appear in the
segments $\theta_c$ of the Fermi surface. A pseudogap exists outside
the segments $\theta_c$, and the Fermi surface is occupied by
excitations of the BCS type in this region. Excitations of both
types have widths of the same order of magnitude, transferring their
energy and momenta into excitations of normal quasiparticles. These
results are in accord with strong indications of the pairing or
the formation of preformed pairs in the pseudogap regime at
temperatures above $T_c$ \cite{pbar,norm,valln,norm1,jhscien,PLA686}.

We now estimate the value of $\gamma$. If the entire Fermi surface
were occupied by the normal state, the width $\gamma$ would be
$\gamma\approx N(0)^3T^2/\varepsilon_d(T)^2$ with the density of
states $N(0)\sim M^*(T)\sim1/T$ [see Eq. (\ref{FL12}]. The
dielectric constant $\varepsilon_d(T)\sim N(0)$ and, hence,
$\gamma\sim T$ \cite{dkss,dkss1}. However, only a part of the Fermi
surface within $\theta_c$ is occupied by normal excitations in our
case. Therefore, the number of states accessible for quasiparticles
and quasiholes is proportional to $\theta_c$, and the factor $T^2$
is replaced by the factor $T^2\theta_c^2$. Taking all this into
account yields $\gamma\sim \theta_c^2 T\sim
T(T-T_c)/T_c\sim(T-T_c)$. Here, we ignored the small contribution
provided by excitations of the BCS type. It is precisely for this
reason that the width $\gamma$ vanishes at $T=T_c$. Moreover, the
resistivity of the normal state $\rho(T)\propto
\gamma\propto(T-T_c)$, because $\gamma\sim(T-T_c)$. Obviously, at
temperatures $T>T^*$, the relation $\rho(T)\propto\gamma\propto T$
remains valid up to $T\sim T_f$, and $T_f$ may be as high as the
Fermi energy if FC occupies a significant part of the Fermi volume.

The temperature $T_{node}$ is determined mainly by the repulsive
interaction, which is part of the pairing interaction
$\lambda_0V({\bf p}_1,{\bf p}_2)$. The value of the repulsive
interaction, in turn, may be determined by the properties of the
materials, such as composition or doping. Because superconductivity
is destroyed at $T_c\simeq T_{node}$, the ratio $2\Delta_1/T_c$ may
vary within broad limits and strongly depends on the properties of
the material \cite{sh,ars,ms1}. For instance, in the case of
Bi$_2$Sr$_2$CaCu$_2$ O$_{6+\delta}$ it is assumed that
superconductivity and the pseudogap are of common origin:
$2\Delta_1/T_c\simeq28$, while $2\Delta_1/T ^*\simeq4$, which agrees
with the experimental data obtained in measurements involving other
high-$T_c$ superconductors \cite{kug}.

We note that Eq. (\ref{SC16}) also provides a good description of
the maximum value of the gap $\Delta_1$ in the case of $d$-wave
superconductivity, because different regions with the maximum
density of states may be considered unrelated \cite{abr}. We may
also conclude that without a strong repulsion, with which $s$-wave
pairing is possible, there can be no pseudogap. Thus, the transition
from the superconducting gap to the pseudogap may proceed only in
the case of $d$-wave pairing, when superconductivity is destroyed at
$T_c\simeq T_{node}$ and the superconducting gap gradually
transforms into a pseudogap, which closes at a certain temperature
$T^*>T_c$ \cite{sh,ars,ms1}. The fact that there is no pseudogap in
the case of $s$-wave pairing agrees with the experimental data
(e.g., see Ref. \cite{chen}).

\subsection{Dependence of the critical temperature $T_c$
of the superconducting phase transition on doping}

We examine the maximum value of the superconducting gap $\Delta_1$
as a function of the number density  $x$ of mobile charge carriers,
which is proportional to the degree of doping. Using Eq.
(\ref{FL15}), we can rewrite Eq. (\ref{SC10}) as
\begin{equation}\frac
{\Delta_1}{\varepsilon_F}\sim\beta\frac{(x_{FC}-x)x}{x_{FC}}.
\label{SC17}\end{equation} where we took into account that the Fermi
level $\varepsilon_F\propto p_F^2$ and that the number density
$x\sim p_F^2/(2M^*)$, with the result that $\varepsilon_F\propto x$.
It is realistic to assume that $T_c\propto\Delta_1$, because the
curve $T_c(x) $ obtained in experiments with high-$T_c$
superconductors \cite{varma} must be a smooth function of $x$.
Hence, we can approximate $T_c(x)$ by a smooth bell-shape function
\cite{ams3}:
\begin{equation}
T_c(x)\propto\beta(x_{FC}-x)x. \label{SC18}
\end{equation}

To illustrate the application of the above analysis, we examine the
main features of a superconductor that can hypothetically exist at
room temperature. Such a superconductor must be a two-dimensional
structure, just as high-$T_c$ superconducting cuprates are. Equation
(\ref{SC10}) implies that $\Delta_1\sim
\beta\varepsilon_F\propto\beta/r_s^2$. Bearing in mind that FCQPT
occurs at $r_s\sim 20$ in 3D systems and at $r_s\sim 8$ in 2D
systems \cite{ksz}, we can expect that in 3D systems $\Delta_1$
amounts to 10\% of the maximum size of the superconducting gap in 2D
systems, which in our case amounts to 60 mV for lightly doped
cuprates with $T_c=70$ K \cite{mzo}. On the other hand, Eq.
(\ref{SC10}) implies that $\Delta_1$ may be even larger,
$\Delta_1\sim 75 $ mV. We can expect that $T_c\sim 300$ K in the
case of $s$-wave pairing, as the simple relation $2T_c\simeq
\Delta_1$ implies. Indeed, we can take $\varepsilon_F\sim500$ mV,
$\beta\sim 0.3$, and $(p_f-p_i)/p_F\sim0.5$.

Thus, the hypothetical superconductor at room temperature must be
an $s$-wave superconductor in order to eliminate the pseudogap
effect, which dramatically decreases the temperature $T_c$ at which
superconductivity is destroyed. We note that the number density $x$
of mobile charge carriers must satisfy the condition $x\leq x_{FC}$
and must be varied to reach the optimum degree of doping
$x_{opt}\simeq x_{FC}/2$.

\subsection{The gap and heat capacity near $T_c$}

We now calculate the gap and heat capacity at temperatures $T\to
T_c$. Our analysis is valid if $T^*\simeq T_c$, since otherwise the
discontinuities in the heat capacity considered below are smeared
over the temperature interval between $T^{*}$ and $T_c$. To simplify
matters, we calculate the leading contribution to the gap and heat
capacity related to FC. We use Eq. (\ref{SC14}) to find the function
$\Delta_1(T\to T_c)$ simply by expanding the first integral on its
right-hand side in powers of $\Delta_1$ and dropping the
contribution from the second integral. This procedure leads to the
equation \cite{ams}
\begin{equation}
\Delta_1(T)\simeq3.4T_c\sqrt{1-\frac{T}{T_c}}.\label{SC19}\end{equation}
Therefore, the gap in the spectrum of single-particle excitations
behaves in the ordinary manner.

To calculate the heat capacity, we can use the standard expression
for the entropy $S$ \cite{bcs}:
\begin{eqnarray}
\nonumber
  S(T)&=& -2\int[f({\bf p})\ln f({\bf p})+(1-f ({\bf
p})) \\
 &\times&\ln(1-f({\bf p}))]\frac{d{\bf p}}{(2\pi)^2},\label{SC20}
\end{eqnarray}
where
\begin{eqnarray}\label{EFT}
\nonumber f({\bf p})&=&(1+\exp[E({\bf p})/T])^{-1}\,\,\,,\\ E({\bf
p})&=&\sqrt{(\varepsilon({\bf
p})-\mu)^2+\Delta_1^2(T)}.\end{eqnarray} The heat capacity $C$ is
given by
\begin{eqnarray}
\nonumber C(T)&=&T\frac {dS} {dT}\simeq\ 4\frac
{N_{FC}}{T^2}\int\limits_0^ {E_0}f(E)(1-f (E))\\
&\times& \left[E^2+T\Delta_1(T)\frac{d\Delta_1(T)}
{dT}\right]d\xi\nonumber \\
&+& 4\frac{N_{L}}{T^2}\int\limits_{E_0}^{\omega_D}f(E)(1-f (E))\nonumber \\
&\times& \left[E^2+T\Delta_1(T)\frac{d\Delta_1(T)}{dT}\right] d\xi.
\label{SC21}
\end{eqnarray}
In deriving Eq. (\ref{SC21}), we again used the variable $\xi$, the
above notation for the density of states, $N_{FC}$ and $N_{L}$, and
the notation $E =\sqrt{\xi^2+\Delta_1^2(T)}$. Equation \eqref{SC21}
describes a jump in heat capacity, $\delta C(T)=C_s(T)-C_n(T)$,
where $C_s(T)$ and $C_n(T)$ are respectively the heat capacities of the
superconducting and normal states at $T_c$; the jump is determined
by the last two terms in the square brackets on the right-hand side
of this equation. Using Eq. \eqref{SC19} to calculate the first term
on the right-hand side of Eq. \eqref{SC21}, we find \cite{ams}
\begin{equation}\delta C(T_c)\simeq\frac3{2\pi^2}(p_f-p_i)p_F^n.
\label{SC22}\end{equation} where $n=1$ in the 2D case and $n=2$ in
the 3D case. This result differs from the ordinary BCS result,
according to which the discontinuity in the heat capacity is a
linear function of $T_c$. The jump $\delta C(T_c)$ is independent of
$T_c$ because, as Eq. (\ref{SC13}) shows, the density of state
varies in inverse proportion to $T_c$. We note that in deriving Eq.
(\ref{SC22}) we took the leading contribution coming from FC into
account. This contribution disappears as $E_0\to0$, and the second
integral on the right-hand side of Eq. (\ref{SC21}) yields the
standard result.

As we will show in Section \ref{SCEL} [see Eq. (\ref{SL4})], the heat
capacity of a system with FC  behaves as $C_n(T)\propto
\sqrt{T/T_f}$. The jump in the heat capacity given by Eq.
(\ref{SC22}) is temperature-independent. As a result, we find that
\begin{equation}\frac {\delta
C(T_c)}{C_n(T_c)}\sim\sqrt{\frac{T_f}{T_c}}\frac{(p_f-p_i)}
{p_F}.\label{SC23}
\end{equation}
In contrast to the case of normal superconductors, in which $\delta
C(T_c)/C_n(T_c)=1.43$ \cite{lanl1}, in our case Eq. \eqref{SC23}
implies that the ratio $\delta C(T_c)/C_n(T_c)$ is not constant and
may be very large when $T_f/T_c\gg 1$ \cite{yakov,ams}. It is
instructive to apply this analysis to $\rm CeCoIn_5$, where
$T_c=$2.3 K \cite{yakov}. In this material \cite{petrov}, $\delta
C/C_n\simeq4.5$ is substantially higher than the BCS value, in
agreement with Eq. \eqref{SC23}.

\section{The dispersion law and lineshape
of single-particle excitations} \label{DISP}

The recently discovered break in the dispersion of quasiparticles
at energies between 40 and 70 mV, resulting in a change in the
quasiparticle speed at this energy \cite{blk,krc,vall,vall1}, can hardly
be explained by the marginal Fermi-liquid theory, because this
theory contains no additional energy scales or parameters that
would allow taking the break into account \cite{var,varm1}. We
could assume that the break, which leads to a new energy scale,
occurs because of the interaction of electrons and collective
excitations, but then we would have to discard the idea of a
quantum protectorate, which would contradict the experimental data
\cite{rlp,pa}.

As shown in Sections \ref{FLFC} and \ref{SC}, a system with FC has
two effective masses: $M^*_{FC}$, which determines the
single-particle spectrum at low energies, and $M^*_L$, which
determines the spectrum at high energies. The fact that there are
two effective masses manifests itself in the form of a break in the
quasiparticle dispersion law. The dispersion law can be approximated
by two straight lines intersecting at a binding energy $E_0/2$ [see
Eqs. (\ref{FL14}) and (\ref{SC8})]. The break in the dispersion law
occurs at temperatures much lower than $T\ll T_f$, when the system
is in the superconducting or normal state. Such behavior is in good
agreement with the experimental data \cite{blk}. It is pertinent to
note that at temperatures below $T<T_c$, the effective mass
$M^*_{FC}$ is independent of the momenta $p_F$, $p_f$, and $p_i$, as
shown by Eqs. (\ref{SC7}) and (\ref{SC10}):
\begin{equation} M^*_{FC}\sim \frac{2\pi}{\lambda_0}.
\label{DI1}\end{equation} This formula implies that $M^*_{FC}$ is
only weakly dependent on $x$ if a dependence of $\lambda_0$ on $x$
is allowed. This result is in good agreement with the experimental
facts \cite{ino,zhou,padil}. The same is true of the dependence of
the Fermi velocity $v_F=p_F/M^*_{FC}$ on $x$ because the Fermi
momentum $p_F\sim\sqrt{n}$ is weakly dependent on the electron
number density $n=n_0(1-x)$ \cite{ino, zhou}; here, $n_0$ is the
single-particle electron number density at half-filling.

Because $\lambda_0$ is the coupling constant that determines the
magnitude of the pairing interaction, e.g., the electron-phonon
interaction, we can expect the break in the quasiparticle dispersion
law to be caused by the electron-phonon interaction. The phonon
scenario could explain the constancy of the break at $T>T_c$ because
phonons are temperature independent. On the other hand, it was found
that the quasiparticle dispersion law distorted by the interaction
with phonons has a tendency to restore itself to the ordinary single
particle dispersion law when the quasiparticle energy becomes higher
than the phonon energy \cite{vald}. However, there is no
experimental evidence that such restoration of the dispersion law
actually takes place \cite{blk}.

The quasiparticle excitation curve $L(q,\omega)$ is a function of
two variables. Measurements at a constant energy $\omega=\omega_0$,
where $\omega_0$ is the single particle excitation energy, determine
the curve $L(q,\omega=\omega_0)$ as a function of the momentum $q$.
We established above that $M^*_{FC}$ is finite and constant at
temperatures not exceeding $T_c$. Hence, at excitation energies
$\omega<E_0$, the system behaves as an ordinary superconducting
Fermi liquid with the effective mass determined by Eq. (\ref{SC7})
\cite{ms, shb, ars}. At Ïðè $T_c\leq T$, the effective mass
$M^*_{FC}$ is also finite and is given by Eq. (\ref{FL12}). In other
words, at $\omega<E_0$, the system behaves as a Fermi liquid whose
single-particle spectrum is well defined and the width of the
single-particle excitations is of the order of $T$
\cite{ms,shb,dkss}. Such behavior has been observed in experiments
in measuring the quasiparticle excitation curve at a fixed energy
\cite{koral,vall,feng}.

The quasiparticle excitation curve can also be described as a
function of $\omega$, at a constant momentum $q=q_0$. For small
values of $\omega$, the behavior of this function is similar to
that described above, with $L(q=q_0,\omega)$ having a
characteristic maximum and width. For $\omega\geq E_0$, the
contribution provided by quasiparticles of mass $M ^*_{L}$ becomes
significant and leads to an increase in the function
$L(q=q_0,\omega)$. Thus, $L(q=q_0,\omega)$ has a certain structure
of maxima and minima \cite{dess} directly determined by the
existence of two effective masses, $M^*_{FC}$ and $M^*_L$
\cite{ms,shb,ars}. We conclude that, in contrast to Landau
quasiparticles, these quasiparticles have a more complicated
spectral lineshape.

We use the Kramers-Kronig transformation to calculate the imaginary
part ${\mathrm{Im}}\Sigma ({\bf p},\varepsilon)$ of the self-energy
part $\Sigma({\bf p},\varepsilon)$. But we begin with the real part
${\mathrm{Re}}\Sigma ({\bf p},\varepsilon)$, which determines the
effective mass $M^*$ \cite{mig},
\begin{equation} \left.
\frac1{M^*}=\left(\frac{1}{m}+\frac{1}{p_F}
\frac{\partial{\mathrm{Re}}\Sigma}{\partial p}\right)\right/
\left(1-\frac{\partial{\mathrm{Re}}\Sigma}{\partial
\varepsilon}\right).\label{DI2} \end{equation} The corresponding
momenta $p$ and energies $\varepsilon$ satisfy the inequalities
$|p-p_F |/p_F\ll 1 $, and $\varepsilon/\varepsilon_F\ll 1$. We take
${\mathrm{Re}}\Sigma({\bf p},\varepsilon)$ in the simplest form
possible that ensures the variation of the effective mass at the
energy $E_0/2$,
\begin{eqnarray}
\nonumber \mbox{Re}\Sigma({\bf p},\varepsilon)&=& -\varepsilon
\frac{M^*_{FC}}m+\left(\varepsilon-\frac{E_0}2\right)
\frac{M^*_{FC}-M^*_L}m\\
&\times&\left[\theta\left( \varepsilon\!-\!\frac{E_0}2\right)
+\theta\left(\mbox{-}\varepsilon\!-\!\frac{E_0}2\right)\right],
\label{DI3}\end{eqnarray} where $\theta (\varepsilon)$ is the step
function. To ensure a smooth transition from the single-particle
spectrum characterized by $M^*_{FC}$ to the spectrum characterized
by $M^*_{L}$, we must replace the step function with a smoother
function. Substituting Eq. \eqref{DI3} in Eq. \eqref{DI2}, we see
that $M^*\simeq M^*_{FC}$ within the interval $(-E_0/2,$ $E_0/2)$,
while $M^*\simeq M^*_{L}$ outside this interval. Applying the
Kramers-Kronig transformation to ${\mathrm{Re}}\Sigma({\bf
p},\varepsilon)$, we express the imaginary part of the self-energy
as \cite{ams}
\begin{eqnarray}
\nonumber& &\mbox{Im}\Sigma({\bf
p},\varepsilon)\sim\varepsilon^2\frac{M^*_{FC}}{\varepsilon_F m}+
\frac{M^*_{FC}-M^*_L}{m} \\
&\times&\left[\varepsilon\ln\left|
\frac{2\varepsilon+E_0}{2\varepsilon-E_0}\right|+
\frac{E_0}2\ln\left|\frac{4\varepsilon^2-E^2_0}{E^2_0}
\right|\right].\label{DI4}
\end{eqnarray}
Clearly, with $\varepsilon/E_0\ll 1 $, the imaginary part is
proportional to $\varepsilon^2$; at $2\varepsilon/E_0\simeq1$, we
have ${\mathrm{Im}} \Sigma\sim \varepsilon$, and for
$E_0/\varepsilon\ll1$, the main contribution to the imaginary part
is approximately constant.

It follows from Eq. (\ref{DI4}) that as $E_0\to0$, the second term
on its right-hand side tends to zero and the single-particle
excitations become well-defined, which resembles the situation with
a normal Fermi liquid, while the pattern of minima and maxima
eventually disappears. Now the quasiparticle
renormalization factor $z({\bf p})$ is given by the equation
\cite{mig}
\begin{equation} \frac1{z({\bf p})}=
1-\frac{\partial\mbox{Re}\Sigma({\bf
p},\varepsilon)}{\partial\varepsilon}.\label{DI5}
\end{equation}

Consequently, from Eqs. \eqref{DI4} and \eqref{DI5} for $T\leq T_c$,
the amplitude of a quasiparticle on the Fermi surface increases as
the characteristic energy $E_0$ decreases. Equations \eqref{SC8} and
(\ref{SC18}) imply that $E_0\sim (x_{FC}-x)/x_{FC}$. When $T>T_c$,
it follows from (\ref{DI3}) and (\ref{DI5}) that the quasiparticle
amplitude increases as the effective mass $M^*_{FC}$ decreases. So,
from Eqs. (\ref{FL12}) and (\ref{FL15})
$M^*_{FC}\sim(p_f-p_i)/p_F\sim(x_{FC}-x)/x_{FC}$. As a result, we
conclude that the amplitude increases with the doping level and the
single-particle excitations are better defined in heavily doped
samples. As $x\to x_{FC}$, the characteristic energy $E_0\to0$ and
the quasiparticles become normal excitations of LFL. We note that
such behavior has been observed in experiments with heavily doped
Bi2212, which demonstrates high-$T_c$ superconductivity with a gap
of about 10 mV \cite{val1}. The size of the gap suggests that the
region occupied by FC is small because $E_0/2\simeq\Delta_1$. For
$x>x_{FC}$ and low temperatures, the HF liquid behaves as LFL (see
Fig. \ref{fig1} and Section \ref{HCEL}). Experimental data show
that, as expected, the LFL state exists in super-heavily doped
nonsuperconducting La$_{1.7}$Sr$_{0.3}$CuO$_4$ \cite{nakam,huss}.

\section{Electron liquid with FC in magnetic fields} \label{HFL}

In this Section, we discuss the behavior of HF liquid with FC in
magnetic field.
We assume that the coupling constant is nonzero, $\lambda_0\neq0$,
but is infinitely small. We found in Section \ref{SC} that at $T=0$
the superconducting order parameter $\kappa({\bf p})$ is finite in
the region occupied by FC and that the maximum value of the
superconducting gap $\Delta_1\propto \lambda_0$ is infinitely small.
Hence, any weak magnetic field $B\neq0$ is critical and destroys
$\kappa({\bf p})$ and FC. Simple energy arguments suffice to
determine the type of rearrangement of the FC state. On the one
hand, because the FC state is destroyed, the gain in energy $\Delta
E_B\propto B^2$ tends to zero as $B\to 0 $. On the other hand, the
function $n_0({\bf p})$, which occupies the finite interval
$(p_f-p_i)$ in the momentum space and is specified by Eq.
(\ref{FL8}) or (\ref{SC8}), leads to a finite gain in the
ground-state energy compared to the ground-state energy of a normal
Fermi liquid \cite{ks}. Thus, the distribution function is to be
reconstructed so that the order parameter is to vanish while a new
distribution function is to deliver the same ground state energy.

\subsection{Phase diagram of electron liquid in magnetic field}

\begin{figure} [! ht]
\begin{center}
\includegraphics [width=0.60\textwidth] {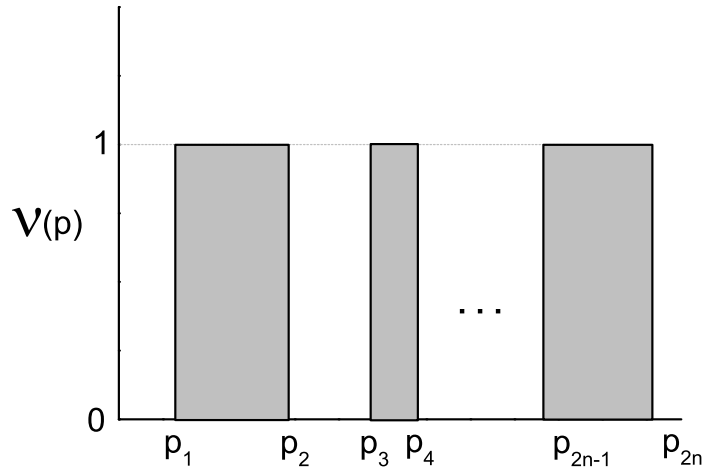}
\end{center}
\caption{The function $\nu({\bf p})$ for the multiply connected
distribution that replaces the function $n_0({\bf p})$ in the region
$(p_f-p_i)$ occupied by FC. The momenta satisfy the inequalities
$p_i<p_F<p_f$. The outer Fermi surface at $p\simeq p_{2n}\simeq p_f$
has the shape of a Fermi step, and therefore the system behaves like
LFL at sufficiently low temperatures.}\label{Fig4}
\end{figure}

Thus, in weak magnetic fields, the new ground state without FC must
have almost the same energy as the state with FC. As shown in
Section \ref{TFT}, such a state is formed by multiply connected
Fermi spheres resembling an onion, in which a smooth distribution
function of quasiparticles, $n_0({\bf p})$, is replaced in the
interval $(p_f-p_i)$ with the distribution function \cite{asp,pogsh}
\begin{equation}
\nu ({\bf p})=\sum\limits_{k=1}^n\theta(p-p_{2k-1})
\theta(p_{2k}-p).\label{HF1}
\end{equation}
where the parameters $p_i\leq p_1<p_2<\ldots<p_{2n}\leq p_f$ are
chosen such that they satisfy the normalization condition and the
condition needed for the conservation of the number of particles:
$$
\int_{p_{2k-1}}^{p_{2k+3}}\nu({\bf p})\frac{d{\bf p}}{(2\pi)^3}=
\int_{p_{2k-1}}^{p_{2k+3}}n_0({\bf p})\frac{d{\bf p}}{(2\pi)^3}.
$$
Figure \ref{Fig4} shows the corresponding multiply connected
distribution. For definiteness, we present the most interesting case
of a three-dimensional system. The two-dimensional case can be
examined similarly. We note that the possibility of the existence of
multiply connected Fermi spheres was studied in e.g.
\cite{khodb,llvp,llvp1,zb}.

We assume that the thickness of each inner slice of the Fermi
sphere, $\delta p\simeq p_{2k+1}-p_{2k}$, is determined by the
magnetic field $B$. Using the well-known rule for estimating errors
in calculating integrals, we find that the minimum loss of the
ground-state energy due to slice formation is approximately $(\delta
p)^4$. This becomes especially clear if we account for the fact that
the continuous FC functions $n_0 ({\bf p})$ ensure the minimum value
of the energy functional $E [n({\bf p})]$, while the approximation
of $\nu({\bf p})$ by steps of width $\delta p$ leads to a minimal
error of the order of $(\delta p)^4$. Recalling that the gain due to
the magnetic field is proportional to $B^2$ and equating the two
contributions, we obtain
\begin{equation} \delta p\propto\sqrt{B}.\label{HF2}
\end{equation}
Therefore, as $T\to 0 $, with $B\to0$, the slice thickness $\delta
p$ also tends to zero and the behavior of a Fermi liquid with FC  is
replaced with that of LFL with the Fermi momentum $p_f$. Equation
(\ref{SC6}) implies that $p_f>p_F$ and the electron number density x
remains constant, with the Fermi momentum of the multiply connected
Fermi sphere $p_{2n}\simeq p_f>p_F$ (see Fig. \ref{Fig4}). We see in
what follows that these observations play an important role in
studying the behavior of the Hall coefficients $R_H(B)$ as a
function of $B$ in heavy-fermion metals at low temperatures.

To calculate the effective mass $M^*(B)$ as a function of the
applied magnetic field $B$, we first note that at $T=0$ the field
$B$ splits the FC state into Landau levels, suppresses the
superconducting order parameter $\kappa({\bf p})$, and destroys FC,
which leads to restoration of LFL \cite{shag4,shag}. The Landau
levels near the Fermi surface can be approximated by separate slices
whose thickness in momentum space is $\delta p$. Approximating the
quasiparticle dispersion law within a single slice,
$\varepsilon(p)-\mu\sim(p-p_f+\delta p)(p-p_f)/M^*$, we find the
effective mass $M^*(B)\sim M^*/(\delta p/p_f)$. The energy increment
$\Delta E_{FC}$ caused by the transformation of the FC state can be
estimated based on the Landau formula \cite{lanl1}
\begin{equation}
\Delta E_{FC}=\int(\varepsilon({\bf p})-\mu)\delta n({\bf p})\frac
{d{\bf p}^3}{(2\pi)^3}.\label{HF3}\end{equation} The region
occupied by the variation $\delta n({\bf p})$ has the thickness
$\delta p$, with $(\varepsilon ({\bf
p})-\mu)\sim(p-p_f)p_f/M^*(B)\sim \delta p p_f/M^*(B)$. As a
result, we find that $\Delta E_{FC}\sim$ $p_f^3\delta p^2/M^*(B)$.
On the other hand, there is the addition $\Delta
E_B\sim(B\mu_{B})^2M^*(B)p_f$ caused by the applied magnetic field,
which decreases the energy and is related to the Zeeman splitting.
Equating $\Delta E_B$ and $\Delta E_{FC}$ and recalling that
$M^*(B)\propto1/\delta p$ in this case, we obtain the chain of
relations
\begin{equation} \frac{\delta
p^2}{M^*(B)}\propto \frac{1}{(M^*(B))^3}\propto
B^2M^*(B).\label{HF4}
\end{equation}
which implies that the effective mass $M^*(B)$ diverges as
\begin{equation} M^*(B)\propto \frac{1}{\sqrt{B-B_{c0}}}.\label{HF5}
\end{equation}
where $B_{c0}$ is the critical magnetic field, which places HF metal
at the magnetic-field-tuned quantum critical point and nullifies the
respective N\`eel temperature, $T_{NL}(B_{c0})=0$ \cite{shag4}. In
our simple model of HF liquid, the quantity $B_{c0}$ is a parameter
determined by the properties of the specific metal with heavy
fermions. We note that in some cases $B_{c0}=0$, e.g., the HF metal
CeRu$_2$Si$_2$ has no magnetic order, exhibits no superconductivity,
and does not behave like a Landau Fermi liquid even at the lowest
reached temperatures \cite{takah}.

Formula (\ref{HF5}) and Fig. \ref{Fig4} show that the application of a
magnetic field $B>B_{c0}$ brings the FC system back to the LFL state
with the effective mass $M^*(B)$ that depends on the magnetic field.
This means that the following characteristic of LFL are restored:
$C/T=\gamma_0(B)\propto M^*(B)$ for the heat capacity and
$\chi_0(B)\propto M^*(B)$ for the magnetic susceptibility. The
coefficient $A(B)$ determines the temperature-dependent part of the
resistivity, $\rho (T)=\rho_0+\Delta\rho$, where $\rho_0$ is the
residual resistivity and $\Delta\rho=A(B)T^2$. Because this
coefficient is directly determined by the effective mass,
$A(B)\propto (M^*(B))^2$ \cite{ksch}, Eq. (\ref{HF5}) yields
\begin{equation}
A(B)\propto\frac{1}{B-B_{c0}}.\label{HF6}\end{equation} Thus, the
empirical Kadowaki-Woods relation \cite{kadw} $K=A/\gamma_0^2\simeq
const$ is valid in our case \cite{ksch}. Furthermore, $K$ may depend
on the degree of degeneracy of the quasiparticles. With this
degeneration, the Kadowaki-Woods relation provides a good
description of the experimental data for a broad class of HF metals
\cite{tky,natphys}. In the simplest case, where HF liquid is formed
by spin-$1/2$ quasiparticles with the degeneracy degree $2$, the
value of $K$ turns out to be close to the empirical value
\cite{ksch} known as the Kadowaki-Woods ratio \cite{kadw}. Hence,
under a magnetic field, the system returns to the state of LFL and
the constancy of the Kadowaki-Woods relation holds.

At finite temperatures, the system remains in the LFL state, but
when $T>T^*(B)$, the NFL behavior is restored. As regards finding
the function $T^*(B)$, we note that the effective mass $M ^*$
characterizing the single-particle spectrum cannot change at
$T^*(B)$ because no phase transition occurs at this temperature. To
calculate $M^*(T)$, we equate the effective mass $M^*(T)$ in Eq.
(\ref{FL12}) to $M^*(B)$ in (\ref{HF5}), $M^*(T)\sim M^*(B)$,
\begin{equation}\frac{1}{M^*(T)}\propto T^*(B)\propto\frac{1}{M^*(B)}
\propto \sqrt{B-B_{c0}},\label{HF7}
\end{equation}
whence
\begin{equation}
T^*(B)\propto\sqrt{B-B_{c0}}.\label{HF8}
\end{equation}
At temperatures $T\geq T^*(B)$, the system returns to the NFL
behavior and the effective mass $M^*$ specified by Eq. \eqref{FL12}.
Thus, expression \eqref{HF8} determines the line in the $T-B$ phase
diagram that separates the region where the effective mass depends
on $B$ and the heavy Fermi liquid behaves like a Landau Fermi liquid
from the region where the effective mass is temperature-dependent.
At $T^*(B)$, the temperature dependence of the resistivity ceases to
be quadratic and becomes linear.

\begin{figure}[!ht]
\begin{center}
\includegraphics [width=0.60\textwidth]{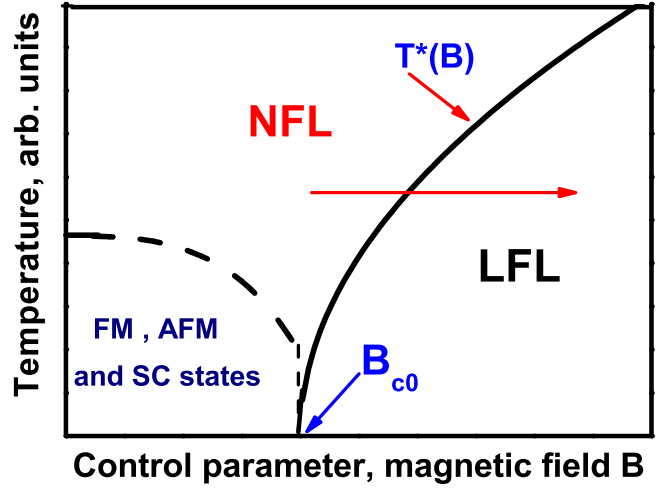}
\vspace*{-0.5cm}
\end{center}
\caption{Schematic $T-B$ phase diagram of heavy electron liquid.
$B_{c0}$ denotes the magnetic field at which the effective mass
diverges as given by \eqref{HF5}. The horizontal arrow illustrates
the system moving in the NFL-LFL direction along $B$ at fixed
temperature. As shown by the dashed curve, at $B<B_{c0}$
the system can be in its
ferromagnetic (FM), antiferromagnetic (AFM) or superconducting (SC)
states. The NFL state is characterized by the entropy $S_{0}$ given
by Eq. \eqref{snfl}. The solid curve $T^*(B)$ separates the NFL state and
the weakly polarized LFL one and represents the transition
regime.}\label{figPD2}
\end{figure}

A schematic $T-B$ phase diagram of HF liquid with FC in magnetic
field is shown in Fig. \ref{figPD2}. At magnetic field $B<B_{c0}$
the FC state can captured by ferromagnetic
(FM), antiferromagnetic (AFM) and
superconducting (SC) states lifting the degeneracy of
the FC state. It follows from
(\ref{HF8}) that at a certain temperature $T^*(B)\ll T_f$, the
heavy-electron liquid transits from its NFL state to LFL one
acquiring the properties of LFL at $(B-B_{c0})\propto(T^*(B))^2$. At
temperatures below $T^*(B)$, as shown by the horizontal arrow in
Fig. \ref{figPD2}, the heavy-electron liquid demonstrates an
increasingly metallic behavior as the magnetic field $B$ increases,
because the effective mass decreases [see Eq. \eqref{HF5}]. Such
behavior of the effective mass can be observed, for instance, in
measurements of the heat capacity, magnetic susceptibility,
resistivity, and Shubnikov-de Haas oscillations. From the $T-B$
phase diagram shown in Fig. \ref{figPD2} and constructed in this
manner, it follows that a unique possibility emerges where a
magnetic field can be used to control the variations in the physical
nature and type of behavior of the electron liquid with FC.

We briefly discuss the case where the system is extremely close to
FCQPT on the ordered size of this transition, and hence $\delta
p_{FC}= (p_f-p_i)/p_F\ll1$. Because $\delta p\propto M^*(B)$, it
follows from Eqs. (\ref{HF2}) and (\ref{HF5}) that
\begin{equation}
\frac{\delta p}{p_F}\sim a_c\sqrt{\frac{B-B_{c0}}{B_{c0}}},
\label{HF9}\end{equation} where $a_c$ is a constant of the order of
unity, $a_c\sim1$. As the magnetic field $B$ increases, $\delta
p/p_F$ becomes comparable to $\delta p_{FC}$, and the distribution
function $\nu({\bf p})$ disappears, being absorbed by the ordinary
Zeeman splitting. As a result, we are dealing with HF liquid located
on the disordered side of FCQPT. We show in Section \ref{SCEL} that
the behavior of such a system differs markedly from that of a system
with FC. Equation \eqref{HF9} implies that the relatively weak
magnetic field $B_{cr}$,
\begin{equation}B_{red}\equiv \frac{B-B_{c0}}
{B_{c0}}=(\delta p_{FC})^2\sim B_{cr}, \label{HF10}\end{equation}
where $B_{red}$ is the reduced field, takes the system from the
ordered side of the phase transition to the disordered if $\delta
p_{FC}\ll1$.

\subsection{Dependence of effective mass on magnetic fields
in HF metals and high-$T_c$ superconductors}

Observations have shown that in the normal state obtained by
applying a magnetic field whose strength is higher than the maximum
critical field $B_{c2}$ that destroys superconductivity, the
heavily doped cuprate (Tl$_2$Ba$_2$CuO$_{6+\delta}$) \cite{cyr} and
the optimally doped cuprate (Bi$_2$Sr$_2$CuO$_{6+\delta}$)
\cite{cyr1} exhibit no significant violations of the
Wiedemann-Franz law. Studies of the electron-doped superconductor
Pr$_{0.91}$LaCe$_{0.09}$Cu0$_{4-y}$ ($T_c$=24 K), revealed that
when a magnetic field destroyed superconductivity in this material,
the spin-lattice relaxation constant $1/T_1$ obeyed the relation
$T_1T=const$, known as the Korringa law, down to temperatures about
$T\simeq 0.2$ K \cite{korr,zheng}. At higher temperatures and in
magnetic fields up to 15.3 T perpendicular to the CuO$_2$ plane,
the ratio $1/T_1T$ remains constant as a function of $T$ for $T\leq
55$ K. In the temperature range from 50 to 300 K, the ratio
$1/T_1T$ decreases as the temperature increases \cite{zheng}.
Measurements involving the heavily doped nonsuperconducting
material La$_{1.7}$Sr$_{0.3}$CuO$_4$ have shown that the
resistivity $\rho$ varies with $T$ as $T^2$ and that the
Wiedemann-Franz law holds \cite{nakam,huss}.

Because the Korringa and Wiedemann-Franz laws strongly indicate the
presence of the LFL state, experiments show that the observed elementary
excitations cannot be distinguished from Landau quasiparticles in
high-$T_c$ superconductors. This places severe restrictions on
models describing hole- or electron-doped high-$T_c$
superconductors. For instance, for a Luttinger liquid
\cite{kane,kane1}, for spin-charge separation \cite{sen}, and in
some $t-J$ models \cite{hough}, a violation of the Wiedemann-Franz
law was predicted, which contradicts experimental evidence and
points to the limited applicability of these models.

If the constant $\lambda_0$ is finite, then a HF liquid with FC is in
the superconducting state. We examine the behavior of the system in
magnetic fields $B>B_{c2}$. In this case, the system becomes LFL
induced by the magnetic field, and the elementary excitations become
quasiparticles that cannot be distinguished from Landau
quasiparticles, with the effective mass $M^*(B)$ given by Eq.
(\ref{HF5}). As a result, the Wiedemann-Franz law holds as $T\to0$,
which agrees with the experimental facts \cite{cyr,cyr1}. The
low-temperature properties of the system depend on the effective
mass; in particular, the resistivity $\rho(T)$ behaves as given by
Eq. \eqref{RESIS}, with $A(B)\propto(M^*(B))^2$. Assuming that the
critical field $B=B_{c0}$ in the case of high-$T_c$ superconductors,
we deduce from Eq. (\ref{HF5}) that
\begin{equation}
\gamma_0\sqrt{B-B_{c0}}=const.\label{HF11}
\end{equation}
Taking Eqs. (\ref{HF6}) and (\ref{HF11}) into account, we find that
\begin{equation}
\gamma_0\sim A(B)\sqrt{B-B_{c0}}.\label{HF12}
\end{equation}
At finite temperatures, the system remains LFL, but for $T>T^*(B)$
the effective mass becomes temperature-dependent, $M^*\propto1/T$,
and the resistivity becomes a linear function of the temperature,
$\rho(T)\propto T$ \cite{khodrho}. Such behavior of the resistivity
has been observed in the high-$T_c$ superconductor
Tl$_2$Ba$_2$CuO$_{6+\delta}$ ($T_c<15$ K) \cite{mac}. At $B<10$ T,
the resistivity is a linear function of the temperature in the range
from 120 mK to 1.2 K, and at $B=10$ T the temperature dependence of
the resistivity can be written in the form $\rho(T)\propto AT^2$ in
the same temperature range \cite{mac,aspla}, clearly
demonstrating that the LFL state is restored
under the application of magnetic fields.

In LFL, the spin-lattice relaxation parameter $1/T_1$ is determined
by the quasiparticles near the Fermi level, whose population is
proportional to $M^* T$, whence $1/T_1T\propto M^*$, and is a
constant quantity \cite{korr,zheng}. When the superconducting state
disappears as a magnetic field is applied, the ground state can be
regarded as a field-induced LFL with the field-dependent effective
mass. As a result, $T_1T=const$, which implies that the Korringa law
holds. According to Eq. (\ref{HF5}), the ratio $1/T_1T\propto
M^*(B)$ decreases as the magnetic field increases at $T<T^*(B)$,
whereas in the case of a Landau Fermi liquid it remains constant, as
noted above. On the other hand, at $T>T^*(B)$, the ratio $1/T_1T$ is
a decreasing function of the temperature,
$1/T_1T\propto M^*(T)$. These results are
in good agreement with the experimental
facts \cite{zheng}. Because $T^*(B)$ is an increasing function of
the magnetic field [see Eq. \eqref{HF8}], the Korringa law remains
valid even at higher temperatures and in stronger magnetic fields.
Hence, at $T_0\leq T^*(B_0)$ and high magnetic fields $B>B_0$, the
system demonstrates distinct metallic behavior, because the
effective mass decreases as $B$ increases, see Eq. \eqref{HF5}.

The existence of FCQPT  can also be verified in experiments, because
at number densities $x>x_{FC}$ or beyond the FCQPT point, the system
must become LFL at sufficiently low temperatures \cite{shag}.
Experiments have shown that such a liquid indeed exists in the
heavily doped non-superconducting compound
La$_{1.7}$Sr$_{0.3}$CuO$_4$ \cite{nakam,huss}. It is remarkable that
for $T<55$ K, the resistivity exhibits a $T^2$-behavior without an
additional linear term and the Wiedemann-Franz law holds
\cite{nakam,huss}. At temperatures above 55 K, experimenters have
observed significant deviations from the LFL behavior. Observations
\cite{obz,pogsh} are in accord with these experimental findings
showing that the system can again be returned to the LFL state by
applying sufficiently strong magnetic fields (also see Section
\ref{HCEL}).

\subsubsection{Common QCP in the high-$T_c$
$\rm Tl_2Ba_2CuO_{6+x}$ and the HF metal $\rm
YbRh_2Si_2$}\label{HTCHF}

Under the application of magnetic fields $B>B_{c2}>B_{c0}$ and at
$T<T^*(B)$, a high-$T_c$ superconductor or HF metal can be driven to
the LFL state with its resistivity given by Eq. \eqref{RESIS}. In that
case measurements of the coefficient $A$ produce information on its
dependence on the applied field. We note that relationships between
critical magnetic fields $B_{c2}$ and $B_{c0}$ are clarified in
Subsection \ref{BC0C2}.

\begin{figure}[!ht]
\begin{center}
\includegraphics [width=0.60\textwidth]{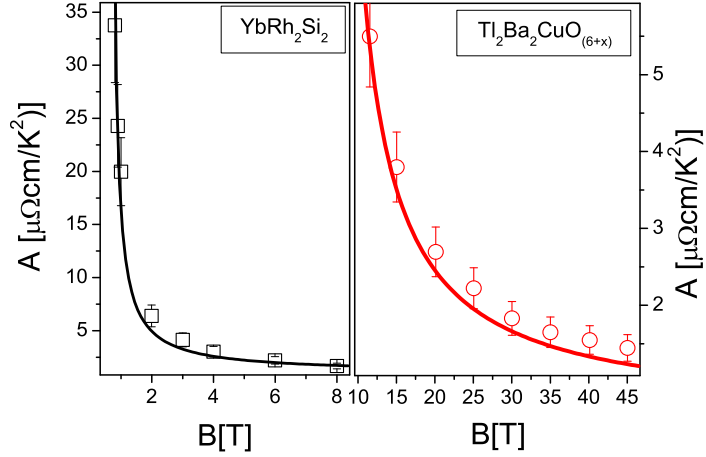}
\vspace*{-0.5cm}
\end{center}
\caption{The charge transport coefficient $A(B)$ as a function of
magnetic field $B$ obtained in measurements on $\rm YbRh_2Si_2$
\cite{geg} and $\rm Tl_2Ba_2CuO_{6+x}$ \cite{pnas}. The different
field scales are clearly seen. The solid curves represent our fit by
using Eq. \eqref{ABDR}}\label{ff2}
\end{figure}

Precise measurements of the coefficient $A(B)$ on high-$T_c$ $\rm
Tl_2Ba_2CuO_{6+x}$ \cite{pnas} allow us to establish relationships
between the physics of both high-$T_c$ superconductors and HF metals
and clarify the role of the extended quasiparticle paradigm. The
$A(B)$ coefficient, being proportional to the
quasiparticle–--quasiparticle scattering cross-section, is found to
be $A\propto(M^*(B))^2$ \cite{geg,ksch}. With respect to Eq.
\eqref{HF5}, this implies that
\begin{equation}
A(B)\simeq A_0+\frac{D}{B-B_{c0}},\label{ABDR}
\end{equation}
where $A_0$ and $D$ are fitting parameters.

\begin{figure}[!ht]\begin{center}
\includegraphics [width=0.60\textwidth]{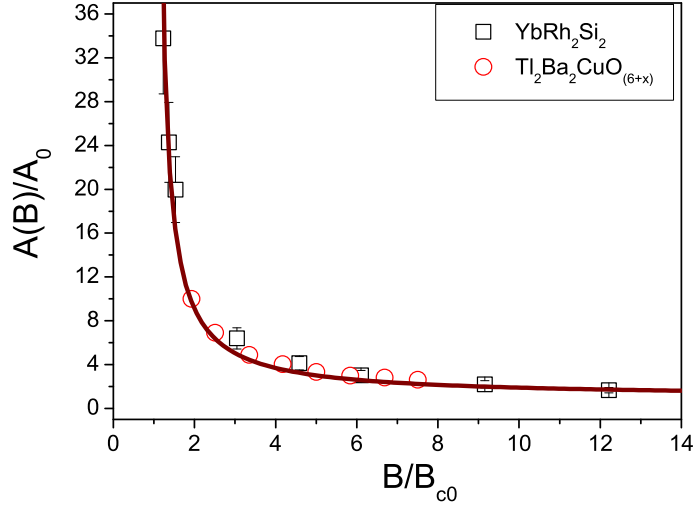}
\end{center}
\caption{Normalized coefficient $A(B)/A_0\simeq 1+D_N/(y-1)$ given
by Eq. \eqref{HFC} as a function of normalized magnetic field
$y=B/B_{c0}$ shown by squares for $\rm YbRh_2Si_2$ and by circles
for high-$T_c$ $\rm Tl_2Ba_2CuO_{6+x}$. $D_N$ is the only fitting
parameter.}\label{ff3}
\end{figure}
Figure \ref{ff2} reports the fit of our theoretical dependence
\eqref{ABDR} to the experimental data for the measurements of the
coefficient $A(B)$ for two different classes of substances: HF metal
$\rm YbRh_2Si_2$ (with $B_{c0}=0.06$ T, left panel) \cite{geg} and
high-$T_c$ $\rm Tl_2Ba_2CuO_{6+x}$ (with $B_{c0}=5.8$ T, right
panel) \cite{pnas}. In Fig. \ref{ff2}, left panel, $A(B)$ is shown
as a function of magnetic field $B$, applied both along and
perpendicular to the $c$ axis. For the latter the $B$ values have
been multiplied by a factor of 11 \cite{geg}. The different scales
of field $B_{c0}$ are clearly seen and demonstrate that $B_{c0}$ has
to be taken as an input parameter. Indeed, the critical field of
$\rm Tl_2Ba_2CuO_{6+x}$ with $B_{c0}=5.8$ T is 2 orders of magnitude
larger than that of $\rm YbRh_2Si_2$ with $B_{c0}=0.06$ T.

Figure \ref{ff2} displays good coincidence of the theoretical dependence
\eqref{HF6} with the experimental facts \cite{pnas,shpo}. This means that the
physics underlying the field-induced reentrance of LFL behavior, is
the same for both classes of substances. To further corroborate this
point, we rewrite Eq. \eqref{ABDR} in the reduced variables $A/A_0$ and
$B/B_{c0}$. Such rewriting immediately reveals the scaling nature
of the behavior of these two substances - both of them are driven to
common QCP related to FCQPT and induced by the application of
magnetic field. As a result,  Eq. \eqref{ABDR} takes the form
\begin{equation} \frac{A(B)}{A_0}\simeq
1+\frac{D_N}{B/B_{c0}-1},\label{HFC}\end{equation} where
$D_N=D/(A_0B_{c0})$ is a constant. From Eq. \eqref{HFC} it is seen that
upon applying the scaling to both coefficients $A(B)$ for $\rm
Tl_2Ba_2CuO_{6+x}$ and $A(B)$ for $\rm YbRh_2Si_2$ they are reduced to a
function depending on the single variable $B/B_{c0}$ thus
demonstrating universal behavior. To support Eq. \eqref{HFC}, we
replot both dependencies in reduced variables $A/A_0$ and $B/B_{c0}$
in Fig. \ref{ff3}. Such replotting immediately reveals the universal
scaling nature of the behavior of these two substances.
It is seen from Fig.
\ref{ff3} that close to the magnetic induced QCP
there are no "external" physical
scales revealing the scaling. Therefore the
normalization by the scales $A_0$ and
$B_{c0}$ immediately reveals the common physical nature of these
substances, allowing us to get rid of the specific
properties of the system that define the values of $A_0$ and
$B_{c0}$.

Based on the above analysis of the $A$ coefficients, we conclude
that there is at least one quantum phase transition inside the
superconducting dome of high-$T_c$ superconductors, and this
transition is FCQPT \cite{PLA686}.

\section{Appearance of FCQPT in Fermi systems} \label{FCDL}

We say that Fermi systems that approach QCP from a disordered state
are highly correlated systems in order to distinguish them from
strongly correlated systems (or liquids) that are already beyond
FCQPT placed at the quantum critical line as shown in Fig.
\ref{fig1}. A detailed description of the properties of highly
correlated systems are given in Section \ref{HCEL}, and the
properties of  strongly correlated systems are discussed in Section
\ref{SCEL}. In the present section, we discuss the behavior of the
effective mass $M^*$ as a function of the density $x$ of the system
as $x\to x_{FC}$.

The experimental facts for high-density 2D $^3$He
\cite{cas1,cas,mor,3he} show that the effective mass becomes
divergent when the value of the density at which the 2D liquid
$^3$He begins to solidify is reached \cite{cas}. Also observed was a
sharp increase in the effective mass in the metallic 2D electron
system as the density $x$ decreases and tends to the critical
density of the metal-insulator transition \cite{skdk}. We note that
there is no ferromagnetic instability in the Fermi systems under
consideration and the corresponding Landau amplitude $F^a_0>-1$
\cite{skdk,cas}, which agrees with the model of nearly localized
fermions \cite{vollh,pfw,vollh1}.

We examine the divergence of the effective mass in 2D and 3D highly
correlated Fermi liquids at $T=0$ as the density $x\to x_{FC}$
approaching FCQPT from the disordered phase. We begin by calculating
$M^*$ as a function of the difference $(x-x_{FC})$ for a 2D Fermi
liquid. For this, we use the equation for $M^*$ derived in
\cite{ksz}, where the divergence of $M^*$ related to the generation
of density wave in various Fermi liquids was predicted. As $x\to
x_{FC}$, the effective mass $M^*$ can be approximately written as
\begin{eqnarray}
\nonumber \frac{1}{M^{*}}&\simeq&\frac{1}{m}+\frac{1}{4\pi^{2}}
\int\limits_{-1}^{1}\int\limits_0^{g_0}\frac{ydydg}{\sqrt{1-y^{2}}} \\
&\times& \frac{v(q(y))}{\left[1-R(q(y),g) \chi_0(q(y))\right]^{2}}.
\label{DL1}\end{eqnarray} Here we use the notation
$p_F\sqrt{2(1-y)}=q(y)$, where $q(y)$ is the momentum transfer,
$v(y)$ is the pair interaction, the integral with respect to the
coupling constant $g$ is taken from zero to the actual value $g_0$,
$\chi_0(q,\omega)$ is the linear response function for the
noninteracting Fermi liquid, and $R(q,\omega)$ is the effective
interaction, with both functions taken at $\omega=0$. The
quantities $R$ and $\chi_0$ determine the response function for the
system,
\begin{equation}
\chi(q,\omega,g)=\frac{\chi_0(q,\omega)}
{1-R(q,\omega,g)\chi_0(q,\omega)}.\label{DL2}
\end{equation}
Near the instability related to the generation of density wave at
the density $x_{cdw}$, the singular part of the response function
$\chi$ has the well-known form, see e.g. \cite{varma}
\begin{equation}
\chi^{-1}(q,\omega,g)\simeq
a(x_{cdw}-x)+b(q-q_c)^2+c(g_0-g),\label{DL3}
\end{equation}
where $a$, $b$, and $c$ are constants and $q_c\simeq 2p_F$ is the
density-wave momentum. Substituting Eq. (\ref{DL3}) in (\ref{DL1})
and integrating, we can represent the equation for the effective
mass $M ^*$ as
\begin{equation}\frac{1}{M^*(x)}=\frac{1}{m}-\frac{c}
{\sqrt{x-x_{cdw}}},\label{DL4} \end{equation} where $c$ is a
positive constant. It follows from Eq. \eqref{DL4} that $M^*(x)$
diverges as a function of the difference $(x-x_{FC})$ and
$M^*(x)\to\infty$ as $x\to x_{FC}$ \cite{khod1,shag1}
\begin{equation}
\frac{M^*(x)}{m}\simeq a_1+\frac{a_2}{x-x_{FC}},\label{DL5}
\end{equation}
where $a_1$ and $a_2$ are constants. We note that Eqs. (\ref{DL4})
and (\ref{DL5}) do not explicitly contain the interaction $v(q)$,
although $v(q)$ affects $a_1$, $a_2$ and $x_{FC}$. This result
agrees with Eq. \eqref{FL7}, which determines the same universal
type of divergence (i.e., a divergence that is independent of the
interaction). Hence, both Eqs. \eqref{FL7} and \eqref{DL5} can be
applied to 2D $^3$He, the electron liquid, and other Fermi liquids.
We also see that FCQPT precedes the formation of density waves (or
charge-density waves) in Fermi systems.

We note that the difference $(x-x_{FC})$ must be positive in both
cases, since the density $x$ approaches $x_{FC}$ when the system is
on the disordered side of FCQPT with the finite effective mass
$M^*(x)>0$. In the case of $\rm ^3He$, FCQPT occurs as the density
increases, when the potential energy begins to dominate the
ground-state energy due to the strong repulsive short ranged part of
the inter-particle interaction. Thus, for the 2D $^3$He liquid, the
difference $(x-x_{FC})$ on the right hand side of Eq. (\ref{DL5})
must be replaced with $(x_{FC}-x)$. Experiments have shown that the
effective mass indeed diverges at high densities in the case of 2D
$\rm ^3He$ and at low densities in the case of 2D electron systems
\cite{skdk,cas}.

\begin{figure} [! ht]
\begin{center}
\includegraphics [width=0.60\textwidth]{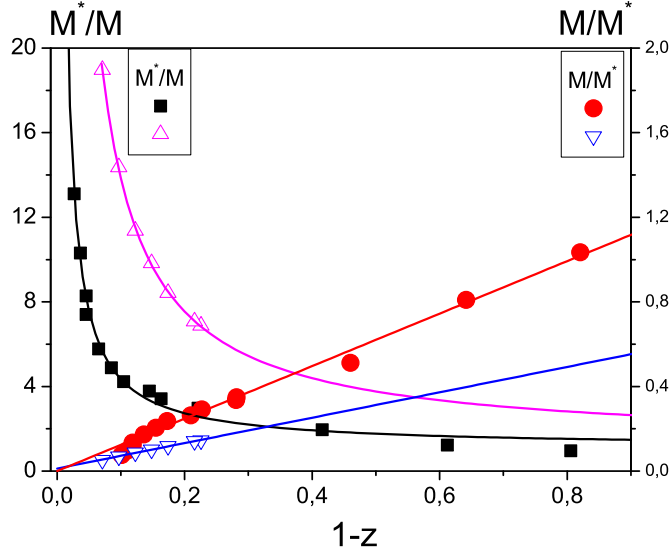}
\end{center}
\vspace*{-1.0cm} \caption{The dependence of the effective mass
$M^*(z)$ on dimensionless density $1-z=1-x/x_{FC}$. Experimental
data from Ref. \cite{cas} are shown by circles and squares and those
from Ref. \cite{3he} are shown by triangles. The effective mass is
fitted as $M^*(z)/m\propto b_1+b_2/(1-z)$ (also see Eq.
\eqref{FL7}), while the reciprocal one as $m/M^*(z)\propto b_3\,z$,
where $b_1,b_2$ and $b_3$ are constants.}\label{MXM}
\end{figure}

In Fig. \ref{MXM}, we report the experimental values of the
effective mass $M^*(z)$ obtained by the measurements on $^3$He
monolayer \cite{cas}. These measurements, in coincidence with those
from Ref. \cite{3he,3he1}, show the divergence of the effective
mass at $x=x_{FC}$. To show, that our FCQPT approach is able to
describe the above data, we represent the fit of $M^*(z)$ by the
fractional expression $M^*(z)/M\propto b_1+b_2/(1-z)$ and the
reciprocal effective mass by the linear fit $m/M^*(z)\propto
b_3\,z$. We note here, that the linear fit has been used to
describe the experimental data for bilayer $^3$He
\cite{3he,pepin3he} and we use this function here for the sake of
illustration. It is seen from Fig. \ref{MXM} that the data of Ref.
\cite{3he} ($^3$He bilayer) can be equally well approximated by
both linear and fractional functions, while the data in Ref.
\cite{cas} cannot. For instance, both fitting functions give for
the critical density in bilayer $x_{FC}\approx 9.8$ nm$^{-2}$,
while for monolayer \cite{cas} these values are different -
$x_{FC}=5.56$ nm$^{-2}$ for linear fit and $x_{FC}=5.15$ nm$^{-2}$
for fractional fit. It is seen from Fig. \ref{MXM}, that linear fit
is unable to properly describe the experiment \cite{cas} at small
$1-z$ (i.e. near $x=x_{FC}$), while the fractional fit describes
the experiment very well. This means that more detailed
measurements are necessary in the vicinity $x=x_{FC}$
\cite{prl3he}.

The effective mass as a function of the electron density $x$ in a
silicon MOSFET  is shown in Fig. \ref{Fig5}. We see that Eq.
\eqref{DL4} provides a good description of the experimental results.
The divergence of the effective mass  $M^*(x)$ discovered in
measurements involving 2D $^3$He \cite{cas1,cas,3he} is illustrated
by Figs. \ref{Fig6} and \ref{MXM}. Figures \ref{Fig5}, \ref{Fig6}
and \ref{MXM} show that the description provided by Eqs.
 \eqref{FL7}, \eqref{DL4} and \eqref{DL5} is in good agreement with
the experimental data.

In the case of 3D systems, as $x\to x_{FC}$, the effective mass is
given by the expression \cite{ksz}
\begin{equation}
\frac {1}{M^{*}}\simeq\frac{1}{m}+\frac{p_F}{4\pi^{2}}\int\limits_
{-1}^{1}\int\limits_0^{g_0}\frac{v(q(y))ydydg}{\left[1-R(q
(y),g)\chi_0(q(y))\right]^{2}}.\label{DL6}
\end{equation}
Comparison of Eqs. (\ref{DL1}) and (\ref{DL6}) shows that there is
no essential difference between them, although they describe
different cases, 2D and 3D. In the 3D case, we can derive equations
similar to (\ref{DL4}) and (\ref{DL5}) just as we did in the 2D
case, but the numerical coefficients are different, because they
depend on the number of dimensions. The only difference between 2D
and 3D electron systems is that FCQPT occurs in 3D systems at
densities much lower than those corresponding to 2D systems. No such
transition occurs in massive 3D $^3$He because the transition is
absorbed by the first-order liquid-solid phase transition
\cite{cas1,cas}.

\section{A highly correlated Fermi liquid
in HF metals} \label{HCEL}

As noted in the Introduction, the challenge for the theories is
to explain the scaling behavior of the normalized effective mass
$M^*_N(y)$ displayed in Fig. \ref{YBRHSIN}; the theories
analyzing only the critical exponents that characterize $M^*_N(y)$
at $y\gg 1$ consider a part of the problem. In this Section we
analyze and derive the scaling behavior of the normalized effective
mass near QCP as depicted in Fig. \ref{YBRHSIN} and show that
numerous facts collected in measurements  of the thermodynamic,
transport and relaxation properties carried out at the transition
regime on HF metals can be explained within the framework of the
extended quasiparticle paradigm describing the scaling behavior.

\subsection{Dependence of the effective mass $M^*$
on magnetic field} \label{HCEL1}

When the system approaches FCQPT from the disordered side, at
sufficiently low temperatures as shown in Fig. \ref{fig1}, it
remains LFL with the effective mass $M^*$ that strongly depends on
the distance $r = (x-x_{FC})/x_{FC}$ and  magnetic field $B$. This
state of the system with $M^*$ that strongly depends on $r$ and $B$
resembles the state of strongly correlated liquid described in
Sections \ref{FLFC} and \ref{SCEL}. But in contrast to a strongly
correlated liquid, the system in question involves no temperature
independent entropy $S_0$ specified by Eq. (\ref{snfl}) and at low
temperatures becomes LFL with effective mass $M^*\propto1/r$ [see
Eqs. (\ref{FL7}) and (\ref{DL5})]. Such a liquid can be called a
highly correlated liquid; as we see shortly, its effective mass
exhibits the scaling behavior. We study this behavior when the
system transits from its LFL to NFL states.

We use the Landau equation to study the behavior of the effective
mass $M^*(T,B)$ as a function of the temperature and the magnetic
field. For the model of homogeneous HF liquid at finite temperatures
and magnetic fields, this equation acquires the form \cite{lanl1}
\begin{eqnarray}
\nonumber \frac{1}{M^*(T,
B)}&=&\frac{1}{m}+\sum_{\sigma_1}\int\frac{{\bf p}_F{\bf
p}}{p_F^3}F_
{\sigma,\sigma_1}({\bf p_F},{\bf p}) \\
&\times&\frac{\partial n_{\sigma_1} ({\bf
p},T,B)}{\partial{p}}\frac{d{\bf p}}{(2\pi)^3}. \label{HC1}
\end{eqnarray}
where $F_{\sigma,\sigma_1}({\bf p_F},{\bf p})$ is the Landau
amplitude dependent on momenta $p_F$, $p$ and spin $\sigma$.
For the sake of definiteness, we assume that the HF liquid is 3D
liquid. As seen in Section \ref{FCDL}, the scaling behavior
calculated within the model of HF liquid does not depend on
dimensionality and on the inter-particle interaction, while the
values of scales like $M^*_M$ and $T_M$ do depend. To simplify
matters, we ignore the spin dependence of the effective mass,
because $M^*(T,B)$ is nearly independent of the spin in weak fields.
The quasiparticle distribution function can be expressed as
\begin{equation} n_{\sigma}({\bf p},T)=\left\{ 1+\exp
\left[\frac{(\varepsilon({\bf p},T)-\mu_{\sigma})}T\right]\right\}
^{-1},\label{HC2}
\end{equation}
where $\varepsilon({\bf p},T)$ is determined by (\ref{FL2}). In our
case, the single-particle spectrum depends on the spin only weakly,
but the chemical potential may depend on the spin due to the Zeeman
splitting. When this is important, we specifically indicate the
spin dependence of physical quantities. We write the quasiparticle
distribution function as $n_{\sigma}({\bf p},T,B) \equiv\delta
n_{\sigma}({\bf p},T,B)+n_{\sigma}({\bf p},T=0,B=0)$. Equation
(\ref{HC1}) then becomes
\begin{eqnarray}
\nonumber
&&\frac{m}{M^*(T,B)}=\frac{m}{M^*(x)}+\frac{m}{p_F^2}\sum_{\sigma_1}
\int\frac{{\bf p}_F{\bf p_1}}{p_F}\\
&\times&F_{\sigma,\sigma_1}({\bf p_F},{\bf
p}_1)\frac{\partial\delta n_{\sigma_1}({\bf p}_1,T,B)}
{\partial{p}_1}\frac{d{\bf p}_1}{(2\pi) ^3}. \label{HC3}
\end{eqnarray}

We assume that the highly correlated HF liquid is close to FCQPT and
the distance $r\to0$, and therefore $M/M^*(x)\to0$, as follows from
Eq. \eqref{FL7}. For normal metals, where the electron liquid
behaves like LFL with the effective mass of several bare electron
masses $M^*/m\sim1$, at temperatures even near 1000 K, the second
term on the right hand side of Eq. (\ref{HC3}) is of the order of
$T^2/\mu^2$ and is much smaller than the first term. The same is
true, as can be verified, when a magnetic field $B\sim100$ T is
applied. Thus, the system behaves like LFL with the effective mass
that is actually independent of the temperature or magnetic field,
while $\rho(T)\propto AT^2$. This means that the corrections to the
effective mass determined by the second term on the right-hand side
of Eq. \eqref{HC3} are proportional to $(T/\mu)^2$ or $(\mu_B
B/\mu)^2$.

Near QCP $x_{FC}$, with $m/M^*(x\to x_{FC})\to0$, the behavior of
the effective mass changes dramatically because the first term on
the right-hand side of Eq. (\ref{HC3}) vanishes, the second term
becomes dominant, and the effective mass is determined by the
homogeneous version of Eq. (\ref{HC3}) as a function of $B$ and $T$.
As a result, the LFL state vanishes and the system demonstrates the
NFL behavior down to lowest temperatures.

We now qualitatively analyze the solutions of Eq. (\ref{HC3}) at
$x\simeq x_{FC}$ and $T=0$. Application of magnetic field leads to
Zeeman splitting of the Fermi surface, and the distance $\delta p$
between the Fermi surfaces with spin up and spin down becomes
$\delta p=p_F^{\uparrow}-p_F^{\downarrow}\sim\mu_{B}BM^*(B)/p_F$. We
note that the second term on the right-hand side of Eq. (\ref{HC3})
is proportional to $(\delta p)^2\propto(\mu_{B}BM^*(B)/p_F)^2$, and
therefore Eq. (\ref{HC3}) reduces to \cite{shag4,ckhz,shag}
\begin{equation}
\frac{m}{M^*(B)}=\frac{m}{M^*(x)}+c\frac{(\mu_{B}BM^*(B))^2}
{p_F^4},\label{HC4}
\end{equation}
where $c$ is a constant. We also note that $M^*(B)$ depends on $x$
and that this dependence disappears at $x=x_{FC}$. At this point,
the term $m/M^*(x)$ vanishes and Eq. (\ref{HC4}) becomes homogeneous
and can be solved analytically \cite{ckhz,shag1,shag}:
\begin{equation}
M^*(B)\propto\frac{1}{(B-B_{c0})^{2/3}}.\label{HC5}
\end{equation}
where $B_{c0}$ is the critical magnetic field, regarded as a
parameter (see remarks to Fig. \ref{ff2}).

Equation (\ref{HC5}) specifies the universal power-law behavior of
the effective mass, irrespective of the interaction type and is
valid in 3D and 2D cases. For densities $x>x_{FC}$, the effective
mass $M^*(x)$ is finite and we deal with the LFL state if the
magnetic field is so weak that
\begin{equation}
\frac{M^*(x)}{M^*(B)}\ll1,\label{HC5a}
\end{equation}
with $M^*(B)$ given by Eq. \eqref{HC5}. The second term on the
right-hand side of Eq. \eqref{HC4}, which is proportional to
$(BM^*(x))^2$, is only a small correction. In the opposite case, at
$T/T^*(B)\ll1$, where
\begin{equation}
\frac{M^*(x)}{M^*(B)}\gg1,\label{HC5b}
\end{equation}
the electron liquid behaves as if it were at the quantum critical
point. In the LFL state, the main thermodynamic and transport
properties of the system are determined by the effective mass. It
therefore follows from Eq. \eqref{HC5} that we have the unique
possibility of controlling the magnetoresistance, resistivity, heat
capacity, magnetization, thermal bulk expansion, etc by varying the
magnetic field $B$. It must be noted that a large effective mass
leads to a high density of states, which causes the emergence of a
large number of competing states and phase transitions. We believe
that such states can be suppressed by applying an external magnetic
field, and we examine the thermodynamic properties of the system
without considering such competition.

\subsection{Dependence of the effective mass $M^*$
on temperature and the damping of quasiparticles} \label{HCEL2}

For a qualitative examination of the behavior of $M^*(T,B,x)$ as the
temperature increases, we simplify Eq. \eqref{HC3} by dropping the
variable $B$ and by imitating the effect of an external magnetic
field by a finite effective mass in the denominator of the first
term on the right hand side of Eq. \eqref{HC3}. Then the effective
mass becomes a function of the distance $r$, $M^*(r)$, determined
also by both the magnitude of the magnetic field $B$ and $x$. In a
zero magnetic field, $r=(x-x_{FC})/x_{FC}$, We integrate the second
term on the right-hand side of Eq. \eqref{HC3} with respect to the
angular variables, then integrate by parts with respect to $p$, and
replace $p$ with $z=(\varepsilon(p)-\mu)/T$. In the case of a flat
and narrow band, we can use the approximation where
$(\varepsilon(p)-\mu) \simeq p_F (p-p_F)/M^*(T)$. The result is
\begin{eqnarray}
\nonumber \frac{M}{M^*(T)}&=&\frac{m}{M^*(r)}-
\alpha\int^{\infty}_{0}\frac{F(p_F,p_F(1 +\alpha z))dz}{1+e^z} \\
&+&\alpha\int^{1/\alpha}_{0}F(p_F, p_F(1-\alpha
z))\frac{dz}{1+e^z}.\label{HC8}
\end{eqnarray}
where we use the notation $$F\sim m\frac{d(F^1p^2)}{dp}, \alpha=
\frac{TM^*(T)}{p_F^2}=\frac{TM^*(T)}{(T_kM^*(r))},$$
$T_k=p_F^2/M^*(r)$, and the Fermi momentum is defined by the
condition $\varepsilon(p_F)=\mu$.

We first consider the case where $\alpha\ll 1 $. Then, discarding
terms of the order $\exp(-1/\alpha) $, we can set the upper limit in
the second integral on the right hand side of Eq. (\ref{HC8}) to
infinity, with the result that the sum of the second and third terms
is an even function of $\alpha$. The resulting integrals are typical
expressions involving the Fermi-Dirac function in the integrand and
can be evaluated by a standard procedure (e.g., see Ref.
\cite{lanl2}). Because we need only an estimate of the integrals, we
write Eq. (\ref{HC8}) as
\begin{eqnarray}
\nonumber \frac{m}{M^*(T)}&\simeq& \frac{m}{M^*(r)}
+a_1\left(\frac{TM^*(T)}{T_kM^*(r)}\right)^2\\
&+&a_2\left(\frac{TM^*(T)} {T_kM^*(r)}\right)^4+... ,\label{HC9}
\end{eqnarray}
where $a_1$ and $a_2$ are constants of the order of unity.

Equation (\ref{HC9}) is a typical equation of the LFL theory. The
only exception is the effective mass $M^*(r)$, which depends
strongly on the distance $r$ and diverges as $r\to0$. Nevertheless,
Eq. (\ref{HC9}) implies that as $T\to0$, the corrections to $M^*(r)$
begin with terms of the order $T^2$ if
\begin{equation} \frac{m}{M^*(r)}\gg\left(\frac{TM^*(T)}{T_kM^*(r)}
\right)^2\simeq\frac{T^2}{T_k^2},\label{HC10} \end{equation} and the
system behaves like LFL. Condition (\ref{HC10}) implies that the
behavior inherent in LFL disappears as $r\to0$ and
$M^*(r)\to\infty$. Then the free term on the right-hand side of Eq.
(\ref{HC8}) is negligible, $m/M^*(r)\to0$, and Eq. (\ref{HC8})
becomes homogeneous and determines the universal behavior of the
effective mass $M^*(T)$. At a certain temperature $T^*\ll T_k$, the
value of the sum on the right-hand side of Eq. (\ref{HC9}) is
determined by the second term and relation (\ref{HC10}) becomes
invalid. Keeping only the second term in Eq. (\ref{HC9}), we arrive
at an equation for determining $M^*(T)$ in the transition region
\cite{ckhz,shag5}:
\begin{equation}
M^*(T)\propto\frac{1}{T^{2/3}}.\label{HC11}
\end{equation}

As regards an estimate of the transition temperature $T^*(B)$ at
which the effective mass becomes temperature-dependent, we note that
the effective mass is a continuous function of the temperature and
the magnetic field: $M^*(B)\sim M^*(T^*)$. With Eqs. (\ref{HC5}) and
(\ref{HC11}), we obtain
\begin{equation}
T^*(B)\simeq \mu_B(B-B_{c0}).\label{HC14a}
\end{equation}
As the temperature increases, the system transfers into another
mode. The coefficient $\alpha$ is then of the order of unity,
$\alpha\sim 1 $, the upper limit in the second integral in Eq.
(\ref{HC8}) cannot be set to infinity, and odd terms begin to play a
significant role. As a result, Eq. (\ref{HC9}) breaks down and the
sum of the first and second integrals on the right-hand side of Eq.
(\ref{HC8}) becomes proportional to $M^*(T)T$. Ignoring the first
term $m/M^*(r)$ and approximating the sum of integrals by $M^*(T)T$,
we obtain from (\ref{HC8}) that
\begin{equation}
M^*(T)\propto\frac{1}{\sqrt{T}}.\label{HC14}
\end{equation}
We note that $M^*(T)$ is also given by Eq. \eqref{HC14} if the
Landau amplitude $F(p)$ is determined by a nonanalytic function,
that is the function cannot be expanded in Tailor series at $p\to0$,
see Section \ref{TFT}.

We therefore conclude that as the temperature increases and the
condition $x\simeq x_{FC}$ is satisfied, the system demonstrates
regimes of three types: (i) the behavior of the Landau Fermi liquid
at $\alpha\ll 1$, when Eq. (\ref{HC10}) is valid and the behavior
of the effective mass is described by Eq. (\ref{HC5}); (ii) the
behavior defined by Eq. (\ref{HC11}), when $M^*(T)\propto T^{-2/3}$
and $S (T)\propto M^*(T)T\propto T^{1/3}$; and (iii) the behavior
at $\alpha\sim 1$, when Eq. (\ref{HC14}) is valid,
$M^*(T)\propto1/\sqrt{T}$, the entropy $S(T)\propto
M^*(T)T\propto\sqrt{T}$, and the heat capacity $C (T)=T(\partial
S(T)/\partial T)\propto\sqrt{T}$.
\begin{figure} [! ht]
\begin{center}
\includegraphics [width=0.60\textwidth] {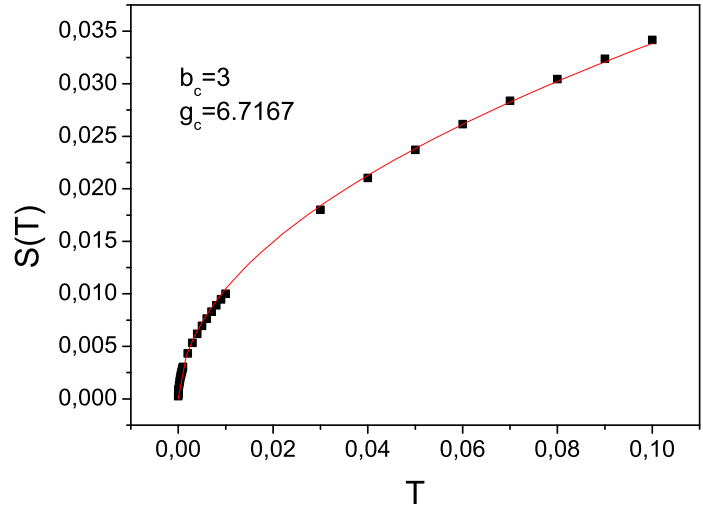}
\end{center}
\caption {The entropy $S(T)$ of a highly correlated Fermi liquid at
the critical point of FCQPT. The solid line represents the function
$S(T)\propto \sqrt{T}$, and the squares mark the results of
calculations.} \label {Fig10}
\end{figure}

We illustrate the behavior of $S(T)$ when
Eq. \eqref{HC14} is valid using calculations
based on the model Landau functional \cite{ksk,ksn}
\begin{eqnarray}
\nonumber E[n(p)]&=&\int\frac{{\bf p}^2}{2M}\frac{d{\bf
p}}{(2\pi)^3}+\frac{1}
{2}\int V({\bf p}_1-{\bf p}_2)\\
&\times&n({\bf p}_1)n({\bf p}_2) \frac{d{\bf p}_1d{\bf
p}_2}{(2\pi)^6},\label{HC6}
\end{eqnarray}
with the nonanalytic Landau amplitude
\begin{equation}
V({\bf p})=g_0\frac{\exp(-\beta_0|{\bf p}|)}{|{\bf p}|}.\label{HC7}
\end{equation}
We normalize the effective mass to $m$, i.e., $M^*=M^*/m$, and the
temperature $T_0$ to the Fermi energy $\varepsilon^0_F$,
$T=T_0/\varepsilon^0_F$, and use the dimensionless coupling constant
$g=(g_0 m)/(2\pi^2)$ and also $\beta=\beta_0p_F$. FCQPT occurs when
these parameters reach the critical values $\beta=b_c$ and $g=g_c$.
On the other hand, a transition of this kind occurs as
$M^*\to\infty$. This condition allows deriving a relation between
$b_c$ and $g_c$ \cite{ksk,ksn}:
\begin{equation}\label{BCGC}
\frac{g_c}{b_c^3}(1+b_c)\exp(-b_c)[b_c\cosh(b_c)-\sinh(b_c)]=1.
\end{equation}
This relation implies that the critical point of FCQPT can be
reached by varying $g_0$ if $\beta_0$ and $p_F$ are fixed, by
varying $p_F$ if $\beta_0$ and $g_0$ are fixed, etc. For
definiteness, we vary $g$ to reach FCQPT or to study the properties
of the system beyond the critical point. Calculations of $M^*(T)$,
$S(T)$, and $C(T)$ based on the model functional \eqref{HC6} with the
parameters $g=g_c=6.7167$ and $\beta=b_c=3$ show that $M^*(T)\propto
1/\sqrt{T}$, $S(T)\propto \sqrt{T}$, and $C(T)\propto\sqrt{T}$. The
temperature dependence of the entropy in this case is depicted in
Fig. \ref{Fig10}.

We now estimate the quasiparticle damping $\gamma(T)$. In the
Landau Fermi-liquid theory, $\gamma(T)$ is given by \cite{lanl1}
\begin{equation} \gamma\sim|\Gamma|^2(M^*)^3T^2, \label{HC15}\end{equation}
where $\Gamma$ is the particle-hole amplitude. In the case of highly
correlated HF system with a high density of states caused by the
enormous effective mass, $\Gamma$ cannot be approximated by the
"bare" interaction between particles but can be estimated within the
"ladder" approximation, which yields $|\Gamma|\sim1/(p_FM^*(T))$
\cite{dkss,dkss1}. As a result, we find that $\gamma(T)\propto T^2$
in the Landau Fermi-liquid regime since $M^*$ is
temperature-independent. Then, $\gamma(T)\propto T^{4/3}$ in the
$T^{-2/3}$-regime, and $\gamma (T)\propto T^{3/2}$ in the
$1/\sqrt{T}$-regime. We note that in all these cases, the width is
small compared to the characteristic quasiparticle energy, which is
assumed to be of the order of $T$, and hence the quasiparticle
concept is meaningful.

The conclusion that can be drawn here is that when the HF liquid is
localized near QCP of FCQPT and is on the disordered side, its
low-energy excitations are quasiparticles with the effective mass
$M^*(B,T)$.We note that at FCQPT, the quasiparticle renormalization
$z$-factor remains approximately constant and the divergence of the
effective mass that follows from Eq. (\ref{FL7}) is not related to
the fact that $z\to0$ \cite{khodb,x18,khod2}. Therefore, the
quasiparticle concept remains valid and it is these quasiparticles
that constitute the extended paradigm and determine the transport,
relaxation and thermodynamic properties of HF liquid.

\subsection{Scaling behavior of the effective mass} \label{HCEL3}

As was mentioned in the Introduction, to avoid difficulties
associated with the anisotropy generated by the crystal lattice of
HF metals, we study the universal behavior of HF metals using the
model of the homogeneous HF (electron) liquid. The model is quite
meaningful because we consider the scaling behavior exhibited by the
effective mass at low temperatures. The scaling behavior of the
effective mass is determined by energy and momentum transfers that
are small compared to the Debye characteristic temperature and
momenta of the order of the reciprocal lattice cell length $a^{-1}$.
Therefore quasiparticles are influenced by the crystal lattice
averaged over large distances compared to the length $a$.  Thus, we
can use the well-known jelly model. We note that the values of such
scales as $M^*_M$, $T_M$, $B_{c0}$ and $B_{c2}$ etc depend on the
properties of a HF metal, its lattice, composition etc. For example,
the critical magnetic field $B_{c0}$ depends even on its orientation
with respect to the lattice.

To explore the scaling behavior of $M^*$, we start with qualitative
analysis of Eq. \eqref{HC1}. At FCQPT the effective mass $M^*$
diverges and Eq. \eqref{HC1} becomes homogeneous determining $M^*$
as a function of temperature as given by Eq. \eqref{HC11}. If the
system is located before FCQPT, $M^*$ is finite, and at low temperatures
the system demonstrates the LFL behavior, that is $M^*(T)\simeq
M^*+a_1T^2$. As we have seen in Subsection \ref{HCEL2}, the LFL
behavior takes place when the second term on the right hand side of
Eq. \eqref{HC1} is small in comparison with the first one. Then, at
increasing temperatures the system enters the transition regime: $M^*$
grows, reaching its maximum $M^*_M$ at $T=T_M$, with subsequent
diminishing. Near temperatures $T\geq T_M$ the last "traces" of LFL
regime disappear, the second term starts to dominate, and again Eq.
\eqref{HC1} becomes homogeneous, and the NFL behavior restores,
manifesting itself in decreasing of $M^*$ as $T^{-2/3}$. When the
system is near FCQPT, it turns out that the solution of Eq.
\eqref{HC1} $M^*(T)$ can be well approximated by a simple universal
interpolating function \cite{shag2}. The interpolation occurs
between the LFL ($M^*\simeq M^*+a_1T^2$) and NFL ($M^*\propto
T^{-2/3}$) regimes, thus describing the above crossover. Introducing
the dimensionless variable $y=T_N=T/T_M$, we obtain the desired
expression
\begin{equation}M^*_N(y)\approx c_0\frac{1+c_1y^2}{1+c_2y^{8/3}}.
\label{UN2}
\end{equation}
Here $M^*_N=M^*/M^*_M$ is the normalized effective mass,
$c_0=(1+c_2)/(1+c_1)$, $c_1$ and $c_2$ are fitting parameters,
parameterizing the Landau amplitude.

It follows from Eq. \eqref{HC5}, that the application of magnetic
field restores the LFL behavior, so that $M^*_M$ depends on $B$ as
\begin{equation}\label{LFLB}
    M^*_M\propto (B-B_{c0})^{-2/3},
\end{equation} while
\begin{equation}T_M\propto\mu_B(B-B_{c0}).\label{LFLT}\end{equation}
Employing Eqs. \eqref{LFLB} and \eqref{LFLT} to calculate $M^*_M$
and $T_M$, we conclude that Eq. \eqref{UN2} is valid to describe
the normalized effective mass in external fixed magnetic fields
with $y=T/(B-B_{c0})$. On the other hand, Eq. \eqref{UN2} is valid
when the applied magnetic field becomes a variable, while
temperature is fixed at $T=T_f$. In that case, it is convenient to
represent the variable as $y=(B-B_{c0})/T_f$.

\subsubsection{Schematic phase diagram of HF metal}\label{PHDS}

\begin{figure}[!ht]
\begin{center}
\vspace*{-0.5cm}
\includegraphics [width=0.90\textwidth]{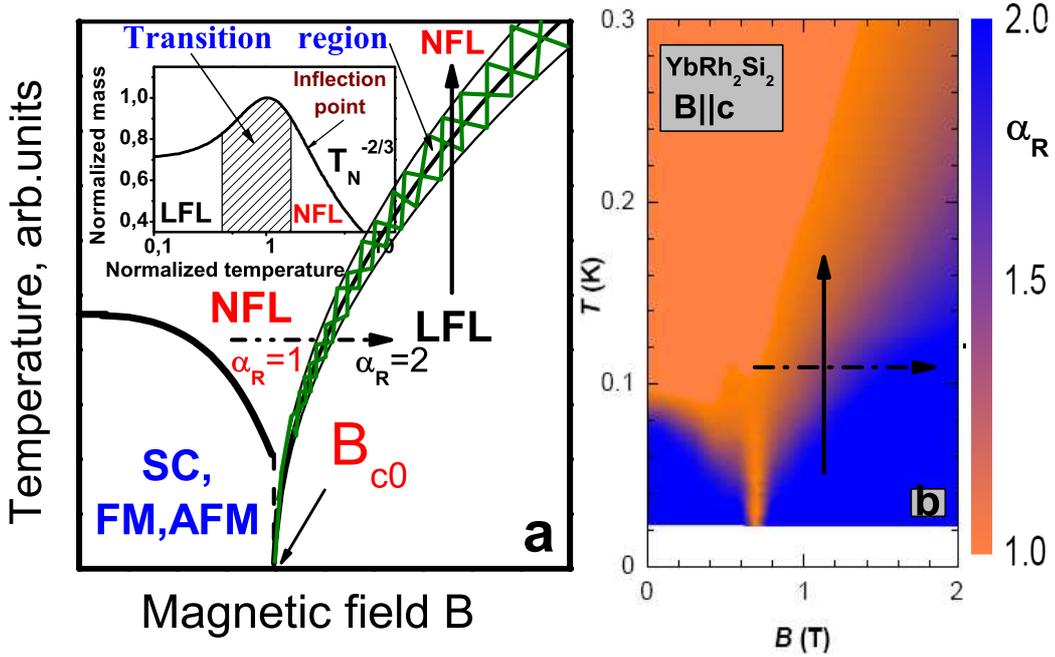}
\vspace*{-1.0cm}
\end{center}
\caption{ The panel {\bf a} represents a schematic phase diagram of
HF metals. $B_{c0}$ is magnetic field at which the effective mass
divergences. $\rm {SC,FM,AFM}$ denote the superconducting (SC),
ferromagnetic (FM) and antiferromagnetic (AFM) states, respectively.
At $B<B_{c0}$ the system can be in SC, FM or AFM states. The
vertical arrow shows the transition from the LFL to the NFL state at
fixed $B$ along $T$ with $M^*$ depending on $T$. The dash-dot
horizontal arrow illustrates the system moving from the NFL to LFL
state along $B$ at fixed $T$. The exponent $\alpha_R$ determines the
temperature dependent part of the resistivity, see Eq.
\eqref{RESIS}. In the LFL state the exponent  $\alpha_R=2$ and in
the NFL $\alpha_R=1$. In the transition regime the exponent evolves
from its LFL value to the NFL one. The inset shows a schematic plot
of the normalized effective mass versus the normalized temperature.
Transition regime, where $M^*_N$ reaches its maximum value $M^*_M$
at $T=T_M$, is shown by the hatched area both in the panel {\bf a}
and in the inset. The arrows mark the position of inflection point
in $M^*_N$ and the transition region. The panel {\bf b} shows the
experimental $T-B$ phase diagram of the exponent $\alpha_R(T,B)$ as
a function of temperature $T$ versus magnetic field $B$ \cite{cust}.
The evolution of $\alpha_R(T,B)$ is shown by the color: the yellow
color corresponds to $\alpha_R(T,B)=1$ (the NFL state) and the blue
color corresponds to $\alpha_R(T,B)=2$ (the LFL state). The NFL
behavior occurs at the lowest temperatures right at QCP tuned by
magnetic field. At rising magnetic fields $B>B_{c0}$ and $T\sim
T^*(B)$, the broad transition regime from the NFL state to the
field-induced LFL state occurs. As in the panel {\bf a}, the both
transitions from the LFL to the NFL state and from the NFL to LFL
state are shown by the corresponding arrows.}\label{PHD}
\end{figure}

The schematic phase diagram of HF metal is reported in Fig.
\ref{PHD}, panel {\bf a}. Magnetic field $B$ is taken as the control
parameter. In fact, the control parameter can be pressure $P$ or
doping (the number density) $x$ etc as well. At $B=B_{c0}$, FCQPT
takes place leading to a strongly degenerate state, where $B_{c0}$
is a critical magnetic field, such that at $B>B_{c0}$ the system is
driven towards the LFL state. We recall, that in our simple model
$B_{c0}$ is a parameter. The FC state is captured by the
superconducting (SC), ferromagnetic (FM), antiferromagnetic (AFM)
etc. states lifting the degeneracy. Below in Subsection \ref{HCEL4}
we consider the HF metal $\rm YbRh_2Si_2$. In that case,
$B_{c0}\simeq 0.06$ T ($B\bot c$) and at $T=0$ and $B<B_{c0}$ the
AFM state takes place with temperature dependent resistivity
$\rho(T)\propto T^2$ \cite{geg}. At elevated temperatures and fixed
magnetic fields the NFL regime occurs, while rising $B$ again drives
the system from the NFL state to the LFL one as shown by the
dash-dot horizontal arrow in Fig. \ref{PHD}. We consider the
transition region when the system moves from the NFL state to LFL
one along the horizontal arrow and also moves from LFL state to NFL
one along the vertical arrow as shown in Fig. \ref{PHD}. The inset
to Fig. \ref{PHD} demonstrates the scaling behavior of the
normalized effective mass $M^*_N=M^*/M^*_M$ versus normalized
temperature $T_N=T/T_M$, where $M^*_M$ is the maximum value that
$M^*$ reaches at $T=T_M$. The $T^{-2/3}$ regime is marked as NFL
since the effective mass depends strongly on temperature. The
temperature region $T\simeq T_M$ signifies the crossover (or the
transition region) between the LFL state with almost constant
effective mass and NFL behavior, given by $T^{-2/3}$ dependence.
Thus temperatures $T\sim T_M$ can be regarded as the crossover
region between the LFL and NFL states.

The transition (crossover) temperature $T^*(B)$ is not really the
temperature of a phase transition. It is
necessarily broad, very much depending on the criteria for
determination of the point of such a crossover, as it is seen from
the inset to Fig. \ref{PHD} {\bf a},
see e.g. Refs. \cite{geg,pnas}. As usual, the temperature $T^*(B)$ is
extracted from the field dependence of charge transport, for example
from the resistivity $\rho(T)$  given by Eq. \eqref{RESIS}. The LFL
state is characterized by the $T^{\alpha_R}$ dependence of the
resistivity with $\alpha_R=2$, see Subsection \ref{HCELR}.
The crossover (that is the transition regime shown
by the hatched area both in the panel {\bf a} of Fig. \ref{PHD} and
in its inset) takes
place at temperatures where the resistance starts to deviate from
the LFL behavior with $\alpha_R=2$ so that the exponent becomes
$1<\alpha_R<2$, see Subsection \ref{HCELR}. As it will be shown in
Subsection \ref{HCELR}, in the NFL state $\alpha_R=1$.

The panel {\bf b} of Fig.  \ref{PHD} represents the experimental
$T-B$ phase diagram of the exponent $\alpha_R(T,B)$ as a function of
temperature $T$ versus magnetic field $B$ \cite{cust}. The evolution
of $\alpha_R(T,B)$ is shown by the color: the yellow color
corresponds to $\alpha_R(T,B)=1$ and the blue color corresponds to
$\alpha_R(T,B)=2$. It is seen from the panel that at the critical
field $B_{c0}\simeq 0.66$ T ($B\| c$) the NFL behavior occurs down
to the lowest temperatures. While $\rm YbRh_2Si_2$ transits from the
NFL to LFL behavior under the application of magnetic field. It is
worthy to note that the phase diagram displayed in Fig. \ref{PHD}
(the panel {\bf a}) coincides with that of shown in the panel {\bf
b}.

A few remarks are in order here. As we shall see, the magnetic field
dependence of the effective mass or of other observable like the
longitudinal magnetoresistance do not have "peculiar points" like
maxima. The normalization is to be performed in the other points
like the inflection point at $T=T_{inf}$ (or at $B=B_{inf}$) shown
in the inset to Fig. \ref{PHD} by the arrow. Such a normalization is
possible since it is established on the scales, $T_{inf}\propto
T_M\propto(B-B_{c0})$. As a result, we obtain
\begin{equation}\label{LFLf}\mu_B(B_{inf}-B_{c0})\propto T_f.\end{equation}
It follows from Eq. \eqref{UN2} that in contrast to the Landau
paradigm of quasiparticles the effective mass strongly depends on
$T$ and $B$. This dependence leads to the extended quasiparticle
paradigm and forms the NFL behavior. Also from Eq. \eqref{UN2} the
scaling behavior of $M^*$ near QCP is revealed by the application of
appropriate physical scales to measure the effective mass, magnetic
field and temperature. At fixed magnetic fields, the characteristic
scales of temperature and of the function $M^*(T,B)$ are defined by
both $T_M$ and $M^*_M$ respectively. At fixed temperatures, the
characteristic scales are $(B_M-B_{c0})$ and $M^*_M$. From Eqs.
\eqref{LFLB} and \eqref{LFLT} it is seen that at fixed magnetic
fields, $T_M\to0$, and $M^*_M\to\infty$, and the width of the
transition region shrinks to zero as $B\to B_{c0}$ when these are
measured in "external" scales like $T$ in K, magnetic field $B$ in T
etc. In the same way, it follows from Eqs. \eqref{HC11} and
\eqref{LFLf} that at fixed temperature $T_f$,
$(B_{inf}-B_{c0})\to0$, and $M^*_M\to\infty$, and the width of the
transition region shrinks to zero as $T_f\to0$. Thus, the
application of the external scales obscure the scaling behavior of
the effective mass and the thermodynamic and transport properties.

\begin{figure} [! ht]
\begin{center}
\vspace*{-0.5cm}
\includegraphics [width=0.60\textwidth]{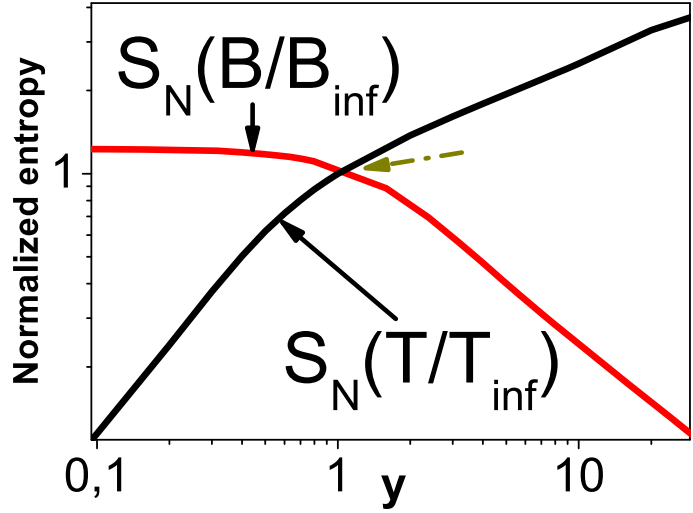}
\end{center}
\vspace*{-1.2cm} \caption{The normalized entropy $S_N(B/B_{inf})$
versus $y=B/B_{inf}$ and the normalized entropy $S_N(T/T_{inf})$
versus $y=T/T_{inf}$ calculated at fixed temperature and magnetic
field, correspondingly, are represented by the solid lines and shown
by the arrows. The inflection point is depicted by the dash-dot
arrow.}\label{STB}
\end{figure}

In what follows, we compute the effective mass using Eq. \eqref{HC1}
and employ Eq. \eqref{UN2} for estimations of the considered values. To
compute the effective mass $M^*(T,B)$, we solve Eq. \eqref{HC1} with
a quite general form of Landau interaction amplitude \cite{ckhz}.
Choice of the amplitude is dictated by the fact that the system has
to be at QCP, which means that the first two $p$-derivatives of the
single-particle spectrum $\varepsilon({\bf p})$ should equal zero.
Since the first derivative is proportional to the reciprocal
quasiparticle effective mass $1/M^*$, its zero just signifies QCP of
FCQPT. The second derivative must vanish; otherwise
$\varepsilon(p)-\mu$ has the same sign below and above the Fermi
surface, and the Landau state becomes unstable before $r\to0$
\cite{khodb,zb}. Zeros of these two subsequent derivatives mean that
the spectrum $\varepsilon({\bf p})$ has an inflection point at $p_F$
so that the lowest term of its Taylor expansion is proportional to
$(p-p_F)^3$. After solution of Eq. \eqref{HC1}, the obtained
spectrum has been used to calculate the entropy $S(B,T)$, which, in
turn, has been used to recalculate the effective mass $M^*(T,B)$ by
virtue of the well-known LFL relation $M^*(T,B)=S(T,B)/T$. Our
calculations of the normalized entropy as a function of the
normalized magnetic field $B/B_{inf}=y$ and as a function of the
normalized temperature $y=T/T_{inf}$ are reported in Fig. \ref{STB}.
Here $T_{inf}$ and $B_{inf}$ are the corresponding inflection points
in the function $S$. We normalize the entropy by its value at the
inflection point $S_N(y)=S(y)/S(1)$. As seen from Fig. \ref{STB},
our calculations corroborate the scaling behavior of the normalized
entropy, that is the curves at different temperatures and magnetic
fields merge into a single one in terms of the variable $y$. The
inflection point $T_{inf}$ in $S(T)$ makes $M^*(T,B)$ have its
maximum as a function of $T$, while $M^*(T,B)$ versus $B$ has no
maximum. We note that our calculations of the entropy confirm the
validity of Eq. \eqref{UN2} and the scaling behavior of the
normalized effective mass shown in Fig. \ref{PHD}.

\subsection{Non-Fermi liquid behavior in $\rm YbRh_2Si_2$} \label{HCEL4}

In this Subsection, we analyze the transition regime and the NFL
behavior of the HF metal $\rm YbRh_2Si_2$. We demonstrate that the
NFL behavior observed in the thermodynamic and transport properties
of $\rm YbRh_2Si_2$ can be described in terms of the scaling
behavior of the normalized effective mass. This allows us to
construct the scaled thermodynamic and transport properties
extracted from experimental facts in a wide range of the variation
of scaled variable and conclude that the extended quasiparticles
paradigm is strongly valid. We show that "peculiar points" of the
normalized effective mass give rise to the energy scales observed in
the thermodynamic and transport properties of HF metals. Our
calculations of the thermodynamic and transport properties are in
good agreement with the heat capacity, magnetization, longitudinal
magnetoresistance and magnetic entropy obtained in remarkable
measurements on the heavy fermion metal $\rm YbRh_2Si_2$
\cite{geg,steg,oesb,oesbs}. For $\rm YbRh_2Si_2$ the constructed
thermodynamic and transport functions extracted from experimental
facts show the scaling over three decades in the variable. The
energy scales in these functions are also explained \cite{dft373}.

\subsubsection{Heat capacity and the Sommerfeld coefficient} \label{HCEL5}

Exciting measurements of $C/T\propto M^*$ on samples of the new
generation of $\rm YbRh_2Si_2$ in different magnetic fields $B$ up
to 1.5 T \cite{oesb} allow us to identify the scaling behavior of
the effective mass $M^*$ and observe the different regimes of $M^*$
behavior such as the LFL regime, transition region from LFL to NFL
regimes, and the NFL regime itself. A maximum structure in
$C/T\propto M^*_M$ at $T_M$ appears under the application of
magnetic field $B$ and $T_M$ shifts to higher $T$ as $B$ is
increased. The value of $C/T=\gamma_0$ is saturated towards lower
temperatures decreasing at elevated magnetic fields.

The transition region corresponds to the temperatures where the
vertical arrow in the main panel {\bf a} of Fig. \ref{PHD} crosses
the hatched area. The width of the region, being proportional to
$T_M\propto (B-B_{c0})$ shrinks,  $T_M$ moves to zero temperature
and $\gamma_0\propto M^*$ increases as $B\to B_{c0}$. These
observations are in accord with the experimental facts \cite{oesb}.

\begin{figure} [! ht]
\begin{center}
\vspace*{-0.2cm}
\includegraphics [width=0.60\textwidth]{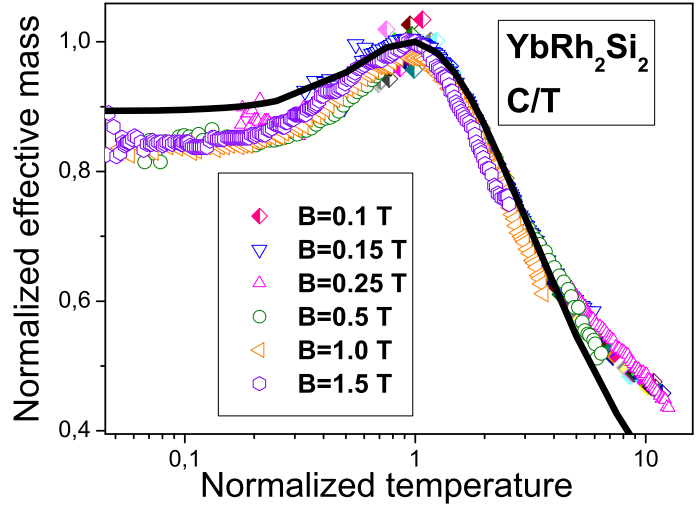}
\end{center}
\vspace*{-0.3cm} \caption{The normalized effective mass $M^*_N$
extracted from the measurements of the specific heat $C/T$ on $\rm
YbRh_2Si_2$ in magnetic fields $B$ \cite{oesb} listed in the legend.
Our calculations are depicted by the solid curve tracing the scaling
behavior of $M^*_N$.}\label{fig2H}
\end{figure}

To obtain the normalized effective mass $M^*_N$, the maximum
structure in $C/T$ was used to normalize $C/T$, and $T$ was
normalized by $T_M$. In Fig. \ref{fig2H} $M^*_N$ as a function of
normalized temperature $T_N$ is shown by geometrical figures, our
calculations are shown by the solid line. Figure \ref{fig2H} reveals
the scaling behavior of the normalized experimental curves - the
scaled curves at different magnetic fields $B$ merge into a single
one in terms of the normalized variable $y=T/T_M$. As seen, the
normalized mass $M^*_N$ extracted from the measurements is not a
constant, as would be for  LFL. The two regimes (the LFL regime and
NFL one) separated by the transition region, as depicted by the
hatched area in the inset to Fig. \ref{PHD} {\bf a}, are clearly
seen in Fig. \ref{fig2H} displaying good agreement between the
theory and experimental facts. It is worthy to note that the
normalization procedure allows us to construct the scaled function
$C/T$ extracted from the experimental facts in wide range variation
of the normalized temperature. Indeed, it integrates measurements of
$C/T$ taken at the application of different magnetic fields into
unique function of the normalized temperature which demonstrates the
scaling behavior over three decades in the normalized temperature as
seen from Fig. \ref{fig2H}. As seen from Figs. \ref{YBRHSI}, the NFL
behavior extends at least to temperatures up to few Kelvins. Thus,
we conclude that the extended quasiparticle paradigm does take into
account the remarkably large temperature ranges over which the NFL
behavior is observed. We note that  at these temperatures the
contribution coming from phonons is still small.

\subsubsection{Magnetization} \label{HCEL6}

Consider now the magnetization $M$ as a function of magnetic field
$B$ at fixed temperature $T=T_f$
\begin{equation}\label{CHIB}
M(B,T)=\int_0^B \chi(b,T)db,
\end{equation}
where the magnetic susceptibility $\chi$ is given by \cite{lanl1}
\begin{equation}\label{CHI}
\chi(B,T)=\frac{\beta M^*(B,T)}{1+F_0^a}.
\end{equation}
Here, $\beta$ is a constant and $F_0^a$ is the Landau amplitude
related to the exchange interaction. In the case of strongly
correlated systems $F_0^a\geq -0.9$ \cite{vollh,pfw,vollh1}.
Therefore, as seen from Eq. \eqref{CHI}, due to the normalization
the coefficients $\beta$ and $(1+F_0^a)$ drops out from the result,
and $\chi\propto M^*$.
\begin{figure} [! ht]
\begin{center}
\vspace*{-0.5cm}
\includegraphics [width=0.60\textwidth]{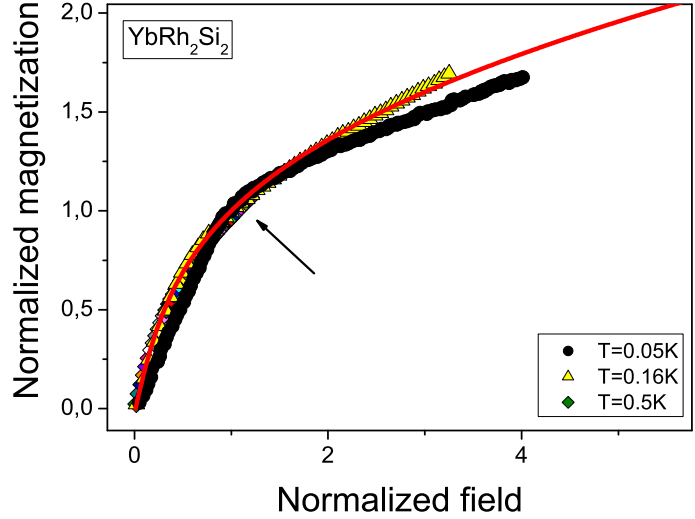}
\end{center}
\vspace*{-0.8cm} \caption{The field dependencies of the normalized
magnetization $M$ collected at different temperatures shown at right
bottom corner are extracted from measurements collected on $\rm{
YbRu_2Si_2}$ \cite{steg,oesbs}. The kink (shown by the arrow) is
clearly seen at the normalized field $B_N=B/B_k\simeq 1$. The solid
curve represents our calculations.}\label{fig3H}
\end{figure}
One might assume that $F_0^a$ can strongly depend on $B$. This is
not the case \cite{dft373,dftjtp}, since the Kadowaki-Woods ratio is
conserved \cite{geg1,kadw,natphys}, $A(B)/\gamma_0^2(B)\propto
A(B)/\chi^2(B)\propto const$, we have $\gamma_0\propto M^*\propto
\chi$. Note that the Sommerfeld coefficient does not depend on
$F_0^a$.

Our calculations show that the magnetization exhibits a kink at some
magnetic field $B=B_k$. The experimental magnetization demonstrates
the same behavior \cite{steg,oesbs}. We use $B_k$ and $M(B_k)$ to
normalize $B$ and $M$ respectively. The normalized magnetization
$M(B)/M(B_k)$ extracted from experimental facts depicted by the
geometrical figures and calculated magnetization shown by the solid
line are reported in Fig. \ref{fig3H}. As seen, the scaled data at
different $T_f$ merge into a single one in terms of the normalized
variable $y=B/T_k$. It is also seen, that these exhibit energy
scales separated by kink at the normalized magnetic field
$B_N=B/B_k=1$. The kink is a crossover point from the fast to slow
growth of $M$ at rising magnetic field. Figure \ref{fig3H} shows
that our calculations are in good agreement with the experimental
facts, and all the data exhibit the kink (shown by the arrow) at
$B_N\simeq 1$ taking place as soon as the system enters the
transition region corresponding to the magnetic fields where the
horizontal dash-dot arrow in the main panel {\bf a} of Fig.
\ref{PHD} crosses the hatched area. Indeed, as seen from Fig.
\ref{fig3H}, at lower magnetic fields $M$ is a linear function of
$B$ since $M^*$ is approximately independent of $B$. Then, Eqs.
\eqref{UN2} and \eqref{LFLB} show that at elevated magnetic fields
$M^*$ becomes a diminishing function of $B$ and generates the kink
in $M(B)$ separating the energy scales discovered in Refs.
\cite{steg,oesbs}. It is seen from Eq. \eqref{LFLf} that the
magnetic field $B_k$ at which the kink appears, $B_k\simeq
B_M\propto T_f$, shifts to lower $B$ as $T_f$ is decreased. This
observation is in accord with experimental facts \cite{steg,oesbs}.

\begin{figure} [! ht]
\begin{center}
\vspace*{-0.5cm}
\includegraphics [width=0.60\textwidth]{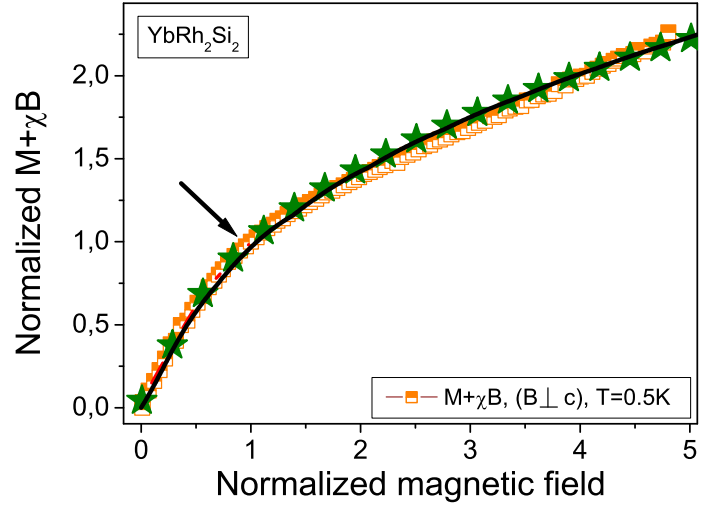}
\end{center}
\vspace*{-0.8cm} \caption{The field dependence of the normalized
``average'' magnetization $\underline{M}\equiv M+B\chi$ is shown by
squares and has been extracted from measurements collected on $\rm{
YbRu_2Si_2}$ \cite{steg}. The kink (shown by the arrow) is clearly
seen at the normalized field $B_N=B/B_k\simeq 1$. The solid curve
and stars (see text) represent our calculations.}\label{fig3HM}
\end{figure}
Consider now an ``average'' magnetization $\underline{M}\equiv
B\chi+M$ as a function of the magnetic field $B$ at fixed
temperature $T=T_f$ \cite{steg}. We again use $B_k$ and
$\underline{M}(B_k)$ to normalize $B$ and $\underline{M}$
respectively. The normalized $\underline{M}$ vs the normalized field
$B_N=B/B_K$ are shown in Fig. \ref{fig3HM}. Our calculations are
depicted by the solid line. The stars trace out our calculations of
$\underline{M}$ with $M^*(y)$ extracted from the data $C/T$ shown in
Fig. \ref{fig2H}. It is seen from Fig. \ref{fig3HM} that our
calculations are in good agreement with the experimental facts, and
all the data exhibit the kink (shown by arrow) at $B_N\simeq 1$
taking place as soon as the system enters the transition region
corresponding to the magnetic fields where the horizontal dash-dot
arrow in the main panel {\bf a} of Fig. \ref{PHD} crosses the
hatched area. Indeed, as seen from Fig. \ref{fig3HM}, at lower
magnetic fields $\underline{M}$ is a linear function of $B$ since
$M^*$ is approximately independent of $B$. It follows from Eq.
\eqref{LFLB} that at elevated magnetic fields $M^*$ becomes a
diminishing function of $B$ and generates the kink in
$\underline{M}(B)$ separating the energy scales discovered in Ref.
\cite{steg}. Then, it seen from Eq. \eqref{LFLf} that the magnetic
field $B_k\simeq B_M$ at which the kink appears shifts to lower $B$
as $T_f$ is decreased.

\subsubsection{Longitudinal magnetoresistance} \label{HCEL7}

\begin{figure} [! ht]
\begin{center}
\vspace*{-0.5cm}
\includegraphics [width=0.60\textwidth]{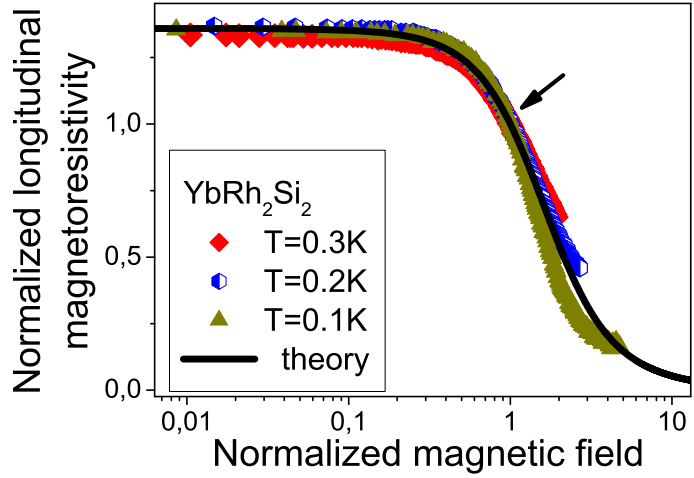}
\end{center}
\vspace*{-0.8cm} \caption{Magnetic field dependence of the
normalized magnetoresistance $\rho_N$ versus normalized magnetic
field. $\rho_N$ was extracted from LMR of $\rm YbRh_2Si_2$ at
different temperatures \cite{steg,oesbs} listed in the legend. The
inflection point is shown by the arrow, and the solid line
represents our calculations.}\label{fig4H}
\end{figure}

Consider a longitudinal magnetoresistance (LMR)
$\rho(B,T)=\rho_0+AT^2$ as a function of $B$ at fixed $T_f$. In that
case, the classical contribution to LMR due to orbital motion of
carriers induced by the Lorentz force is small, while the
Kadowaki-Woods relation \cite{geg1,kadw,ksch,tky,natphys},
$K=A/\gamma_0^2\propto A/\chi^2=const$, allows us to employ $M^*$ to
construct the coefficient $A$, since $\gamma_0\propto\chi\propto
M^*$. Omitting the classical contribution to LMR, we obtain that
$\rho(B,T)-\rho_0\propto(M^*)^2$. Fig. \ref{fig4H} reports the
normalized magnetoresistance
\begin{equation}\label{rn}
\rho_N(y)\equiv\frac{\rho(y)-\rho_0}{\rho_{inf}}=(M_N^*(y))^2
\end{equation}
versus normalized magnetic field $y=B/B_{inf}$ at different
temperatures, shown in the legend. Here $\rho_{inf}$ and $B_{inf}$
are LMR and magnetic field respectively taken at the inflection
point marked by the arrow in Fig. \ref{fig4H}. Both theoretical
(shown by the solid line) and experimental (marked by the
geometrical symbols) curves have been normalized by their inflection
points, which also reveal the scaling behavior - the scaled curves
at different temperatures merge into a single one as a function of
the variable $y$ and show the scaling behavior over three decades in
the normalized magnetic field. The transition region at which LMR
starts to decrease is shown in the inset to Fig. \ref{PHD} {\bf a}
by the hatched area. Obviously, as seen from Eq. \eqref{LFLf}, the
width of the transition region being proportional to $B_M\simeq
B_{inf}\propto T_f$ decreases as the temperature $T_f$ is lowered.
In the same way, the inflection point of LMR, generated by the
inflection point of $M^*$ shown in the inset to Fig. \ref{PHD} by
the arrow, shifts to lower $B$ as $T_f$ is decreased. All these
observations are in good agreement with the experimental facts
\cite{steg,oesbs}.

\subsubsection{Magnetic entropy} \label{HCEL8}

The evolution of the derivative of magnetic entropy $dS(B,T)/dB$ as
a function of magnetic field $B$ at fixed temperature $T_f$ is of
great importance since it allows us to study the scaling behavior of
the derivative of the effective mass $TdM^*(B,T)/dB\propto
dS(B,T)/dB$. While the scaling properties of the effective mass
$M^*(B,T)$ can be analyzed via LMR, see Fig. \ref{fig4H}.
\begin{figure} [! ht]
\begin{center}
\includegraphics [width=0.60\textwidth]{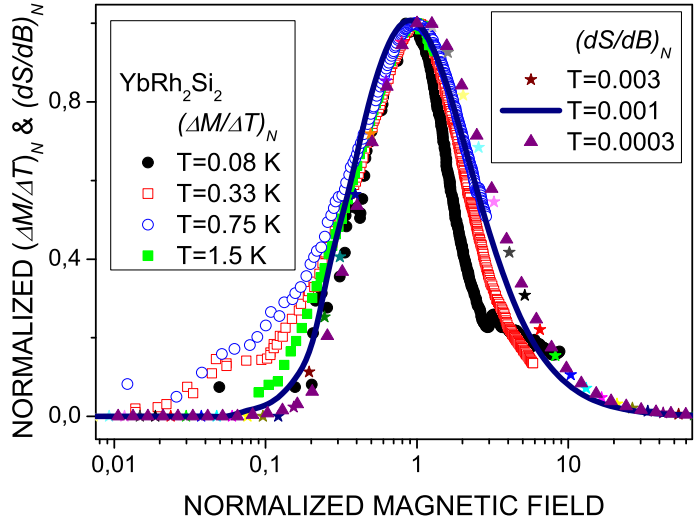}
\vspace*{-0.6cm}
\end{center}
\caption{ Normalized magnetization difference divided by temperature
increment $(\Delta M/\Delta T)_N$ vs normalized magnetic field at
fixed temperatures (listed in the legend in the upper left corner)
is extracted from the facts collected  on $\rm YbRh_2Si_2$
\cite{gegtok}. Our calculations of the normalized derivative
$(dS/dB)_N\simeq (\Delta M/\Delta T)_N$ vs normalized magnetic field
are given at fixed dimensionless temperatures $T/\mu$ (listed in the
legend in the upper right corner). All the data are shown by the
geometrical figures depicted in the legend at the upper left
corner.} \label{fig5H}
\end{figure}
As seen from Eqs. \eqref{UN2} and \eqref{LFLf}, at $y\leq 1$
the derivative $$-\frac{dM_N(y)}{dy}\propto y$$ with
$y=(B-B_{c0})/(B_{inf}-B_{c0})\propto (B-B_{c0})/T_f$. We note that
the effective mass as a function of $B$ does not have the maximum.
At elevated $y$ the derivative $-dM_N(y)/dy$ possesses a maximum at
the inflection point and then becomes a diminishing function of $y$.
Upon using the variable $y=(B-B_{c0})/T_f$, we conclude that at
decreasing temperatures, the leading edge of the function
$-dS/dB\propto -TdM^*/dB$ becomes steeper and its maximum at
$(B_{inf}-B_{c0})\propto T_f$ is higher. These observations are in
quantitative agreement with striking measurements of the
magnetization difference divided by temperature increment, $-\Delta
M/\Delta T$, as a function of magnetic field at fixed temperatures
$T_f$ collected on $\rm YbRh_2Si_2$ \cite{gegtok}. We note that
according to the well-known thermodynamic equality $dM/dT=dS/dB$, and
$\Delta M/\Delta T\simeq dS/dB$. To carry out a quantitative
analysis of the scaling behavior of $-dM^*(B,T)/dB$, we calculate as
described above the entropy $S(B,T)$ shown in Fig. \ref{STB} as a
function of $B$ at fixed dimensionless temperatures $T_f/\mu$ shown
in the upper right corner of Fig. \ref{fig5H}. This figure reports
the normalized $(dS/dB)_N$ as a function of the normalized magnetic
field. The function $(dS/dB)_N$ is obtained by normalizing
$(-dS/dB)$ by its maximum taking place at $B_M$, and the field $B$
is scaled by $B_M$. The measurements of $-\Delta M/\Delta T$ are
normalized in the same way and depicted in Fig. \ref{fig5H} as
$(\Delta M/\Delta T)_N$ versus normalized field. It is seen from
Fig. \ref{fig5H} that our calculations are in good agreement with
the experimental facts and both the experimental functions $(\Delta M/\Delta
T)_N$ and the calculated $(dS/dB)_N$ show the scaling behavior
over three decades in the normalized magnetic field.

\subsubsection{Energy scales} \label{HCEL9}

\begin{figure} [! ht]
\begin{center}
\vspace*{-0.6cm}
\includegraphics [width=0.60\textwidth]{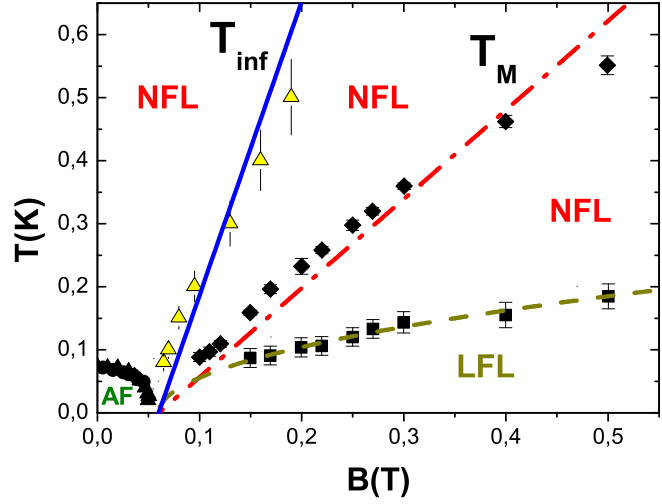}
\end{center}
\vspace*{-0.8cm} \caption{Temperature versus magnetic field $T-B$
phase diagram for $\rm YbRh_2Si_2$. Solid circles represent the
boundary between AF and NFL states. The solid squares  denote the
boundary of the NFL and LFL regime \cite{geg,steg,oesbs} shown by
the dashed line which is approximated by $\sqrt{B-B_{c0}}$
\cite{obz}. Diamonds mark the maximums $T_M$ of $C/T$ \cite{oes}
shown in Fig. \ref{fig2H}. The dash-dot line is approximated by
$T_M\propto a(B-B_{c0})$, $a$ is a fitting parameter, see Eq.
\eqref{LFLT}. Triangles along the solid line denote $T_{inf}$ in LMR
\cite{steg,oesbs} sown in Fig. \ref{fig5H}, and the solid line
represents the function $T_{inf}\propto b(B-B_{c0})$, $b$ is a
fitting parameter, see Eq. \eqref{LFLf}.}\label{fig6H}
\end{figure}
Fig. \ref{fig6H} reports $T_{inf}$ and $T_M$ versus $B$ depicted by
the solid and dash-dotted lines, respectively. The boundary between
the NFL and LFL regimes is shown by the dashed line, and AF marks
the antiferromagnetic state. The corresponding data are taken from
Ref. \cite{steg,oesbs,gegtok}. It is seen that our calculations are
in good agreement with the experimental facts. In Fig. \ref{fig6H},
the solid and dash-dotted lines corresponding to the functions
$T_{inf}$ and $T_M$, respectively, represent the positions of the
kinks separating the energy scales in $C$ and $M$ reported in Ref.
\cite{steg,gegtok}. Furthermore, our calculations are in accord with
experimental facts, and we conclude that the energy scales are
reproduced by Eqs. \eqref{LFLT} and \eqref{LFLf} and related to the
peculiar points $T_{inf}$ and $T_M$ of the normalized effective mass
$M^*_N$ which are shown by the arrows in the inset to Fig.
\ref{PHD}.

At $B\to B_{c0}$ both $T_{inf}\to 0$ and $T_{M}\to 0$, thus the LFL
and the transition regimes of both $C/T$ and $M$ as well as those of
LMR and the magnetic entropy are shifted to very low temperatures.
Therefore due to experimental difficulties these regimes cannot
often be observed in experiments on HF metals. As it is seen from
Figs. \ref{fig2H}, \ref{fig3H}, \ref{fig4H}, \ref{fig5H} and
\ref{fig6H}, the normalization allows us to construct the unique
scaled thermodynamic and transport functions extracted from the
experimental facts in a wide range of the variation of the scaled
variable $y$. As seen from the mentioned Figures, the constructed
normalized thermodynamic and transport functions show the scaling
behavior over three decades in the normalized variable.

\subsection{Electric resistivity of HF metals}\label{HCELR}

The electric resistivity of strongly correlated Fermi systems,
$\rho(T)=\rho_0+\Delta\rho_1(B,T)$, is determined by the
effective mass, because of the Kadowaki-Woods relation
$\Delta\rho_1(B,T)=A(B,T)T^2\propto(M^*(B,T)T)^2$, see
Subsection \ref{HCEL7} and Refs.
\cite{ksch,tky,natphys}, and therefore the temperature dependence of
the effective mass discussed above can be observed in measurements
of the resistivity of HF metals. At temperatures $T\ll
T^*(B)$, the system is in the LFL state, the behavior of the
effective mass as $x\to x_{FC}$ is described by Eq. \eqref{HC5}, and
the coefficient $A(B)$ can be represented as
\begin{equation}
A(B)\propto\frac{1}{(B-B_{c0})^{4/3}}.\label{HC16}
\end{equation}
In this regime, the resistivity behaves as
$\Delta\rho_1=c_1T^2/(B-B_{c0})^{4/3}\propto T^2$. The second
regime, a highly correlated Fermi liquid determined by Eq.
(\ref{HC11}), is characterized by the resistivity dependence
$\Delta\rho_1=c_2T^2/(T^{2/3})^2\propto T^{2/3}$. The third regime
at $T>T^*(B)$ is determined by Eq. (\ref{HC14}). In that case we
obtain $\Delta\rho_1=c_3T^2/(T^{1/2})^2\propto T$. If the system is
above the quantum critical line as shown in Fig. \ref{fig1}, the
dependence of the effective mass on temperature is given by Eq.
\eqref{FL12}, so we obtain from Eq. \eqref{HC15} that the
quasiparticle damping $\gamma(T)\propto T$ \cite{dkss}. As a
result, we see that the resistivity dependence on temperature is
$\Delta\rho_1=c_4T$ \cite{khodrho}. Here, $c_1$, $c_2$, $c_3$ and
$c_4$ are constants. If the system at the transition regime, as
shown by the arrows in Fig. \ref{PHD}, the dependence of the
effective mass on temperature cannot be characterized be a single
exponent as it is clearly seen from the inset to Fig. \ref{PHD}
{\bf a}. So we have that $\Delta\rho_1\propto T^{\alpha_R}$ with
$1<\alpha_R<2$. We note that all temperature dependencies
corresponding to these regimes have been observed in measurements
involving the heavy-fermion metals $\rm CeCoIn_5$, $\rm YbRh_2Si_2$
and $\rm YbAgGe$ \cite{geg,pag,pag2,bud,movsh}.

\subsection{Magnetic susceptibility and magnetization measured on $\rm CeRu_2Si_2$}

Experimental investigations of the magnetic properties of $\rm
CeRu_2Si_2$ down to the lowest temperatures (down to 170 mK) and
ultrasmall magnetic fields (down to 0.21 mT) have shown neither
evidence of the magnetic ordering, superconductivity  nor
conventional LFL behavior \cite{takah}. These results imply a
magnetic quantum critical point in $\rm CeRu_2Si_2$ is absent and
the critical field $B_{c0}=0$. Even if the
magnetic quantum critical point were there it should maintain the NFL
behavior over four decades in temperature. Such a strong influence
can hardly exist within the framework of conventional quantum phase
transitions.

\begin{figure} [! ht]
\begin{center}
\includegraphics [width=0.60\textwidth] {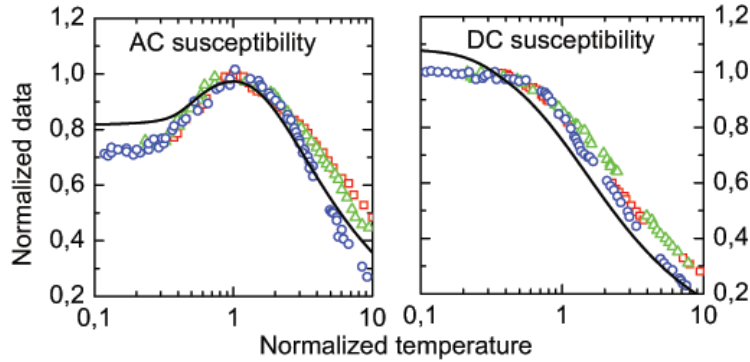}
\end {center}
\caption {The normalized magnetic susceptibility
$\chi(B,T)/\chi(B,T_M)$ (the left panel) and normalized
magnetization ${M_B}(B,T)/{M_B}(B,T_M)$ (DC susceptibility, the
right panel) for $\rm CeRu_2Si_2$ in magnetic fields 0.20 mT
(squares), 0.39 mT (triangles), and 0.94 mT (circles) as functions
of the normalized temperature $T/T_M$ \cite{takah}. The solid lines
depict the calculated scaling behavior \cite{ckhz} as described
in Subsection \ref{PHDS}.}\label{Fig12}
\end{figure}

Temperature dependence on a logarithmic scale of the normalized $AC$
susceptibility $\chi(B,T)$ is shown at different applied magnetic
fields $B$ as indicated in the left panel of Fig. \ref{Fig12} versus
normalized temperature. The right panel of the Figure shows the
normalized static magnetization $M_B(B,T)$
(DC susceptibility) in the same normalized
temperature range. The temperature is normalized to $T_M$ (the
temperature at which the susceptibility reaches its peak value), the
susceptibility is normalized to the peak value $\chi(B,T_M)$, and
the magnetization is normalized to $M_B(B,T\to0)$, for each value of
the field \cite{takah}.
If we use Eq. \eqref{CHIB} and the definition of susceptibility
\eqref{CHI}, we conclude that the susceptibility and magnetization
also demonstrate the scaling behavior and can be represented by the
universal function \eqref{UN2} of the single variable $y$, if they are
respectively normalized as discussed above.
We see from Fig. \ref{Fig12} that at finite field strengths $B$, the
curve describing $\chi(B,T)/\chi(B,T_M)$ has a peak at a certain
temperature $T_M$, while $M_B(B,T)/{M_B}(B,T_M)$
has no such peak \cite{shag4,ckhz,shag5}.
This behavior agrees well with the
experimental results \cite{shag4,ckhz,shag5} obtained in
measurements on $\rm CeRu_2Si_2$ \cite{takah}. We note
that such behavior of the susceptibility is not typical of ordinary
metals and cannot be explained within the scope of theories that
take only ordinary quantum phase transitions into account
\cite{takah}.
\begin{figure}[!ht]
\begin{center}
\includegraphics [width=0.60\textwidth]{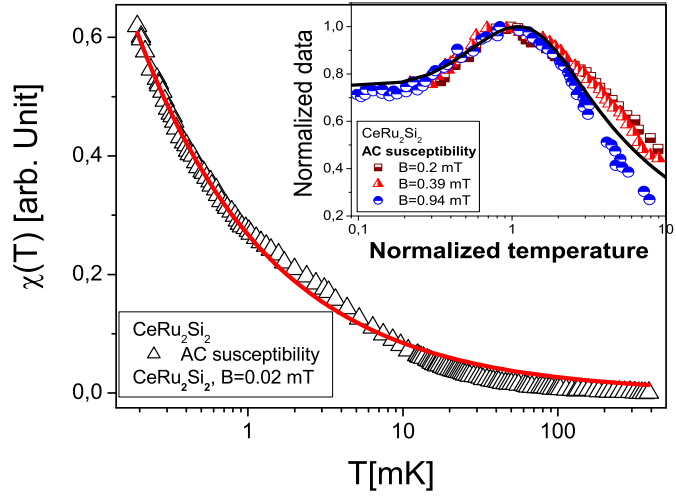}
\end{center}
\caption{Temperature dependence of the $AC$ susceptibility
$\chi_{AC}$ for $\rm{CeRu_2Si_2}$. The solid curve is a fit for the
data shown by the triangles at $B=0.02$ mT \cite{takah} and
represented by the function $\chi(T)=a/\sqrt{T}$ given by Eq.
\eqref{HC14} with $a$ being a fitting parameter. Inset shows the
normalized effective mass versus normalized temperature $T_N$
extracted from $\chi_{AC}$ measured at different fields as indicated
in the inset \cite{takah}. The solid curve traces the universal
behavior of $M^*_N(T_N)$ determined by Eq. \eqref{UN2}. Parameters
$c_1$ and $c_2$ are adjusted to fit the average behavior of the
normalized effective mass $M^*_N$.}\label{MRM}
\end{figure}

To verify Eq. \eqref{HC14} and illustrate the transition from LFL
behavior to NFL one, we use measurements of $\chi_{AC}(T)$ in
$\rm{CeRu_2Si_2}$ at magnetic field $B=0.02$ mT at which this HF
metal demonstrates the NFL behavior down to lowest temperatures
\cite{takah}. Indeed, in this case we expect that LFL regime to start
to form at temperatures lower than $T_M\sim \mu_B B\sim 0.01$ mK as
it follows from Eq. \eqref{LFLT}. It is seen from Fig. \ref{MRM}
that Eq. \eqref{HC14} gives good description of the
experimental facts in the
extremely wide range of temperatures: the susceptibility $\chi_{AC}$
as a function of $T$, is not a constant upon cooling, as would be
for a Fermi liquid, but shows a $1/\sqrt{T}$ divergence over almost
four decades in temperature. The inset of Fig. \ref{MRM} exhibits a
fit for $M^*_N$ extracted from measurements of $\chi_{AC}(T)$ at
different magnetic fields, clearly indicating the change from LFL
behavior at $T_N<1$ to NFL one at $T_N>1$ when the system moves
along the vertical arrow as shown in Fig. \ref{PHD}. It seen from
Figs. \ref {Fig12} and \ref{MRM} that
the function given by Eq. \eqref{UN2} represents
a good approximation for $M^*_N$ within the extended paradigm. In
Subsection \ref{HCEL4} we have seen that the same is true in the
case of $\rm YbRh_2Si_2$ with the AF quantum critical point. We
conclude that both alloys $\rm{CeRu_2Si_2}$ and $\rm YbRh_2Si_2$
demonstrate the universal NFL thermodynamic behavior, independent of
the details of the HF metals such as their lattice structure,
composition and magnetic ground state. This conclusion implies also
that numerous QCPs related to conventional quantum phase transitions
assumed to be responsible for the NFL behavior of different HF
metals can be well reduced to a single QCP related to FCQPT and
accounted for within the extended quasiparticle paradigm
\cite{epl2}.

\subsection{Transverse magnetoresistance in the HF metal $\rm CeCoIn_5$}\label{MAGRES}

Our comprehensive theoretical study of both
the longitudinal and transverse magnetoresistance (MR)
shows that it is (similar
to other thermodynamic characteristics like magnetic susceptibility,
specific heat, etc) governed by the scaling behavior of the
quasiparticle effective mass. The crossover from negative to
positive MR occurs at elevated temperatures and fixed magnetic
fields when the system transits from the LFL behavior to NFL one and
can be well captured by this scaling behavior.

By definition, MR is given by
\begin{equation}
\rho_{mr}(B,T)=\frac{\rho(B,T)-\rho(0,T)}{\rho(0,T)},\label{HC23}
\end{equation}
We apply Eq. (\ref{HC23}) to study MR of strongly correlated
electron liquid versus temperature $T$ as a function of magnetic
field $B$. The resistivity $\rho(B,T)$ is
\begin{equation}
\rho(B,T)=\rho_0+\Delta\rho(B,T)+\Delta\rho_{L}(B,T), \label{RBT}
\end{equation}
where $\rho_0$ is a residual resistance, $\Delta\rho=c_1AT^2$, $c_1$
is a constant. The classical
contribution $\Delta\rho_{L}(B,T)$ to MR due to orbital motion of
carriers induced by the Lorentz force obeys the Kohler's rule
\cite{zim}. We note that $\Delta\rho_{L}(B)$ $\ll\rho(0, T)$ as it
is assumed in the weak-field approximation. To calculate $A$, we
again use the quantities $\gamma_0=C/T\propto M^*$ and/or
$\chi\propto M^*$ as well as employ the fact that the Kadowaki-Woods
ratio $K=A/\gamma_0^2\propto A/\chi^2=const$. As a result, we obtain
$A\propto (M^*)^2$, so that $\Delta\rho(B,T)=c(M^*(B,T))^2T^2$ and
$c$ is a constant. Suppose that the temperature is not very low, so
that $\rho_0\leq \Delta\rho(B=0,T)$, and $B\geq B_{c0}$.
Substituting (\ref{RBT}) into (\ref{HC23}), we find that
\cite{plamr}
\begin{equation}
\rho_{mr}\simeq \frac{\rho_0+\Delta\rho_{L}(B,T)}{\rho(0,T)}+cT^2
\frac{(M^*(B,T))^2-(M^*(0, T))^2}{\rho(0,T)}\label{HC25}.
\end{equation}

Consider the qualitative behavior of MR described by Eq.
(\ref{HC25}) as a function of $B$ at a certain temperature $T=T_0$.
In weak magnetic fields, when the system exhibits NFL (see Fig.
\ref{PHD}), the main contribution to MR is made by the term
$\Delta\rho_{L}(B)$, because the effective mass is independent of
the applied magnetic field. Hence, $|M^*(B, T)-M^*(0,T)|/M^*(0,
T)\ll1$ and the leading contribution is made by
$\Delta\rho_{L}(B)$. As a result, MR is an increasing function of
$B$. When $B$ becomes so high that $T^*(B)\sim \mu_B(B-B_{c0})\sim
T_0$, the difference $(M^*(B, T)-M^*(0, T))$ becomes negative
because $M^*(B, T)$ is now the diminishing function of $B$ given by
Eq. \eqref{LFLB}. Thus, MR as a function of $B$ reaches its maximal
value at $T^*(B)\sim T_N(B)\sim  T_0$. At further increase of
magnetic field, when $T_M(B)>T_0$, the effective mass $M^*(B,T)$
becomes a decreasing function of $B$. As $B$ increases,
\begin{equation}
\frac{(M^*(B,T)-M^*(0,T))} {M ^*(0, T)}\to -1,\label{HC25a}
\end{equation}
and the magnetoresistance, being a decreasing function of $B$, can
reach its negative values.

Now we study the behavior of MR as a function of $T$ at fixed value
$B_0$ of magnetic field. At low temperatures $T\ll T^*(B_0)$, it
follows from Eqs. \eqref{UN2} and \eqref{HC5} that
$M^*(B_0,T)/M^*(0,T)\ll1$, and it is seen from Eq. \eqref{HC25a}
that $\rho_{mr}(B_0,T)\sim-1$, because
$\Delta\rho_{L}(B_0,T)/\rho(0,T)\ll1$. We note that $B_0$ must be
relatively high to guarantee that $M^*(B_0,T)/M^*(0,T)$ $\ll1$. As
the temperature increases, MR increases, remaining negative. At
$T\simeq T^*(B_0)$, MR is approximately zero, because
$\rho(B_0,T)\simeq\rho(0,T)$ at this point. This allows us to
conclude that the change of the temperature dependence of
resistivity $\rho(B_0,T)$ from quadratic to linear manifests itself
in the transition from negative to positive MR. One can also say
that the transition takes place when the system goes from the LFL
behavior to the NFL one. At $T\geq T^*(B_0)$, the leading
contribution to MR is made by $\Delta\rho_{L}(B_0,T)$ and MR reaches
its maximum. At $T_M(B_0)\ll T$, MR is a decreasing function of the
temperature, because
\begin{equation}
\frac{|M^*(B,T)-M^*(0, T)|}{M^*(0,T)} \ll1,\label{HC25b}
\end{equation}
and $\rho_{mr}(B_0,T)\ll1$. Both transitions (from positive to
negative MR with increasing $B$ at  fixed temperature $T$ and from
negative to positive MR with increasing $T$ at fixed $B$ value) have
been detected in measurements of the resistivity of $\rm{ CeCoIn_5}$
in a magnetic field \cite{pag}.
\begin{figure} [! ht]
\begin{center}
\vspace*{-0.5cm}
\includegraphics [width=0.60\textwidth]{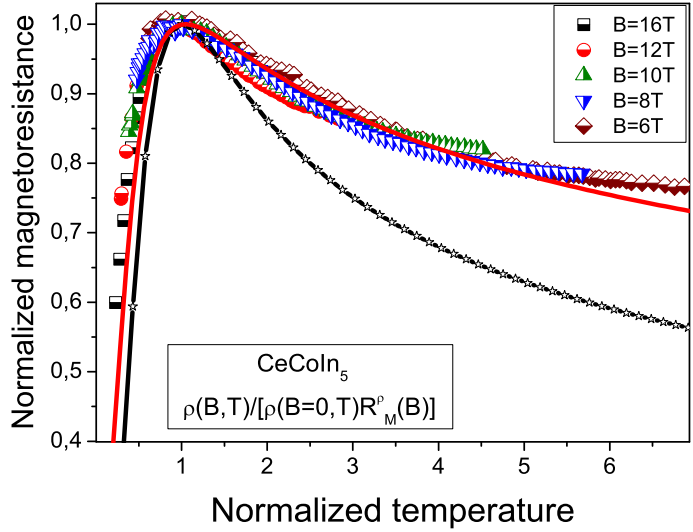}
\end {center}\vspace{-0.8cm}
\caption {The normalized magnetoresistance $R^{\rho}_N(y)$ given by
Eq. \eqref{HC27} versus normalized temperature $y=T/T_{\rm Rm}$.
$R^{\rho}_N(y)$ was extracted from MR shown in Fig. \ref{MRT} and
collected on $\rm{ CeCoIn_5}$ at fixed magnetic fields $B$
\cite{pag} listed in the right upper corner. The starred  line
represents our calculations based on Eqs. \eqref{UN2} and
\eqref{HC27} with the parameters extracted from $AC$ susceptibility
of $\rm CeRu_2Si_2$ (see the caption to Fig. \ref{MRM}). The solid
line displays our calculations based on Eqs. \eqref{HC28} and
\eqref{HC27}; only one parameter was used to fit the data, while the
other were extracted from the $AC$ susceptibility measured on $\rm{
CeRu_2Si_2}$.}\label{MRTU}
\end{figure}

Let us turn to quantitative analysis of MR \cite{plamr}. As it was
mentioned above, we can safely assume that the classical
contribution $\Delta\rho_{L}(B,T)$ to MR is small as compared to
$\Delta\rho(B,T)$. Omission of $\Delta\rho_{L}(B,T)$ allows us to
make our analysis and results transparent and simple while the
behavior of $\Delta\rho_{L}(B_0,T)$ is not known in the case of HF
metals. Consider the ratio $R^{\rho}=\rho(B,T)/\rho(0,T)$ and assume
for a while that the residual resistance $\rho_0$ is small in
comparison with the temperature dependent terms. Taking into account
Eq. \eqref{RBT} and $\rho(0,T)\propto T$, we obtain from Eq.
\eqref{HC25} that
\begin{equation}
R^{\rho}=\rho_{mr}+1= \frac{\rho(B,T)}{\rho(0, T)}\propto
T(M^*(B,T))^2,\label{HC26}
\end{equation}
and consequently, from Eqs.  \eqref{UN2} and \eqref{HC26} that the ratio
$R^{\rho}$ reaches its maximal value $R^{\rho}_M$ at some
temperature $T_{\rm Rm}\sim T_M$. If the ratio is measured in units
of its maximal value $R^{\rho}_M$ and $T$ is measured in units of
$T_{\rm Rm}\sim T_M$ then it is seen from Eqs. \eqref{UN2} and
\eqref{HC26} that the normalized MR
\begin{equation}
R^{\rho}_N(y)= \frac{R^{\rho}(B,T)}{R^{\rho}_M(B)}\simeq
y(M^*_N(y))^2\label{HC27}
\end{equation}
becomes a function of the only variable $y=T/T_{\rm Rm}$.
To verify Eq. \eqref{HC27}, we use MR obtained in measurements on
CeCoIn$_5$, see Fig. 1(b) of Ref. \cite{pag}. The results of the
normalization procedure of MR are reported in Fig. \ref{MRTU}. It is
clearly seen that the data collapse into the same curve, indicating
that the normalized magnetoresistance $R^{\rho}_N$ obeys the
scaling behavior well given by Eq. \eqref{HC27}. This scaling behavior
obtained directly from the experimental facts is a vivid evidence
that MR behavior is predominantly governed by the effective mass
$M^*(B,T)$.

Now we are in position to calculate $R^{\rho}_N(y)$ given by Eq.
\eqref{HC27}. Using Eq. \eqref{UN2} to parameterize $M^*_N(y)$, we
extract parameters $c_1$ and $c_2$ from measurements of the magnetic
$AC$ susceptibility $\chi$ on $\rm CeRu_2Si_2$ \cite{takah} and
apply Eq. \eqref{HC27} to calculate the normalized ratio. It is seen
that the calculations shown by the starred line in Fig. \ref{MRTU}
start to deviate from experimental points at elevated temperatures.
To improve the coincidence, we employ  Eq. \eqref{HC14} which
describes the behavior of the effective mass at elevated
temperatures and ensures that at these temperatures the resistance
behaves as $\rho(T)\propto T$. To correct the behavior of $M^*_N(y)$
at rising temperatures $M^*\sim T^{-1/2}$, we add a term to Eq.
\eqref{UN2} and obtain
\begin{equation}
M^*_N(y)\approx\frac{M^*(x)}{M^*_M}\left[\frac{1+c_1y^2}{1+c_2y^{8/3}}
+c_3\frac{\exp(-1/y)}{\sqrt{y}}\right], \label{HC28}\end{equation}
where $c_3$ is a parameter. The last term on the right hand side of
Eq. \eqref{HC28} makes $M^*_N$ satisfy Eq. \eqref{HC14} at
temperatures $T/T_M>2$.
In Fig. \ref{MRTU}, the fit of $R^{\rho}_N(y)$ by Eq. \eqref{HC28}
is shown by the solid line. Constant $c_3$ is taken as a fitting
parameter, while the other were extracted from $AC$ susceptibility
of $\rm{ CeRu_2Si_2}$ as described in the caption to Fig. \ref{MRM}.
\begin{figure} [! ht]
\begin{center}
\vspace*{-0.5cm}
\includegraphics [width=0.60\textwidth]{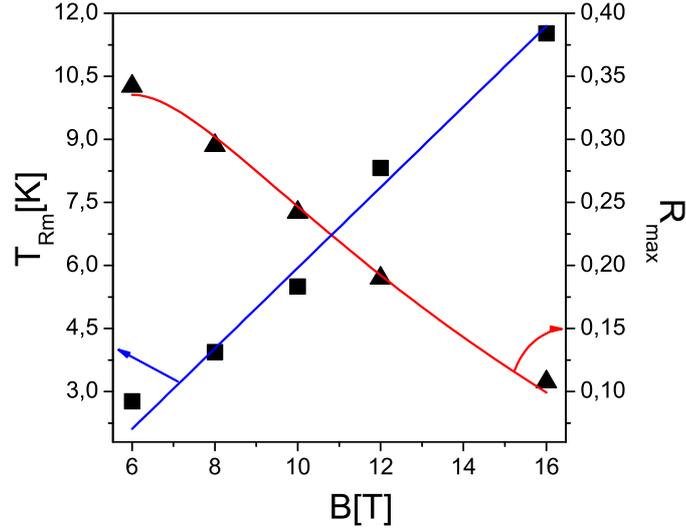}
\end {center}\vspace{-1.0cm}
\caption {The peak temperatures $T_{\rm Rm}$ (squares) and the peak
values $R_{\rm max}$ (triangles) versus magnetic field $B$ extracted
from measurements of MR \cite{pag}. The solid lines represent our
calculations based on Eqs. \eqref{RTB1} and \eqref{RTB2}.}\label{TB}
\end{figure}

Before discussing  the magnetoresistance $\rho_{mr}(B,T)$ given by
Eq. \eqref{HC23}, we consider the magnetic field dependence of
both the MR peak value $R_{\rm max}(B)$ and the corresponding peak
temperature $T_{\rm Rm}(B)$. It is possible to use Eq. \eqref{HC26}
which relates the position and value of the peak with the function
$M^*(B,T)$. Since $T_{\rm Rm}\propto \mu_B(B-B_{c0})$,  $B$ enters Eq.
\eqref{HC26} only as tuning parameter of QCP, as both $\Delta\rho_L$
and $\rho_0$ were omitted. At $B\to B_{c0}$ and $T\ll T_{\rm
Rm}(B)$, this omission is not correct since $\Delta\rho_L$ and
$\rho_0$ become comparable with $\Delta\rho(B,T)$. Therefore, both
$R_{\rm max}(B)$ and $T_{\rm Rm}(B)$ are not characterized by any
critical field, being continuous functions at the quantum critical
field $B_{c0}$, in contrast to $M^*(B,T)$ whose peak value diverges
and the peak temperature tends to zero at $B_{c0}$ as seen
from Eqs. \eqref{LFLB} and \eqref{LFLT}. Thus, we have to take into
account $\Delta\rho_{L}(B,T)$ and $\rho_0$ which prevent $T_{\rm
Rm}(B)$ from vanishing and make $R_{\rm max}(B)$ finite at $B\to
B_{c0}$. As a result, we have to replace $B_{c0}$ by some effective
field $B_{eff}<B_{c0}$ and take $B_{eff}$ as a parameter which
imitates the contributions coming from both $\Delta\rho_{L}(B,T)$
and $\rho_0$. Upon modifying Eq. \eqref{HC26} by taking into account
$\Delta\rho_{L}(B,T)$ and $\rho_0$, we obtain
\begin{figure} [! ht]
\begin{center}
\vspace*{-0.6cm}
\includegraphics [width=0.60\textwidth]{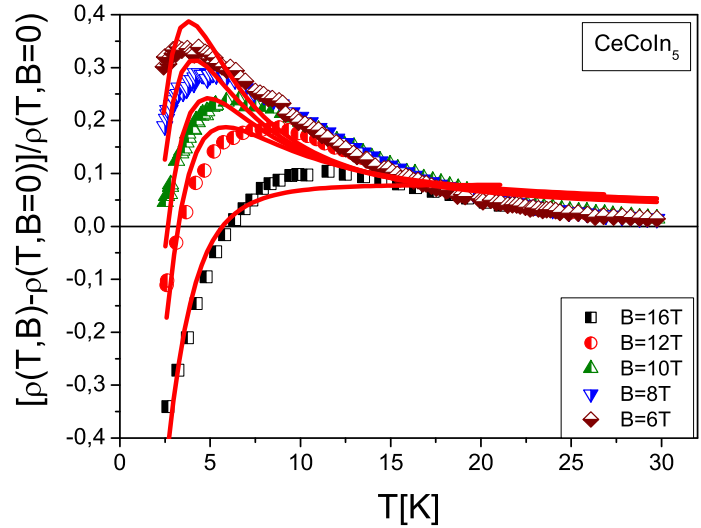}
\end {center}\vspace{-1.0cm}
\caption {MR versus temperature $T$ as a function of magnetic field
$B$. The experimental data on MR were collected on $\rm{ CeCoIn_5}$
at fixed magnetic field $B$ \cite{pag} shown in the right bottom
corner of the Figure. The solid lines represent our calculations,
Eq. \eqref{UN2} is used to fit the effective mass entering Eq.
\eqref{HC27}.}\label{MRT}
\end{figure}
\begin{eqnarray}
&&T_{\rm Rm}(B)\simeq b_1(B-B_{eff}),\label{RTB1}\\
&&R_{\rm max}(B)\simeq
\frac{b_2(B-B_{eff})^{-1/3}-1}{b_3(B-B_{eff})^{-1}+1}.\label{RTB2}
\end{eqnarray}
Here $b_1$, $b_2$, $b_3$ and $B_{eff}$ are the fitting parameters.
It is pertinent to note that while deriving Eq. \eqref{RTB2} we use
Eq. \eqref{RTB1} with substitution $(B-B_{eff})$ for $T$. Then, Eqs.
\eqref{RTB1} and \eqref{RTB2} are not valid at $B\lesssim B_{c0}$.
In Fig. \ref{TB}, we show the field dependence of both $T_{\rm Rm}$
and $R_{\rm max}$, extracted from measurements of MR \cite{pag}.
Clearly both $T_{\rm Rm}$ and $R_{\rm max}$ are well described by
Eqs. \eqref{RTB1} and \eqref{RTB2} with $B_{eff}=$3.8 T. We note
that this value of $B_{eff}$ is in good agreement with observations
obtained from the $T-B$ phase diagram of $\rm{ CeCoIn_5}$, see the
position of the MR maximum shown by the filled circles in Fig. 3 of
Ref. \cite{pag}.

To calculate $\rho_{mr}(B,T)$, we apply Eq. \eqref{HC27} to describe
its universal behavior, Eq. \eqref{UN2} for the effective mass along
with Eqs. \eqref{RTB1} and \eqref{RTB2} for MR parameters. Figure
\ref{MRT} shows the calculated MR versus temperature as a function
of magnetic field $B$ together with the experimental points from
Ref. \cite{pag}. We recall that the contributions coming from
$\Delta\rho_{L}(B,T)$ and $\rho_0$ were omitted. As seen from Fig.
\ref{MRT}, our description of the experiment is good.

\subsection{Magnetic-field-induced reentrance of Fermi-liquid behavior
and spin-lattice relaxation rates in ${\rm
YbCu_{5-x}Au_x}$}\label{TT1}

One of the most interesting and puzzling issues in the research on
HF metals is their anomalous dynamic and relaxation properties. It
is important to verify whether quasiparticles with effective mass
$M^*$ still exist and determine the physical properties of the muon
and $\rm ^{63}Cu$ nuclear spin-lattice relaxation rates $1/T_1$ in
HF metals throughout their temperature - magnetic field phase
diagram, see Fig. \ref{PHD}. This phase diagram comprises both LFL
and NFL regions as well as NFL-LFL transition or the crossover
region, where magnetic-field-induced LFL reentrance occurs.
Measurements of the muon and $^{63}$Cu nuclear spin-lattice
relaxation rates $1/T_1$ in ${\rm {YbCu_{4.4}Au_{0.6}}}$ have shown
that it differs substantially from ordinary Fermi liquids obeying
the Korringa law \cite{osn}. Namely, it was reported that for $T\to
0$ reciprocal relaxation time diverges as $1/T_1T\propto T^{-4/3}$
following the behavior predicted by the self-consistent
renormalization (SCR) theory \cite{kn}. The static uniform
susceptibility $\chi$ diverges as $\chi\propto T^{-2/3}$ so that
$1/T_1T$ scales with $\chi^2$. Latter result is at variance with
SCR theory \cite{osn}. Moreover, the application of magnetic field
$B$ restores the LFL behavior from initial the NFL one,
significantly reducing $1/T_1$. These experimental findings are
hard to explain within both the conventional LFL approach and in
terms of other approaches like SCR theory \cite{osn,kn}.

In this Subsection we show that the above anomalies along with
magnetic-field-induced reentrance of LFL properties are indeed
determined by the dependence of the quasiparticle effective mass
$M^*$ on magnetic field $B$ and temperature $T$ and demonstrate that
violations of the Korringa law also come from $M^*(B,T)$ dependence.
Our theoretical analysis of experimental data on the base of FCQPT
approach permits not only to explain the above two experimental
facts in a unified manner, but to unveil their universal properties,
relating the peculiar features of both longitudinal
magnetoresistance and specific heat in $\rm YbRh_2Si_2$ to the
behavior of spin-lattice relaxation rates.

To discuss the deviations from the Korringa law in light of NFL
properties of ${\rm {YbCu_{4.4}Au_{0.6}}}$, we notice that in LFL
theory the spin-lattice relaxation rate $1/T_1$ is determined by the
quasiparticles near the Fermi level. The above relaxation rate is
related to the decay amplitude of the quasiparticles, which in turn
is proportional to the density of states at the Fermi level
$N(E_F)$. Formally, the spin-lattice relaxation rate is determined by
the imaginary part $\chi''$ of the low-frequency dynamical magnetic
susceptibility $\chi({\bf q}, \omega \to 0)$, averaged over momentum
${\bf q}$
\begin{equation}\label{chi1}
\frac{1}{T_1}=\frac{3T}{4\mu_B^2}\sum _{\bf q}A_{\bf q}A_{-{\bf
q}}\frac{\chi''({\bf q},\omega)}{\omega},
\end{equation}
where $A_{\bf q}$ is the hyperfine coupling constant of the muon (or
nuclei) with the spin excitations at wave vector $\bf q$ \cite{kn}.
If $A_{\bf q}\equiv A_0$ is independent of $q$, then standard LFL
theory yields the relation
\begin{equation}\label{chi3}
\frac{1}{T_1T}=\pi A^2_0N^2(E_F).
\end{equation}
Equation \eqref{chi3} can be viewed as Korringa law. Since in our
FCQPT approach the physical properties of the system under
consideration are determined by the effective mass $M^*(T,B,x)$, we
express $1/T_1T$ in Eq. \eqref{chi3} via it. This is accomplished
with the standard expression \cite{lanl1} $N(E_F)=M^*p_F/\pi^2$,
rendering Eq. \eqref{chi3} to the form
\begin{equation}\label{chi5}
\frac{1}{T_1T}=\frac{A^2_0p_F^2}{\pi^3}M^{*2}\equiv\eta
\left[M^*(T,B,x)\right]^2,
\end{equation}
where $\eta=(A^2_0p_F^2)/\pi^3=$const. The empirical expression
\begin{equation}\label{chi5a}
\frac{1}{T_1T}\propto \chi^2(T),
\end{equation}
extracted from experimental data in ${\rm {YbCu_{5-x}Au_x}}$
\cite{osn}, follows explicitly from Eq. \eqref{chi5} and well-known
LFL relations $M^*\propto \chi \propto C/T$.

\begin{figure} [! ht]
\begin{center}
\vspace*{-0.3cm}
\includegraphics [width=0.60\textwidth]{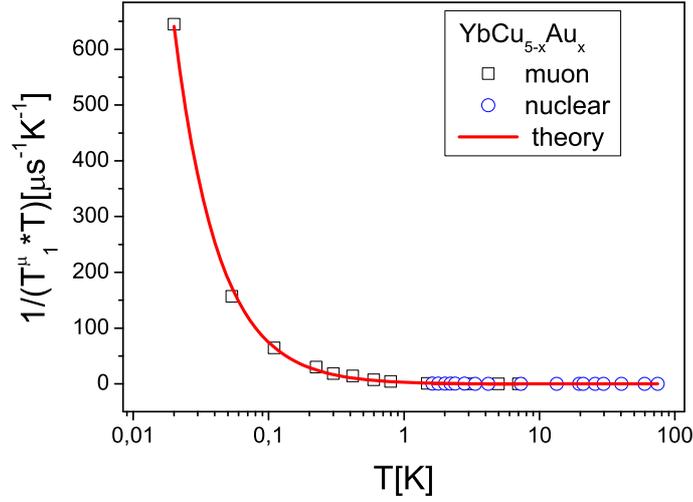}
\end{center}
\vspace*{-0.3cm} \caption{Temperature dependence of muon (squares)
and nuclear (circles) spin-lattice relaxation rates (divided by
temperature) for ${\rm {YbCu_{4.4}Au_{0.6}}}$ at zero magnetic field
\cite{osn}. The solid curve represents our calculations based on Eq.
\eqref{t43}. }\label{fi3}
\end{figure}

In what follows, we compute the effective mass as it was explained
in Subsection \ref{PHDS} and employ Eq. \eqref{UN2} for estimations
of obtained values \cite{T1T}. The decay law given by Eq.
\eqref{HC11} along with Eq. \eqref{chi5} permits to express
the relaxation rate in this temperature range as
\begin{equation}\label{t43}
\frac{1}{T_1T}=a_1+a_2T^{-4/3}\propto \chi^2(T),
\end{equation}
where $a_{1}$ and $a_{2}$ are fitting parameters. The dependence
\eqref{t43} is reported in Fig. \ref{fi3} along with experimental
points for the muon and nuclear spin-lattice relaxation rates in ${\rm
{YbCu_{4.4}Au_{0.6}}}$ at zero magnetic field \cite{osn}. It is seen
from Fig. \ref{fi3} that Eq. \eqref{t43} gives good
description of the experiment in the extremely wide temperature
range. This means that the extended paradigm is valid and
quasiparticles survive in close vicinity of FCQPT, while the
observed violation of Korringa law comes from the dependence of the
effective mass on temperature.

\begin{figure} [! ht]
\begin{center}
\vspace*{-0.2cm}
\includegraphics [width=0.60\textwidth]{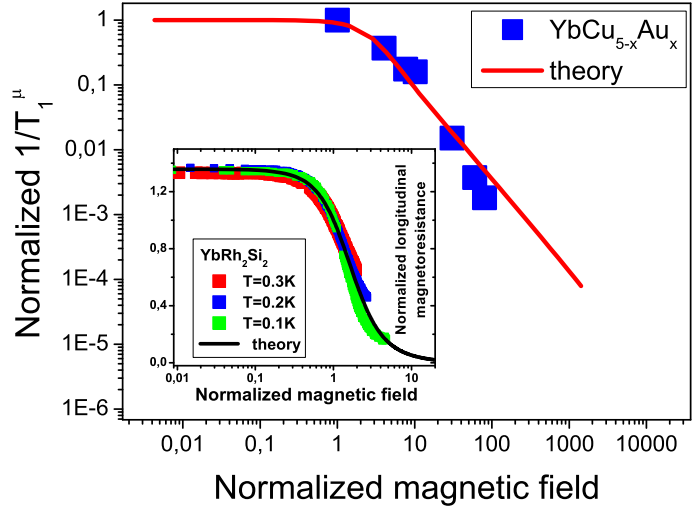}
\vspace*{-0.20cm}
\end{center}
\caption{Magnetic field dependence of normalized at the inflection
point muon spin-lattice relaxation rate $1/T_{1N}^\mu$ extracted from
measurements \cite{osn} on ${\rm {YbCu_{4.4}Au_{0.6}}}$ along with
our calculations of $B$-dependence of the quasiparticle effective
mass. Inset shows the normalized LMR $R^{\rho}_N(y)$ versus
normalized magnetic field. $R^{\rho}_N(y)$ was extracted from LMR of
$\rm YbRh_2Si_2$ at different temperatures \cite{steg} listed in the
legend. The solid curves represent our calculations. }\label{fig:fi2}
\end{figure}

Figure \ref{fig:fi2} displays magnetic field dependence of
normalized  muon spin-lattice relaxation rate $1/T_{1N}^\mu$ in ${\rm
{YbCu_{5-x}Au_x}}$ (x=0.6) along with our theoretical
$B$-dependence. To obtain the latter theoretical curve we (for fixed
temperature and in magnetic field $B$) employ Eq. \eqref{chi5} and
solve the Landau integral equation to calculate $M^*(T,B)$ as it was
described in Subsection \ref{PHDS}. We note that
the normalized effective mass $M_N^*(y)$ was
obtained by normalizing $M^*(T,B)$ at
its infection point shown in the inset to Fig. \ref{PHD}.

It is instructive to compare the LMR analyzed in Subsection
\ref{HCEL7} and $1/T_1^\mu$. LMR $\rho(B,T)=\rho_0+\rho_B+A(B,T)T^2$
is as a function of $B$ at fixed $T$, where $\rho_0$ is the residual
resistance, $\rho_B$ is the contribution to LMR due to orbital
motion of carriers induced by the Lorentz force, and $A$ is the
coefficient. As we see in Subsection \ref{HCEL7}, $\rho_B$ is small
and we omit this contribution. The Kadowaki-Woods relation allows us
to employ $M^*$ to calculate $A(B,T)$. As a result,
$\rho(B,T)-\rho_0\propto(M^*)^2$, and $1/T_{1N}^\mu\propto(M^*)^2$
as seen from Eq. \eqref{chi5}. As a result, we see that that LMR and
the magnetic field dependence of normalized  muon spin-lattice
relaxation rate $1/T_{1N}^\mu$ can be evaluated from the same equation
\begin{equation}\label{rn}
R_N^\rho(y)=\frac{\rho(y)-\rho_0}{\rho_{\rm{inf}}}=\frac1{T_{1N}^\mu}
=(M_N^*(y))^2.
\end{equation}

Inset to Fig. \ref{fig:fi2} reports the normalized LMR vs normalized
magnetic field $y=B/B_{\rm{inf}}$ at different temperatures, shown
in the legend. Here $\rho_{\rm{inf}}$ and $B_{\rm{inf}}$ are
respectively LMR and magnetic field taken at the inflection point.
The inflection points of both LMR and $1/T_{1N}$ are generated by
the inflection point of $M^*$ shown in the inset to Fig. \ref{PHD}
{\bf a} by the arrow. The transition region where LMR starts to
decrease is shown in the inset by the hatched area and takes place
when the system moves along the horizontal dash-dot arrow. We note
that the same normalized effective mass has been used to calculate
both $1/T_{1N}^\mu$ in ${\rm {YbCu_{4.4}Au_{0.6}}}$ and the
normalized LMR in $\rm YbRh_2Si_2$. Thus, Eq. \eqref{rn} determines
the close relationship between the quite different dynamic
properties, showing the validity of the quasiparticle extended
paradigm. In Fig. \ref{fig:fi2}, both theoretical and experimental
curves have been normalized by their inflection points, which also
reveals the scaling behavior - the curves at different
temperatures merge into a single one in terms of the scaled variable
$y$. Figure \ref{fig:fi2} shows clearly that both normalized
magnetoresistance $R^{\rho}_N$ and reciprocal spin-lattice
relaxation time obey well the scaling behavior given by Eq.
\eqref{rn}. This fact obtained directly from the experimental
findings is vivid evidence that the behavior of both the above
quantities is predominantly governed by the field and temperature
dependence of the effective mass.

We remark that the same normalized effective mass determines the
behavior of the thermodynamic and transport properties in $\rm
YbRh_2Si_2$, see Subsection \ref{HCEL4}. It is seen from the Figures
presented in Subsection \ref{HCEL4} that our calculations of the
effective mass offer good descriptions of such different quantities
as the relaxation rates $(1/T_1T)$ and the transport (LMR) and
thermodynamic properties in such different
HF metals as$\rm YbCu_{5-x}Au_x$
and $\rm YbRh_2Si_2$. It is pertinent to note
that the obtained good
description makes an impressive case in
favor of the reliability of the
quasiparticle extended paradigm.

\subsection{Relationships between critical magnetic fields $B_{c0}$
and $B_{c2}$ in HF metals and high-$T_c$
superconductors}\label{BC0C2}

Recently, in high-$T_c$ superconductors, exciting measurements
revealing their physics have been performed. One type of the
measurements demonstrate the existence of Bogoliubov quasiparticles
(BQ) in the superconducting state \cite{mat,norm,valln}. While in
the pseudogap regime at temperatures above $T_c$ when the
superconductivity vanishes, a strong indication of the pairing of
electrons or the formation  of preformed pairs of electrons was
observed, while the gap continues to follow the simple d-wave form
\cite{norm,valln}. Another type of the measurement explored the
normal state induced by the application of magnetic field, when the
transition from the NFL behavior to LFL one occurs \cite{pnas}. As
we have mentioned in Subsection \ref{HTCHF}, there are the
experimental relationships between the critical fields $B_{c2}\geq
B_{c0}$, where $B_{c2}$ is the field destroying the superconducting
state, and $B_{c0}$ is the critical field at which the magnetic
field induced QCP takes place. Now we show that $B_{c2}\geq
B_{c0}$. We note that to study the aforementioned transition
experimentally in high-$T_c$ superconductors, strong magnetic
fields of $B\geq B_{c2}$ are required; earlier, such investigation
was technically inaccessible. An attempt to study the transition
experimentally had already been made \cite{mac}.

\begin{figure}[!ht]
\begin{center}
\includegraphics [width=0.60\textwidth]{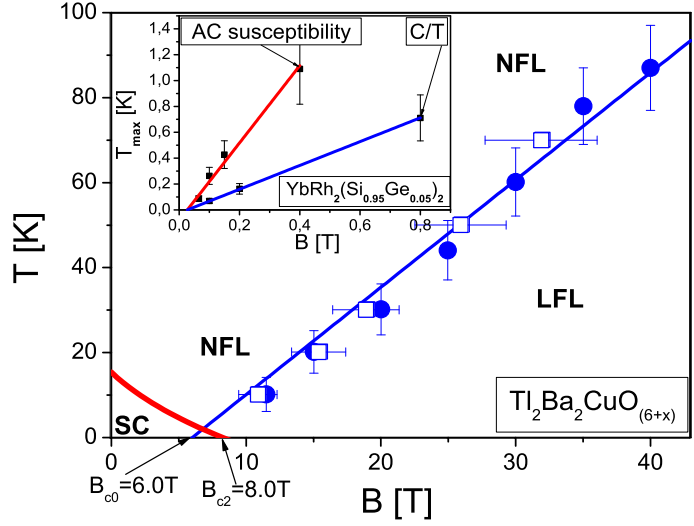}
\end{center}
\caption{$T-B$ phase diagram of the superconductor
Tl$_2$Ba$_2$CuO$_{6+x}$. The crossover (from LFL to NFL regime) line
$T^*(B)$ is given by the Eq. \eqref{LFLT}. Open squares and solid
circles are experimental values \cite{pnas}. Thick line represents
the boundary between the superconducting and normal phases. Arrows
near the bottom left corner indicate the critical magnetic field
$B_{c2}$ destroying the superconductivity and the critical field
$B_{c0}$. Inset displays the peak temperatures $T_{\rm max}(B)$,
extracted from measurements of $C/T$ and $\chi_{AC}$ on
YbRh$_2$(Si$_{0.95}$Ge$_{0.05}$)$_2$ \cite{cust,geg3} and
approximated by straight lines Eq. \eqref{LFLT}. The lines intersect
at $B\simeq 0.03$ T.}\label{TMM}
\end{figure}

Let us now consider the $T-B$ phase diagram of the high-$T_c$
superconductor Tl$_2$Ba$_2$CuO$_{6+x}$ shown in Fig. \ref{TMM}. The
substance is a superconductor with $T_c$ from 15 K to 93 K, being
controlled by oxygen content \cite{pnas}. In Fig. \ref{TMM} open
squares and solid circles show the experimental values of the
crossover temperature from the LFL to NFL regimes \cite{pnas}. The
solid line given by Eq. \eqref{HC14a} shows our fit with $B_{c0}=6$
T that is in good agreement with $B_{c0}=5.8$ T obtained from the
field dependence of the charge transport \cite{pnas}.

\begin{figure} [! ht]
\begin{center}
\vspace*{-0.5cm}
\includegraphics [width=0.60\textwidth]{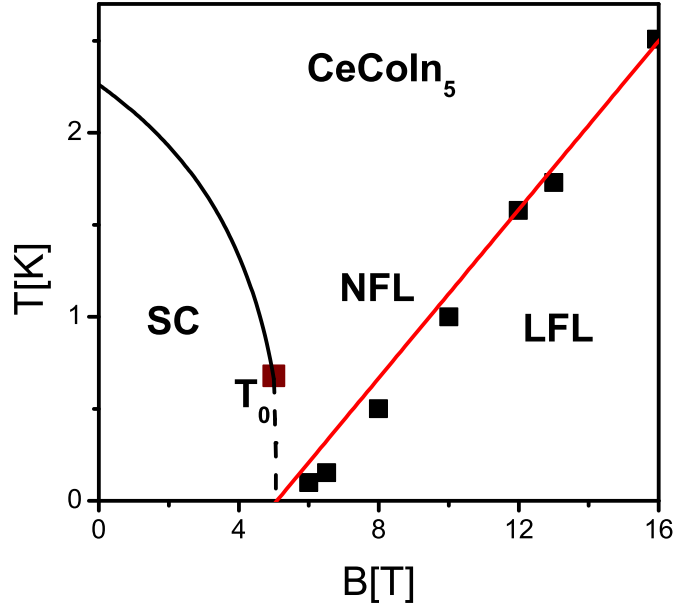}
\end{center}
\vspace*{-0.7cm} \caption{$T-B$ phase diagram of the $\rm CeCoIn_5$
heavy fermion metal. The interface between the superconducting and
normal phases is shown by the solid line to the square where the
phase transition becomes a first-order phase transition. At $T<T_0$,
the phase transition is a first-order phase transition \cite{bian}.
The interface between the superconducting and normal phases is shown
by the dashed line. The solid straight line represented by Eq.
\eqref{LFLT} with the experimental points \cite{pag2} shown by
squares is the interface between the LFL and NFL
states.}\label{CeCo}
\end{figure}
As it is seen from Fig. \ref{TMM}, the linear behavior agrees well
with the experimental data \cite{pnas,{plamr}}. The peak
temperatures $T_{\rm max}$ shown in the inset to Fig. \ref{TMM},
depict the maxima of $C(T)/T$ and $\chi_{AC}(T)$ measured on
YbRh$_2$(Si$_{0.95}$Ge$_{0.05}$)$_2$ \cite{cust,geg3}. From Fig.
\ref{TMM}, $T_{\rm max}$ is seen to shift to higher values with
increase of the applied magnetic field and both functions can be
represented by straight lines intersecting at $B\simeq 0.03$ T.
This observation is in good agreement with experiments
\cite{cust,geg3}. Clearly from Fig. \ref{TMM} the critical field
$B_{c2}=8$ T destroying the superconductivity is close to
$B_{c0}=6$ T. We now show that this is more than a simple
coincidence, and $B_{c2}\gtrsim B_{c0}$. Indeed, at $B>B_{c0}$ and
low temperatures $T<T^*(B)$, the system is in its LFL state. The
superconductivity is then destroyed since the superconducting gap
is exponentially small as we have seen in Subsection \ref{SCFT}. At
the same time, there is the FC state at $B<B_{c0}$ and this
low-field phase has large prerequisites towards superconductivity
as in this case the gap is a linear function of the superconducting
coupling constant $\lambda_0$ as it was shown in Subsection
\ref{SCFT}. We note that this is exactly the case in $\rm CeCoIn_5$
where $B_{c0}\simeq B_{c2}\simeq 5$ T \cite{pag} as seen from Fig.
\ref{CeCo}, while the application of pressure makes $B_{c2}>B_{c0}$
\cite{ronn}. However, if the superconducting coupling constant is
rather weak then antiferromagnetic order wins the competition. As a
result, $B_{c2}=0$, while $B_{c0}$ can be finite as in $\rm
YbRh_2Si_2$ and $\rm{YbRh_2(Si_{0.95}Ge_{0.05})_2}$
\cite{geg,geg3}.

Comparing the phase diagram of Tl$_2$Ba$_2$CuO$_{6+x}$  with that of
$\rm CeCoIn_5$ shown in Figs. \ref{TMM} and \ref{CeCo} respectively,
it is possible to conclude that they are similar in many respects.
Further, we note that the superconducting boundary line $B_{c2}(T)$
at decreasing temperatures acquires a step, i.e. the corresponding
phase transition becomes first order \cite{shagstep,bian}. This
leads us to speculate that the same may be true for
Tl$_2$Ba$_2$CuO$_{6+x}$. We expect that in the NFL state the
tunneling conductivity is asymmetrical function of the applied
voltage, while it becomes symmetrical at the application of
increased magnetic fields when Tl$_2$Ba$_2$CuO$_{6+x}$ transits to
the LFL behavior, as it predicted to be in $\rm CeCoIn_5$
\cite{shagpopov}.

It follows from Eq. \eqref{ABDR} that it is impossible to observe
the relatively high values of $A(B)$ since in our case
$B_{c2}>B_{c0}$. We note that Eq. \eqref{ABDR} is valid when
the superconductivity is destroyed by the application of magnetic
field, otherwise the effective mass is also finite being given by
Eq. \eqref{SC7}. Therefore, as was mentioned above, in high-$T_c$
QCP is poorly accessible to experimental observations being
"hidden in superconductivity". Nonetheless, thanks to the
experimental facts \cite{pnas}, we have
seen in Subsection \ref{HTCHF} that it is possible to
study QCP of high-$T_c$ \cite{PLA686}.
As seen from Fig. \ref{ff3}, the facts give evidences that the
physics underlying the field-induced reentrance of LFL behavior, is
the same for both HF metals and hight-$T_c$ superconductors.

\subsection{Scaling behavior of the HF $\rm CePd_{1-x}Rh_x$
ferromagnet} \label{CePdRh}

QCP can arise by suppressing the transition temperature $T_{NL}$ of a
ferromagnetic (FM) (or antiferromagnetic (AFM)) phase to zero by
tuning some control {parameter $\zeta$} other than temperature, such
as pressure $P$, magnetic field $B$, or doping $x$ as it takes place
in the HF ferromagnet $\rm{CePd_{1-x}Rh_x}$ \cite{sereni,pikul} or
the HF metal $\rm{CeIn_{3-x}Sn_x}$ \cite{kuch}.

The HF metal $\rm {CePd_{1-x}Rh_x}$ evolves from ferromagnetism at
$x=0$ to a non-magnetic state at some critical concentration
$x_{FC}$. Utilizing the extended quasiparticle paradigm picture and
the concept of FCQPT, we address the question about the NFL behavior
of the ferromagnet $\rm {CePd_{1-x}Rh_x}$ and show that it coincides
with that of the antiferromagnets $\rm
{YbRh_2(Si_{0.95}Ge_{0.05})_2}$ and $\rm {YbRh_2Si_2}$, and
paramagnets $\rm {CeRu_2Si_2}$ and $\rm {CeNi_2Ge_2}$. We again
conclude that the NFL behavior, being independent of the
peculiarities of a specific alloy, is universal. Incidentally,
numerous quantum critical points assumed to be responsible for the
NFL behavior of different HF metals can be well reduced to the only
quantum critical point related to FCQPT \cite{cepdrh,physb}.

\begin{figure} [! ht]
\begin{center}
\includegraphics [width=0.60\textwidth]{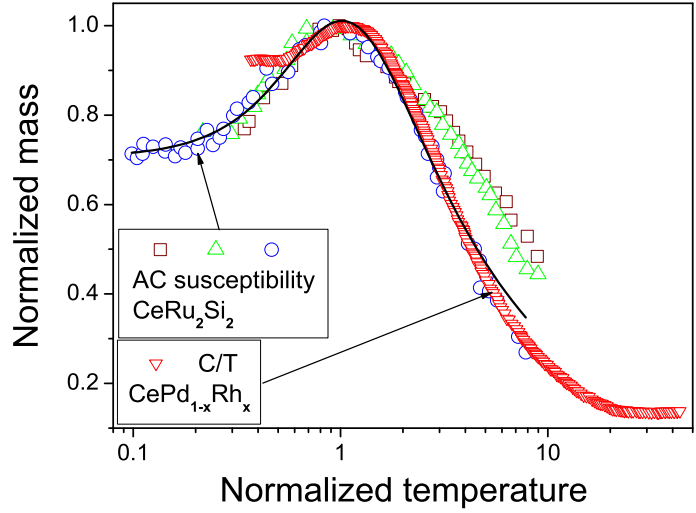}
\vspace*{-1.0cm}
\end{center}
\caption{Normalized magnetic susceptibility
$\chi_{N}(T_N,B)=\chi_{AC}(T/T_M,B)/\chi_{AC}(1,B)=M^*_N(T_N)$ for
$\rm{CeRu_2Si_2}$ in magnetic fields 0.20 mT (squares), 0.39 mT
(upright triangles) and 0.94 mT (circles) versus normalized
temperature $T_N=T/T_M$ \cite{takah}. The susceptibility reaches its
maximum $\chi_{AC}(T_M,B)$ at $T=T_M$. The normalized specific heat
$(C(T_N)/T_N)/C(1)$  of the HF ferromagnet $\rm{CePd_{1-x}Rh_x}$
with $x = 0.8$ versus $T_N$ is shown by downright triangles
\cite{pikul}. Here $T_M$ is the temperature at the peak of $C(T)/T$.
The solid curve traces the universal behavior of the normalized
effective mass determined by Eq. (\ref{UN2}). Parameters $c_1$ and
$c_2$ are adjusted for $\chi_{N}(T_N,B)$ at $B=0.94$ mT. }\label{UB}
\end{figure}

As we have seen above, the effective mass $M^*(T,B)$ can be measured
in experiments on HF metals. For example, $M^*(T,B)\propto
C(T)/T\propto \alpha(T)/T$ and $M^*(T,B)\propto \chi_{AC}(T)$ where
$\chi_{AC}(T)$ is ac magnetic susceptibility. If the corresponding
measurements are carried out at fixed magnetic field $B$ (or at
fixed both the concentration $x$ and $B$) then the effective mass
reaches its maximum at some temperature $T_M$. Upon normalizing both
the effective mass by its peak value at each field $B$  and the
temperature by $T_M$, we observe that all the curves merge into a
single one, given by Eq. (\ref{UN2}), thus demonstrating a scaling
behavior.

It is seen from Fig. \ref{UB}, that the  behavior of the normalized
ac susceptibility
$\chi_{AC}^N(y)=\chi_{AC}(T/T_M,B)/\chi_{AC}(1,B)=M^*_N(T_N)$
obtained in measurements on the HF paramagnet $\rm {CeRu_2Si_2}$
\cite{takah} agrees with both the approximation given by Eq.
(\ref{UN2}) and the normalized specific heat
$(C(T_N)/T_N)/C(1)=M^*_N(T_N)$ obtained in measurements on
$\rm{CePd_{1-x}Rh_x}$ \cite{pikul}. Also, from Fig. \ref{UB}, we see
that the curve given by Eq. (\ref{UN2}) agrees perfectly with the
measurements on ${\rm CeRu_2Si_2}$ whose electronic system is placed
at FCQPT \cite{epl2}.

\begin{figure} [! ht]
\begin{center}
\includegraphics [width=0.60\textwidth]{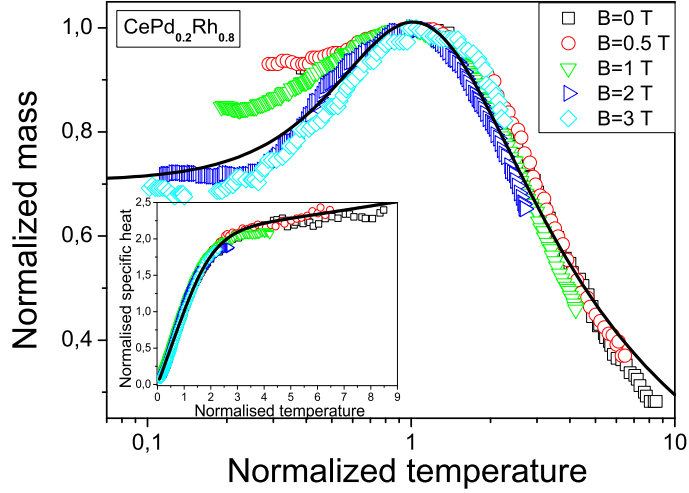}
\end{center}
\caption{The normalized effective mass at elevated magnetic fields
as a function of $y=T/T_M$. The mass taken from the specific heat
$C/T$ of the HF ferromagnet $\rm CePd_{1-x}Rh_x$ with $x = 0.8$
(Ref. \cite{pikul}) is shown at different magnetic fields $B$
depicted at the right upper corner. At $B\geq 1$ T, $M^*_N(y)$
coincides with that of $\rm{CeRu_2Si_2}$ (solid curve, see the
caption to Fig. \ref{UB}). The normalized specific heat
$C(y)/C(T_M)$ of $\rm CePd_{1-x}Rh_x$ at different magnetic fields
$B$ is shown in the inset. The kink in the specific heat is clearly
seen at $y\simeq 2$. The solid curve represents the function
$yM^*_N(y)$ with parameters $c_1$ and $c_2$ adjusted for the
magnetic susceptibility of $\rm{CeRu_2Si_2}$ at $B=0.94$
mT.}\label{UBB}
\end{figure}

Now we consider the behavior of $M^*_N(T)$, extracted from
measurements of the specific heat on $\rm{CePd_{1-x}Rh_x}$ under the
application of magnetic field \cite{pikul} and shown in Fig.
\ref{UBB}. It is seen from Fig. \ref{UBB} that for $B\geq 1$ T
$M^*_N$ describes the normalized specific heat almost perfectly,
coinciding with that of $\rm{CeRu_2Si_2}$ and is in accord with the
universal behavior of the normalized effective mass given by Eq.
(\ref{UN2}). Thus, we conclude that the thermodynamic properties of
$\rm{CePd_{1-x}Rh_x}$ with $x=0.8$ are determined by quasiparticles
rather than by the critical magnetic fluctuations. On the other
hand, one could expect the growth of the critical fluctuations
contribution as $x\to x_{FC}$ so that the behavior of the normalized
effective mass would deviate from that given by Eq. (\ref{UN2}).
This is not the case as observed from Fig. \ref{UBB}. It is also
seen that at increasing magnetic fields $B$ all the curves
corresponding to the normalized effective masses extracted from $\rm
CePd_{1-x}Rh_x$ with $x = 0.8$ merge into a single one, thus
demonstrating a scaling behavior in accord with equation
\eqref{UN2}. We note that existing theories based on the quantum and
thermal fluctuations predict that magnetic and thermal properties of
the ferromagnet $\rm CePd_{1-x}Rh_x$ differ from those of the
paramagnet $\rm CeRu_2Si_2$, Refs.
\cite{vojta,col3,pikul,kir,butch}. Clearly, from the inset of Fig.
\ref{UBB}, there is the kink in the temperature dependence of the
normalized specific heat $C(T_N)/C(T_M)$ of $\rm CePd_{1-x}Rh_x$
appearing at $T_N\simeq 2$. In the inset, the solid line depicts the
function $T_NM^*_N(T_N)$ with parameters $c_1$ and $c_2$ which are
adjusted for the magnetic susceptibility at $B=0.94$ mT. Since the
function $T_NM^*_N(T_N)$ describes the normalized specific heat very
well and its bend (or kink) comes from the crossover from the LFL
regime to the NFL one, we safely conclude that the kink emerges at
temperatures when the system transits from the LFL behavior to the
NFL one. As shown in Subsection \ref{MAGRES}, the magnetoresistance
changes from positive values to negative ones at the same
temperatures. One may speculate that there is an energy scale which
could make the kink coming from fluctuations of the order parameter
\cite{steg}. In that case we must to concede that such different HF
metals as $\rm CePd_{1-x}Rh_x$, $\rm{CeRu_2Si_2}$ and $\rm CeCoIn_5$
with different magnetic ground states have the same fluctuations
which exert coherent influence on the heat capacity, susceptibility
and transport properties. Indeed, as we have seen above and will
also see below in this Subsection, that Eq. \eqref{UN2} allows us to
describe quantitatively all the mentioned quantities.

\begin{figure}[!ht]
\begin{center}
\includegraphics[width=0.60\textwidth]{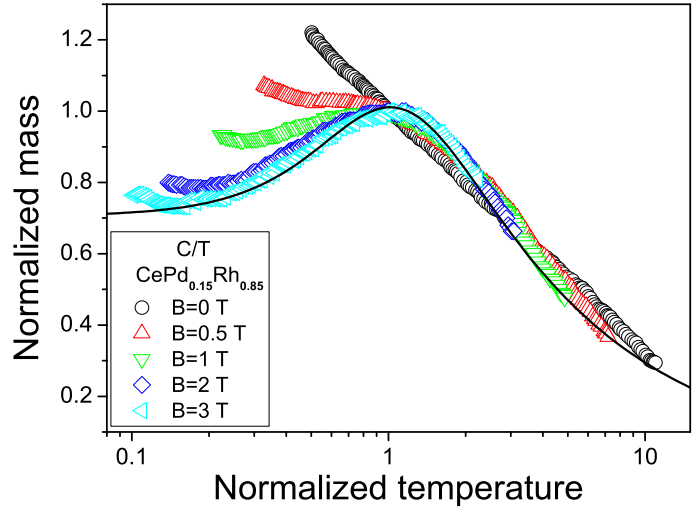}
\vspace*{-1.0cm}
\end{center}
\caption{Same as in Fig. \ref{UBB} but $x=0.85$ \cite{pikul}. At
$B\geq 1$ T, $M^*_N(T_N)$ demonstrates the universal behavior (solid
curve, see the caption to Fig. \ref{UB}).}\label{Ce1}
\end{figure}
In Fig. \ref{Ce1}, the effective mass $M^*_N(T_N)$ at fixed $B$'s is
shown. Since the curve shown by circles and extracted from
measurements at $B=0$ does not exhibit any maximum down to 0.08 K
\cite{pikul}, we conclude that in this case $x$ is very close to
$x_{FC}$ and the maximum is shifted to very low
temperatures. As seen from Fig. \ref{Ce1}, the application of
magnetic field restores the scaling behavior given by Eq.
(\ref{UN2}). Again, this permits us to conclude that the
thermodynamic properties of $\rm{CePd_{1-x}Rh_x}$ with $x=0.85$ are
determined by quasiparticles rather than by the critical magnetic
fluctuations.

\begin{figure} [! ht]
\begin{center}
\includegraphics [width=0.60\textwidth] {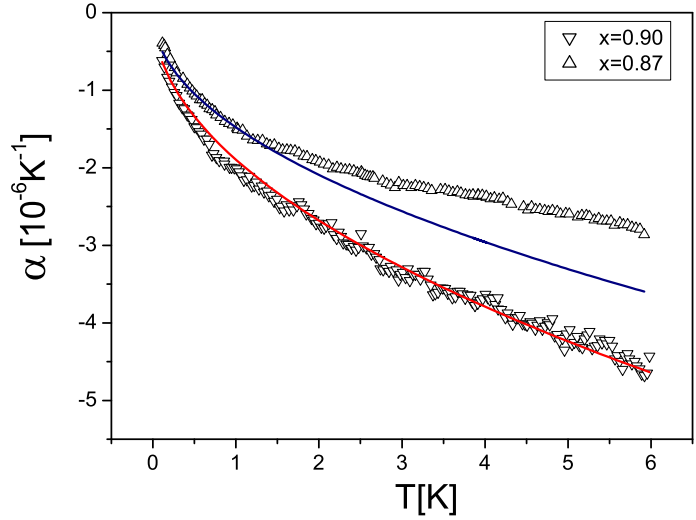}
\end {center}
\caption{Thermal expansion coefficient $\alpha(T)$ as a function of
temperature in the interval $0.1\leq T\leq 6$ K. The experimental
values for doping levels $x=0.90, 0.87$ are taken from Ref.
\cite{sereni}. The solid lines represent approximations of the
experimental values of $\alpha(T)=c_1\sqrt{T}$, where $c_1$ is a
fitting parameter.} \label {Fig11}
\end{figure}
The thermal expansion coefficient $\alpha(T)$  is given by $
\alpha(T) \simeq M^*T/(p_F^2K)$ \cite{lanl2}. The compressibility
$K(\rho)$ is not expected to be singular at FCQPT and is
approximately constant \cite{nozz}. Taking into account Eq.
\eqref{HC14}, we find that $\alpha(T)\propto \sqrt{T}$ and the
specific heat $C(T)=TM^*\propto\sqrt{T}$. Measurements of the
specific heat $C(T)$ on $\rm{CePd_{1-x}Rh_x}$ with $x=0.9$ show a
power-law temperature dependence. It is described by the expression
$C(T)/T=AT^{-q}$ with $q\simeq 0.5$ and $A$=const \cite{sereni}.

Figure  \ref{Fig11} shows that at the critical point $x=0.90$ at
which the critical temperature of the ferromagnetic phase transition
vanishes, the thermal expansion coefficient is well approximated by
the dependence $\alpha(T)\propto\sqrt{T}$ as the temperature varies
by almost two orders of magnitude. However, even a small deviation
of the system from the critical point destroys the correspondence
between this approximation and the experimental data. We note that
it is possible to describe the critical behavior of two entirely
different heavy-fermion metals (one is a paramagnet and the other a
ferromagnet) by the function $\alpha(T)=c_1\sqrt{T}$ with only one
fitting parameter $c_1$. This fact vividly shows that fluctuations
do not determine the behavior of $\alpha(T)$. Heat-capacity
measurements for $\rm CePd_{1-x}Rh_x$ with $x=0.90$ have shown that
$C(T)\propto\sqrt{T}$ \cite{sereni}. Thus, the electron systems of
both metals can be interpreted as being highly correlated electron
liquids. Hence, we conclude that the behavior of the effective mass
given by Eq. \eqref{HC14} agrees with experimental facts.

\begin{figure} [! ht]
\begin{center}
\vspace*{-0.5cm}
\includegraphics [width=0.60\textwidth]{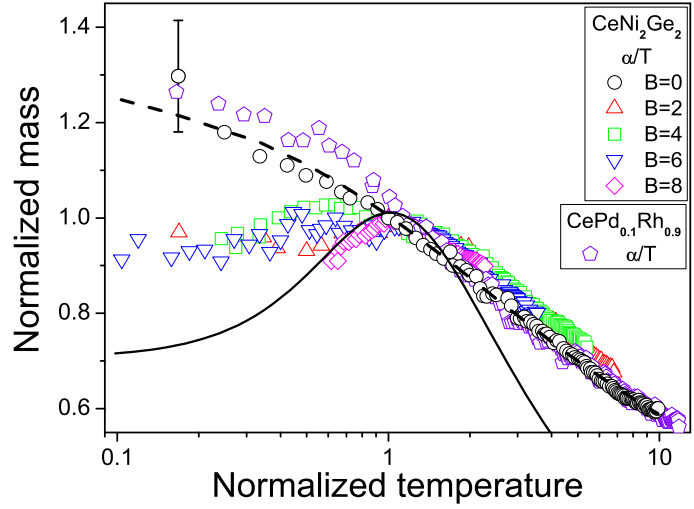}
\vspace*{-1.0cm}
\end{center}
\caption{The normalized thermal expansion coefficient
$(\alpha(T_N)/T_N)/\alpha(1)=M^*_N(T_N)$ for $\rm{CeNi_2Ge_2}$
\cite{geg1} and for $\rm{CePd_{1-x}Rh_x}$ with $x = 0.90$
\cite{pikul} versus $T_N=T/T_M$. Data obtained in measurements on
$\rm{CePd_{1-x}Rh_x}$ at $B=0$ are multiplied by some factor to
adjust them at one point to the data for $\rm{CeNi_2Ge_2}$. Dashed
line is a fit to the data shown by the circles and pentagons at
$B=0$; it is represented by the function $\alpha(T)=c_3\sqrt{T}$
with $c_3$ being a fitting parameter. The solid curve traces the
universal behavior of the normalized effective mass determined by
Eq. (\ref{UN2}), see the caption to Fig. \ref{UB}.}\label{ALPB}
\end{figure}
Measurements of $\alpha(T)/T$ on both $\rm{CePd_{1-x}Rh_x}$ with
$x=0.9$ \cite{sereni} and $\rm{CeNi_2Ge_2}$ \cite{geg1} are shown in
Fig. \ref{ALPB}. It is seen that the approximation
$\alpha(T)=c_3\sqrt{T}$ is in good agreement with the results of
measurements of $\alpha(T)$ in $\rm{CePd_{1-x}Rh_x}$ and
$\rm{CeNi_2Ge_2}$ over two decades in $T_N$. It is noted that
measurements on $\rm{CeIn_{3-x}Sn_x}$ with $x=0.65$ \cite{kuch}
demonstrate the same behavior $\alpha(T)\propto\sqrt{T}$ (not shown
in Fig. \ref{ALPB}). As a result, we suggest that
$\rm{CeIn_{3-x}Sn_x}$ with $x=0.65$, $\rm{CePd_{1-x}Rh_x}$ with
$x\simeq 0.9$, and $\rm{CeNi_2Ge_2}$ are located at FCQPT; recall
that $\rm{CePd_{1-x}Rh_x}$ is a three dimensional FM
\cite{sereni,pikul}, $\rm{CeNi_2Ge_2}$ exhibits a paramagnetic
ground state \cite{geg1} and $\rm{CeIn_{3-x}Sn_x}$ is AFM cubic
metal \cite{kuch}.

\begin{figure} [! ht]
\begin{center}
\includegraphics [width=0.60\textwidth]{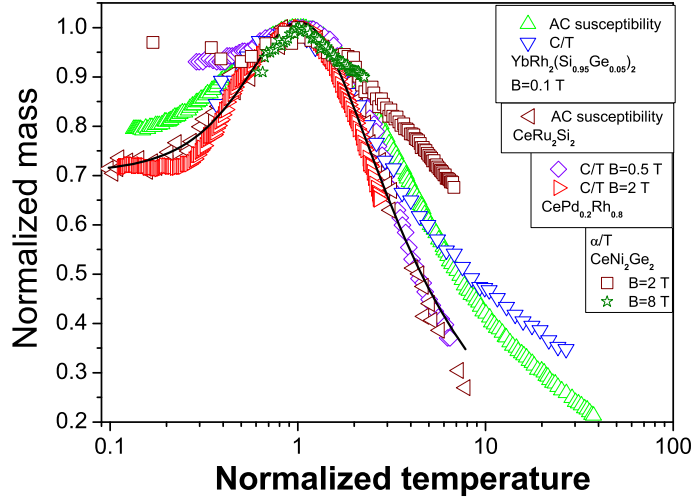}
\vspace*{-1.0cm}
\end{center}
\caption{The universal behavior of $M^*_N(T_N)$, extracted from
$\chi_{AC}(T,B)/\chi_{AC}(T_M,B)$ for both
$\rm{YbRh_2(Si_{0.95}Ge_{0.05})_2}$ and $\rm{CeRu_2Si_2}$
\cite{takah,geg3}, $(C(T)/T)/(C(T_M)/T_M)$ for both
$\rm{YbRh_2(Si_{0.95}Ge_{0.05})_2}$ and $\rm{CePd_{1-x}Rh_x}$ with
$x=0.80$ \cite{pikul,geg3}, and $(\alpha(T)/T)/(\alpha(T_M)/T_M)$
for $\rm{CeNi_2Ge_2}$ \cite{geg1}. All the measurements were
performed under the application of magnetic field as shown in the
insets. The solid curve gives the universal behavior of $M^*_N$
determined by Eq.(\ref{UN2}), see the caption to Fig.
\ref{UB}.}\label{UnB}
\end{figure}

The normalized effective mass $M^*_N(T_N)$ extracted from
measurements on the HF metals $\rm{YbRh_2(Si_{0.95}Ge_{0.05})_2}$,
$\rm{CeRu_2Si_2}$, $\rm{CePd_{1-x}Rh_x}$ and $\rm{CeNi_2Ge_2}$ is
reported in Fig. \ref{UnB}. Clearly, the scaling behavior of the
effective mass given by Eq. (\ref{UN2}) is in accord with the
experimental facts and $M^*_N(T_N)$, shown by inverted triangles
and collected on the AFM phase of
$\rm{YbRh_2(Si_{0.95}Ge_{0.05})_2}$ \cite{geg3}, coincides with
that collected on the PM phase (shown by upright triangles) of
$\rm{YbRh_2(Si_{0.95}Ge_{0.05})_2}$ \cite{geg3}. We note that in
the case of LFL theory the corresponding normalized effective mass
$M^*_{NL}\simeq1$ is independent of both $T$ and $B$ as shown in
Fig. \ref{YBRHSIN}.

\begin{figure} [! ht]
\begin{center}
\vspace*{-0.3cm}
\includegraphics [width=0.60\textwidth]{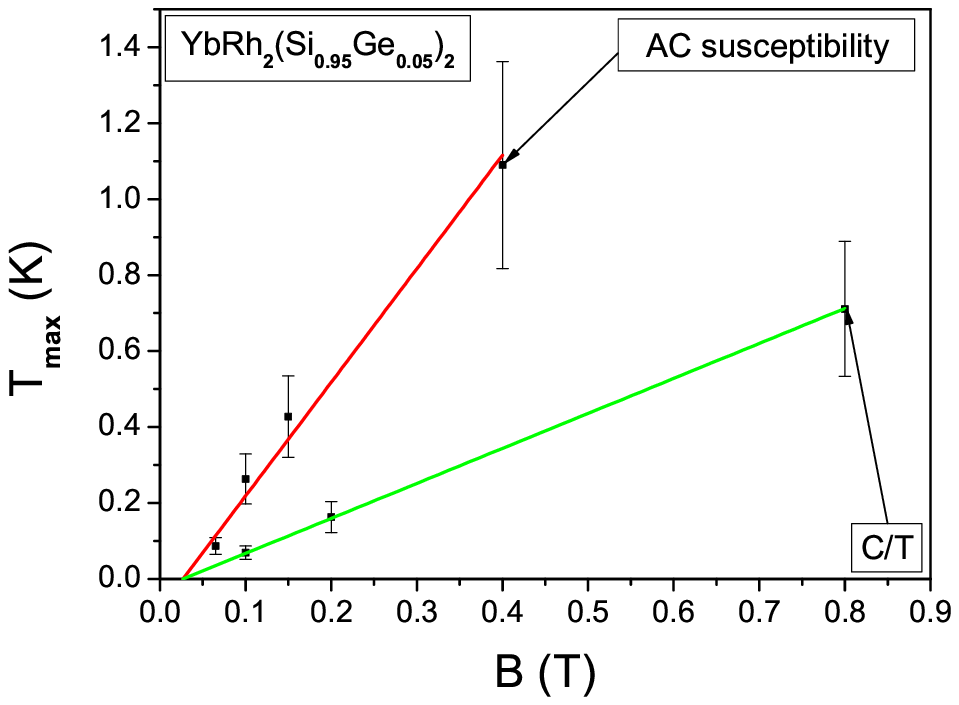}
\end{center}
\vspace*{-1.0cm} \caption{The peak temperatures $T_{\rm max}(B)$,
extracted from measurements of $\chi_{AC}$ and $C/T$ on
$\rm{YbRh_2(Si_{0.95}Ge_{0.05})_2}$ \cite{gegtok,geg3} and
approximated by straight lines given by Eq. \eqref{LFLT}. The lines
intersect at $B\simeq 0.03$ T.}\label{TM}\end{figure}

The peak temperatures $T_{\rm max}$, where the maxima of $C(T)/T$,
$\chi_{AC}(T)$ and $\alpha(T)/T$  occur, shift to higher values
with increase of the applied magnetic field. In Fig. \ref{TM},
$T_{\rm max}(B)$ are shown for $C/T$ and $\chi_{AC}$, measured on
$\rm{YbRh_2(Si_{0.95}Ge_{0.05})_2}$. It is seen that both functions
can be represented by straight lines intersecting at $B\simeq 0.03$
T. This observation \cite{gegtok,geg3} as well as the measurements
on $\rm{CePd_{1-x}Rh_x}$, $\rm{CeNi_2Ge_2}$ and $\rm{CeRu_2Si_2}$
demonstrate similar behavior \cite{geg1,takah,pikul} which is well
described by Eq. \eqref{LFLT}.

We conclude, that subjecting the different experimental data (like
$C(T)/T$, $\chi_{AC}(T)$, $\alpha(T)/T$ etc) collected
in measurements on different
HF metals ($\rm{YbRh_2 (Si_{0.95} Ge_{0.05})_2}$, $\rm{Ce Ru_2Si_2}$,
$\rm{CePd_{1-x}Rh_x}$, $\rm{CeIn_{3-x}Sn_x}$ and $\rm{CeNi_2Ge_2})$
to the above normalized form immediately reveals their universal
scaling behavior \cite{physb}. This is because all the above
experimental quantities are indeed proportional to the normalized
effective mass exhibiting the scaling behavior. Since the effective
mass determines the thermodynamic properties, we further conclude
that the above HF metals demonstrate the same scaling behavior,
independent of the details of HF metals such as their lattice
structure, magnetic ground states, dimensionality etc
\cite{epl2,physb}.

\section{Metals with a strongly correlated
electron liquid} \label{SCEL}

For $T\ll T_f$, the function $n_0({\bf p})$ given by Eq. (\ref{FL8})
determines the entropy $S_{NFL}(T)$ given by Eq. \eqref{FL3} of the
HF liquid located above the quantum critical line shown in Fig.
\ref{fig1}. From Eqs. \eqref{FL3} and \eqref{snfl}, the entropy
contains a temperature-independent contribution,
\begin{equation}
 S_0\sim\frac{p_f-p_i}{p_F}\sim|r|,\label{SL1a}
\end{equation}
where $r=(x-x_{FC})/x_{FC}$. Another specific contribution is
related to the spectrum $\varepsilon({\bf p})$, which ensures a link
between the dispersionless region $(p_f-p_i)$ occupied by FC and the
normal quasiparticles in the regions $p<p_i$ and $p>p_f$. This
spectrum has the form $\varepsilon({\bf
p})\propto(p-p_f)^2\sim(p_i-p)^2$. Such a shape of the spectrum,
corroborated by exactly solvable models for systems with FC , leads
to a contribution to the heat capacity $C\sim\sqrt{T/T_f}$
\cite{ks}. Therefore, for $0<T\ll T_f$, the entropy can be
approximated by the function \cite{alp}
\begin{equation}
S_{NFL}(T)\simeq
S_0+a\sqrt{\frac{T}{T_f}}+b\frac{T}{T_f},\label{SL1}
\end{equation}
where $a$ and $b$ are constants. The third term on the right-hand
side of Eq. (\ref{SL1}), which emerges because of the contribution
of the temperature-independent part of the spectrum
$\varepsilon({\bf p})$, yields a relatively small addition to the
entropy. As we will see shortly, the temperature-independent term
$S_0$ determines the universal transport and thermodynamic
properties of the heavy-electron liquid with FC, which we call a
strongly correlated Fermi liquid. The properties of this liquid
differ dramatically from those of highly correlated Fermi liquid
that at $T\to0$ becomes LFL liquid. As a result, we can think of QCP
of FCQPT as the phase transition that separates highly correlated
and strongly correlated Fermi liquids. Because the highly correlated
liquid behaves like LFL as $T\to0$, QCP separates LFL from a
strongly correlated Fermi liquid. On the other hand, as was shown in
Subsection \ref{PHDFC}, at elevated temperatures the properties of
both liquids become indistinguishable. Thus, as shall be seen below,
both systems can be discriminated at diminishing temperatures when
the impact of both QCP and the quantum critical line on the
properties become more vivid.

\begin{figure} [! ht]
\begin{center}
\includegraphics [width=0.60\textwidth] {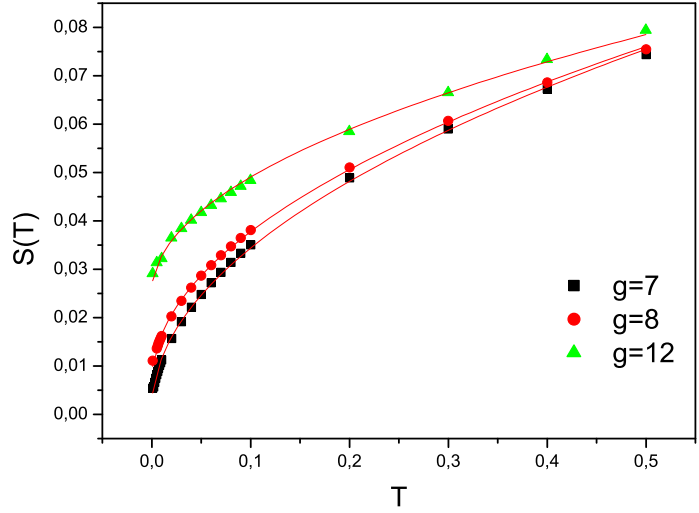}
\end {center}
\caption {Entropy $S(T)$ as a function of temperature. The lines
represent the approximation for $S(T)$ based on Eq. (\ref{SL1}), the
symbols mark the results of calculations based on (\ref{HC6}).}
\label {Fig13}
\end{figure}
Figure \ref{Fig13} shows the temperature dependence of $S(T)$
calculated on the basis of the model functional (\ref{HC6}). The
calculations were done with $g=7,8,12$ and $\beta=b_c=3$. We recall
that the critical value of $g$ is $g_c=6.7167$. We see in Fig.
\ref{Fig13} that in accord with Eq. \eqref{SL1a} $S_0$ increases as
the system moves away from QCP along the quantum critical line, see
Fig. \ref{fig1}. Obviously, the term $S_0$ on the right-hand side of
Eq. (\ref{SL1}), which is temperature-independent, contributes
nothing to the heat capacity; the second term in (\ref{SL1}) makes a
contribution so that the heat capacity behaves as
$C(T)\propto\sqrt{T}$.

\subsection{Entropy, linear expansion, and Gr\"uneisen's
law}\label{ENTR}

The unusual temperature dependence of the entropy of a strongly
correlated electron liquid specified by Eq. (\ref{SL1}) determines
the unusual behavior of the liquid. The existence of a
temperature-independent term $S_0$ can be illustrated by
calculating the thermal expansion coefficient $\alpha(T)$
\cite{alp,zver}, which is given by \cite{lanl1}
\begin{equation}
\alpha(T)=\frac13\left(\frac{\partial(\log V)}{\partial T}\right)
_P=-\frac{1}{3V}\left(\frac{\partial(S/x)}{\partial
P}\right)_T,\label{SL2}
\end{equation}
where $P$ is the pressure and $V$ is the volume. We note that the
compressibility $K=d\mu/d(Vx)$ does not develop a singularity at
FCQPT and is approximately constant in systems with FC  \cite{nozz}.
Substituting (\ref{SL1}) in Eq. (\ref{SL2}), we find that
\cite{alp,zver}
\begin{equation}\frac{\alpha_{FC}(T)}{T}\simeq
\frac{a_0}{T}\sim\frac{M_{FC}^{*}}{p_F^2K},\label{SL3}
\end{equation}
where $a_0\sim \partial S_0/\partial P$ is temperature-independent.
In (\ref{SL3}), we took only the leading contribution related to
$S_0$ into account. We recall that
\begin{equation} C(T)=T\frac{\partial S(T)}{\partial
T}\simeq\frac{a}{2} \sqrt{\frac{T}{T_f}},\label{SL4}
\end{equation}
and obtain from Eqs. \eqref{SL3} and \eqref{SL4} that the
Gr\"uneisen ratio $\Gamma(T)$ diverges as
\begin{equation} \Gamma(T)=\frac{\alpha(T)}{C(T)}
\simeq2\frac{a_0}{a}\sqrt{\frac{T_f}{T}},\label{SL5}
\end{equation}
from which we conclude that Gr\"uneisen's law does not hold in
strongly correlated Fermi systems.

Measurements that have been conducted with ${\rm
YbRh_2(Si_{0.95}Ge_{0.05})_2}$ show that $\alpha/T\propto1/T$ and
that the Gr\"uneisen ratio diverges as $\Gamma(T)\simeq T^{-q}$,
$q\simeq0.33$, which allows classifying the electron system of this
compound as strongly correlated liquid \cite{geg1}. Our estimate
$q=0.5$, which follows from Eq. (\ref{SL5}), is in satisfactory
agreement with this experimental value. The behavior of
$\alpha(T)/T$ given by Eq. \eqref{SL3} contradicts the LFL theory,
according to which the thermal expansion coefficient $\alpha(T)/T=
M^*=const$ as $T\to0$.  The $1/T$-dependence of the ration
$\alpha/T$ predicted in Ref. \cite{zver} is in good agreement with
facts collected on  ${\rm YbRh_2(Si_{0.95}Ge_{0.05})_2}$
\cite{geg1}.

Equation (\ref{FL12}) implies that $M^*(T\to0)\to\infty$ and that
the strongly correlated electron system behaves as if it were placed
at the quantum critical point. Actually, as we have seen in
Subsection \ref{PHDFC} the system is at the quantum critical line
$x/x_{FC}\leq1$, and critical behavior is observed for all $x\leq
x_{FC}$ as $T\to0$. It was shown in Section \ref{SC} that as
$T\to0$, the strongly correlated electron liquid undergoes the
first-order quantum phase transition, because the entropy becomes a
discontinuous function of the temperature: at finite temperatures,
the entropy is given by Eq. (\ref{SL1}), while $S(T=0)=0$. Hence,
the entropy has a discontinuity $\delta S=S_0$ as $T\to0$. This
implies that, as a result of the first-order phase transition, all
critical fluctuations are suppressed along the quantum critical
curve and the respective divergences, e.g., the divergence of
${\rm\Gamma}(T)$, are determined by quasiparticles and not critical
fluctuations, as could be expected in the case of an ordinary
quantum phase transition \cite{voj}. We note that according to the
well-known inequality \cite{lanl2} $q\leq T\delta S$, in our case
the heat $q$ of the first order transition tends to zero as its
critical temperature  $T_{NL}\to 0$.

\subsection{The $T-B$ phase diagram of $\rm YbRh_2Si_2$, Hall
coefficient and magnetization}

To study the $T-B$ phase diagram of strongly correlated electron
liquid, we examine the case where NFL behavior emerges when the AF
phase is suppressed by an external magnetic field $B$, as it is in
the HF metals $\rm YbRh_2(Si_{0.95} Ge_{0.05})_2$ and $\rm
YbRh_2Si_2$ \cite{geg1,geg}.

\begin{figure} [! ht]
\begin{center}
\includegraphics [width=0.60\textwidth] {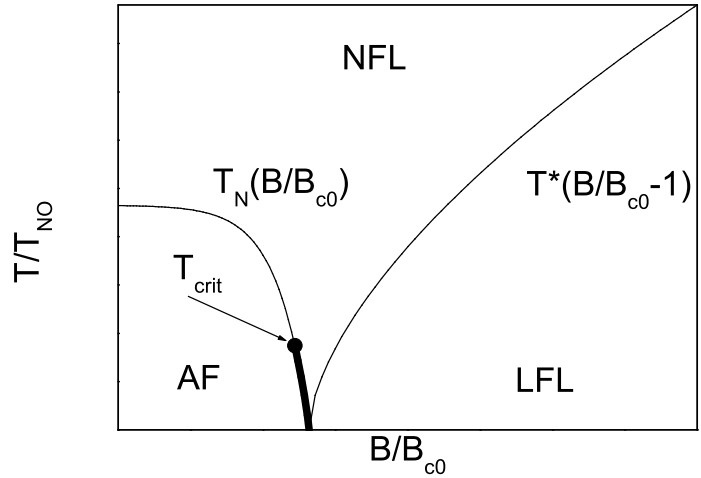}
\end{center}
\caption {The $T-B$ phase diagram of a strongly correlated electron
liquid. The line $T_{N}(B/B_{c0})$ represents the dependence of the
N\'eel temperature on the field strength $B$. The black dot at
$T=T_{crit}$ marks the critical temperature at which the
second-order AF phase transition becomes a first-order one. For
$T<T_{crit}$, the heavy solid line represents the function $T_N
(B/B_{c0})$, when the AF phase transition becomes a first-order one.
The strongly correlated liquid in the NFL region is characterized by
the entropy $S_{NFL}$ given by Eq. (\ref{SL1}). The line separating
the strongly correlated liquid (NFL) from the weakly polarized
electron liquid, which behaves like the Landau Fermi liquid, is
described by the function $T^*(B/B_{c0}-1)\propto\sqrt {B/B_{c0}-1}$
[see Eq. \eqref{HF8}].} \label{Fig16}
\end{figure}

The antiferromagnetic phase is LFL with the entropy vanishing as
$T\to0$. For magnetic fields higher than the critical value $B_{c0}$
at which the N\'eel temperature $T_{NL}(B\to B_{c0})\to0$, the
antiferromagnetic phase transforms into a weakly polarized
paramagnetic strongly correlated electron liquid \cite{geg1,geg}. As
shown in Section \ref{HFL}, a magnetic field applied to the system
with $T=0$ splits the FC state occupying the interval $(p_f-p_i)$
into Landau levels and suppresses the superconducting order
parameter $\kappa({\bf p})$. The new state is specified by a
multiply connected Fermi sphere, on which a smooth quasiparticle
distribution function $n_0({\bf p})$ in the interval $(p_f-p_i)$ is
replaced with a distribution $\nu({\bf p})$ as seen from Fig.
\ref{Fig4}. Hence, the behavior of LFL is restored and characterized
by quasiparticles with the effective mass $M^*(B)$ given by Eq.
\eqref{HF5}. When the temperature increases so high that $T>T^*(B)$
with $T^*(B)$  given by Eq. \eqref{HF8}, the entropy of the electron
liquid is determined by Eq. (\ref{SL1}). The described behavior of
the system is shown in the $T-B$ diagram in Fig. \ref{Fig16}.

In accordance with the experimental data, we assume that at
relatively high temperatures, such that $T/T_{N0}\sim 1$, where
$T_{N0}$ is the N\'eel temperature in a zero magnetic field, the
antiferromagnetic phase transition is a second-order one \cite{geg}.
In this case, the entropy and other thermodynamic functions at the
transition temperature $T_{NL}$ are continuous. This means that the
entropy $S_{AF}$ of the antiferromagnetic phase coincides with the
entropy $S_{NFL}$ of the strongly correlated liquid given by Eq.
\eqref{SL1}:
\begin{equation}
S_{AF}(T\to T_{NL}(B))=S_{NFL}(T\to T_{NL}(B))\label{SL6}.
\end{equation}
Since the antiferromagnetic phase behaves like LFL, with its entropy
$S_{AF}(T\to0)\to0$, Eq. (\ref{SL6}) cannot be satisfied at
sufficiently low temperatures $T\leq T_{crit}$ because of the
temperature-independent term $S_0$. Hence, the second order
antiferromagnetic phase transition becomes the first order one at
$T=T_{crit}$ \cite{spa,sopt} as shown by the arrow in Fig.
\ref{Fig16}. A detailed consideration of this item is given in
Section \ref{OPT}.

At $T=0$, the critical magnetic field $B_{c0}$ in which the
antiferromagnetic phase becomes LFL is determined by the condition
that the ground-state energy of the antiferromagnetic phase be equal
to the ground-state energy $E[n_0({\bf p})]$ of the HF liquid with
FC, since, as it was shown in Subsection \ref{ENTR}, the heat of the
transition $q=0$. This means that the ground state of the
antiferromagnetic phase is degenerate at $B=B_{c0}$. Hence, at $B\to
B_{c0}$ the N\'eel temperature $T_{NL}$ tends to zero and the
behavior of the effective mass $M^*(B\geq B_{c0})$ is determined by
Eq. (\ref{HF5}), so that $M^*(B)$ diverges as $B\to B_{c0}$ from
top. As a result, at $T=0$, the phase transition separating the
antiferromagnetic phase existing at $B\leq B_{c0}$ from LFL taking
place at $B\geq B_{c0}$ is the first order quantum phase transition.
The driving parameter of this phase transition is the magnetic field
strength $B$. We note that the respective quantum and thermal
critical fluctuations disappear at $T<T_{crit}$ because the
first-order antiferromagnetic phase transition occurs at such
temperatures.

We now examine the jump in the Hall coefficient detected in
measurements involving $\rm YbRh_2Si_2$ \cite{pash}. The Hall
coefficient $R_H(B)$ as a function of $B$ experiences a jump as
$T\to0$ when the applied magnetic field reaches its critical value
$B=B_{c0}$, and then becomes even higher than the critical value at
$B=B_{c0}+\delta B$, where $\delta B$ is an infinitely small
magnetic field strength \cite{pash}. As shown in Section \ref{HFL},
when $T=0$, the application of the critical magnetic field $B_{c0}$,
which suppresses the antiferromagnetic phase with the Fermi momentum
$p_F$ restores LFL with the Fermi momentum $p_f>p_F$. When
$B<B_{c0}$, the ground-state energy of the antiferromagnetic phase
is lower than that of the LFL state induced by the application of
magnetic field, but for $B>B_{c0}$ we are confronted with the
opposite case, where the LFL state has the lower energy. At
$B=B_{c0}$ and $T=0$, both phases have the same ground state energy
and $T_{NL}=0$, because the phases are degenerate, being separated
by the first order phase transition as shown in  Fig. \ref{Fig16}.

Thus, at $T=0$ and $B=B_{c0}$, an infinitely small increase $\delta
B$ in the magnetic field leads to a finite discontinuity in the
Fermi momentum. This is because the distribution function becomes
multiply connected (see Fig. \ref{Fig4}) and the number of mobile
electrons does not change. Thus, the antiferromagnetic ground state
can be viewed as having a "small" Fermi surface characterized by the
Fermi momentum $p_F$, correspondingly the paramagnetic ground state
at $B>B_{c0}$ has a "large" Fermi surface with $p_f>p_F$. As a
result, the Hall coefficient experiences a sharp jump because
$R_H(B)\propto1/p_F^3$ in the antiferromagnetic phase and
$R_H(B)\propto1/p_f^3$ in the paramagnetic phase. Assuming that
$R_H(B)$ is a measure of the Fermi momentum \cite{pash} (as is the
case with a simply connected Fermi volume), we obtain
\begin{equation}
\frac {R_H(B=B_{c0}-\delta)}{R_H(B=B_{c0}+\delta)}\simeq1+3\frac
{p_f-p_F}{p_F}\simeq1+d\frac{S_0}{x_{FC}},\label{SL7}
\end{equation}
where $S_0/x_{FC}$ is the entropy per heavy electron and $d$ is a
constant $d\sim 5$. It follows from Eq. (\ref{SL7}) that the
discontinuity in the Hall coefficient is determined by the anomalous
behavior of the entropy, which can be attributed to $S_0$. Hence,
the discontinuity tends to zero as $r\to0$ and disappears when the
system is on the disordered side of FCQPT, where the entropy has no
temperature-independent term \cite{spa}.

\begin{figure} [! ht]
\begin{center}
\includegraphics [width=0.60\textwidth] {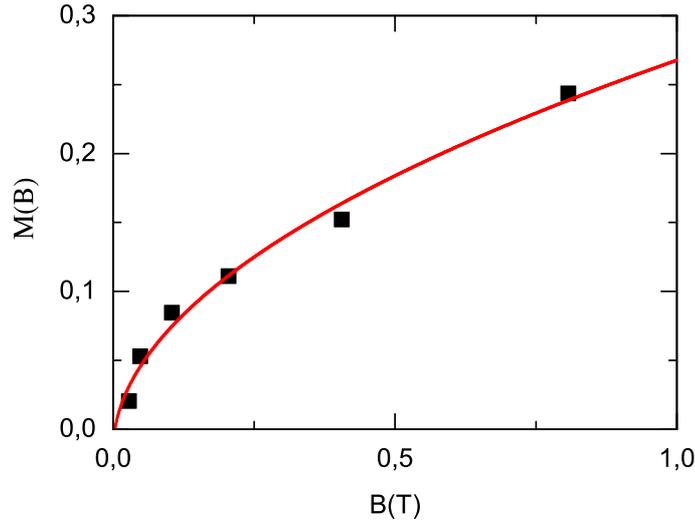}
\end{center}
\caption {The values of magnetization $M(B)$ obtained in
measurements involving $\rm YbRh_2(Si_{0.95} Ge_{0.05})_2$ (black
squares) \cite{cust}. The curve represents the field-dependent
function $M(B)=a_M\sqrt{B}$ given by Eq. \eqref{SL9}, where $a_M$ is
a fitting parameter.} \label{Fig17}
\end{figure}
We now turn to the magnetization which is determined by Eq.
(\ref{CHIB}). For $T\ll T^*(B)$, the effective mass is given by Eq.
\eqref{HF5} and the static magnetization is
\begin{equation}
M(B)\simeq a_M\sqrt{B-B_{c0}}.\label{SL9}
\end{equation}
Figure \ref{Fig17} shows that the function $M(B)$ determined by Eq.
\eqref{SL9} is in good agreement with the data obtained in
measurements on $\rm YbRh_2(Si_{0.95} Ge_{0.05})_2$ \cite{cust}. We
note that $B_{c0}\simeq0$ in this case.

\begin{figure} [! ht]
\begin{center}
\includegraphics [width=0.60\textwidth] {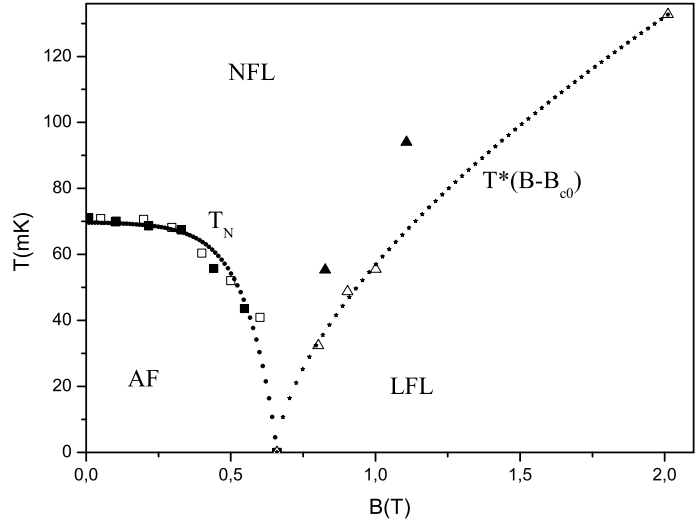}
\end{center}
\caption {$T-B$ phase diagram for ${\rm YbRh_2Si_2}$; the symbols
denote the experimental data \cite{cust,geg}. The line $T_N$ depicts
the field dependence of the N\'eel temperature $T_{NL}(B)$. In the
NFL region, the behavior of the strongly correlated liquid is
characterized by the entropy $S_{NFL}$ determined by Eq.
(\ref{SL1}). The line separating the NFL region from the LFL region
is approximated by the function $T^*(B-B_{c0})=c\sqrt{B-B_{c0}}$
given by Eq. \eqref{HF8} where $c$ is a fitting parameter.}
\label{Fig18}
\end{figure}

We examine the experimental $T-B$ diagram of the heavy-fermion
metal ${\rm YbRh_2Si_{2}}$ \cite{cust,geg} shown in Fig.
\ref{Fig18}. In the LFL state, the behavior of the metal is
characterized by the effective mass $M^*(B)$, which diverges as
$1/\sqrt{B-B_{c0}}$ \cite{geg}. It is quite evident that Eq.
(\ref{HF5}) provides a good description of this experimental fact:
$M^*(B)$ diverges as $B\to B_{c0}$ at $T_N(B=B_{c0})=0$ and, as
Fig. \ref{Fig17} shows, the calculated behavior of the
magnetization agrees with the experimental data. The magnetic-field
dependence of the coefficient $A(B)$ shown in the left panel of
Fig. \ref{ff2} is also in good agreement with experimental facts
collected on $\rm YbRh_2Si_{2}$ \cite{geg}. Figure \ref{Fig18}
shows that in accordance with (\ref{HF8}), the curve separating the
LFL region from the NFL region can be approximated by the function
$c\sqrt{B-B_{c0}}$ with a fitting parameter $c$. Bearing in mind
that the behavior of $\rm YbRh_2Si_{2}$ is like that of $\rm
YbRh_2(Si_{0.95}Ge_{0.05})_2$ \cite{cust,geg1,geg3,pepin}, we also
conclude that the thermal expansion coefficient $\alpha(T)$ is
temperature-independent and that the Gr\"uneisen ratio diverges as
a function of $T$ in the NFL state \cite{geg1}. We conclude that
the entropy in the NFL state is determined by Eq. (\ref{SL1}).
Since the antiferromagnetic phase transition is the second order at
relatively high temperatures \cite{geg}, we can predict that as the
temperature decreases, the phase transition becomes the first
order. The above description of the behavior of the Hall
coefficient $R_H(B)$ also agrees with the experimental facts
\cite{pash}.

Thus, we conclude that the $T-B$ phase diagram of the strongly
correlated electron liquid shown in Fig. \ref{Fig16} agrees with the
experimental $T-B$ diagram obtained from experiments involving the
heavy-fermion metals $\rm YbRh_2Si_2$ and $\rm YbRh_2(Si_{0.95}
Ge_{0.05})_2$ and shown in Fig. \ref{Fig18}.

\subsection{Heavy-fermion metals in the immediate vicinity
of QCP}\label{CeCoIn}

We now consider the case where $\delta p_{FC}=(p_f-p_i)/p_F\ll1$ and
the electron system of HF metal is in a state close to QCP while
remaining on the ordered side that is at the quantum critical line,
see Fig. \ref{fig1}. It follows from Eq. \eqref{HF10} that when the
system is placed in a magnetic field $(B-B_{c0})/B_{c0}\geq B_{cr}$,
the system passes from the ordered side of FCQPT to the disordered
side, or the strongly correlated liquid transforms into the highly
correlated one. As a result, when $T\leq T^*(B)$, the effective mass
$M^*(B)$ is determined by Eqs. (\ref{HC5}) and (\ref{HC11}); thus
both the Kadowaki-Woods relation and the Wiedemann-Franz law remain
valid, and there are quasiparticles in the system. The resistivity
then behaves as described in Subsection \ref{HCELR}.

In magnetic field with $B\simeq B_{c0}$ and at temperatures $T_f\gg
T>T^*(B)$, the system behaves like the strongly correlated Fermi
liquid, the effective mass $M^*(T)$ is given by Eq. (\ref{FL12}),
and the entropy is determined by Eq. (\ref{SL1}). The thermal
expansion coefficient $\alpha(T)$ is temperature-independent [as
follows from Eq. (\ref{SL3})], and the Gr\"uneisen ratio diverges,
as follows from Eq. (\ref{SL5}). It follows from Eq. (\ref{FL12})
that  the width $\gamma(T)\propto T$ (see also Section
\ref{pseudogap}). Hence, at $T_f\gg T\gg T^*(B)$, the
temperature-dependent part of the resistivity behaves as
$\Delta\rho(T)\propto\gamma(T)\propto T$ in either case, when the
electron system is in the highly correlated state or in the strongly
correlated state.

We assume that the system becomes superconducting at a certain
temperature $T_c$. In contrast to the jump $\delta C(T_c)$ of the
heat capacity at $T_c$ in ordinary superconductors, which is a
linear function of $T_c$, the value of $\delta C(T_c)$ is
independent of $T_c$ in our case. Equations (\ref{SC22}) and
(\ref{SC23}) show that both $\delta C(T_c)$ and the ratio $\delta
C(T_c)/C_n(T_c)$ can be very large compared to the corresponding
quantities in the ordinary BCS case as it was observed in the HF
metal $\rm CeCoIn_5$ \cite{yakov,ams,bau}. Experiments show that the
electron system in $\rm CeCoIn_5$ can be considered as a strongly
correlated electron liquid. Indeed, for $T>T^*(B)$, the linear
thermal expansion coefficient $\alpha(T)\propto const$ and the
Gr\"uneisen ratio diverges \cite{oes} [see Eqs. (\ref{SL3}) and
(\ref{SL5})], so we may assume that the entropy is given by
(\ref{SL1}).

A finite magnetic field takes the system to the disordered side of
FCQPT; for $T<T^*(B)$, the system behaves like the highly correlated
liquid with the effective mass given by Eq. \eqref{HC5}. Estimates
of $\delta p_{FC}$ based on calculations of the magnetic
susceptibility show that $\delta p_{FC}\simeq0.044$ \cite{yakov}. We
conclude that $B_{cr}\sim 0.01$, as follows from Eq. (\ref{HF10}),
and the electron system of the heavy-fermion metal $\rm CeCoIn_5$
passes, in relatively weak magnetic fields, to the disordered side
of FCQPT and acquires the behavior characteristic of highly
correlated liquid. We note that the estimated value of $\delta
p_{FC}$ provides an explanation for the relatively large jump
$\delta C(T_c)$ \cite{yakov} observed at $T_c=2.3$ K in experiments
with $\rm CeCoIn_5$ \cite{bau}.

\begin{figure} [! ht]
\begin{center}
\includegraphics [width=0.60\textwidth] {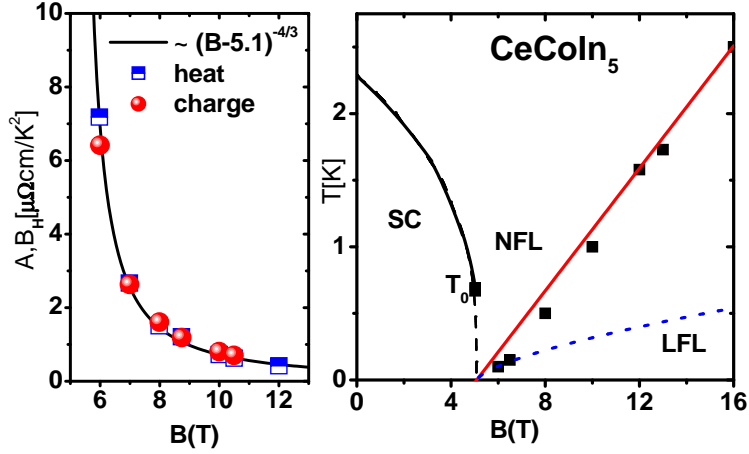}
\end{center}
\caption {$T-B$ phase diagram for $\rm CeCoIn_5$. In the left panel,
shown are $A(B)$ and $B_H(B)$ that determine the $T^2$-dependence of
the resistance and heat transfer in the LFL state induced by the
magnetic field; the symbols mark the experimental data. The right
panel depicts the curves of phase transitions in a magnetic field;
the line separates the normal (NFL) state from the superconducting
(SC) state \cite{bi}; the solid curve corresponds to the
second-order phase transitions, the dashed curve corresponds to the
first-order phase transitions, the black square (at $T_0$) is the
point where second-order transitions become first-order transitions.
The dotted line represents the function $T^*(B)$ calculated in
accordance with (\ref{tr1}) for the transition region between the
LFL and NFL states. The light solid line represents the function
$T^*(B)$ calculated according to Eq. \eqref{HC14a} for the transition
region (when $B>B_{cr}$) between the highly correlated and strongly
correlated liquids; the black squares mark the experimental results
obtained from resistivity measurements \cite{pag,pag2}.}
\label{Fig14}
\end{figure}

As Fig. \ref{Fig14} shows, the behavior $A(B)\propto B_H(B)\propto
M^*(B)\propto(B-B_{c0})^{-4/3}$ specified by Eq. (\ref{HC16}) is in
good agreement with the experimental results \cite{pag,pag2}. The
coefficient $B_H(B)$ determines the $T^2$-dependence of the thermal
resistance, and the ratio $A(B)/B_H(B)$ is field-independent, with
$A/B_H\simeq 0.70$ \cite{pag2,pag}. In the LFL state, the
Kadowaki-Woods relation and the Wiedemann-Franz law hold, and the
behavior of the system is determined by quasiparticles
\cite{pag,pag2,bi}. Thus, we conclude that our description is in
good agreement with the experimental facts.

At low temperatures and in magnetic fields $B_{red}\sim B_{cr}$ [see
Eq. \eqref{HF10}], the electron system is in its LFL state. As the
temperature increases, the behavior of the strongly correlated
liquid determined by the entropy $S_0$ is restored at $T^*(B)$, and
the effective mass becomes temperature-dependent, according to Eq.
(\ref{FL12}). To calculate $T^*(B)$, we use the fact that the
behavior of the effective mass is given by Eq. \eqref{HC5} for
$T<T^*(B)$ and by Eq. \eqref{FL12} for $T>T^*(B)$. Since the
effective mass cannot change at $T=T^*(B)$, we can estimate $T^*(B)$
by equating these two values of the effective mass. As a result, we
obtain
\begin{equation}\label{tr1}
T^*(B)\propto (B-B_{c0})^{2/3}.
\end{equation}
The function $T^*(B)$ \eqref{tr1} is shown by the dotted line in
Fig. \ref{Fig14}. As the magnetic field becomes stronger, $B\gg
B_{cr}$, the system becomes the highly correlated liquid in which
the behavior of $M^*(T)$ is given by Eq. \eqref{HC11} and that of
$M^*(B)$ by Eq. \eqref{HC5}. Comparison of these two types of
behavior yields Eq. \eqref{HC14a}.
The function $T^*(B)$ given be Eq. \eqref{HC14a} is depicted by the light solid
line in Fig. \ref{Fig14}. Clearly, both lines match the experimental
results.

Using Eq. (\ref{SL6}) to study the superconducting phase transition,
we can explain the main universal properties of the $T-B$ phase
diagram of the HF metal $\rm CeCoIn_5$ shown in Fig. \ref{Fig14}.
The latter substance is a $d$-wave superconductor with $T_c=2.3$ K,
while field tuned QCP with a critical field of $B_{c0}=5.1$ T
coincides with $B_{c2}$, the upper critical field where
superconductivity vanishes \cite{pag,pag2,bi}. Under the application
of magnetic fields $B_{c0}$, $\rm CeCoIn_5$ demonstrates the NFL
behavior \cite{oes}. It also follows from the above consideration
given in Subsection \ref{BC0C2} that $B_{c2}\geq B_{c0}$. Therefore,
the approximate equality $B_{c2}\simeq B_{c0}$ observed in $\rm
CeCoIn_5$ is an accidental coincidence that has to disappear under
the application of external factors. Indeed, $B_{c2}$ is determined
by $\lambda_0$ which in turn is given by the coupling of electrons
with magnetic, phonon, etc excitations rather than by $B_{c0}$. As a
result, under the application of pressure influencing differently
the coupling constant $\lambda_0$ and $B_{c0}$, the above
coincidence is lifted in complete agreement with experimental facts,
so that $B_{c2}>B_{c0}$ \cite{ronn} as has been shown in Subsection
\ref{BC0C2}. At relatively high temperatures, the
superconducting-normal phase transition in $\rm CeCoIn_5$ shown by
the solid line in the right panel of Fig. \ref{Fig14} is of the
second order \cite{bian,izawa} so that $S$ and the other
thermodynamic quantities are continuous at the transition
temperature $T_c(B)$. Since $B_{c2}\simeq B_{c0}$, upon the
application of magnetic field, the HF metal transits to its NFL
state down to lowest temperatures as it is seen from Fig.
\ref{Fig14}. As long as the phase transition is of the second order,
the entropy of the superconducting phase $S_{\rm SC}(T)$ coincides
with the entropy $S_{NFL}(T)$ of the NFL state and Eq. (\ref{SL6})
becomes
\begin{equation}\label{pq17b}
S_{SC}(T\to T_c(B))=S_{NFL}(T\to T_c(B)).
\end{equation}
Since $S_{\rm SC}(T\to 0)\to 0$, Eq. (\ref{pq17b}) cannot be
satisfied at sufficiently low temperatures due to the presence of
the temperature independent term $S_0$. Thus, in accordance with
experimental results \cite{bian,izawa}, the second order phase
transition converts to the first order one below some temperature
$T_{0}(B)$ \cite{shagstep}. To estimate $T_{0}(B)$, we use the
scaling idea of Volovik (see Ref. \cite{volvol} for details), who
derived the interpolation formula for the entropy of a $d$-wave
superconductor in a magnetic field $B$, while $S_{NFL}$ has been
estimated in \cite{yakov}. As a result, we obtain $T_0(B)/T_c\simeq
0.3$. This point coincides well with the experimental value, shown
on the Fig. \ref{Fig14}. Note that the prediction that the
superconducting phase transition may change its order had been made
in the early 1960-s \cite{maki}. Since our consideration is based on
purely thermodynamic reasoning, it is robust and can be generalized
to the cases when the superconducting phase is replaced by another
ordered state, e.g. ferromagnetic state or antiferromagnetic one.

Under constant entropy (adiabatic) conditions, there should be a
temperature step as a magnetic field crosses the phase boundary due
to the above thermodynamic inequality. Indeed, the entropy jump
would release the heat, but since $S=const$ the heat $q$ is
absorbed, causing the temperature to decrease in order to keep the
constant entropy of the NFL state. Note that the minimal jump is
given by the temperature-independent term $S_0$, and $q$ can be
quite large so that the corresponding HF metal can be used as an
effective cooler at low temperatures.

\section{Scaling behavior of heavy fermion systems}\label{UBH}

As we have seen in Section \ref{FLFC1} and Subsection \ref{HCEL1}
the core of Landau Fermi liquid theory, the effective mass $M^*_L$
practically does not depend on temperature $T$, magnetic field $B$
etc, $M^*_L(T,B)=M^*_L=const$ \cite{landau}. The thermodynamic functions
such as the entropy $S$, heat capacity $C$, magnetic susceptibility
$\chi$ behave as in the case of noninteracting Fermi gas, namely low
temperatures $S/T\propto C/T\propto \chi\propto M^*_L$. In other
words, when the inter-particle interaction is switching on and its
strength $\lambda$ is increasing, a noninteracting Fermi gas
continuously transforms into LFL with $S(\lambda)$, $M^*_L(\lambda)$
etc. becoming functions of $\lambda$, while the main scaling
behavior of LFL, $S\propto M^*_L(\lambda)T$, remains untouched. This
fact imposes strict conditions on the low temperature thermodynamic
properties causing LFL exhibit the scaling behavior, which could
be represented by some reference LFL with a normalized effective
mass $M^*_{NL}=M^*_L(T,B)/M^*_L\simeq 1$. As seen
from Fig. \ref{YBRHSIN}, in the case of HF metals the scaling behavior of
$M^*_N$ is different from that of $M^*_L$.

Here we show that despite of the very different microscopic nature of 2D
$^3$He and HF metals with various ground state magnetic properties
their NFL behavior is universal and can be captured well within the
framework of FCQPT \cite{obz,ks,ksk,volovik2,vol,epl2} that supports
the extended quasiparticles paradigm. We concentrate on the NFL
behavior observed when heavy fermion systems transit from their LFL
to NFL states. This area is mostly puzzling and important because
the behavior of the system in its transition state strongly depends
on the scenario shaping the corresponding QCP. For example, if the
transition region is described by theories based on quantum and
thermal critical fluctuations there are no theoretical grounds to
expect that these systems with different magnetic ground states
could exhibit a universal scaling behavior \cite{ste,vojta,geg,col2,col3}.

There are many measurements of the heat capacity $C(T,B)$, thermal
expansion coefficient $\alpha(T,B)$ and the magnetic AC
susceptibility $\chi(T,B)$ on strongly correlated Fermi systems such
as HF metals, high-$T_c$ superconductors and 2D $\rm^3He$ carried out
at different temperatures $T$, fixed magnetic fields $B$ and the
number density (or doping) $x$. Many of these measurements allow to
explore the systems at their transition from the LFL state to the NFL
one. Due to the equation
\begin{equation}\label{MM}C/T\propto S/T\propto
\sqrt{A}\propto\chi\propto\alpha/T\propto M^*,\end{equation}
relating all the above quantities to the effective mass,
these can be regarded as the
effective mass $M^*(T,B,x)$ measurements producing information about
the scaling behavior of the normalized effective mass $M^*_N$.

Experimental facts show that the effective mass extracted from
numerous measurements on different strongly correlated Fermi systems
upon using Eq. \eqref{MM} depends on magnetic field, temperature,
number density and composition. As we have seen and shall see, a
4D function describing the normalized effective mass is reduced to a
function of a single variable. Indeed, the normalized effective mass
depends (as the effective mass does) on magnetic field, temperature,
number density and the composition of a strongly correlated Fermi
system such as HF metals and 2D Fermi systems, and all these
parameters can be merged into the single variable $y$ by means of Eq.
\eqref{UN2}.

\subsection{Quantum criticality in {\rm 2D} $\rm ^3He$}

We now discuss how the scaling behavior of the normalized
effective mass $M^*_N$ given by Eq. \eqref{UN2} describes the
quantum criticality observed in 2D $^3$He \cite{cas,mor,3he}. This
quantum criticality is extremely significant as it allows us to check
the possibility of the scaling behavior in the 2D system formed by
$\rm ^3He$ atoms which are essentially different from electrons.
Namely, the neutral atoms of $\rm ^3He$ are fermions interacting
with each other by Van der Waals forces with strong hardcore
repulsion and a weakly attractive tail. The different character of
the inter-particle interaction along with the fact that the mass of
the $\rm ^3He$ atom is 3 orders of magnitude larger than that of an
electron, makes $\rm ^3He$ systems have drastically different
properties than those of HF metals. Because of this difference
nobody can be sure that the macroscopic physical properties of these
systems will be more or less similar to each
other at their QCP. The 2D
$^3$He has a very important feature: a change in the total density
of $^3$He film drives it towards QCP at which the quasiparticle
effective mass $M^*$ diverges \cite{cas,mor,3he} as seen from
Figs. \ref{Fig6} and \ref{MXM}. This peculiarity permits to plot the
experimental dependence of the normalized effective mass versus
temperature as a function of the number density $x$, which can be
directly compared with $M_N^*$ given by Eq. \eqref{UN2}. As a
result, 2D $\rm ^3He$, being an intrinsically isotropic Fermi-liquid
with negligible spin-orbit interaction becomes an ideal system to
test a theory describing the NFL behavior. Note that the bulk liquid
$\rm ^3He$ is historically the first object to which the LFL theory
had been applied \cite{landau}. One may speculate that at a
sufficiently high pressure the liquid $\rm ^3He$ would exhibits the
NFL behavior. Unfortunately, the application of pressure causes 3D
$\rm ^3He$ to solidify.

Let us consider HF liquid at $T=0$ characterized by the effective
mass $M^*$. As it was shown in Section \ref{FCDL}, at QCP $x=x_{FC}$
the effective mass diverges at $T=0$ and the system undergoes FCQPT.
The leading term of this divergence reads
\begin{equation}\label{zui22}
\frac{M^*(x)}{m}=a_1+\frac{a_2}{1-z},\ z=\frac {x}{x_{FC}},
\end{equation}
where $m$ is the bare mass. Equation \eqref{zui22} is valid in both 3D
and 2D cases, while the values of the factors $a_1$ and $a_2$ depend on
dimensionality and inter-particle interaction \cite{obz}. At
$x>x_{FC}$ (or $z>1$) FCQPT takes place. Here we confine ourselves
to the case $x<x_{FC}$. It is seen from Eq. \eqref{zui22} that FCQPT
takes place in 2D $\rm ^3He$ at elevated densities due to Van der
Waals forces with strong hardcore repulsion. This strong hardcore
repulsion makes the potential energy produce the main contribution to
the ground state energy resulting in strong rearrangement of the
single-particle spectrum and FCQPT. We recall that in the heavy
electron liquid FCQPT occurs at diminishing densities due to Coulomb
interaction.

When the system approaches QCP, the dependence of quasiparticle
effective mass on temperature and number density $x$ is governed by
Eq. \eqref{HC1}. It follows from Fig. \ref{PHD} that the effective
mass $M^*(T)$ as a function of $T$ at fixed $x$ reveals three
different regimes at growing temperature. At the lowest
temperatures we have the LFL state.  The effective mass grows,
reaching its maximum $M^*_M(T,x)$ at some temperature $T_M(x)$ and
subsequently diminishing as $T^{-2/3}$ as seen from Eq.
\eqref{HC11}. Moreover, the closer is the number density $x$ to its
threshold value $x_c$, the higher is the rate of the growth. The
peak value $M^*_M$ grows also, but the maximum temperature $T_M$
diminishes. Near the $T_M$ temperature the last "traces" of the LFL
state disappear, manifesting themselves in substantial growth of
$M^*(x)$. The temperature region beginning near $T_M(x)$ signifies
the crossover between the LFL state with almost constant effective
mass and the NFL behavior with the $T^{-2/3}$ dependence. Thus the
$T_M$ point can be regarded as crossover between the LFL and NFL
states or regimes.

As we have seen, $M^*(T,x)$ in the $T$ and $x$ range can be well
approximated by a simple universal interpolating function. The
interpolation occurs between the LFL ($M^*\propto T^2$) and NFL
($M^*\propto T^{-2/3}$) states, thus describing the above crossover.
Substituting $T$ by the dimensionless variable $y=T/T_M$, we obtain the
desired expression \eqref{UN2}. It is possible to calculate $T_M$ as
a function of $z$. Equation \eqref{zui22} shows that $M^*_M\propto
1/(1-z)$ and it follows from \eqref{HC11} that $M^*_M\propto
T^{-2/3}$. As a result, we obtain \cite{prl3he}
\begin{equation}\label{zui4}
T_M\propto(1-z)^{3/2}.
\end{equation}

Equation \eqref{MM} demonstrates that $M^*(T)$ can be measured in
experiments on strongly correlated Fermi systems.  Upon normalizing
both $M^*(T)$ by its peak value at each $x$ and the temperature by
$T_M$, we see from Eq. (\ref{UN2}) that all the curves merge into
a single one demonstrating a scaling behavior.

\begin{figure} [! ht]
\begin{center}
\includegraphics [width=0.60\textwidth]{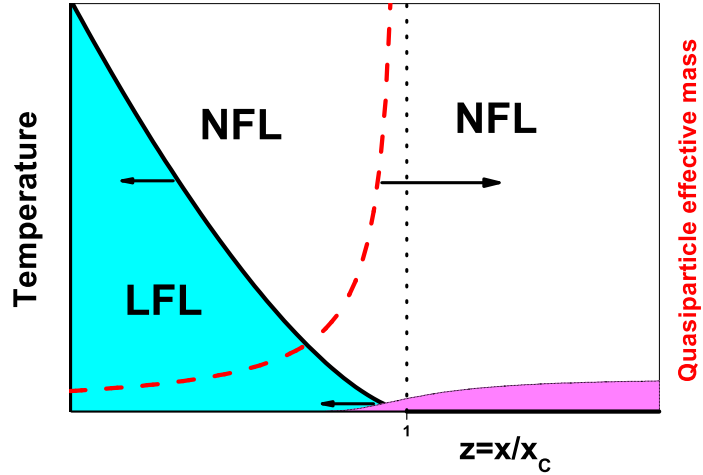}
\end{center}
\vspace*{-1.0cm} \caption{The phase diagram of the 2D $\rm ^3He$
system. The part for $z<1$ corresponds to HF behavior divided into
the LFL and NFL parts by the line $T_M(z)\propto (1-z)^{3/2}$,
where $T_M$ is the temperature at which the effective mass reaches
its maximum. The exponent $3/2=1.5$ coming from Eq. \eqref{zui4} is
in good agreement with the experimental value $1.7\pm 0.1$
\cite{3he}. The dependence $M^*(z)\propto (1-z)^{-1}$ is shown by
the dashed line. The regime for $z\geq 1$ consists of the LFL piece
(the shadowed region, beginning in the intervening phase $z\leq 1$
\cite{3he}, which is due to the substrate inhomogeneities, see
text) and NFL regime at higher temperatures.}\label{fd}
\end{figure}

In Fig. \ref{fd}, we show the phase diagram of 2D $^3$He in the
variables $T$ and $z$ (see Eq. \eqref{zui22}). For the sake of
comparison, the plot of the effective mass versus $z$ is shown by
the dashed line. The part of the diagram where $z<1$ corresponds to HF
behavior and consists of LFL and NFL parts, separated by the line
$T_M(z)\propto (1-z)^{3/2}$. We note here that our
exponent $3/2=1.5$ is exact as compared to that from Ref. \cite{3he}
$1.7\pm 0.1$. The good agreement between the theoretical and
experimental exponents supports our FCQPT
description of the NFL behavior of both 2D $^3$He and HF metals;
the former system is in great detail similar to the latter. The regime for
$z>1$ consists of a low-temperature LFL piece (shadowed region,
beginning in the intervening phase $z\leq 1$ \cite{3he}) and the NFL
state at higher temperatures. The former LFL piece is related to
the peculiarities of the substrate on which the 2D $\rm ^3He$ film is
placed. Namely, it is related to weak substrate heterogeneity (steps
and edges on its surface) so that quasiparticles, being localized
(pinned) on it, give rise to the LFL behavior \cite{3he}. The
competition between thermal and pinning energies returns the system
back to NFL state and hence restores the NFL behavior. Note, that the
presence of the substrate can be considered as the main difference between
2D $\rm ^3He$ and HF metals. Namely, the latter metals do not have a
substrate, the above LFL piece would be absent or very thin if some
3D disorder (like point defects, dislocations etc) is present in a
HF metals.

\begin{figure} [! ht]
\begin{center}
\includegraphics [width=0.60\textwidth]{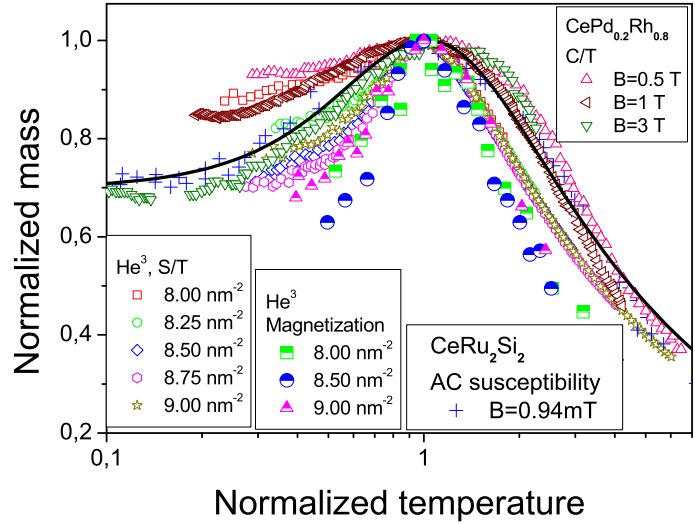}
\end{center}
\vspace*{-1.0cm} \caption{The normalized effective mass $M^*_N$ as a
function of the normalized temperature $T/T_M$ at densities shown in
the left lower corner. The behavior $M^*_N$ is extracted from
experimental data for the entropy in 2D $^3$He \cite{3he} and 3D HF
compounds with different magnetic ground states such as
$\rm{CeRu_2Si_2}$ and $\rm CePd_{1-x}Rh_x$ \cite{takah,pikul},
fitted by the universal function \eqref{UN2}.}\label{f2}
\end{figure}

In Fig. \ref{MXM}, we report the experimental values of the effective
mass $M^*(z)$ obtained by the measurements on $^3$He monolayer
\cite{cas}. These measurements, in coincidence with those from Ref.
\cite{3he}, show the divergence of the effective mass at $x=x_c$. To
show, that our FCQPT approach is able to describe the above data, we
represent the fit of $M^*(z)$ by the fractional expression coming from
Eq. \eqref{zui22} and the reciprocal effective mass by the linear
fit $M/M^*(z)\propto a_1z$.
We apply the universal dependence \eqref{UN2} to fit the
experimental data not only in 2D $^3$He but in 3D HF metals as well.
$M^*_N(y)$ extracted from the entropy measurements on the $^3$He
film \cite{3he,3he1} at different densities $x<x_c$ smaller then the
critical point $x_c=9.9 \pm 0.1$ nm$^{-2}$ is reported in Fig.
\ref{f2}. In the same figure, the data extracted from the heat capacity
of the ferromagnet CePd$_{0.2}$Rh$_{0.8}$
\cite{pikul} and the $AC$ magnetic
susceptibility of the paramagnet CeRu$_2$Si$_2$ \cite{takah} are plotted
for different magnetic fields. It is seen that the scaling
behavior of the normalized effective mass given by Eq. (\ref{UN2}) is in accord
with the experimental facts.
\begin{figure} [! ht]
\begin{center}
\includegraphics [width=0.60\textwidth]{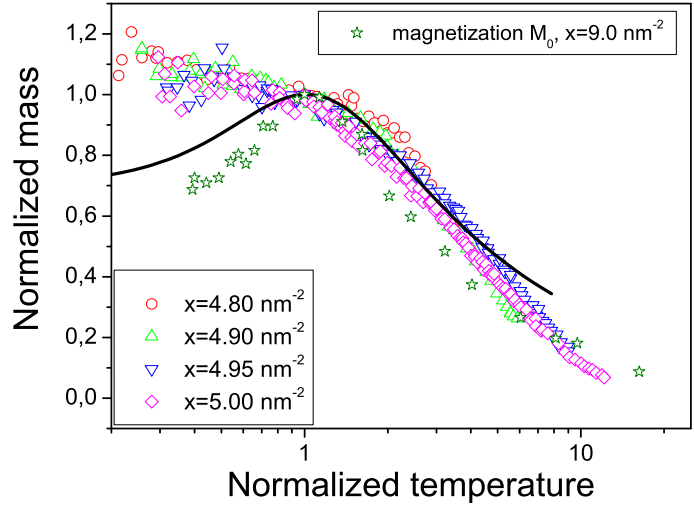}
\end{center}
\vspace*{-1.0cm} \caption{The dependence of
$M^*_N(T/T_M)$ on $T/T_M$ at densities
shown in the left lower corner. The behavior $M^*_N$ is extracted
from experimental data for $C(T)/T$ in 2D $^3$He \cite{cas} and for
the magnetization $M_0$ in 2D $^3$He \cite{3he}. The solid curve
shows the universal function, see the caption to Fig.
\ref{f2}.}\label{PRLC}
\end{figure}
All substances are located at QCP, where the system progressively
disrupts its LFL behavior at elevated temperatures. In that case the
control parameter, which drives the system towards its QCP $x_{FC}$
is represented merely by a number density $x$. It is seen that the
behavior of the effective mass $M^*_N(y)$, extracted from $S(T)/T$
in 2D $^3$He (the entropy $S(T)$ is reported in Fig. S8 A of Ref.
\cite{3he1}) looks very much like that in 3D HF compounds as was
shown in Sections \ref{HCEL}.

The attempt to fit the available experimental data for $C(T)/T$ in
$\rm ^3He$ \cite{cas} by the universal function $M^*_N(y)$ is
reported in Fig. \ref{PRLC}. Here, the data extracted from heat
capacity $C(T)/T$ for $^3$He monolayer \cite{cas} and magnetization
$M_0$ for bilayer \cite{3he,3he1}, are reported. It is seen that
the normalized effective mass extracted from these thermodynamic
quantities can be well described by Eq. \eqref{UN2}. We note the
qualitative similarity between the double layer \cite{3he} and
monolayer \cite{cas} of $^3$He seen from Fig. \ref{PRLC}.

\begin{figure} [! ht]
\begin{center}
\includegraphics [width=0.60\textwidth]{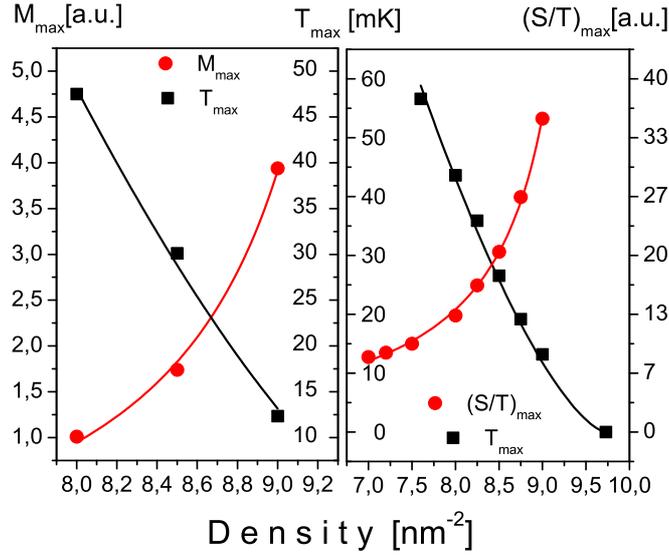}
\end{center}
\vspace*{-1.0cm} \caption{Left panel, the peak values $M_{\rm max}$
and the peak temperatures $T_{\rm max}$ extracted from measurements
of the magnetization $M_0$ in $^3$He \cite{3he,3he1} are plotted
versus density. Right panel shows $T_{\rm max}$ and the peak values
$(S/T)_{\rm max}$ extracted from measurements of $S(T)/T$ in $^3$He
\cite{3he,3he1} also versus density. We approximate $T_{\rm
max}=a_1(1-z)^{3/2}$, Eq. \eqref{zui4}, and $(S/T)_{\rm max}\propto
M_{\rm max}=a_2/(1-z)$, Eq. \eqref{zui22}, with $a_1$ and $a_2$
fitting parameters.}\label{STM}
\end{figure}
On the left panel of Fig. \ref{STM}, we show the density dependence
of $T_{\rm max}$, extracted from measurements of the magnetization
$M_0(T)$ on $^3$He bilayer \cite{3he,3he1}. The peak temperature is
fitted by Eq. \eqref{zui4}. On the same figure, we have also
reported the maximal magnetization $M_{\rm max}$. It is seen that
$M_{\rm max}$ is well described by the expression $M_{\rm max}
\propto (S/T)_{\rm max}\propto (1-z)^{-1}$, see Eq. (\ref{zui22}).
The right panel of Fig. \ref{STM} reports the peak temperature
$T_{\rm max}$ and the maximal entropy $(S/T)_{\rm max}$ versus the
number density $x$. They are extracted from the measurements of
$S(T)/T$ on $^3$He bilayer \cite{3he,3he1}. The fact that both the
left and right panels  extracted from $M_0(T)$ and $S/T$
demonstrate the same behavior shows once more that there are indeed
the quasiparticles, which determine the thermodynamic behavior of
2D $^3$He (and also 3D HF compounds \cite{epl2}) near the point of
their effective mass divergence.

As seen from Fig. \ref{STM}, the amplitude and positions of the
maxima of magnetization $M_0(T)$ and $S(T)/T$ in 2D $^3$He follow
well Eqs. \eqref{zui22} and \eqref{zui4}, while Eq. \eqref{UN2}
describes the scaling behavior of the normalized thermodynamic
functions. We recall that we can calculate only relative values of
the effective mass, that is the normalized effective mass, since the
real values of $T_M$ and $M^*_M$ are determined by the specific
properties of the system in question. Thus, with only two
values defining both the real value, for example, of the entropy and
the corresponding temperature, it is possible to calculate the
thermodynamic or transport properties of HF metals or 2D $\rm ^3He$.
We conclude that Eq. \eqref{UN2} allows us to reduce a 4D function
describing the normalized effective mass to a function of a single
variable. Indeed, the normalized effective mass depends on magnetic
field, temperature, number density and the composition of a strongly
correlated Fermi system such as HF metals and 2D Fermi systems, and
as we have seen above, all these parameters can be merged into the
single variable by means of Eq. \eqref{UN2} \cite{prl3he}.  We note
that the validity of Eq. \eqref{UN2} is confirmed by numerical
calculations as described in Subsection \ref{PHDS}.

In conclusion of this Subsection, we have
described the diverse experimental facts
related to temperature and number density (2D number density)
dependencies of different thermodynamic characteristics of 2D $^3$He
by the single universal function of one argument. The above
universal behavior is also inherent to HF metals with different
magnetic ground state properties. The amplitude and positions of the
maxima of the magnetization $M_0(T)$ and $S(T)/T$ in 2D $^3$He are
also well described. We have shown that bringing the different
experimental data collected on strongly correlated Fermi systems to
the above form immediately reveals their universal scaling behavior.

\subsection{Kinks in the thermodynamic functions}

\begin{figure} [! ht]
\begin{center}
\vspace*{-0.8cm}
\includegraphics [width=0.80\textwidth]{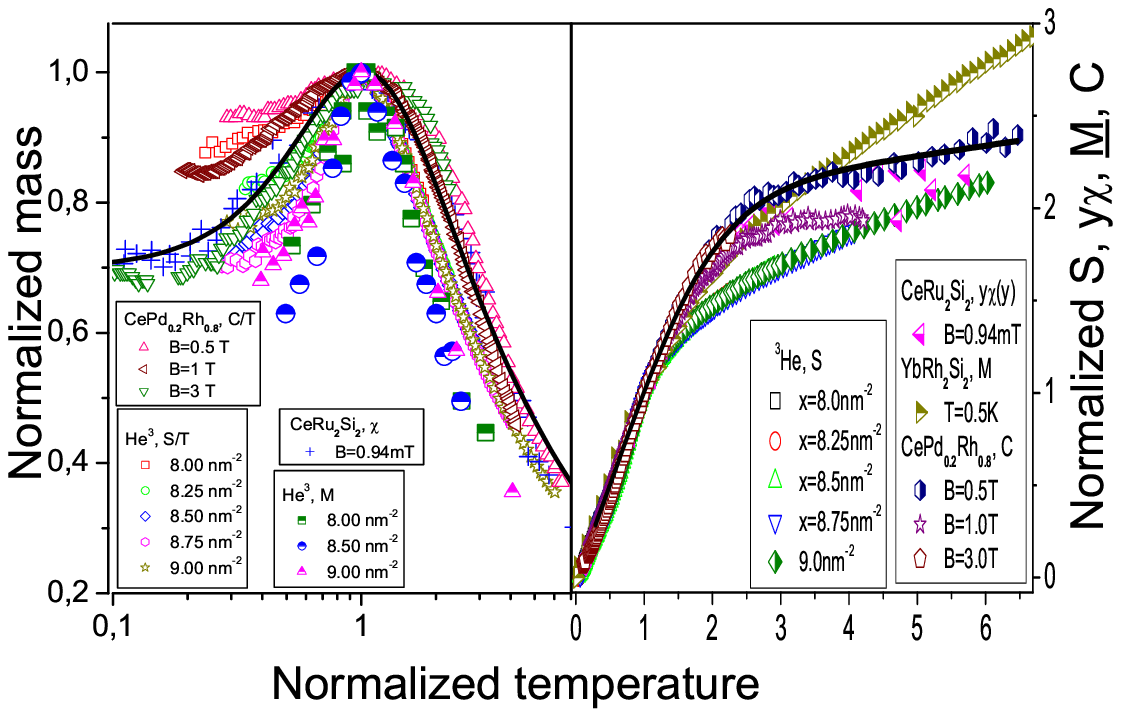}
\end{center}
\vspace*{-0.8cm} \caption{Energy scales in HF metals and 2D
$\rm^3He$. The left panel. The normalized effective mass $M^*_N$
versus the normalized temperature $y=T/T_M$. The dependence $M^*_N
(y)$ is extracted from measurements of $S(T)/T$ and magnetization
$M$ on 2D $\rm{^3He}$ \cite{3he}), from $AC$ susceptibility
$\chi(T)$ collected on $\rm{ CeRu_2Si_2}$ \cite{takah} and from
$C(T)/T$ collected on both $\rm{ CePd_{1-x}Rh_x}$ \cite{pikul} and
$\rm CeCoIn_5$ \cite{bbian}. The data are collected for different
densities and magnetic fields shown in the left bottom corner. The
solid curve traces the universal behavior of the normalized
effective mass determined by Eq. (\ref{UN2}). Parameters $c_1$ and
$c_2$ are adjusted for $\chi_{N}(T_N,B)$ at $B=0.94$ mT. The right
panel. The normalized specific heat $C(y)$ of $\rm{ CePd_{1-x}Rh_x}$
and $\rm CeCoIn_5$ at different magnetic fields $B$, normalized
entropy $S(y)$ of $\rm{ ^3He}$ at different number densities $x$,
and the normalized $y\chi(y)$ at $B=0.94$ mT versus normalized
temperature $y$ are shown. The upright triangles depict the
normalized "average" magnetization $\underline{M}=M+B\chi$ collected
on $\rm{ YbRu_2Si_2}$ \cite{steg}. The kink (shown by the arrow) in
all the data is clearly seen in the transition region $y\geq 1$. The
solid curve represents $yM^*_N(y)$ with parameters $c_1$ and $c_2$
adjusted for the magnetic susceptibility of $\rm{ CeRu_2Si_2}$ at
$B=0.94$ mT.}\label{kink3}
\end{figure}

To illuminate kinks or energy scales observed in the thermodynamic
functions measured on HF metals \cite{steg} and 2D $\rm^3He$, we
present in Fig. \ref{kink3} the normalized effective mass $M^*_N$
extracted from the thermodynamic functions versus normalized
temperature (the left panel) and the normalized thermodynamic functions
proportional to $T_NM^*_N$ (the right panel) as a function of the
normalized temperature $T_N$ \cite{plamr}.

$M^*_N(y)$ extracted from the entropy $S(T)/T$ and magnetization $M$
measurements on the $^3$He film \cite{3he} at different densities
$x$ is reported in the left panel of Fig. \ref{kink3}. In the same
panel, the data extracted from the heat capacity of the ferromagnet
$\rm{ CePd_{0.2}Rh_{0.8}}$ \cite{pikul}, $\rm CeCoIn_5$ \cite{bbian}
and the AC magnetic susceptibility of the paramagnet $\rm{
CeRu_2Si_2}$ \cite{takah} are plotted for different magnetic fields.
It is seen that the universal behavior of the normalized effective
mass given by Eq. (\ref{UN2}) and shown by the solid curve is in
accord with the experimental facts. It is seen that the behavior of
$M^*_N(y)$, extracted from $S(T)/T$ and magnetization $M$ of 2D
$\rm{ ^3He}$ looks very much like that of 3D HF compounds. In the
right panel of Fig. \ref{kink3}, the normalized data on $C(y)$,
$S(y)$, $y\chi(y)$ and $\underline{M}=M(y)+y\chi(y)$ extracted from
data collected on $\rm{ CePd_{1-x}Rh_x}$ \cite{pikul} , $\rm{ ^3He}$
\cite{3he}, $\rm{ CeRu_2Si_2}$ \cite{takah}, $\rm CeCoIn_5$
\cite{bbian} and $\rm{ YbRu_2Si_2}$ \cite{steg} respectively are
presented. Note that in the case of $\rm{ YbRu_2Si_2}$, the variable
$y=(B-B_{c0})\mu_B/T_M$ can be viewed as effective normalized
temperature. We remark that in Subsection \ref{HCEL6} we calculate
$\underline{M}$ as a function of magnetic field.

It is seen from the right panel of Fig. \ref{kink3} that all the
data exhibit the kink (shown by arrow) at $y\geq 1$ taking place as
soon as the system enters the transition region from the LFL state
to the NFL one. This region corresponds to the temperatures where the
vertical arrow in Fig. \ref{PHD} {\bf a}
crosses the hatched area separating the LFL
from NFL behavior. It is also seen that the low temperature LFL
scale of the thermodynamic functions (as a function of $y$) is
characterized by the fast growth, and the high temperature scale
related to the NFL behavior is characterized by the slow growth. As
a result, we can identify the energy scales near QCP, discovered in
Ref. \cite{steg}: the thermodynamic characteristics exhibit the
kinks (crossover points from the fast to slow growth at elevated
temperatures) which separate the low temperature LFL scale and high
temperature one related to the NFL state.

\subsection{Heavy-fermion metals at metamagnetic phase transitions}

A Fermi system can be driven to FCQPT when narrow bands situated
close to the Fermi surface are formed by the application of a high
critical magnetic field $B_m$. The emergence of such state is known
as metamagnetism that occurs when this transformation comes abruptly
at $B_m$ \cite{grig}.

\begin{figure}[!ht]
\begin{center}
\includegraphics[width=0.60\textwidth]{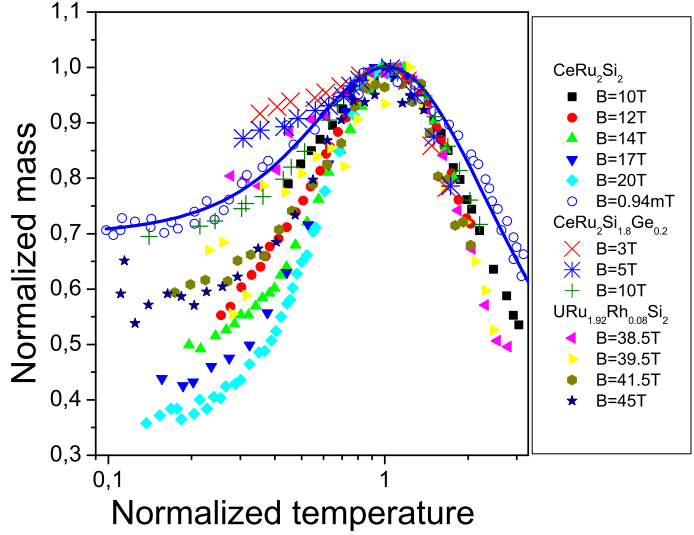}
\end{center}
\caption{The normalized effective mass  as
a function of magnetic field
versus the normalized temperature. $M^*_N(T_N)$ is extracted from
measurements of $C/T$ collected on $\rm URu_{1.92}Rh_{0.08}Si_2$,
CeRu$_2$Si$_2$ and $\rm CeRu_2Si_{1.8}Ge_{0.2}$ at different
magnetic fields \cite{silh,kim} shown in the right panel. The solid
curve gives the universal behavior of $M^*_N$ given by Eq.
\eqref{UN2}, see also the caption to Fig. \ref{UB}.} \label{meta}
\end{figure}
Let us assume that the magnetic field $B_m$ is similar to that of
$B_{c0}$ driving a HF metal to its magnetic field tuned QCP. In our
simple model both $B_{c0}$ and $B_m$ are taken as parameters. To
apply equation \eqref{UN2} when the critical magnetic field is not
zero, we have to replace $B$ by $(B-B_{m})$. Acting as above, we can
extract the normalized effective mass $M^*_N(T_N)$ from data
collected on HF metals at their metamagnetic QCP. In Fig. \ref{meta}
the extracted normalized mass is displayed. $M^*_N(T_N)$ is
extracted from measurements of $C/T$ collected on $\rm
URu_{1.92}Rh_{0.08}Si_2$, $\rm CeRu_2Si_2$ and $\rm
CeRu_2Si_{1.8}Ge_{0.2}$ at their metamagnetic QCP with $B_m\simeq$
35 T, $B_m\simeq$ 7 T and $B_m\simeq$ 1.2 T respectively
\cite{silh,kim}. As seen from Fig. \ref{meta}, the effective mass
$M^*_N(T_N)$ in different HF metals reveals the same form both in
the high magnetic field and in low ones as soon as the corresponding
bands become flat, that is the electronic system of HF metals is
driven to FCQPT. This observation is extremely significant as it
allows us to check the universal behavior in HF metals when these
are under the application of essentially different magnetic fields.
Namely, the magnitude of the applied field ($B\sim 10$ T) at the
metamagnetic point is four orders of magnitude larger than that of
the field applied to tune CeRu$_2$Si$_2$ to the LFL behavior ($B\sim
1$ mT). Relatively small values of $M^*_N(T_N)$ observed in $\rm
URu_{1.92}Rh_{0.08}Si_2$ and CeRu$_2$Si$_2$ at the high fields and
small temperatures can be explained by taking into account that the
narrow band is completely polarized \cite{silh}. As a result, at low
temperatures the summation over the spins "up" and "down" reduces to
a single direction producing the coefficient $1/2$ in front of the
normalized effective mass. At high temperatures the summation is
restored. As seen from Fig. \ref{meta}, these observations are in
accord with the experimental facts.

\section{ Asymmetric conductivity
in HF metals and high-$T_c$ superconductors} \label{TUN}

The main subjects of investigation in experiments on HF
metals are the thermodynamic properties. Therefore, it seems reasonable to
study the properties of HF liquids that are determined by the
quasiparticle distribution function $n({\bf p},T)$ and not only by
the density of states or by the behavior of the effective mass $M^*$
\cite{obz,shagpopov,tun,spsk,zag}. As we shall see in this Section, the
FC solutions $n_0({\bf p})$ leads to the NFL behavior and violate
the particle-hole symmetry inherent in LFL and generate dramatic
changes in transport properties of HF metals, particularly, the
differential conductivity becomes asymmetric. As was shown in
Section \ref{HFL}, the LFL behavior is restored under the
application of magnetic field. Thus, we expect that in magnetic
fields the asymmetric part of the differential conductivity is
suppressed. Scanning tunnel microscopy and point-contact spectroscopy
closely related to the Andreev reflection are
sensitive to both the density of states and
the probability of the population of quasiparticle
states determined by the function $n({\bf p},T)$  \cite{guy,andr}.
Thus, scanning tunnel microscopy and point-contact
spectroscopy are ideal tools for studying specific features of the
NFL behavior of HF metals and high-$T_c$ superconductors.

\subsection{Normal state}

The tunnel current $I$ running through a point contact of two
ordinary metals is proportional to the applied voltage $V$ and to
the square of the absolute value of the quantum mechanical
transition amplitude $t$ times the difference $N_1(0)N_2(0)(n_1(p,
T)-n_2(p,T))$ \cite{zag}, where $N_{1}(0)$
$N_{2}(0)$ are the density of
states of the respective metals and $n_2(p,T))$ and
$n_2(p,T)$ are respectively the distribution functions
of the respective metals. On the other hand, in the
semiclassical approximation, the wave function that determines the
amplitude $t$ is proportional to $(N_1(0)N_2(0))^{-1/2}$. Therefore,
the density of states drops out from the final result and the tunnel
current becomes independent of $N_1(0)N_2(0)$. Because the
distribution $n(p,T\to0)\to\theta(p_F-p)$ as $T\to0$, where
$\theta(p_F-p)$ is the step function, it can be verified that the
differential tunnel conductivity $\sigma_d (V)=dI/dV$ is a symmetric
or even function of $V$ in the Landau Fermi-liquid theory. Actually,
the symmetry of $\sigma_d(V)$ is obeyed if there is the
hole-quasiparticle symmetry (which is present in the LFL theory).
Hence, the fact that $\sigma_d(V)$ is symmetric is obvious and is
natural in the case of metal-metal contacts for ordinary metals that
are in the normal or superconducting state.

\begin{figure}[!ht]
\begin{center}
\includegraphics[width=0.60\textwidth]{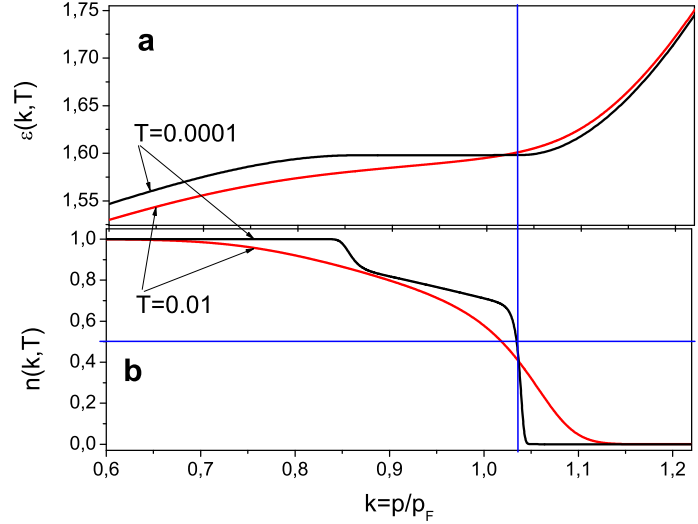}
\end{center}\vspace*{-0.7cm}
\caption{The single particle energy $\varepsilon({\bf k},T)$ ({\bf
a}) and the distribution function $n({\bf k},T)$ ({\bf b}) at
finite temperatures as functions of the dimensionless variable
$k=p/p_F$. The arrows show temperature measured in $T/E_F$. At
$T=0.0001$ the vertical line shows the position of the Fermi level
$E_F$ at which $n({\bf k},T)=0.5$ as depicted by the horizontal
line. At diminishing temperatures $T\to0$, the single particle
energy $\varepsilon({\bf k},T)$ becomes more flat in the region
$(p_f-p_i)$ and the distribution function $n({\bf k},T)$ in this
region becomes more asymmetrical with respect to the Fermi level
$E_F$ producing the particle-hole asymmetry related to the NFL
behavior.} \label{figfc}
\end{figure}
We study the tunnel current at low temperatures, which for ordinary
metals is given by the expression \cite{zag,guy}
\begin{equation}\label{tun1}
I(V)=2|t|^2\int\left[n(\varepsilon-V)-
n(\varepsilon)\right]d\varepsilon.
\end{equation}
where we use the atomic system of units $e=m=\hbar =1$ and
normalize the transition amplitude to unity, $|t|^2=1$. Since the
temperatures are low, we can approximate the distribution function
$n(\varepsilon)$ by the step function $\theta(\mu-\varepsilon)$;
Eq. (\ref{tun1}) then yields $I(V)=a_1V$, and hence the
differential conductivity $\sigma_d(V)=dI/dV=a_1=const$ is a
symmetric function of the applied voltage $V$.

To quantitatively examine the behavior of the asymmetric part of
the conductivity $\sigma_d(V)$, we find the derivatives of both
sides of Eq. (\ref{tun1}) with respect to $V$. The result is the
following equation for $\sigma_d(V)$:
\begin{equation}
\sigma_d=\frac{1}{T}\int n(\varepsilon(z)-V,T)
(1-n(\varepsilon(z)-V,T)) \frac{\partial \varepsilon}{\partial
z}dz,\label{tun2}
\end{equation}
In the integrand in Eq. \eqref{tun2}, we used the dimensionless momentum
$z=p/p_F$ instead of $\varepsilon$ for the variable, because
$n$ is no longer a function of $\varepsilon$ in the
case of a strongly correlated electron liquid; it depends on the
momentum as shown in Figs. \ref{Fig1} and \ref{figfc}. Indeed, the
variable $\varepsilon$ in the interval $(p_f-p_i)$ is equal to
$\mu$, and the quasiparticle distribution function varies within
this interval. It is seen from Eq. \eqref{tun2} that the violation
of the particle-hole symmetry makes $\sigma_d(V)$ asymmetric as a
function of the applied voltage $V$  \cite{obz,shagpopov,tun,spsk}.

The single particle energy $\varepsilon({\bf k},T)$
shown in Fig. \ref{figfc} ({\bf a}) and the corresponding $n({\bf k},T)$
shown in the panel ({\bf b}) evolve
from the FC state characterized by
$n_0({\bf k},T=0)$ determined by Eq. \eqref{FL8}. It is seen
from Fig. \ref{figfc} ({\bf a}), that at elevated temperatures the
dispersion $\varepsilon({\bf k},T)$ becomes more inclined since the
effective mass $M^*(T)$ diminishes as seen from Eq.
\eqref{FL12}. At the Fermi level $\varepsilon(p,T)=\mu$, then
from Eq. \eqref{FL4} the distribution function $n(p,T)=1/2$.
The vertical line in Fig. \ref{figfc} crossing the distribution
function at the Fermi level illustrates the asymmetry of the distribution
function with respect to the Fermi level at $T=0.0001$. It is clearly
seen that the FC state strongly violates the particle-hole symmetry
at diminishing temperatures. As a result, at low temperatures the
asymmetric part of the differential conductivity becomes larger.
Under the application of magnetic fields the system transits to the
LFL state that strongly supports the particle hole symmetry.
Therefore, the application of magnetic fields restoring the symmetry
suppresses the asymmetric part of the differential
conductivity.

\begin{figure} [! ht]
\begin{center}
\includegraphics [width=0.60\textwidth] {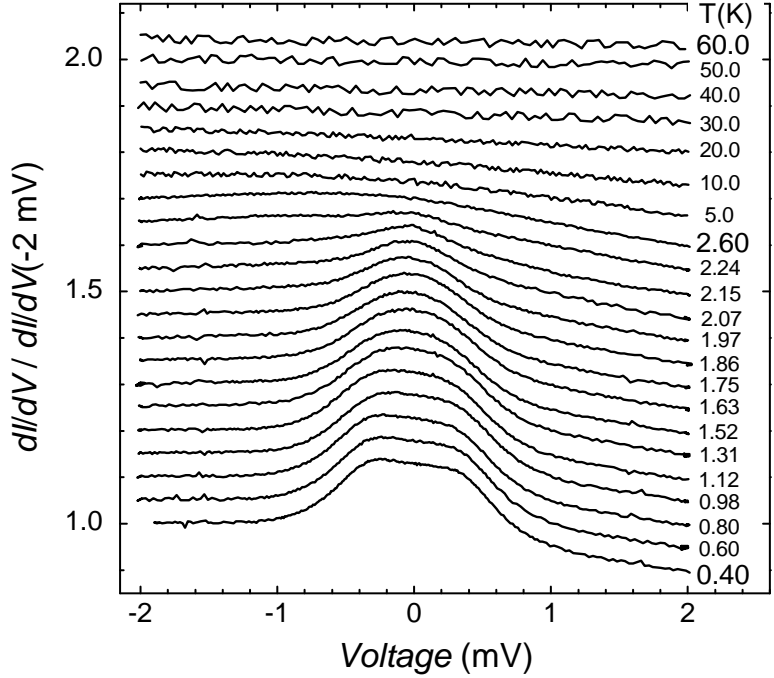}
\end {center}
\caption{Differential conductivity $\sigma_d(V)$ measured in the
case of point contacts Au/CeCoIn$_5$. The curves $\sigma_d(V)$ are
displaced along the vertical axis by 0.05. The conductivity is
normalized to its value at $V=-2$ mV. The asymmetry becomes
noticeable at $T<45$ K and increases as the temperature decreases
\cite{park}.} \label{Fig1_t}
\end{figure}

After performing fairly simple transformations in Eq. (\ref{tun2}),
we find that the asymmetric part
$$\Delta \sigma_d(V)=(\sigma_d(V)-\sigma_d(-V))/2$$
of the differential conductivity can be expressed as
\begin{eqnarray}
&&\Delta\sigma_d(V)=\frac{1}{2}\int\frac{\alpha(1-\alpha^2)}
{[n(z,T)+\alpha[1-n(z,T)]^2} \nonumber\\
&\times&\frac{\partial n(z,T)}{\partial z}\frac{1-2n(z,T)}{[\alpha
n(z,T)+[1-n(z,T)]]^2}dz,\label{tun3}
\end{eqnarray}
where $\alpha=\exp(-V/T)$.

\begin{figure} [! ht]
\begin{center}
\includegraphics [width=0.60\textwidth] {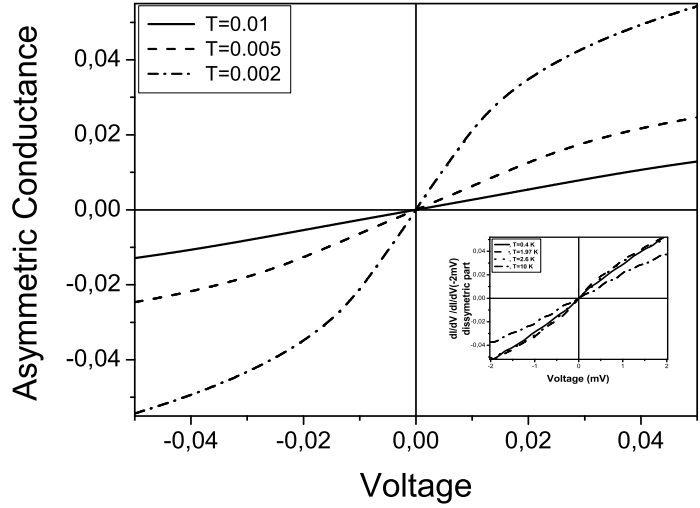}
\end {center}
\caption{The asymmetric conductivity $\Delta \sigma_d(V)$ as a
function of $V/\mu$ for three values of the temperature $T/\mu$
(normalized to $\mu$). The inset shows the behavior of the
asymmetric conductivity extracted from the data in Fig.
\ref{Fig1_t}.} \label{Fig2_t}
\end{figure}

Asymmetric tunnel conductivity can be observed in measurements
involving metals whose electron system is located near FCQPT or
behind it. Among such metals are high-$T_c$
superconductors and heavy-fermion metals, e.g.,
YbRh$_2$(Si$_{0.95}$Ge$_{0.05}$)$_2$, $\rm CeCoIn_5$, $\rm
YbCu_{5-x}Al_{x}$ or YbRh$_2$Si$_2$. The measurements must be
conducted when the heavy-fermion metal is in the superconducting or
normal state. If the metal is in its normal state, measurements of
$\Delta\sigma_d(V)$ can be done in a magnetic field $B>B_{c0}$ at
temperatures $T^*(B)<T\leq T_f$ or in a zero magnetic field at
temperatures higher than the corresponding critical temperature when
the electron system is in the paramagnetic state and its behavior is
determined by the entropy $S_0$.

Recent measurements of the differential conductivity in $\rm CeCoIn_5$
carried out using by the point-contact
spectroscopy technique \cite{park} have
vividly revealed the asymmetry in the differential conductivity in
the superconducting ($T_c=2.3$ K) and normal states. Figure
\ref{Fig1_t} shows the results of these measurements. Clearly,
$\Delta\sigma_d(V)$ is nearly constant when the heavy-fermion metal is in
the superconducting state, experiencing no substantial variation near
$T_c$, see also Fig. \ref{CEasym} below.
Then it monotonically decreases as the temperature increases
\cite{park}.

Figure \ref{Fig2_t} shows the results of calculations of the
asymmetric part $\Delta\sigma_d(V)$ of the conductivity
$\sigma_d(V)$ obtained from Eq. (\ref{tun3}) \cite{shagpopov}.
In calculating
the distribution function $n(z,T)$, we used the functional (\ref{HC6})
(with the parameters $\beta=3$ and $g=8$). In this case,
$(p_f-p_i)/p_F\simeq 0.1$. Figure \ref{Fig2_t} also shows that the
asymmetric part $\Delta \sigma_d(V)$ of the conductivity is a linear
function of $V$ for small voltages. Consistent with the Fig.
\ref{figfc} showing that the asymmetry of $n({\bf k},T)$ diminishes
at elevated temperatures, the asymmetric part decreases with
increasing temperature, which agrees with the behavior of the
experimental curves in the inset in Fig. \ref{Fig2_t}.

We now derive an estimate formula for analyzing the asymmetric
part of the differential conductivity. It follows from Eq.
(\ref{tun3}) that for small values of $V$, the asymmetric part
behaves as $\Delta\sigma_d(V)\propto V$. Here, it is appropriate to
note that the asymmetric part of the tunnel conductivity is an odd
function of $V$, and therefore $\Delta\sigma_d(V)$ must change sign
when $V$ changes sign. The natural unit for measuring voltage is
$2T$, because this quantity determines the characteristic energy for
FC, as shown by Eq. (\ref{FL14}). Actually, the asymmetric part must
be proportional to the size $(p_f-p_i)/p_F$ of the region occupied
by FC:
\begin{equation}
\Delta\sigma_d(V)\simeq c\frac{V}{2T}\frac{p_f-p_i} {p_F}\simeq
c\frac{V}{2T}\frac{S_0}{x_{FC}}.\label{tun4}
\end{equation}
where $S_0/x_{FC}\sim (p_f-p_i)/p_F$ is the temperature-independent
part of the entropy [see Eq. \eqref{SL1a}] and $c$ is a constant of
the order of unity. For instance, calculations of $c$ using the
distribution function displayed in Fig. \ref{figfc} yield $c\sim 1$.
From Eq. (\ref{tun4}) we see that when $V\simeq 2T$ and FC
occupies a sizable part of the Fermi volume, $(p_f-p_i)/p_F\simeq1$,
the asymmetric part becomes comparable to the differential tunnel
conductivity $\Delta\sigma_d (V)\sim V_d(V)$.

\subsubsection{Suppression of the asymmetrical differential
resistance in $\rm
YbCu_{5-x}Al_{x}$ in magnetic fields} \label{suppb}

\begin{figure} [! ht]
\begin{center}
\includegraphics [width=0.65\textwidth] {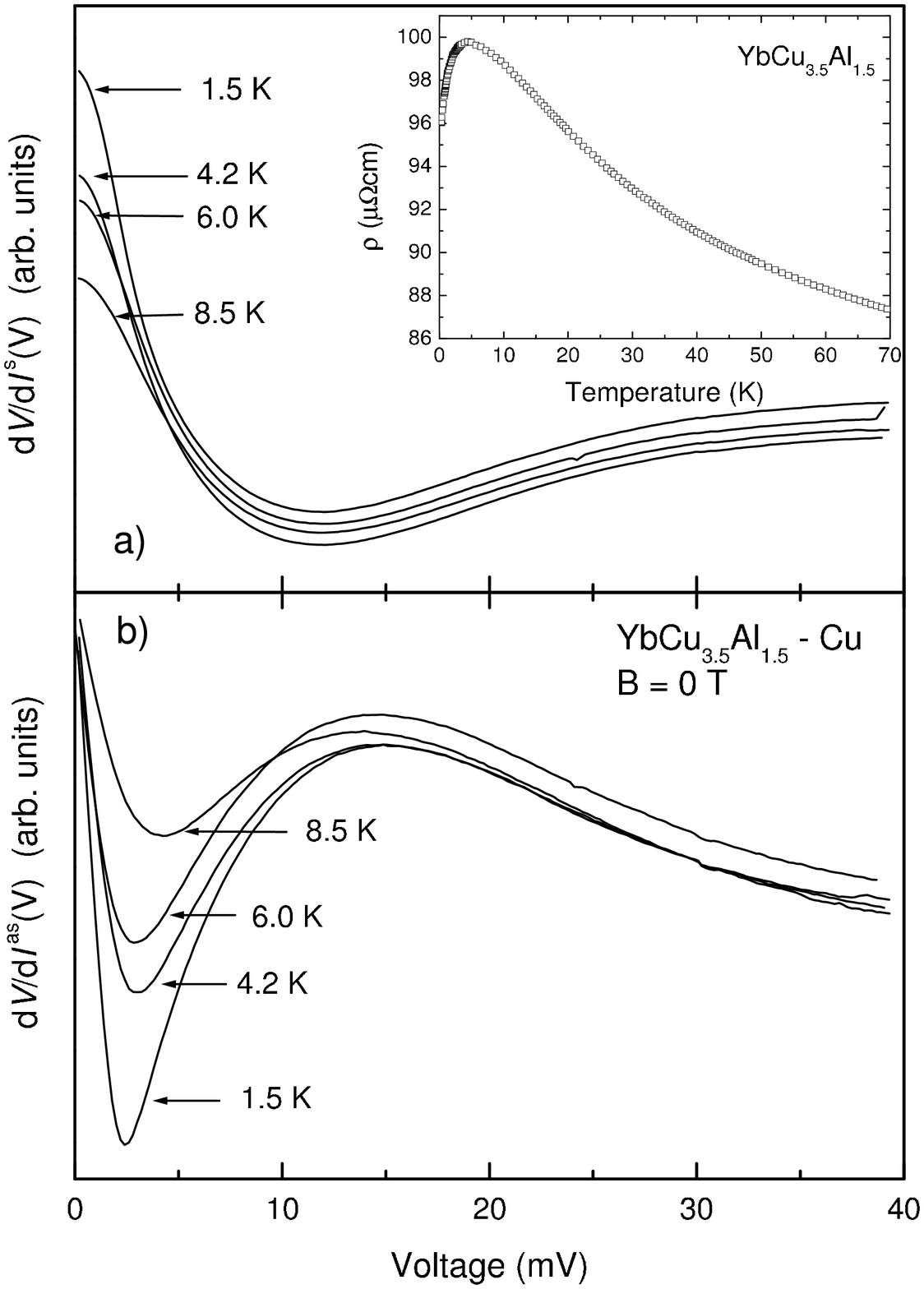}
\end {center}
\caption{Characteristic temperature behavior of (a) symmetric
$dV/dI^{s}(V)$ and (b) asymmetric $dV/dI^{as}(V)$ parts of
$dV/dI(V)$ for heterocontact $\rm YbCu_{3.5}Al_{1.5}-Cu$ at $B=0$ T
and different temperatures shown by the arrows. The inset shows the
bulk resistivity $\rho(T)$ of $\rm YbCu_{3.5}Al_{1.5}$
\cite{asymc}.} \label{at1}
\end{figure}

Now consider the behavior of the asymmetric part of the differential
conductivity $\Delta \sigma_d(V)$ under the application of a magnetic
field $B$. Obviously, the differential conductivity being a scalar
should not to depend on the direction of current $I$. Thus, the
non-zero value of $\Delta \sigma_d(V)$ manifests the violation of the
particle-hole symmetry on a macroscopic scale.
As we have seen in Section \ref{HFL} and Subsection \ref{HCEL1}, at
sufficiently low temperatures $T<T^*(B)$, the application of a
magnetic field $B>B_{c0}$ leads to restoration of the LFL behavior
eliminating the particle-hole asymmetry, and therefore
the asymmetric part of the differential conductivity disappears
\cite{shagpopov,tun}. This prediction is in accord with the
experimental facts collected in measurements on $\rm
YbCu_{5-x}Al_{x}$ of the differential resistance $dV/dI(V)$ under
the application of magnetic fields \cite{asymc}. Representing the
differential resistance as the sum of its symmetrical $dV/dI^s(V)$
and the asymmetrical part $dV/dI^{as}(V)$,
$$dV/dI(V)=dV/dI^s(V)+dV/dI^{as}(V),$$ we obtain the equation
\begin{equation}\label{asymm}
\Delta \sigma_d(V)\simeq -\frac{dV/dI^{as}(V)}{[dV/dI^{s}(V)]^2}.
\end{equation}
Deriving Eq. \eqref{asymm}, we assume that $dV/dI^s(V)\gg
dV/dI^{as}(V)$. Figure \ref{at1} \cite{asymc} shows the temperature
evolution of (a) the symmetric $dV/dI^s(V)$ and (b) the asymmetric
$dV/dI^{as}(V)$ parts at zero applied magnetic field. Also for the
case of a heterocontact, the behavior of the symmetric part does not
show a decrease in $\rho(T)$, while the asymmetric part decreases at
elevated temperatures \cite{asymc}. It seen from Fig. \ref{at1} that
the behavior of the asymmetric part of the differential resistance
given by  Eqs. \eqref{tun4} and \eqref{asymm} is in accord with the
experimental facts.

\begin{figure} [! ht]
\begin{center}
\includegraphics [width=0.65\textwidth] {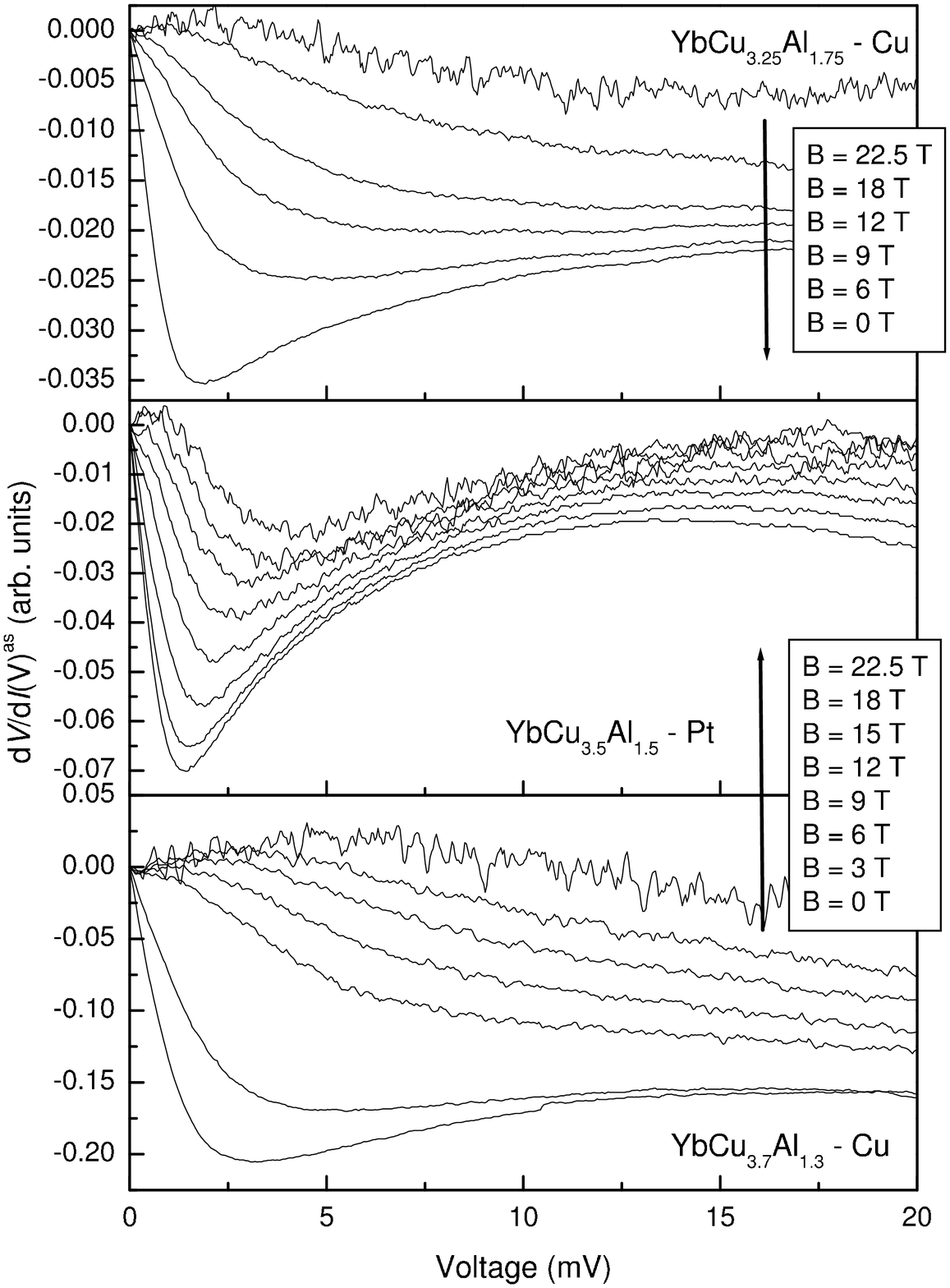}
\end {center}
\caption{Characteristic magnetic-field behavior of the asymmetric
part $dV/dI^{as}(V)$ of the differential conductivity is shown
versus magnetic fields displayed in the legends for
heterocontacts with different $x=1.3$, $1.5$, and $1.75$ at $1.5$ K
\cite{asymc}.} \label{at2}
\end{figure}
It is seen from Fig. \ref{at2} \cite{asymc} that increasing magnetic
fields suppress the asymmetric part. Thus, the application of
magnetic fields destroys the NFL behavior and recovers both the LFL
state and the particle-hole symmetry.
Correspondingly, we conclude that the particle-hole symmetry is
macroscopically broken in the absence of applied magnetic fields,
while the application of magnetic fields restores both the
particle-hole symmetry and the LFL state. It is seen from Figs.
\ref{at1} and \ref{at2} that the asymmetric part shows a linear
behavior as function of the voltage below about 1 mV \cite{asymc} as
predicted \cite{shagpopov}.

\subsection{Superconducting state}

Tunnel conductivity may remain asymmetric as a high-$T_c$
superconductor or a HF metal pass into the
superconducting state from the normal state. The reason is that the
function $n_0({\bf p})$ again determines the differential
conductivity. As we saw in Section \ref{SC}, $n_0({\bf p})$ is not
noticeably distorted by the pairing interaction, which is relatively
weak compared to the Landau interaction, which forms the
distribution function $n_0({\bf p})$. Hence, the asymmetric part of
the conductivity remains practically unchanged for $T\leq T_c$,
which agrees with the results of experiments (see Fig.
\ref{Fig1_t}). In calculating the conductivity using the results
of measurements with a tunneling microscope, we must bear in mind
that the density of states in the superconducting state
\begin{equation}
N_S(E)=N(\varepsilon-\mu)\frac{E}{\sqrt{E^2-\Delta^2}}, \label{tun6}
\end{equation}
determines the conductivity, which is zero for $E\leq|\Delta|$.
Here, $E$ is the quasiparticle energy given by Eq. (\ref{SC3.1}),
and $\varepsilon-\mu=\sqrt{E^2-\Delta^2}$.
Equation (\ref{tun6}) implies that the tunnel conductivity may be
asymmetric if the density of states in the normal state
$N(\varepsilon)$ is asymmetric with respect to the Fermi level
\cite{pand}, as is the case with strongly correlated Fermi systems
with FC. Our calculations of the density of states based on model
functional (\ref{HC6}) with the same parameters as those used in
calculating $\Delta \sigma_d(V)$ shown in Fig. \ref{Fig2_t}
corroborate this conclusion.

\begin{figure} [! ht]
\begin{center}
\includegraphics [width=0.60\textwidth] {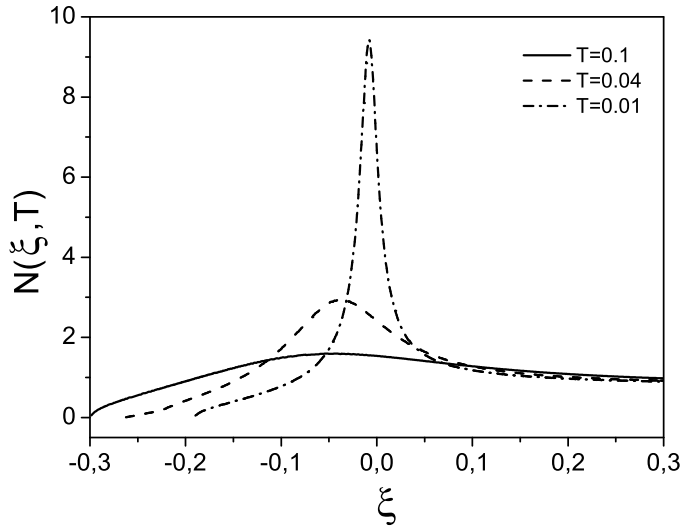}
\end{center}
\caption {Density of states $N(\xi,T)$ as a function of
$\xi=(\varepsilon-\mu)/\mu$, calculated for three values of the
temperature $T$ (normalized to $\mu$).}\label {Fig7}
\end{figure}

Figure \ref{Fig7} shows the results of calculations of the density
of states $N(\xi,T)$. Clearly, $N(\xi,T)$ is strongly asymmetric
with respect to the Fermi level. If the system is in the
superconducting state, the values of the normalized temperature
given in the upper right corner of the diagram can be related to
$\Delta_1$. With $\Delta_1\simeq 2T_c$, we find that $2T/\mu\simeq
\Delta_1/\mu$. Because $N(\xi,T)$ is asymmetric, the first
derivative $\partial N(\xi,T)/\partial\xi$ is finite at the Fermi
level, and the function $N(\xi,T)$ can be written as
$N(\xi,T)\simeq a_0+a_1\xi$ for small values of $\xi$. The
coefficient $a_0$ contributes nothing to the asymmetric part.
Obviously, the value of $\Delta \sigma_d(V)$ is determined by the
coefficient $a_1\propto M^*(\xi=0)$. In turn, $ M^*(\xi=0)$ is
determined by Eq. \eqref{SC7}. As a result, Eq. (\ref{tun6}) yields
\begin{equation}\Delta \sigma_d(V)\sim c_1\frac{V}{|\Delta|}\frac{S_0}{x_{FC}},
\label{tun7}\end{equation} because $(p_f-p_i)/p_F\simeq
S_0/x_{FC}$, the energy $E$ is replaced by the voltage $V$, and
$\xi=\sqrt{V^2-\Delta^2}$. The entropy $S_0$ here refers to the
normal state of a heavy-fermion metal.
\begin{figure}
\begin{center}
\includegraphics[width=0.60\textwidth]{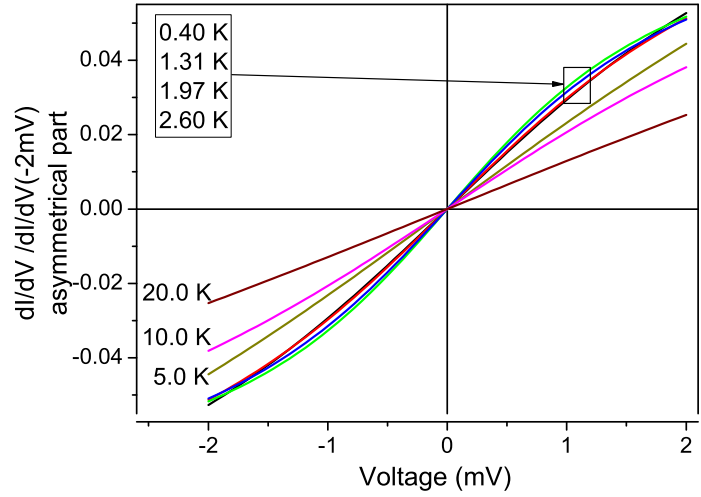}
\end{center}
\caption{ Temperature dependence of the asymmetric parts
$\Delta\sigma_d(V)$ of the conductance spectra extracted from
measurements on  $\rm CeCoIn_5$ \cite{park}. The temperatures are
boxed and shown by the arrow for $T\leq 2.60$ K, otherwise
by numbers near the curves.} \label{CEasym}
\end{figure}

\begin{figure} [! ht]
\begin{center}
\includegraphics [width=0.60\textwidth] {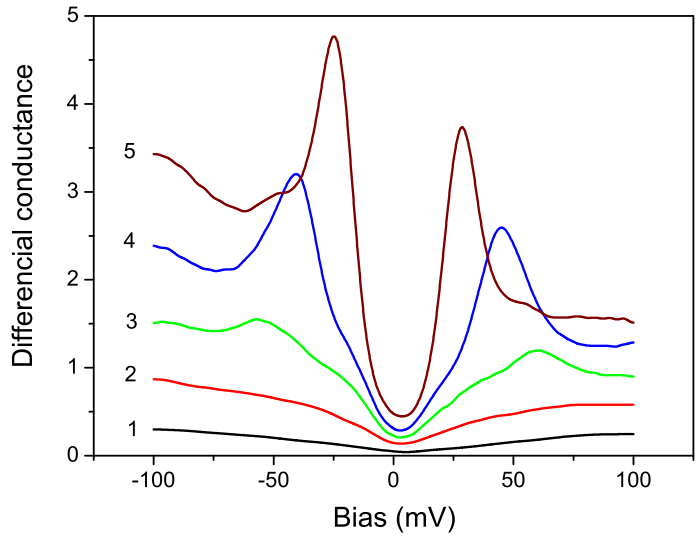}
\end {center}
\caption{Spatial variation of the spectra of the differential
tunnel conductivity measured in $\rm Bi_2Sr_2CaCu_2O_{8+x}$. Lines
1 and 2 belong to regions in which the integrated local density of
states is very low. Low differential conductivity and the absence
of a gap are indications that we are dealing with an insulator.
Line 3 corresponds to a large gap (65 meV) with mildly pronounced
peaks. The integrated value of the local density of states for
curve 3 is small, but is larger than that for lines 1 and 2. Line 4
corresponds to a gap of about 40 meV, which is close to the average
value. Line 5 corresponds to the maximum integrated local density
of states and the smallest gap about of 25 meV, and has two sharp
coherent peaks \cite{pan}.}\label{Fig3_t}
\end{figure}
Actually, Eq. (\ref{tun7}) coincides with Eq. (\ref{tun4}) if we use
the fact that the characteristic energy of the superconducting state
is determined by Eq. (\ref{SC8}) and is temperature-independent. In
studies of the universal behavior of the asymmetric conductivity,
Eq. (\ref{tun7}) has proved to be more convenient than (\ref{tun6}).
It follows from Eqs. (\ref{tun4}) and (\ref{tun7}) that measurements
of the transport properties (the asymmetric part of the
conductivity) allow the determination
of the thermodynamic properties of the
normal phase that are related to the entropy $S_0$. Equation
\eqref{tun7} clearly shows that the asymmetric part of the
differential tunnel conductivity becomes comparable to the
differential tunnel conductivity at $V\sim 2|\Delta|$ if FC occupies
a substantial part of the Fermi volume, $(p_f-p_i)/p_F\simeq1$. In
the case of the $d$-wave symmetry of the gap, the right-hand side of
Eq. \eqref{tun7} must be averaged over the gap distribution
$\Delta(\phi)$, where $\phi$ is the angle. This simple procedure
amounts to redefining the gap size or the constant $c_1$. As a
result, Eq. (\ref{tun7}) can also be applied when $V<\Delta_1$,
where $\Delta_1$ is the maximum size of the  $d$-wave gap
\cite{tun}. For the Andreev reflection, where the current is finite
for any small value of $V$, Eq. (\ref{tun7}) also holds for
$V<\Delta_1$ in the case of the $s$-wave gap.

\begin{figure} [! ht]
\begin{center}
\includegraphics [width=0.60\textwidth] {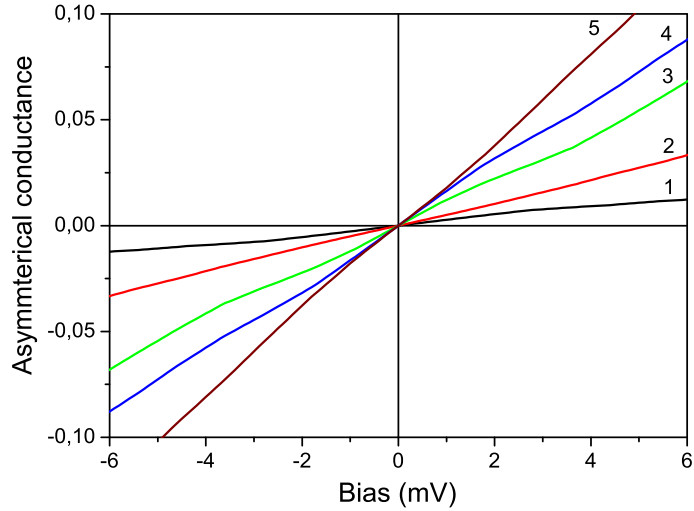}
\end {center}
\caption{The asymmetric part $\Delta\sigma_d(V)$ of the differential
tunnel conductivity in the high-$T_c$ superconductor $\rm
Bi_2Sr_2CaCu_2O_{8+x}$, extracted from the data in Fig.
\ref{Fig3_t}, as a function of the voltage $V$ (mV). The lines are
numbered consistent with the numbers of the lines in Fig.
\ref{Fig3_t}.}\label{Fig4_t}
\end{figure}
It is seen from Fig. \ref{CEasym} that the asymmetrical part
$\Delta\sigma_d(V)$ of the conductivity remains constant up to
temperatures of about $T_c$ and persists up to temperatures well
above $T_c$. At small voltages the asymmetric part is a linear
function of $V$ and starts to diminish at $T\geq T_c$. It follows
from Fig. \ref{CEasym} that the description of the asymmetric part
given by Eqs. (\ref{tun4}) and (\ref{tun7}) coincides with the facts
obtained in measurements on $\rm CeCoIn_5$.

Low-temperature measurements with tunneling microscopy and spectroscopy
techniques were used in \cite{pan} to detect an inhomogeneity in the
electron density distribution in $\rm Bi_2Sr_2CaCu_2O_{8+x}$. This
inhomogeneity manifests itself as spatial variations in the local
density of states in the low-energy part of the spectrum and in the
size of the superconducting gap. The inhomogeneity observed in the
integrated local density of states is not caused by impurities but
is inherent in the system. Observation facilitated
relating the value of
the integrated local density of states to the concentration $x$ of
local oxygen impurities.

\begin{figure} [! ht]
\begin{center}
\includegraphics [width=0.60\textwidth] {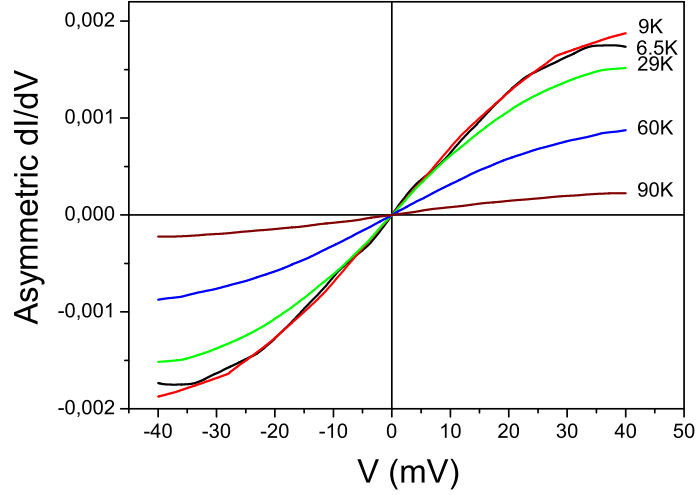}
\end {center}
\caption{Temperature dependence of the asymmetric part
$\Delta\sigma_d(V)$ of the conductivity spectra obtained in
measurements for $\rm YBa_2Cu_3O_{7-x}/La_{0.7}Ca_{0.3}MnO_3$ by the
contact spectroscopy method; the critical temperature $T_c\simeq 30$
K \cite{samanta}.} \label{Fig5_t}
\end{figure}
Spatial variations in the differential tunnel conductivity spectrum
are shown in Fig. \ref{Fig3_t}. Clearly, the differential tunnel
conductivity is highly asymmetric in the superconducting state of
$\rm Bi_2Sr_2CaCu_2O_{8+x}$. The differential tunnel conductivity
shown in Fig. \ref{Fig3_t} may be interpreted as measured at
different values of $\Delta_1(x)$ but at the same temperature,
which allows studying the $\Delta \sigma_d(V)$ dependence on
$\Delta_1(x)$.
Figure  \ref{Fig4_t} shows the asymmetric conductivity diagrams
obtained from the data in Fig. \ref{Fig3_t}. Clearly, for small
values of $V$, $\Delta\sigma_d(V)$ is a linear function of voltage
consistent with (\ref{tun7}) and the slope of the respective
straight lines $\Delta\sigma_d(V)$ is inversely proportional to the
gap size $\Delta_1$.

Figure \ref{Fig5_t} shows the variation in the asymmetric part of
the conductivity $\Delta\sigma_d(V)$ as the temperature increases.
The measurements were done on $\rm
YBa_2Cu_3O_{7-x}/La_{0.7}Ca_{0.3}MnO_3$ with $T_c\simeq 30$ K
\cite{samanta}. Clearly, at $T<T_c$ in the region of the linear
dependence on
$V$, the asymmetric part $\Delta\sigma_d(V)$ of the conductivity
depends only weakly on the temperature; such behavior agrees with
(\ref{tun7}). When $T>T_c$, the slope of the straight line
sections of the $\Delta\sigma_d(V)$ diagrams decreases as the
temperature increases; this behavior is described by Eq.
(\ref{tun4}). We conclude that the description of the universal
behavior of $\Delta\sigma_d(V)$ based on the FCQPT is in good
agreement with the results of the experiments presented in Figs.
\ref{Fig2_t}, \ref{at1}, \ref{at2}, \ref{CEasym}, \ref{Fig4_t}, and
\ref{Fig5_t} and is valid for both high-$T_c$ superconductors and
heavy-fermion metals.

\section{Violation of the Wiedemann-Franz law in HF metals}

As early as in 1853, German physicists Gustav Wiedemann and Rudolph
Franz \cite{widfr} discovered the empirical law stating that for a
metal at a constant temperature the ratio of its thermal
conductivity $\kappa (T)$ to its electrical conductivity $\sigma
(T)$ is a constant, $\kappa(T)/\sigma(T)=$const. Later on, the
Danish physicist Ludvig Valentin Lorenz showed that the above
ratio is proportional to the temperature $T$,
$\kappa(T)/\sigma(T)=LT$, the proportionality constant $L$ is known
as the Lorenz number. What is called Wiedemann-Franz (WF) law is
indeed an independence of the Lorenz number $L$ on temperature.
However, it was firmly established that the WF law is obeyed both at
room temperatures and for low ones (several Kelvins); at the
intermediate temperatures $L=L(T)$.

Strictly speaking, the Lorenz number is temperature-independent
only at low temperatures; its theoretical value
\begin{equation}\label{wf1}
L_0=\lim _{T\to 0}=\frac{\kappa(T)}{T\sigma(T)}=\frac{\pi
^2}{3}\frac{k_B}{e^2}
\end{equation}
($k_B$ and $e$ are Boltzmann constant and electron charge,
respectively) had been calculated by Sommerfeld in 1927 \cite{som1}
in the model of noninteracting electrons, obeying Fermi-Dirac
statistics. The same result is obtained in LFL theory and reflects
merely the fact that both thermal and electrical conductivities of a
metal are determined by Landau quasiparticles. Due to this fact,
possible deviations from the WF law can be regarded as a signature of
NFL behavior in a sample.

Actually, Eq. \eqref{wf1} is usually referred to as the
Wiedemann-Franz (WF) law. It was shown that at $T=0$ Eq. \eqref{wf1}
remains valid for arbitrarily strong scattering \cite{osd1},
disorder \cite{osd2} and interactions \cite{osd3}. This law holds
for ordinary metals \cite{x5,x6,x7,x8} and does not
hold for HF metals \cite{pag2,x10,x12} CeNiSn and CeCoIns,
the electron-doped material \cite{x13} Pr$_{2-x}$Ce$_x$CuO$_{4-y}$, and
the underdoped compound \cite{x14} YbBa$_2$Cu$_3$O$_y$.
In $\rm CeNiSn$, the
experimental value of the reduced Lorenz number $L(T)/L_0 \sim 1.5$
changes little at $T<1$ K. This rules out the phonon contribution
to the violation of the WF law. In the electron-doped compound
Pr$_{2-x}$Ce$_x$CuO$_{4-y}$ the departure of $L(T)$ from $L_0$ at $T
> 0.3$ K is also by more then unity and even larger \cite{x13} than
that in CeNiSn. Other experimental tests of the WF
law have been undertaken in the
normal state of cuprate superconductors. The phase diagram of these
compounds shows evolution from Mott insulator for undoped materials
towards metallic Fermi liquid behavior for overdoped cases. Upward
shift $L/L_0 \simeq 2 - 3$ was measured in underdoped cuprates at
the lowest temperatures \cite{x13,x14,tand23}. In strongly
overdoped cuprates, the WF law was found to be obeyed perfectly
\cite{cyr}.

The physical mechanism for the WF law violation is usually attributed to
the NFL behavior like in Luttinger and Laughlin liquids
\cite{kane,kane1,tand18,tand19} or in the case of a marginal Fermi
liquid \cite{var}. Yet another possibility
for the LFL theory and the WF law
\eqref{wf1} violation occurs near QCPs where the effective mass
$M^*$ of a quasiparticle diverges. This is because at the QCP the
Fermi liquid spectrum with finite Fermi velocity $v_F=p_F/M^*$
becomes meaningless as in this case $v_F\to 0$. In a standard
scenario of the QCP \cite{col2,col1,x16} the divergence of the
effective mass is attributed to the vanishing of the quasiparticle
weight $z$ in the single-particle states close to second-order phase
transitions, implying that the quasiparticles disappear in this
region. A conventional scenario of the WF law violation, associated with
critical fluctuations in the vicinity of the second order phase
transition, has recently been suggested in \cite{x17}.
However, it has been shown in several works \cite{obz,khodb,x18},
that the standard scenario of the QCP is flawed so that to describe
the deviations of $L$ from the WF value $L_0$, we apply a scenario of
the QCP, where the NFL peculiarities are due to FCQPT. That
is related to the rearrangement of the single-particle spectrum of
strongly correlated electron liquid with the conservation of
quasiparticle picture within the extended paradigm.

Therefore, to describe theoretically the violation of the WF law within
the FCQPT formalism, it is sufficient to use the well-known LFL formulas
for thermal and electrical conductivities with the substitution of the
modified single particle spectrum into them. Such theory has been
advanced in Refs. \cite{xopi,xopi1}. The authors showed that close
to the QCP the Lorenz number $L_{\rm QCP}(T=0)=1.81\,L_0$, i.e.
almost two times larger than that from the LFL theory \eqref{wf1}.
This result agrees well with the experimental values \cite{x10,x13}.
Furthermore, the dependence $L(T)/L_0$ has been calculated for
two topologically distinct phases (see Section \ref{TFT}) -
"iceberg" phase and FC phase \cite{xopi,xopi1}. Theoretical
calculations have shown that in both phases the largest departure
from the WF law occurs near QCP \cite{xopi,xopi1}. Deep in the
"iceberg" phase we have the reentrance of the "classical" WF law in a
sense that $L=L_0$ while in the deep FC phase the Lorenz number is
temperature independent at low temperatures, but its value is
slightly larger than $L_0$. This is due to the particle-hole symmetry
violation in FC phase \cite{shagpopov,tun,physbmy}.

Recently, the anisotropy of the WF law violation near the QCP has
been experimentally observed in the HF metal
$\rm CeCoIn_5$ \cite{x12}. In that paper,
the above HF compound has been studied
experimentally in external magnetic fields, close to the critical
value $H_{c2}$, suppressing the superconductivity. Under these
conditions, the WF law was found to be violated. The violation is
anisotropic and cannot be attributed to the standard scenario of
quasiparticle collapse. At the same time, close to the QCP,
sufficiently large external magnetic fields reveal
the anisotropy of the electrical conductivities
$\sigma _{ik}\propto <v_iv_k>$ ($v_i$ are the components
of the group velocity vector) and thermal conductivities $\kappa_{ik}
\propto <\varepsilon ({\bf p})v_iv_k>$ of a substance. This is
because the magnetic field does not affect the $z$-components of the
group velocity ${\bf v}$ so that the QCP $T$-dependence of the
transport coefficients holds, triggering the violation of the WF
relation $L_{zz} = \sigma _{zz}/T\kappa_{zz} =\pi^2k_B/3e^2$. On the
other hand, the magnetic field ${\bf B}$ alters substantially the
electron motion in the perpendicular direction, yielding considerable
increase of the $x$ and $y$ components of the group velocity so that the
corresponding components $L_{ik}$ do not depart from their WF value.

Therefore, the flattening of the single particle spectrum $\varepsilon
(\bf p)$ of strongly correlated electron systems considerably
changes their transport properties, especially beyond the point of
FCQPT due to breaking of the particle-hole symmetry. Also, in
topologically different "iceberg" phases the WF law is also violated
near its QCP. The results of theoretical \cite{xopi} and
experimental investigations demonstrate that the FCQPT scenario with
further occurrence of both "iceberg" and FC phases give natural
and universal explanation of the NFL changes of the transport
properties of HF compounds and the WF law violation in particular.

\section{The impact of FCQPT on ordinary continuous phase transitions in HF
metals}\label{OPT}

The microscopic nature of quantum criticality determining the NFL
behavior in strongly correlated fermion systems of different types
is still unclear. Many puzzling and common experimental features of
such seemingly different systems as two-dimensional (2D) electron
systems and liquid $^3$He as well as 3D heavy-fermion metals and
high-$T_c$ superconductors suggest that there is a hidden
fundamental law of nature, which remains to be recognized. To reveal
this hidden law "the projection" of microscopic properties of the above
materials on their observable, macroscopic characteristics is
needed. One such "projections" is the impact of the FCQPT phenomenon
on the ordinary phase transitions in  HF metals. As we have seen in
Subsection \ref{CeCoIn}, the main peculiarity here is the continuous
magnetic field evolution of the superconductive phase transition
from the second order to the first one \cite{bian,izawa,maki}. The
same changing of the order is valid for magnetic phase transitions.

Exciting measurements on $\rm YbRh_2Si_2$ at antiferromagnetic (AF)
phase transition revealed a sharp peak in low-temperature specific
heat, which is characterized by the critical exponent $\alpha=0.38$
and therefore differs drastically from those of the conventional
fluctuation theory of second order phase transitions \cite{TNsteg},
where $\alpha\simeq0.1$ \cite{lanl2}. The obtained large value of
$\alpha$ casts doubts on the applicability of the conventional
theory and sends a real challenge for theories describing the second
order phase transitions in HF metals \cite{TNsteg}, igniting strong
theoretical effort to explain the violation of the critical
universality in terms of the tricritical point
\cite{sap,misav,imada,kling}.

The striking feature of FCQPT is that it has profound influence on
thermodynamically driven second order phase transitions provided
that these take place in the NFL region formed by FCQPT. As a
result, the curve of second order phase transitions passes into a
curve of the first order ones at the tricritical point leading to a
violation of the critical universality of the fluctuation theory.
For example, as we have seen in Subsection \ref{BC0C2} the second
order superconducting phase transition in $\rm CeCoIn_5$ changes to
the first one in the NFL region. It is this feature that provides the
key to resolve the challenge.

\subsection{$T-B$ phase diagram for $\rm YbRh_2Si_2$ versus
one for $\rm CeCoIn_5$}

\begin{figure}[!ht]
\begin{center}
\includegraphics [width=0.60\textwidth]{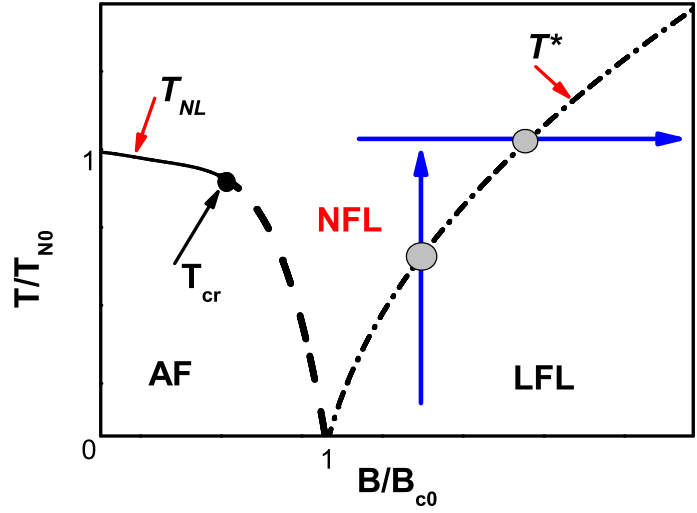}
\vspace*{-0.5cm}
\end{center}
\caption{ Schematic $T-B$ phase diagram for $\rm YbRh_2Si_2$. The
solid and dashed $T_{NL}$ curves separate the AF and NFL states
representing the field dependence of the N\'eel temperature. The
black dot at $T=T_{cr}$ shown by the arrow  in the dashed curve is
the tricritical point, at which the curve of second order AF phase
transitions shown by the solid line passes into the curve of the
first ones. At $T<T_{cr}$, the dashed line represents the field
dependence of the N\'eel temperature when the AF phase transition is
of the first order. The NFL state is characterized by the entropy
$S_{0}$ given by Eq. \eqref{SL1a}. The dash-dot line separating the
NFL state and the weakly polarized LFL state is represented by $T^*$
given by Eq. \eqref{BC0}. The horizontal solid arrow represents the
direction along which the system transits from the NFL behavior to
the LFL one at elevated magnetic field and fixed temperature. The
vertical solid arrow represents the direction along which the system
transits from the LFL behavior to the NFL one at elevated
temperature and fixed magnetic field. The hatched circles depict the
transition temperature $T^*$ from the NFL to LFL
behavior.}\label{fig2}
\end{figure}

In Fig. \ref{fig2}, the $T_{NL}$ line represents temperature
$T_{NL}(B)/T_{N0}$ versus field $B/B_{c0}$ in the schematic phase
diagram for $\rm YbRh_2Si_2$, with $T_{N0}=T_{NL}(B=0)$. There
$T_{NL}(B)$ is the N\'eel temperature as a function of the magnetic
field $B$. The solid and dashed curves indicate the boundary of the
AF phase at $B/B_{c0}\leq 1$ \cite{geg}. For $B/B_{c0}\geq 1$, the
dash-dot line marks the upper limit of the observed LFL behavior.
This dash-dot line coming from Eq. \eqref{HF8} separates the NFL
state and the weakly polarized LFL state, and in that case is
represented by
\begin{equation}\label{BC0}
\frac{T^*}{T_{NL}}=a_1\sqrt{\frac{B}{B_{c0}}-1},
\end{equation}
where $a_1$ is a parameter. We note that Eq. \eqref{BC0} is in good
agreement with experimental facts \cite{geg}.
Thus, $\rm YbRh_2Si_2$ demonstrates
two different LFL states, where the temperature-dependent electrical
resistivity $\Delta\rho$ follows the LFL behavior $\Delta\rho\propto
T^2$, one being weakly AF ordered ($B\leq B_{c0}$ and $T<T_{NL}(B)$)
and the other being the weakly polarized ($B\geq B_{c0}$ and $T<T^*(B)$)
\cite{geg}.
At elevated temperatures and fixed magnetic field, during which the
system moves along the vertical arrow shown in Fig. \ref{fig2}, the
NFL state occurs which is separated from the AF phase by the curve
$T_{NL}$ of the phase transitions. Consistent with the experimental
facts we assume that at relatively high temperatures
$T/T_{NL}(B)\simeq 1$ the AF phase transition is of the second order
\cite{geg,TNsteg}. In that case, the entropy and the other
thermodynamic functions are continuous functions at the line of the
phase transitions $T_{NL}$ shown in Fig. \ref{fig2}. This means that
the entropy of the AF phase $S_{AF}(T)$ coincides with the entropy
$S_{NFL}(T)$ of the NFL state. Since the AF phase demonstrates the
LFL behavior, that is $S_{AF}(T\to 0)\to0$, while $S_{NFL}(T)$
contains the temperature-independent term given by Eq. \eqref{snfl}.
Thus, in the NFL region formed by FCQPT Eq. \eqref{SL6} cannot be
satisfied at diminishing temperatures and the second order AF phase
transition inevitably becomes the first order one at the tricritical
point with $T=T_{cr}$, as shown in Fig. \ref{fig2}. At $T=0$, the
heat of the transition $q=0$ as was shown in Subsection \ref{ENTR},
thus the critical field $B_{c0}$ is determined by the condition that
the ground state energy of the AF phase coincides with the ground
state energy of the weakly polarized LFL, and the ground state of
$\rm YbRh_2Si_2$ becomes degenerate at $B=B_{c0}$. Therefore, the
N\'eel temperature $T_{NL}(B\to B_{c0})\to 0$. That means that at
$T=0$ the system moving along the horizontal arrow shown in Fig.
\ref{fig2} transits to its paramagnetic state when the applied
magnetic field reaches its critical value $B=B_{c0}$, and becomes
even higher $B=B_{c0}+\delta B$, where $\delta B$ is an
infinitesimal magnetic field increment, while the Hall coefficient
experiences the jump as seen from Eq. \eqref{SL6} \cite{ybrhsi}.

Upon comparing the phase diagram of $\rm YbRh_2Si_2$  depicted in
Fig.  \ref{fig2} with that of $\rm CeCoIn_5$ shown in Fig.
\ref{CeCo}, it is possible to conclude that they are similar in many
respects. Indeed, the line of the second order superconducting phase
transitions changes to the line of the first ones at the tricritical
point shown by the square in Fig. \ref{CeCo}. This transition takes
place under the application of magnetic fields $B>B_{c2}\geq B_{c0}$
(see Subsections \ref{BC0C2} and \ref{CeCoIn}), where $B_{c2}$ is
the critical field destroying the superconducting state, and
$B_{c0}$ is the critical field at which the magnetic field induced
QCP takes place \cite{bian,oes}. We note that the superconducting
boundary line $B_{c2}(T)$ at lower temperatures acquires the
tricritical point due to Eq. \eqref{SL6} that cannot be satisfied at
diminishing temperatures $T\leq T_{cr}$, i.e. the corresponding
phase transition becomes first order \cite{bian}. This permits us to
conclude that at lower temperatures, in the NFL region formed by
FCQPT the curve of any second order phase transition passes into the
curve of the first order one at the tricritical point.

\subsection{The tricritical point in the ${B-T}$ phase diagram of
$\rm YbRh_2Si_2$}

The Landau theory of the second order phase transitions is
applicable as the tricritical point is approached, $T\simeq T_{cr}$,
since the fluctuation theory can lead only to further logarithmic
corrections to the values of the critical indices. Moreover, near
the tricritical point, the difference $T_{NL}(B)-T_{cr}$ is a second
order small quantity when entering the term defining the divergence
of the specific heat \cite{lanl2}. As a result, upon using the
Landau theory we obtain that the Sommerfeld coefficient
$\gamma_0=C/T$ varies as $\gamma_0\propto |t-1|^{-\alpha}$ where
$t=T/T_{NL}(B)$ with the exponent being $\alpha\simeq0.5$ as the
tricritical point is approached at fixed magnetic field
\cite{lanl2}. We will see that $\alpha=0.5$ gives a good description
of the facts collected in measurements of the specific heat on $\rm
YbRh_2Si_2$. Taking into account that the specific heat increases in
going from the symmetrical to the asymmetrical AF phase
\cite{lanl2}, we obtain
\begin{equation} \gamma_0(t)=A_1+\frac{B_1}{\sqrt{|t-1|}}.\label{TNN}
\end{equation} Here, $B_1=B_{\pm}$ are the proportionality factors which
are different for the two sides of the phase transition. The
parameters $A_1=A_{\pm}$, related to the corresponding specific heat
$(C/T)_{\pm}$, are also different for the two sides, and ``$+$''
stands for $t>1$, while ``$-$'' stands for $t<1$.

\begin{figure} [! ht]
\begin{center}
\vspace*{-0.8cm}
\includegraphics [width=0.60\textwidth]{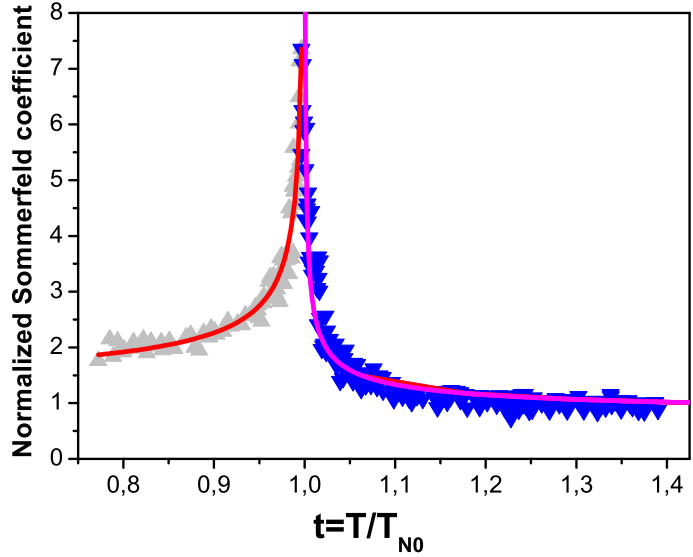}
\end{center}
\vspace*{-0.8cm} \caption{The normalized Sommerfeld coefficient
$\gamma_0/A_{+}$ as a function of the normalized temperature
$t=T/T_{N0}$ given by the formula \eqref{TNN} is shown by the solid
curve. The normalized Sommerfeld coefficient is extracted from the
facts obtained in measurements on $\rm YbRh_2Si_2$ at the AF phase
transition \cite{TNsteg} and shown by the triangles.}\label{fig3}
\end{figure}

The attempt to fit the available experimental data for
$\gamma_0=C(T)/T$ in $\rm YbRh_2Si_2$ at the AF phase transition in
zero magnetic fields \cite{TNsteg} by the function \eqref{TNN} is
reported in Fig. \ref{fig3}. We show there the normalized Sommerfeld
coefficient $\gamma_0/A_{+}$ as a function of the normalized
temperature $t=T/T_{N0}$. It is seen that the normalized Sommerfeld
coefficient $\gamma_0/A_{+}$ extracted from $C/T$ measurements on
$\rm YbRh_2Si_2$ \cite{TNsteg} is well described in the entire
temperature range around the AF phase transition by the formula
\eqref{TNN} with $A_{+}=1$.

\begin{figure} [! ht]
\begin{center}
\vspace*{-0.8cm}
\includegraphics [width=0.80\textwidth]{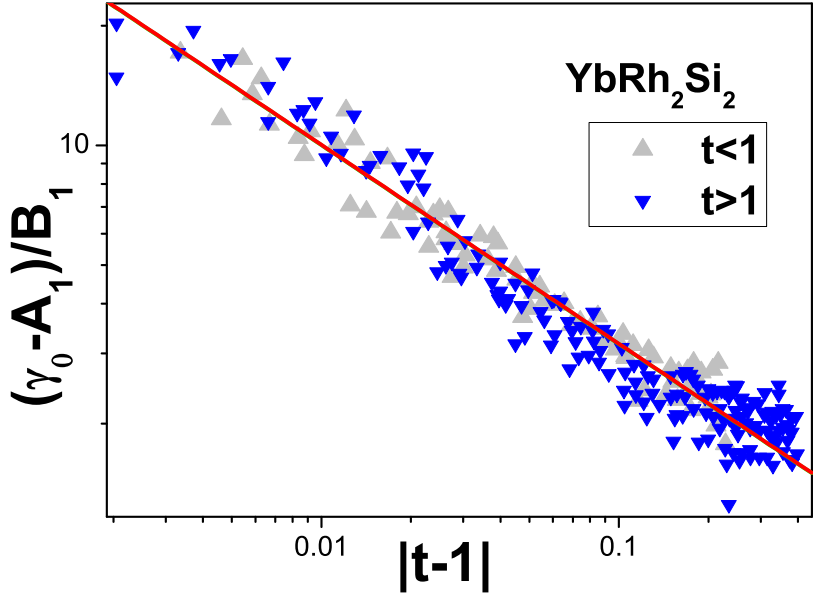}
\end{center}
\vspace*{-0.9cm} \caption{The temperature dependence of the ratios
$(\gamma_0-A_1)/B_1$  for $t<1$ and $t>1$ versus $|1-t|$ given by
the formula \eqref{TN1} is shown by the solid line. The ratios are
extracted from the facts obtained in measurements of $\gamma_0$ on
$\rm YbRh_2Si_2$ at the AF phase transition \cite{TNsteg} and
depicted by the triangles as shown in the legend.}\label{fig4}
\end{figure}

Now transform Eq. \eqref{TNN} to the form
\begin{equation} \frac{\gamma_0(t)-A_1}{B_1}=\frac{1}{\sqrt{|t-1|}}.\label{TN1}
\end{equation} It follows from Eq. \eqref{TN1} that the
ratios $(\gamma_0-A_1)/B_1$  for $t<1$ and $t>1$ versus $|1-t|$
collapse into a single line on logarithmic$\times$logarithmic plot. The
extracted from experimental facts \cite{TNsteg} ratios are depicted
in Fig. \ref{fig4}, the coefficients $A_1$ and $B_1$ are taken from
fitting $\gamma_0$ shown in Fig. \ref{fig3}. It is seen from Fig.
\ref{fig4} that the ratios $(\gamma_0-A_1)/B_1$ shown by the upward
and downwards triangles for $t<1$ and $t>1$, respectively, do collapse
into the single line shown by the solid straight line.

A few remarks are in order here. The good fitting shown in Figs.
\ref{fig3} and \ref{fig4} of the experimental data by the functions
\eqref{TNN} and \eqref{TN1} with the critical exponent $\alpha=1/2$
allows us to conclude that the specific-heat measurements on $\rm
YbRh_2Si_2$ \cite{TNsteg} are taken near the tricritical point and
to predict that the second order AF phase transition in $\rm
YbRh_2Si_2$ changes to the first order under the application of
magnetic field as it is shown by the arrow in Fig. \ref{fig2}
\cite{sopt}. It is seen from Fig. \ref{fig3} that at $t\simeq1$ the
peak is sharp, while one would expect that anomalies in the specific
heat associated with the onset of magnetic order are broad
\cite{TNsteg,PTsteg,lohn}. Such a behavior represents fingerprints
that the phase transition is to be changed to the first order one at
the tricritical point, as it is shown in Fig. \ref{fig2}. As seen
from Fig. \ref{fig3}, the Sommerfeld coefficient is larger below the
phase transition than above it. This fact is in accord with the
Landau theory that states that the specific heat increases when
passing from $t>1$ to $t<1$ \cite{lanl2}.

\subsection{Entropy in $\rm YbRh_2Si_2$ at low temperatures}

It is instructive to analyze the evolution of magnetic entropy in
$\rm YbRh_2Si_2$ at low temperatures. We start with considering the
derivative of magnetic entropy $dS(B,T)/dB$ as a function of
magnetic field $B$ at fixed temperature $T_f$ when the system
transits from the NFL behavior to the LFL one as shown by the
horizontal solid arrow in Fig. \ref{fig2}. Such a behavior is of
great importance since exciting experimental facts \cite{gegtok} on
measurements of the magnetic entropy in $\rm YbRh_2Si_2$ allow us to
analyze the reliability of the theory employed and to study the scaling
behavior of the entropy when the system is in its NFL, transition and
LFL states, correspondingly.

The quantitative analysis of the scaling behavior of $dS(B,T)/dB$ is
given in Subsection \ref{HCEL8}. Fig. \ref{fig5H} reports the
normalized $(dS/dB)_N$ as a function of the normalized magnetic
field. It is seen from Fig. \ref{fig5H} that our calculations shown
by the solid line are in good agreement with the measurements and the
scaled functions $(\Delta M/\Delta T)_N$
extracted from the experimental facts
show the scaling behavior in a wide range variation of the normalized
magnetic field $B/B_M$. The other thermodynamic and transport
properties of $\rm YbRh_2Si_2$ analyzed in Subsection \ref{HCEL4}
are also in good agreement with the measurements.
These developments make
our analysis of the AF phase transition quite substantial.

Now we are in a position to evaluate the entropy $S$ at temperatures
$T\lesssim T^*$ in $\rm YbRh_2Si_2$. At $T<T^*$ the system in its
LFL state, the effective mass is independent of $T$, is a
function of the magnetic field $B$. As a result, Eq. \eqref{HF5} reads
\begin{equation}\label{MB}
\frac{m}{M^*(B)}=a_2\sqrt{\frac{B}{B_{c0}}-1},
\end{equation} where $a_2$ is a parameter.
In the LFL state at $T<T^*$ when the system moves along the
vertical arrow shown in Fig. \ref{fig2}, the entropy is given by
the well-known relation, $S=M^*T\pi^2/p_F^2=\gamma_0T$
\cite{lanl2}. Taking into account Eqs. \eqref{BC0} and \eqref{MB}
we obtain that at $T\simeq T^*$ the entropy is independent of both
the magnetic field and temperature, $S(T^*)\simeq \gamma_0T^*\simeq
S_0\simeq a_1mT_{NL}\pi^2/(a_2p_F^2)$. Upon using the data
\cite{geg}, we obtain that for fields applied along the hard
magnetic direction $S_0(B_{c0}\,\|\,c)\sim0.03R\ln2$, and for
fields applied along the easy magnetic direction $S_0(B_{c0}\,\bot
\,c)\sim0.005R\ln2$. Thus, as it follows from Fig. \ref{Fig18} and
in accordance with the data collected on $\rm YbRh_2Si_2$
\cite{geg} we conclude that the entropy contains the
temperature-independent part $S_0$ \cite{obz,yakov} which gives
rise to the tricritical point.

To summarize this Section, we remark
that a theory is an important tool in
understanding what we observe; we have demonstrated that the
obtained value of $\alpha$ is in good agreement with the
specific-heat measurements on $\rm YbRh_2Si_2$ and conclude that the
critical universality of the fluctuation theory is violated at the
AF phase transition since the second order phase transition is about
to change to the first order one, making $\alpha\to1/2$. We have also
shown that in the NFL region formed by FCQPT the curve of any second
order phase transitions passes into a curve of the first order ones
at the tricritical point leading to the violation of the critical
universality of the fluctuation theory. This change is generated by
the temperature-independent entropy $S_0$ formed behind FCQPT.

\section{Topological phase transitions related to
FCQPT}\label{TFT}

We have now investigated the structure of the Fermi surface beyond QCP
within the extended quasiparticle paradigm. We have shown that
at $T=0$ there is a scenario that entails the formation of FC,
manifested by the emergence of a completely flat portion of the
single-particle spectrum.

In this Section we consider different kinds of instabilities of
normal Fermi liquids relative to several perturbations of initial
quasiparticle spectrum $\varepsilon(p)$ and occupation numbers
$n(p)$ associated with the emergence of a multi-connected Fermi
surface, see e.g. \cite{khodb,asp,pogshag,llvp,zb,zvbld}. Depending
on the parameters and analytical properties of the Landau
amplitude, such instabilities lead to several possible types of
restructuring of initial Fermi liquid ground state. This
restructuring generates topologically distinct phases. One of them
is the FC discussed above, another one belongs to a class of
topological transitions (TT) and will be called "iceberg" phase,
where the sequence of rectangles ("icebergs") $n(p)=0$ and $n(p)=1$
is realized at $T=0$.

In such considerations, we analyze stability of a fermion system
with model repulsive Landau amplitude allowing us to carry out an
analytical consideration of the emergence of a multi-connected
Fermi surface \cite{asp,pogshag}. We show, in particular, that the
Landau amplitude given by the screened Coulomb law does not
generate FC phase, but rather iceberg TT phase. For this model, we
plot a phase diagram in the variables ''screening parameter -
coupling constant'' displaying two kinds of TT: a 5/2-kind similar
to the known Lifshitz transitions in metals, and a 2-kind
characteristic for a uniform strongly interacting system.

\begin{figure}
\begin{center}
\vspace*{-0.2cm}
\includegraphics [width=0.60\textwidth]{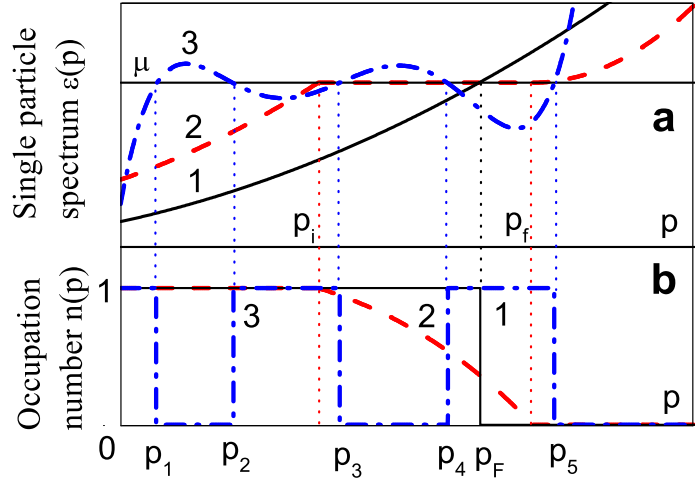}
\vspace*{-0.6cm}
\end{center}
\caption{Schematic plot of the single particle spectrum $\varepsilon(p)$
({\bf a}) and occupation numbers $n(p)$ ({\bf b}), corresponding to
LFL (curves 1), FC (curves 2) and iceberg (curves 3) phases at
$T=0$. For LFL the equation, $\varepsilon (p)=\mu$, has a single
root equal to Fermi momentum $p_F$. For iceberg phase, the above
equation has countable set of the roots $p_1...p_N...$, for FC phase
the roots occupy the whole segment $(p_f-p_i)$. We note that
$p_i<p_F<p_f$ and the states, where $\varepsilon (p)<\mu $ are
occupied (n=1), while those with $\varepsilon(p)>\mu$ are empty
(n=0).}\label{top0}
\end{figure}

The common ground state of isotropic LFL with density $\rho_x$ is
described at zero temperature by the stepwise Fermi function
$n_F(p)=\theta(p_F-p)$, dropping discontinuously from 1 to 0 at the
Fermi momentum $p_F$. The LFL theory states that the quasiparticle
distribution function $n(p)$ and its single particle spectrum $
\varepsilon (p)$ are in all but name similar to those of an ideal
Fermi gas with the substitution of real fermion mass $m$  by the
effective one $M^*$ \cite{landau}. These $n_F(p)$ and $\varepsilon
(p)$ can become unstable under several circumstances. The best known
example is Cooper pairing at arbitrarily weak attractive interaction
with subsequent formation of the pair condensate and gapped
quasiparticle spectrum \cite{bcs}. However, a sufficiently strong
repulsive Landau amplitude can also generate non-trivial ground
states. The first example of such restructuring for a Fermi system
with model repulsive interaction is FC \cite{ks}. It reveals the
existence of a critical value $\alpha_{\rm cr}$ of the interaction
constant $\alpha$ such that at $\alpha=\alpha_{\rm cr}$ the
stability criterion $s(p)=(\varepsilon(p)
-E_F)/(p^2-p_F^2)>0$ fails at the Fermi surface $s(p_F)=0$
($p_F$-instability). We recall that in the case of this instability
the single particle spectrum $\varepsilon(p)$ possesses
the inflection point at the Fermi surface, see Subsection \ref{PHDS}.
Then at $\alpha>\alpha _{\rm cr}$ an exact
solution of a variational equation for $n(p)$ (following from the
Landau functional $E(n(p))$) exists, exhibiting some finite interval
$(p_f-p_i)$ around $p_F$ where the distribution function $n(p)$
varies continuously taking intermediate values between 1 and 0,
while the single-particle excitation spectrum $\varepsilon(p)$ has a
flat plateau. Equation \eqref{FL8} means actually that the roots of
the equation $\varepsilon (p)=\mu$ form an uncountable set in the
range $p_i\leq p\leq p_f$, see Fig. \ref{top0}. It is seen from Eq.
\eqref{FL8} that the occupation numbers $n(p)$ become variational
parameters, deviating from the Fermi step function to minimize the
energy $E$.

The other type of phase transition the so-called iceberg phase
occurs when the equation $\varepsilon (p)=\mu$ has discrete
countable number of roots, either finite or infinite. This is
reported on Fig. \ref{top0} and related to the situation when the
Fermi surface becomes multi-connected. Note that the idea of
multi-connected Fermi surface, with the production of new, interior
segments, had already been considered \cite{llvp,llvp1,zb}.

Let us take the Landau functional $E(n(p))$ of the form
\begin{eqnarray}
\nonumber
E(n(p))&=&\int \frac{p^2}{2M}n(p)\frac{d{\bf p}} {(2\pi )^3}\\
&+&\frac 12\int\int n(p)U(|{\bf p}-{\bf p'}|)n(p') \frac{d{\bf
p}d{\bf p'} }{(2\pi )^6}, \label{energ}
\end{eqnarray}
which, by virtue of Eq. \eqref{FL2}, leads to derivation of the
quasiparticle dispersion law:
\begin{equation}
\varepsilon (p)=\frac{p^2}{2M}+\int U(|{\bf p} -{\bf p'}|)n(p')
\frac{d{\bf p'}}{(2\pi)^3}. \label{disp}
\end{equation}
The angular integration with subsequent change to the dimensionless
variables $x=p/p_F$, $y=y(x)=2M\varepsilon(p)/p_F^2$,
$z=2\pi^2ME/p_F^5$, leads to simplification of the equations \eqref{energ}
and \eqref{disp}
\begin{eqnarray}
z[\nu(x)]&=&\int[x^4+\frac 12x^2V(x)]\nu (x)dx, \label{zet}\\
y(x)&=&x^2+V(x), \label{disp1}
\end{eqnarray}
where
\begin{eqnarray}
&&V(x)=\frac 1x\int x'\nu (x')u(x,x')dx',\nonumber \\
&&u(x,x')=\frac M{\pi^2p_F}\int \limits_{|x-x'|}^{x+x'} u(t)tdt.
\label{kern}
\end{eqnarray}
Here $u(x)\equiv U(p_Fx)$ and the distribution function
$\nu(x)\equiv n(p_Fx)$ is positive, obeys the normalization
condition
\begin{equation}
\int x^2\nu(x)dx=1/3, \label{norm}
\end{equation}
and the Pauli principle limitation $\nu(x)\leq 1$. The latter can be
lifted using, e.g., the ansatz: $\nu (x)=$$[1+\tanh \eta(x)]/2$. In
the latter case the system ground state gives a minimum to the
functional\begin{eqnarray} \nonumber
f[\eta(x)]&=&\int[1+\tanh\eta(x)]\{x^4-\mu x^2\\
&+&x'[1+\tanh\eta(x')]u(x,x')dx'\}dx, \label{funct}
\end{eqnarray}
containing a Lagrange multiplier $\mu$, with respect to an
arbitrary variation of the auxiliary function $\eta(x)$. This
allows us to represent the necessary condition of extremum $\delta f=0$
in the form
\begin{equation}
x^2\nu(x)[1-\nu(x)][y(x)-\mu]=0. \label{extr}
\end{equation}
This means that either $\nu(x)$ takes only the values 0 and 1 or the
dispersion law is flat: $y(x)=\mu$ \cite{ks}, in accordance with Eq.
(\ref{FL8}). The former possibility corresponds to iceberg phase,
while the latter to FC. As it is seen from Eq. (\ref{FL8}), the
spectrum $\varepsilon (p)$ in this case cannot be an analytic function
of complex $p$ in any open domain, containing the FC interval
$(p_f-p_i)$. In fact, all the derivatives of
$\varepsilon(p)$ with respect to $p$ along
the strip $(p_f-p_i)$ should be zero,
while this is not the case outside $(p_f-p_i)$. For instance, in the
FC model with $U(p)=U_0/p$ \cite{ks} the kernel, Eq. (\ref{kern}),
is non-analytic
\begin{equation}
u(x,x')=\frac{MU_0}{\pi^2p_F}(x+x' +|x-x'|),
\end{equation} which eventually
causes non-analyticity of the potential $V(x)$. It follows from Eq.
(\ref{disp1}), that the single particle spectrum is an analytic function
on the whole real axis if $ V(x)$ is such a function. In this case
FC is forbidden and the only alternative to the Fermi ground state
(if the stability criterion gets broken) is iceberg phase
corresponding to TT between the topologically unequal states with
$\nu (x)=0,1$ \cite{vol}.

On the other hand, applying the technique of Poincar\'e mapping, it
is possible to analyze the sequence of iterative maps generated by
Eq. \eqref{EM} for the single-particle spectrum at zero temperature
\cite{khodb}. If the sequence of maps converges, the multi-connected
Fermi surface is formed. If it fails to converge, the Fermi surface
swells into a volume that provides a measure of entropy associated
with the formation of an exceptional state of the system
characterized by partial occupation of single-particle states and
dispersion of their spectrum proportional to temperature as seen
from Eq. \eqref{FL12}.

\begin{figure}
\begin{center}
\includegraphics [width=0.60\textwidth]{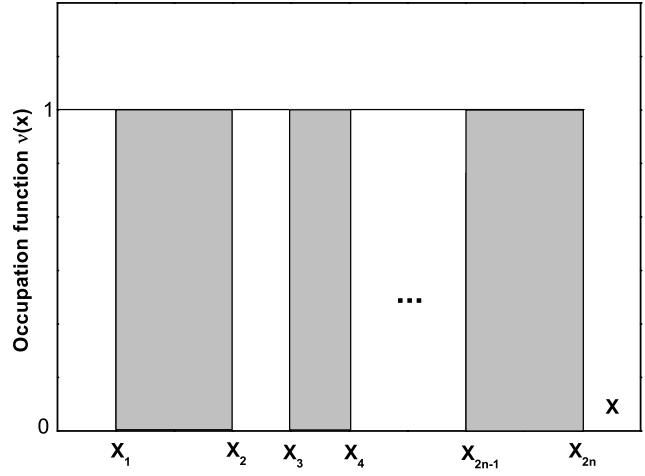}
\end{center}
\caption{Occupation function for a multiconnected
distribution.}\label{top1}
\end{figure}
Generally, all such states related to the formation
of iceberg phases are classified by the indices of
connectedness (known as Betti numbers in algebraic topology
\cite{lifsh,naka}) for the support of $\nu (x)$. In fact, for an
isotropic system, these numbers simply count the separate
(concentric) segments of the Fermi surface. Then the system ground
state corresponds to the following multi-connected distribution
shown in Fig. \ref{top1}
\begin{equation}
\nu(x)=\sum\limits_{i=1}^n\theta (x-x_{2i-1})\theta (x_{2i}-x),
\label{dist}
\end{equation}
where the parameters $0\leq x_1<x_2<\ldots <x_{2n}$ obey the
normalization condition
\begin{equation}
\sum\limits_{i=1}^n(x_{2i}^3-x_{2i-1}^3)=1.  \label{normm}
\end{equation}
The function $z$, Eq. (\ref{zet}),
\begin{equation}
z=\frac 12\sum\limits_{i=1}^n\int
\limits_{x_{2i-1}}^{x_{2i}}x^2[x^2+y(x)]dx, \label{zetm}
\end{equation}
has the absolute minimum with respect to $x_1,\ldots,x_{2n-1}$ and
to $n\geq 1$. To obtain the necessary condition of extremum, we use
the relations
\begin{equation}
\frac{{\partial }x_{2n}}{{\partial }x_k}
=(-1)^{k-1}\left(\frac{x_k}{x_{2n}}\right)^2,\,\,\,1 \leq k\leq
2n-1, \label{deriv}
\end{equation}
following from Eq. (\ref{normm}) and the dependence of the potential
$V(x)$ in the dispersion law $y(x)$ on the parameters
$x_1,\ldots,x_{2n-1}$
\begin{equation}
V(x)=\frac 1x\sum\limits_{i=1}^n\int\limits_{x_{2i-1}}^{x_{2i}}
x'u(x,x')dx'. \label{vm}
\end{equation}
Subsequent differentiation of Eq. (\ref{zetm}) with respect to the parameters
$x_1,\ldots,x_{2n-1}$ and the use of Eqs. \eqref{deriv} and
\eqref{vm} yield the necessary conditions of extremum
in the following form
\begin{equation}
\frac{{\partial z}}{{\partial }x_k}
=(-1)^kx_k^2[y(x_k)-y(x_{2n})]=0,\,\,\,1 \leq k\leq 2n-1.
\label{level}
\end{equation}
This means that a multi-connected ground state is controlled by the
evident rule of unique Fermi level $y(x_k)=y(x_{2n})$ for all
$1\leq k\leq 2n-1$ (except for $x_1=0$). In principle, given the
dispersion law $y(x)$ all the $2n-1$ unknown parameters $x_k$ can be
found from Eq. (\ref{level}). Then, the sufficient stability
conditions $\partial ^2z/\partial x_i\partial x_j=\gamma _i\delta
_{ij}$, $\gamma _i>0$ generate the generalized stability criterion.
Namely, the dimensionless function
\begin{equation}
\sigma (x)=2Ms(p)=\frac{y(x)-y \left( x_{2n}\right) }{x^2-x_{2n}^2},
\label{stab}
\end{equation}
should be positive within filled and negative within empty
intervals, turning to zero at their boundaries in accordance with Eq.
(\ref{level}). It can be proved rigorously that, for given analytic
kernel $u(x,x')$, Eq. (\ref{stab}) uniquely defines the system
ground state.

In what follows we shall label each multi-connected state, Eq.
(\ref{dist}), by an entire number related to the binary sequence of
empty and filled intervals read from $x_{2n}$ to 0. Thus, the Fermi
state with a single filled interval $(x_2=1,x_1=0)$ reads as unity,
the state with a void at the origin (filled $[x_2,x_1]$ and empty
$[x_1,0]$) reads as $(10)=2$, the state with a single gap:
$(101)=3$, etc. Note that all even phases have a void at the origin
and odd phases have not.

\begin{figure}
\begin{center}
\includegraphics [width=0.60\textwidth]{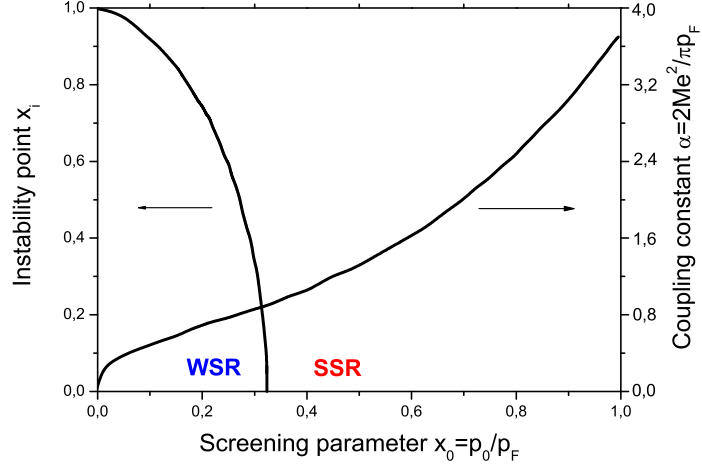}
\end{center}
\caption{Instability point
$x_{i}$ and critical coupling $\alpha^{*}$ as functions of screening.
The regions of weak screening
(WSR) and strong screening (SSR) are separated by the threshold
value $x_{th}$. Note that a $x_{th}$, $\alpha_{th}$ is the triple
point between the phases 1, 2, 3 in Fig. \ref{top33}.}\label{top2}
\end{figure}

For free fermions $V(x)=0$, $y(x)=x^2$, Eq. (\ref{level}) only yields
the trivial solution corresponding to the Fermi state 1. To obtain
non-trivial realizations of TT, we choose $U(p)$ to correspond to
the common screened Coulomb potential:
\begin{equation}
U(p)=\frac{4\pi e^2}{p^2+p_0^2}. \label{utilde}
\end{equation}
The related explicit form of the kernel,
\begin{equation}
u(x,x')=\alpha \ln \frac{(x+x')^2+x_0^2} {(x-x')^2+x_0^2},
\label{uxx}
\end{equation}
with the dimensionless screening parameter $x_0=p_0/p_F$ and the
coupling constant $\alpha =2Me^2/\pi p_F$, evidently displays the
necessary analytical properties for existence of iceberg phase.
Equations (\ref{vm}) and (\ref{uxx}) permit to express the potential
$V(x)$ in elementary functions \cite{asp}.  Then, the
straightforward analysis of Eq. (\ref{level}) shows that their
nontrivial solutions appear only when the coupling parameter $\alpha
$ exceeds a certain critical value $\alpha ^*$. This corresponds to
the situation when the stability criterion \cite{ks} $\sigma
(x)=(y_F (1)-y_F(x))/(1-x^2)>0$ calculated with the Fermi
distribution, $y_F(x)=x^2+V(x;0,1)$, fails in a certain point $0\leq
x_i<1$ within the Fermi sphere: $\sigma(x_i)\to 0$. There are two
different types of such instabilities depending on the screening
parameter $x_0$ (Fig. \ref{top2}). For $x_0$ below a certain threshold
value $x_{\rm th}\approx 0.32365$ (weak screening regime, WSR) the
instability point $x_i$ sets rather close to the Fermi surface:
$1-x_i\ll 1$, while it drops abruptly to zero at $x_0\to x_{\rm th}$
and equals zero for all $x_0>x_{\rm th}$ (strong screening regime,
SSR). The critical coupling $\alpha ^*(x_0)$ results in a monotonously
growing function of $x_0$ with the asymptotic $\alpha^*\approx(\ln
2/x_0-1)^{-1}$ at $x_0\to 0$ and staying analytic at $\alpha_{\rm
th}=\alpha^{*}(x_{\rm th})\approx 0.91535$, where it only exhibits
an inflection point.

\begin{figure}
\begin{center}
\includegraphics [width=0.60\textwidth]{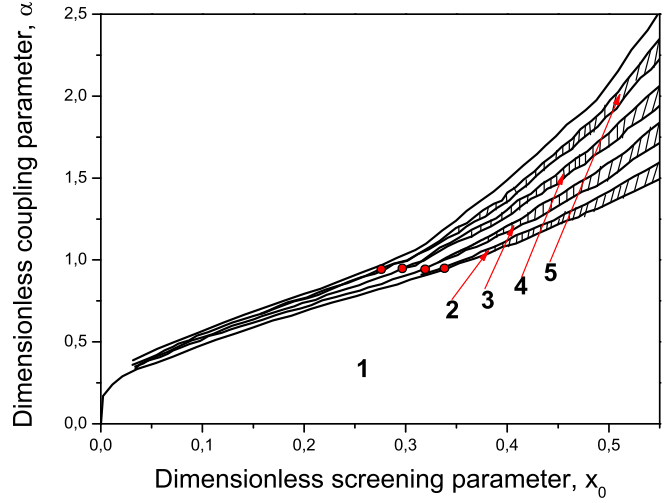}
\end{center}
\caption{Phase diagram in variables ``screening-coupling''. Each
phase with certain topology is labeled by the total number of filled
and empty regions (see Fig. \ref{top1}). Even phases (shaded) are
separated from odd ones by ``$^{5}/_{2}$-kind'' topological
transition (TT) lines, while 2-kind TT lines separate odd phases.
Triple points, where two $^{5}/_{2}$-TT and one 2-TT meet, are shown
by circles.}\label{top33}
\end{figure}

These two types of instabilities give rise to different types of TT
from the state 1 at $\alpha >\alpha^*$: at SSR a void appears around
$x=0$ ($1\to 2$ transition), and at WSR a gap opens around $x_i$
($1\to 3$ transition).  Further analysis of Eq. (\ref{level}) shows
that the point $x_{\rm th},\alpha_{\rm th}$ represents a triple
point in the phase diagram in the variables $x_0,\alpha$ (Fig.
\ref{top2}) where the phases 1, 2, and 3 meet one another. Similarly
to the onset of instability in the Fermi state 1,
each evolution of TT to higher
order phases with growing $\alpha$ is manifested by a zero of $\sigma
(x)$, Eq. (\ref{stab}), at some point $0\leq x_i<x_{2n}$ different
from the existing interfaces. If this occurs at the very origin,
$x_i=0$, the phase number rises at TT by 1, corresponding to the opening
of a void (passing from odd to even phase) or to emerging ''island''
(even $\to$ odd). For $x_i>0$, either a thin spherical gap opens
within a filled region or a thin filled spherical sheet emerges
within a gap, so that the phase number rises by 2, living the parity
unaltered. A part of the whole diagram shown in Fig. \ref{top33}
demonstrates that with decreasing of $x_0$ (screening weakening) all
even phases terminate at certain triple points. This, in particular,
agrees with numerical studies of the considered model along the line
$x_0=0.07$ at growing $\alpha$ \cite{zb}, where only the sequence of
odd phases $1\to 3\to 5\to \ldots $ has been revealed (shown by the
arrow in Fig. \ref{top2}).
The energy gain $\Delta\left(\tau\right)$ at TT as a function of
small parameter $\tau =\alpha /\alpha^*-1$ is evidently proportional
to $\tau $ times the volume of a new emerging phase region (empty or
filled). Introducing a void radius $\delta$ and expanding the energy
gain $\Delta(\delta)=z[n(x,\delta)]-z[n_F(x)]$ in $\delta$, one gets
$\Delta=-\beta_1\tau\delta^3 +\beta_2\delta^5+O(\delta^6)$,
$\beta_1,\beta_2>0$. As a result, the optimum void radius is
$\delta\sim\sqrt\tau$. Consequently we have $\Delta(\tau)\sim\tau
^{5/2}$ indicating a resemblance to the known ''$^5/_2$-kind'' phase
transitions in the theory of metals \cite{lifsh}. The peculiar
feature of our situation is that the new segment of the Fermi surface
opens at very small momentum values, which can dramatically change
the system response to, e.g., electron-phonon interaction. On the
other hand, this segment may have a pronounced effect on the
thermodynamical properties of $^3$He at low temperatures, especially
in the case of $P$-pairing, producing excitations with extremely
small momenta.

For a TT with unchanged parity, the width of a gap (or a sheet) is
found to be $\sim\tau$ so that the energy gain is
$\Delta\left(\tau\right)\sim\tau^2$ and such TT can be related to
the second kind. It follows from the above consideration that each
triple point in the $x_0-\alpha $ phase diagram is a point of
confluence of two $^5/_2$-kind TT lines into one 2-kind line. The
latter type of TT has already been discussed in the literature
\cite{llvp,zb}. Here we only mention that its occurrence on a whole
continuous surface in the momentum space is rather specific for
systems with strong fermion-fermion interaction, while the known
TT's in metals, under the effects of crystalline field, occur
typically at separate points in the quasi-momentum space. It is
interesting to note that in the limit $x_0\to 0,\alpha \to 0$,
reached along a line $\alpha=kx_0$, we attain the exactly solvable
model: $U(p)\to(2\pi)^3U_0\delta(p)$ with $U_0=k/(2Mp_F)$, which is
known to display FC at all $U_0>0$ \cite{ks}. The analytic mechanism
of this behavior is the disappearance of the poles of $U(p)$, Eq.
(\ref{utilde}), as $p_0\to 0$, restoring the analytical properties
necessary for FC. Otherwise, the FC regime corresponds to the phase
order $\to \infty $, when the density of infinitely thin filled
(separated by empty) regions approaches some continuous function
$0<\nu (x)<1$ \cite{zb} and the dispersion law turns flat according
to Eq. (\ref{level}). A few remarks should be made at this point.

First, the considered model formally treats $x_0$ and $\alpha $ as
independent parameters, though in fact a certain relation between
them can be imposed. Under such restriction, the system ground state
should depend on a single parameter, say the particle density
$\rho_x $, along a certain trajectory $\alpha (x_0)$ in the above
suggested phase diagram. For instance, with the simplest
Thomas-Fermi relation for a free electron gas $\alpha
(x_0)=x_0^2/2$, this trajectory stays fully within the Fermi state 1
over all the physically reasonable range of densities. Hence a
faster growth of $\alpha (x_0)$ is necessary for realization of TT
in any fermionic system with the interaction, Eq. (\ref{utilde}).

Second, at increasing temperatures, the stepwise form of the
quasiparticle distribution is melting. Therefore, as temperature
moves away from its zero value, the concentric Fermi spheres are
taken up by FC. In fact, these arguments do not work in the case of
a few icebergs. Thus, it is quite possible to observe the two
separate Fermi sphere regimes related to the FC and iceberg states.

There is a good reason to mention that neither in the FC phase nor
in the other TT phases, the standard Kohn-Sham scheme
\cite{wks,wks1} is no longer valid. This is because in the systems
with FC or TT phase transitions the occupation numbers of
quasiparticles are indeed variational parameters. Thus, to get a
reasonable description of the system, one has to consider the ground
state energy as a functional of the occupation numbers $E[(n(p)]$
rather than a functional of the density $E[\rho_x]$
\cite{dft,dft269,sh}.

\section{Conclusions}

In this review, we have described the effect of FCQPT on the
properties of various Fermi systems and presented substantial
evidence in favor of the existence of such a transition. We have
demonstrated that FCQPT supporting the extended quasiparticle
paradigm forms strongly correlated Fermi systems with their unique
NFL behavior. Vast body of experimental facts gathered in studies of
various materials, such as high-$T_c$ superconductors, heavy-fermion
metals, and correlated 2D Fermi liquids, can be explained by a
theory based on the concept of FCQPT.

We have established that the physics of systems with heavy fermions is
determined by the extended quasiparticle paradigm. In contrast to
the stated Landau paradigm that the quasiparticle effective mass is a
constant, within the extended quasiparticle paradigm the effective mass
of new quasiparticles strongly
depends on the temperature, magnetic field, pressure, and other
parameters. The quasiparticles and the order parameter are well
defined and can be used to describe the scaling behavior of the
thermodynamic, relaxation and transport properties of high-$T_c$
superconductors, HF metals, 2D electronic and $\rm ^3He$ systems and
other correlated Fermi systems. The quasiparticle system determines
the conservation of the Kadowaki-Woods relation and the restoration
of the LFL behavior under the application of magnetic fields.

We have also shown both analytically and using arguments based
entirely on the experimental grounds  that the data collected on
very different strongly correlated Fermi systems reveal their
universal scaling behavior. This is because all above
experimental quantities are indeed proportional to the normalized
effective mass exhibiting the scaling behavior. Since the effective
mass determines the thermodynamic, transport and relaxation properties,
we conclude again that HF metals placed near their QCP demonstrate the
same scaling behavior, independent of the details of HF metals such as
their lattice structure, magnetic ground state, dimensionality etc. In
other words, materials with strongly correlated fermions
can unexpectedly be uniform despite their prominent diversity.

We have also
shown that in finite magnetic fields, in the
NFL region formed by FCQPT the curve of any second
order phase transitions passes into a curve of the first order ones
at the tricritical point leading to the violation of the critical
universality of the fluctuation theory. This change is generated by
the temperature-independent term of the entropy formed behind FCQPT.
The quantum and thermal critical fluctuations
corresponding to second-order phase transitions disappear and have
no effect on the behavior of the system at low temperatures, and the
low temperature thermodynamics of heavy-fermion metals is determined
by quasiparticles.

We have found that the differential conductivity between a metal
point contact and a HF metal or a high-$T_c$ superconductor can be
highly asymmetric as a function of the applied voltage. This
asymmetry is observed when a strongly correlated metal is in its
normal or superconducting state. We have shown that the application
of magnetic field restoring the LFL behavior suppresses the
asymmetry. Correspondingly, we conclude that the particle-hole
symmetry is macroscopically broken in the absence of applied
magnetic fields, while the application of magnetic fields restores
both the LFL state and the particle-hole symmetry. The above
features determine the universal behavior of strongly correlated
Fermi systems and are related to the anomalous low-temperature
behavior of the entropy, which contains the temperature independent
term.

In the future, the realm of problems should be broadened and certain
efforts should be made to describe the other
macroscopic features of FCQPT,
such as the propagation of zero-sound, sonic  and shock waves. In
addition to the already known materials whose properties not only
provide information on the existence of FC but also almost cry aloud
for such a condensate, there are other materials of enormous
interest which could serve as possible objects for studying the
phase transition in question. Among such objects are neutron stars,
atomic clusters, ultra cold gases in traps, nuclei, and quark
plasma. Another possible area of research is related to the
structure of the nucleon, in which the entire "sea" of non-valence
quarks may be in FCQPT. The combination of quarks and the gluons
that hold them together is especially interesting because gluons,
quite possibly, can be in the gluon-condensate phase, which could be
qualitatively similar to the pion condensate proposed by A.B. Migdal
long ago. We believe that FC can be observed in traps, where there
is the possibility of controlling the emergence of a quantum phase
transition accompanied by the formation of FCQPT by changing the
particle number density.

Overall, the ideas associated with a new phase transition in one
area of research stimulates intensive studies of the possible
manifestation of such a transition in other areas. This has happened
in the case of metal superconductivity, whose ideas were
successfully used in describing atomic nuclei and in a possible
explanation of the origin of the mass of elementary particles. This,
quite possibly, could be the case with FCQPT.

Finally, our general discussion shows that FCQPT
develops unexpectedly simple, yet completely good
description of the NFL behavior of strongly correlated
Fermi systems, while the extended
quasiparticle paradigm constitute the properties inherent in
strongly correlated fermion systems. Moreover, the extended paradigm
can be considered as the universal reason for the NFL behavior
observed in  various HF metals, liquids, and other Fermi systems.

\section*{Acknowledgments}

We grateful to V.A. Khodel and V.A. Stephanovich for valuable
discussions. This work was partly supported by the RFBR \#
09-02-00056 and U.S. DOE, Division of Chemical Sciences,
Office of Basic Energy Sciences, Office of Energy Research.

\section*{References}

\bibliographystyle{elsart-num}

\end{document}